\catcode`\@=11
\def\citen#1{\if@filesw \immediate\write \@auxout {\string\citation{#1}}\fi%
\@tempcntb\m@ne \let\@h@ld\relax \def\@citea{}%
\@for \@citeb:=#1\do {\@ifundefined {b@\@citeb}%
    {\@h@ld\@citea\@tempcntb\m@ne{\bf ?}%
    \@warning {Citation `\@citeb ' on page \thepage \space undefined}}%
    {\@tempcnta\@tempcntb \advance\@tempcnta\@ne
    \setbox\z@\hbox\bgroup\ifcat0\csname b@\@citeb \endcsname \relax
    \egroup \@tempcntb\number\csname b@\@citeb \endcsname \relax
    \else \egroup \@tempcntb\m@ne \fi \ifnum\@tempcnta=\@tempcntb
    \ifx\@h@ld\relax \edef \@h@ld{\@citea\csname b@\@citeb\endcsname}%
    \else \edef\@h@ld{\hbox{--}\penalty\@highpenalty
    \csname b@\@citeb\endcsname}\fi
    \else \@h@ld\@citea\csname b@\@citeb \endcsname \let\@h@ld\relax \fi}%
\def\@citea{,\penalty\@highpenalty\hskip.13em plus.13em minus.13em}}\@h@ld}
\def\@citex[#1]#2{\@cite{\citen{#2}}{#1}}%
\def\@cite#1#2{\leavevmode\unskip\ifnum\lastpenalty=\z@\penalty\@highpenalty\fi%
  \ [{\multiply\@highpenalty 3 #1%
  \if@tempswa,\penalty\@highpenalty\ #2\fi}]}   %
\makeatother 
\catcode`\@=12

\newcounter{defthm}

\def\AA            {{\!A}}
\def\alg           {algebra}
\def\Alpha         {\aleph}
\newcommand\Ann[2] {{{\rm A}_{#1}}^{\!\!#2}}
\def\AutO          {{{\rm Aut}\,\mathcal{O}}}
\def\Bb            {\ensuremath{\Langle\vhi\vsi\Rangle}}
\def\bbb           {\ensuremath{\Langle\vsi\vsi\vsi\Rangle}}
\def\BBB           {\ensuremath{\Langle\vhi\vhi\vhi\Rangle}}
\def\BBmat         {{\sf B}}
\newcommand\Bmat[9]{{\sf B}_{\!{\sss#6}#4{\sss#7},{\sss#8}#5{\sss#9}}
                   ^{\,({#1}\,{#2})\,{#3}}}
\newcommand\Bsmat[6]{{\sf B}_{\,{#5}\,{#6}}^{\,({#1}\,{#2}\,{#3})\,{#4}}} 
\def\bc            {boundary condition}
\def\Bc            {Boundary condition}
\def\be            {\begin{equation}}
\def\bea           {\begin{equation}\begin{array}l}
\def\beaa          {\begin{equation}\begin{array}{ll}}
\def\bearl         {\begin{array}{l}}
\def\bearll        {\begin{array}{ll}}
\def\bndb          {\mathrm{b}}
\def\BX            {\ensuremath{\Langle\vhi\Rangle_\times}}
\def\CA            {{\mathfrak V}}
\def\calb          {\mathcal{B}}
\def\calc          {\ensuremath{\mathcal C}}
\def\calca         {\ensuremath{{\mathcal C}_{\!A}}}
\def\calcaa        {\ensuremath{{}_A\!{\,}{\mathcal C}_{\!A}}}
\def\cald          {{\mathcal D}}
\def\calh          {\ensuremath{\mathcal H}}

\def\calm          {\mathcal{M}}
\def\calp          {\mathcal{P}}
\def\cats          {categories}
\def\Cft           {Conformal field theory}
\def\cft           {conformal field the\-o\-ry}
\newcommand\cbs    {{\hat\calb}}
\def\cfts          {conformal field theories}
\def\Cfts          {Conformal field theories}

\def\cir           {\,{\circ}\,}
\def\coborc        {\mbox{{\sf 3}\hy{\sf cobord}$(\calc)$}}
\def\cobord        {\mbox{{\sf 3}\hy{\sf cobord}}}
\def\compac        {compactification}
\def\complex       {\mathbbm C}
\def\con           {conformal }
\def\Con           {Conformal }
\def\corfu         {correlation function}
\def\Corfu         {Correlation function}
\def\cua           {current algebra}
\def\DDD           {\ensuremath{\Langle\Thetha\Thetha\Thetha\Rangle}}
\def\dftd          {\mathrm{d}}
\def\dim           {{\rm dim}}

\def\dsty          {\displaystyle }
\def\DX            {\ensuremath{\Langle\Thetha\Rangle_\times}}

\def\eE            {{\rm e}}
\def\ee            {\end{equation}}
\def\eear          {\end{array}}

\def\End           {{\rm End}}
\newcommand\epicture[2] {\end{picture}\\{}\\[#1.#2em]\end{array}}
\def\eps           {\varepsilon}

\def\eq            {\,{=}\,}

\newcommand\erf[1] {(\ref{#1})}
\newcommand\erp[2] {(\ref{pic4-#1#2})}
\newcommand\Erp[3] {(\ref{pic4-#1#2#3})}
\newcommand\F[9]   {{\sf F}_{\!{\sss#6}#4{\sss#7},{\sss#8}#5{\sss#9}}
                   ^{\,({#1}\,{#2})\,{#3}}}

\def\ff            {{\mathscr F}}
\def\FF            {{\sf F}}

\newcommand{\fose}[1] {{[\![#1]\!]}}
\newcommand\Frac[2]{\mbox{\large$\frac{#1}{#2}$}}
\def\frob          {Fro\-be\-ni\-us algebra}
\newcommand\Fs[6]  {{\sf F}_{\,{#5}\,{#6}}^{\,({#1}\,{#2}\,{#3})\,{#4}}}

\def\ft            {field theory}

\def\gemo          {Gepner model}
\def\GG            {{\sf G}}

\newcommand\Gs[6]  {{\sf G}_{\,{#5}\,{#6}}^{\,({#1}\,{#2}\,{#3})\,{#4}}}

\def\hatpsi        {\varPsi}
\def\Hom           {{\rm Hom}}
\def\HomA          {{\rm Hom}_{\!A}}
\def\HomAA         {{\rm Hom}_{\!A|A}}
\def\hopf          {Hopf algebra}
\newcommand\hsp[1] {\mbox{\hspace{#1 em}}}
\def\hy            {$\mbox{-\hspace{-.66 mm}-}$}

\def\id            {\mbox{\sl id}}

\def\ii            {{\rm i}}
\def\II            {\mathcal I}

\def\iN            {\,{\in}\,}
\def\In            {\prec}

\newcommand\includeourbeautifulpicture[2] {{\begin{picture}(0,0)(0,0)
            \scalebox{.38}{\includegraphics{pic4_#1#2.eps}} \end{picture}}}
\newcommand\Includeourbeautifulpicture[3] {{\begin{picture}(0,0)(0,0)
            \scalebox{.38}{\includegraphics{pic4_#1#2#3.eps}} \end{picture}}}
\newcommand\INcludeourbeautifulpicture[4] {{\begin{picture}(0,0)(0,0)
            \scalebox{.38}{\includegraphics{pic4_#1#2#3#4.eps}} \end{picture}}}
\newcommand\includeournicelargepicture[2] {{\begin{picture}(0,0)(0,0)
            \scalebox{.28}{\includegraphics{pic4_#1#2.eps}} \end{picture}}}
\newcommand\Includeournicelargepicture[3] {{\begin{picture}(0,0)(0,0)
            \scalebox{.28}{\includegraphics{pic4_#1#2#3.eps}} \end{picture}}}
\newcommand\includeourniceLargepicture[2] {{\begin{picture}(0,0)(0,0)
            \scalebox{.14}{\includegraphics{pic4_#1#2.eps}} \end{picture}}}
\newcommand\includeournicehugepicture[2] {{\begin{picture}(0,0)(0,0)
            \scalebox{.10}{\includegraphics{pic4_#1#2.eps}} \end{picture}}}
\newcommand\Includeournicehugepicture[3] {{\begin{picture}(0,0)(0,0)
            \scalebox{.10}{\includegraphics{pic4_#1#2#3.eps}} \end{picture}}}

\def\Intro         {Introduction }

\def\Itemize       {\def\leftmargini{1.8em}\begin{itemize}\addtolength\itemsep{-3pt}} 
\def\ITemize       {\def\leftmargini{1.0em}\begin{itemize}\addtolength\itemsep{-3pt}} 
\def\J             {\mbox{$\JJ$}}
\def\JJ            {\mathcal J}
\def\K             {{\rm K}}

\def\KK            {\mathcal{K}} 

\def\KZ            {Knizh\-nik\hy Za\-mo\-lod\-chi\-kov }
\newcommand\labl[1]{\label{#1}\ee}
\newcommand\labp[2]{\label{pic4-#1#2}\ee}
\newcommand\Labp[3]{\label{pic4-#1#2#3}\ee}
\def\Langle        {\langle}
\def\lhs           {left hand side}

\def\M             {{\dot M}}
\def\mathcalD      {{\mathscr{D}}}

\def\mimo          {minimal model}

\def\modinv        {modular invarian}

\def\M             {{\rm M}}
\def\MX            {{\rm M}_{\rm X}}

\newcommand\N[3]   {{N_{#1#2}}^{\!\!#3}}
\def\nxt           {\raisebox{.08em}{\rule{.44em}{.44em}}\hsp{.4}}
\def\Nxt           {\raisebox{.08em}{\rule{.44em}{.44em}}}

\def\oa            {operator algebra}
\def\obj           {{\mathcal O}bj}
\def\objc          {{\mathcal O}bj(\calc)}

\def\one           {{\bf1}}

\def\ope           {operator product expansion}

\def\orr           {{\rm or}}
\newcommand\ot[1]  {\,{\otimes^{#1}}\,}
\newcommand\oT[1]  {\,{\otimes^{#1}}}

\def\oti           {\,{\otimes}\,}
\def\Oti           {{\otimes}}
\def\otim          {\,{\otimes^-}\,}
\def\otip          {\,{\otimes^+}\,}
\def\otiP          {\,{\otimes^+}}

\def\PHI           {{[\vPhi]}}
\newcommand\ppmatrix[4]{\mbox{{\Large(}$\!\!\begin{array}{cc}{}\\[-1.8em]
                   \scs #1\!&\!\!\scs #2\\[-.39em]\scs #3\!&\!\!\scs #4
                   \\[-.19em]\eear\!\!${\Large)}}}
\newcommand\ppvector[2]{\mbox{{\Large(}$\!\!\begin{array}{cc}{}\\[-1.8em]
                   \!\scs #1\!\\[-.39em]\!\scs #2\!\\[-.19em]\eear\!\!${\Large)}}}
\def\q             {quantum }
\def\Q             {Quantum }
\def\qed           {\hfill\checkmark\medskip}
\def\qft           {quantum field theory}
\def\qg            {quantum group}
\long\def\query#1{\hskip 0pt{\vadjust{\everypar={}\small\vtop to 0pt{\hbox{}%
     \vskip -13pt\rlap{\hbox to 47.0pc{\hfil{\vtop{\hsize=8pc\tolerance=6000%
     \hfuzz=.5pc\rightskip=0pt plus 3em\noindent#1}}}}\vss}}}}%
\newcommand\R[5]   {{\sf R}^{(#1\,#2)\,#3}_{#4\,#5}}
\def\Rangle        {\rangle}
\def\RCA           {{\mathrm{R}}} 
\def\rcfts         {rational conformal field theories}
\def\rep           {representation}
\def\Rep           {{\mathcal R}ep}

\def\rhs           {right hand side}
\def\RIsu          {Riemann surface}
\newcommand\Rm[5]  {{\sf R}^{-\,(#1\,#2)\,#3}_{\;#4\,#5}}

\def\reals         {\mbox{$\mathbb R$}}

\newcommand\RP     {\mathbb{RP}^3}
\def\RR            {{\sf R}}
\newcommand\Rs[4]  {{\sf R}^{#1\,(#2\,#3)\,#4}}

\def\SCA           {{\mathrm{S}}} 
\def\scs           {\scriptstyle}
\newcommand\sect[1]{\section{#1}\setcounter{equation}0\setcounter{defthm}0}
\def\sigrp         {\sigma_{\mathbb{RP}^2}}

\def\sss           {\scriptscriptstyle}
\def\Stn           {S^3{}_{\!\!\!\! {\rm n}.}}
\def\stt           {string theory}
\def\stts          {string theories}

\def\susic         {supersymmetric }
\def\sym           {symmetry}
\def\tcs           {tensor categories}
\newcommand\Tfrac[2]{\textstyle{\frac{#1}{#2}}}
\def\tft           {topological field theory}

\def\Thetha        {\varTheta}
\def\tildeF        {\tilde{\ff}}
\def\Times         {\,{\times}\,}
\def\To            {\,{\to}\,}

\def\twodim        {two-di\-men\-si\-o\-nal}

\def\vhi           {\vPhi}
\def\Vir           {{\mathrm{Vir}}}
\newcommand\VO[5]  {{\mathrm V}^{#1}_{#2#3}(#4;#5)}
\def\voa           {vertex operator algebra}
\newcommand\VOb[6] {{\mathrm V}^{#1,#4}_{#2#3}(#5;#6)}
\def\vop           {vertex operator}
\def\VOs           {{\mathrm V}}
\def\vPhi          {\varPhi}
\def\vsi           {\hatpsi}
\def\WZ            {Wess\hy Zu\-mi\-no }
\def\WZW           {Wess\hy Zu\-mi\-no\hy Wit\-ten }
\def\wzwm          {WZW model}
\def\wzwt          {WZW theory}

\def\X             {{\rm X}}
\newcommand\xlabp[2]{}
\newcommand\Xlabp[3]{}

\def\zet           {\ensuremath{\mathbb Z}}

\documentclass[12pt]{article}

\usepackage{latexsym,amsmath,amssymb,amsfonts,epsf,color,colordvi}
\usepackage{bbm}

\usepackage[dvips]{graphics} 
\usepackage[mathscr]{eucal}
\usepackage{rotating}

\newtheorem{defthm}{\whattheorem}[section]
\newcommand\dt[1]  {\noindent\def\whattheorem{#1}\pagebreak[0]\begin{defthm}{}%
                   \samepage{$\!\!${\rm:}\nopagebreak\\[-1.91em]{}}\end{defthm}}
\newcommand\dtl[2] {\noindent\def\whattheorem{#1}\pagebreak[0]\begin{defthm}{}%
                   \samepage{$\!\!${\rm:}\label{#2}\nopagebreak\\[-1.91em]{}}%
                   \end{defthm}}

\begin{document}

\begin{flushright}  {~} \\[-12mm]
{\sf hep-th/0412290}\\[1mm]{\sf AEI-2004-124}\\[1mm]
{\sf Hamburger$\;$Beitr\"age$\;$zur$\;$Mathematik$\;$Nr.$\;$209}\\[1mm]
{\sf December 2004} \end{flushright}
\begin{center} \vskip 13mm
{\Large\bf TFT CONSTRUCTION OF RCFT CORRELATORS}\\[4mm]
{\Large\bf IV: STRUCTURE CONSTANTS}\\[4mm]
{\Large\bf AND CORRELATION FUNCTIONS}\\[18mm]
{\large
J\"urgen Fuchs$\;^1$ \ \ \ Ingo Runkel$\;^2$ \ \ \ Christoph Schweigert$\;^{3}$}
\\[8mm]
$^1\;$ Institutionen f\"or fysik, \ Karlstads Universitet\\
Universitetsgatan 5, \ S\,--\,651\,88\, Karlstad\\[5mm]
$^2\;$ Max-Planck-Institut f\"ur Gravitationsphysik,
Albert-Einstein-Institut\\
Am M\"uhlenberg 1, \ D\,--\,14\,476\, Golm\\[5mm]
$^3\;$ Fachbereich Mathematik, \ Universit\"at Hamburg\\
Schwerpunkt Algebra und Zahlentheorie\\
Bundesstra\ss e 55, \ D\,--\,20\,146\, Hamburg
\end{center}
\vskip 12mm

\begin{quote}{\bf Abstract}\\[1mm]
We compute the fundamental correlation functions in two-dimensional rational
conformal field theory, from which all other correlators can be obtained by 
sewing: the correlators of three bulk fields on the sphere,
one bulk and one boundary field on the disk, three boundary fields on the disk,
and one bulk field on the cross cap. We also consider conformal defects and
calculate the correlators
of three defect fields on the sphere and of one defect field on the cross cap.
\\
Each of these correlators is presented as the product of a structure constant 
and the appropriate conformal two- or three-point block. The structure 
constants are expressed as invariants of ribbon graphs in three-manifolds.

\end{quote} \vfill \newpage

\tableofcontents



\sect{Introduction and summary}

This paper constitutes part IV of the series announced in \cite{fuRs}, in which
the relation \cite{witt27,frki2,fffs3} between rational two-dimensional \cft\
(RCFT) and three-dimensional topological quantum field theory (TFT) is 
combined with non-commutative algebra in modular tensor categories to obtain 
a universal, model independent, construction of CFT correlation functions.
In the previous parts of the series, we concentrated on the description of 
correlators without field insertions, i.e.\ partition functions \cite{fuRs4},
on the special features that arise when the world sheet is unoriented
\cite{fuRs8}, and on the case that the CFT is obtained as a simple current
construction \cite{fuRs9}. In the present part we give the fundamental 
correlation functions of the CFT, i.e.\ a finite set of correlators from 
which all others can be obtained \cite{sono2,lewe3,fips} via sewing
and chiral Ward identities.

\subsection{Correlators from sewing of world sheets}\label{sec:in-1}

The correlators of a CFT should be single-valued functions of the world sheet 
moduli, like the positions of the field insertions or, more generally, the 
world sheet metric. Further, the correlators are subject to factorisation, or 
sewing, constraints. These constraints arise when a world sheet is cut along a
circle or an interval and a sum over intermediate states is inserted on the cut 
boundaries. This procedure allows one to express a correlator on a complicated 
world sheet (e.g.\ of higher genus) in terms of simpler building blocks. 
The fundamental building blocks are \cite{sono2,lewe3,fips} the correlators 
\BBB\ for three bulk fields on the sphere,
\Bb\ for one bulk field and one boundary field on the disk,
\bbb\ for three boundary fields on the disk, and
\BX\ for one bulk field on the cross cap. Actually, the requirement of 
covariance under global conformal transformations determines each of these 
correlators up to the choice of a non-zero vector in the relevant coupling 
space. The components of these vectors are called {\em structure constants\/}.

Depending on what class of surfaces one allows as world sheets for the CFT,
different sets of these
fundamental correlators (or, equivalently, structure constants) are required:
\\[-1.1em]
\begin{center}
\begin{tabular}{rll}
       \multicolumn2l {class of world sheets}
     & \multicolumn1l {fundamental correlators} \\[4pt]
 (1) & oriented, with empty boundary                 & \BBB            \\[2pt]
 (2) & unoriented, with empty boundary               & \BBB, \BX       \\[2pt]
 (3) & oriented, with empty or non-empty boundary    & \BBB, \Bb, \bbb \\[2pt]
 (4) & unoriented, with empty or non-empty boundary  & \BBB, \Bb, \bbb, \BX
\end{tabular}  
\end{center}

\smallskip

The factorisation of a given correlator into the fundamental building blocks
is not unique. Requiring that the different ways to express a correlator
in terms of sums over intermediate states agree gives rise to an infinite
system of constraints for the data \BBB, \Bb, \bbb, \BX. 
One way to define a CFT is thus to provide a collection of candidates for the 
fundamental correlators that are required in (1)\,--\,(4) and show that they 
solve these factorisation constraints and lead to single-valued correlators. 

\medskip

Holomorphic bulk fields (like the component $T$ of the stress tensor) lead
to conserved charges and chiral Ward identities, which constrain the form of the
correlation functions. Solutions to the chiral Ward identities are called
conformal blocks. In rational CFT the space of conformal blocks for a given 
correlator is finite-dimensional. 
Another important consequence of rationality is that the fundamental correlators
have to be given only for a finite collection of fields; for all other
fields they are then determined by the chiral Ward identities.

The conformal blocks that are relevant for the correlators on a
given world sheet \X\ are actually to be constructed on the double $\hat\X$ of
\X. The surface $\hat\X$ is obtained as the orientation bundle over $\X$, 
divided by the equivalence relation that identifies the two possible 
orientations above points on the boundary of $\X$: 
  \be
  \hat\X = {\rm Or}(\X){/}{\sim} \qquad {\rm with}
  \qquad (x,{\rm or}) \sim (x,-{\rm or}) ~{\rm for}~ x \iN \partial\X \,.
  \labl{eq:double-def}
For example, if $\X$ is orientable and has empty boundary, then $\hat \X$
just consists of two copies of $\X$ with opposite orientation.
A correlator $C(\X)$ on the world sheet $\X$ is an element in the space
of conformal blocks on the double,
  \be 
  C(\X) \in \calh(\hat \X) \,.
  \labl{eq:CX-HX}
This is known as the principle of {\em holomorphic factorisation} \cite{witt39}.

The conformal blocks on a surface $\Sigma$ are generically multivalued functions
of the moduli of $\Sigma$. Upon choosing bases in the spaces $\calh(\Sigma)$, 
the behaviour of the blocks under analytic continuation and factorisation can 
be expressed through braiding and fusing matrices $\BBmat$ and $\FF$ 
\cite{Mose,frki,bakI}.
These matrices are in fact already determined by the four-point blocks on the 
Riemann sphere. One can then express the constraints on the fundamental
correlators coming from factorisation and single-valuedness in terms of 
fusing and braiding matrices. It turns out that there is a {\em finite\/} set of
equations which the structure constants must obey in order to
yield a consistent CFT. These are the so-called {\em sewing constraints\/} 
\cite{sono2,lewe3}.

\medskip

The first class of CFTs for which explicit solutions for the structure
constants have been calculated are the A-series Virasoro minimal models.
The constants for \BBB, \Bb, \bbb\ and \BX\ are given in 
\cite{dofa2,cale,runk,fips}, respectively. By now many more solutions are 
known; pertinent references will be given at the end of sections 
\ref{sec:rc-bnd3pt}\,--\,\ref{sec:rc-1bulk-xcap}, in which the calculation 
of the individual structure constants in the TFT framework is carried out.

Remarkably, one can also obtain exact expressions for some of the fundamental 
correlation functions in a few examples of non-rational CFTs in which the 
chiral algebra $\CA$ is too small to make all spaces of conformal blocks 
finite-dimensional. The most complete answers are known for Liouville theory, 
see e.g.\ \cite{pons3,tesc12,naKA'} and references cited therein. The structure 
of the expressions found in this case is very similar to the Cardy case of a 
rational CFT. This similarity might serve as a guideline when trying to extend 
aspects of the TFT-based construction beyond the rational case.

\subsection{Correlators from 3-d TFT and algebras in tensor categories}

The sewing constraints form a highly overdetermined system of polynomial
equations, which is difficult to solve directly. Instead, one can try to
identify an underlying algebraic structure that encodes all information
about a solution. In our TFT-based construction, this underlying structure
is an algebra $A$ in a certain tensor category.
The associativity of this algebra is in fact a sewing constraint, applied
to a disk with four insertions on the boundary. This turns out to be the
only non-linear constraint to be solved in the TFT approach; all other
relevant equations are linear.

We will consider CFTs that can be defined on world sheets of type (3)
or type (4). Each of these two cases requires slightly different properties of
the algebra $A$, but they can, and will, be treated largely in parallel.

\medskip\noindent

Our construction of CFT correlators takes two pieces of data as an input:

\Itemize
\item[(i)~] The {\em chiral algebra} $\CA$. 
\\
This fixes the minimal amount of symmetry present in the CFT to be constructed. 
More specifically, the holomorphic (and anti-holomorphic) bulk fields contain 
as a subset the fields in $\CA$, and all boundary conditions and defect lines 
preserve $\CA$ (with trivial gluing automorphism). 
\end{itemize}
We demand $\CA$ to be a rational conformal vertex algebra and denote by 
${\Rep}(\CA)$ its representation category. The properties\,\footnote{%
  ~To be precise, $\CA$ must obey certain conditions on its
  homogeneous subspaces, be self-dual, have a semi-simple representation
  category, and fulfill Zhu's $C_2$ cofiniteness condition \cite{huan21}.} 
of $\CA$ then imply that ${\Rep}(\CA)$ is a modular tensor category
(see section I:2.1 for definition, references and our notational conventions).
To any modular tensor category \calc\ one can associate a three-dimensional 
topological field theory TFT($\calc$), see e.g.\ \cite{TUra,BAki} as well as
section I:2.4 and section 3.1 below. The TFT supplies two assignments.
The first, $E \,{\mapsto}\, \calh(E)$ associates to an (extended) surface
$E$ the space $\calh(E)$ of states, a finite-dimensional vector space. The 
second assignment takes
a cobordism $E{\overset{\M}{\longrightarrow}}E'$, i.e.\ a three-manifold
with `in-going' boundary $E$ and `out-going' boundary $E'$ together with
an embedded ribbon graph, and assigns to it a linear map
$Z(\M){:}\; \calh(E)\To\calh(E')$. In the case of Chern\hy Simons theory, the 
state spaces of the TFT can be identified with the spaces of conformal blocks 
of the corresponding WZW model \cite{witt27}. 
In the case of three points on the sphere, for a general RCFT the
identification of the state spaces of the TFT with spaces of conformal
blocks is a consequence of the definition of the tensor product;
for more general situations it follows by the hypothesis of factorisation.

According to the principle \erf{eq:CX-HX}, in order to specify a CFT we must assign
to every world sheet $\X$ a vector $C(\X)$ in $\calh(\hat \X)$. This will be
done with the help of TFT($\calc$), by specifying for every $\X$ a cobordism 
$\emptyset\,{\overset{\MX\,}{\longrightarrow}}\,\hat \X$ and setting 
$C(\X)\,{:=}\, Z(\MX)$. Since $\calc \eq {\Rep}(\CA)$ encodes all the monodromy
properties of the conformal blocks, the question whether the collection $\{C(\X)\}$
of candidate correlators satisfies the requirements
of single-valuedness and factorisation can be addressed solely at the level of 
the category \calc. In particular, the analysis does not require the knowledge 
of the explicit form of $\CA$ or of the conformal blocks. The first input
in the TFT construction is thus
   \def\leftmargini{2.3em}
   \begin{itemize}
   \item
a modular tensor category $\calc$ 
   \end{itemize}
which determines in particular also TFT($\calc$). 

For the construction 
of the vector $C(\X) \iN \calh(\hat \X)$ it is actually not necessary that
\calc\ is the representation category of a vertex algebra; only
the defining properties of a modular tensor category are used. 
The vertex algebra and its conformal blocks are relevant only when
one wishes to determine the correlators as actual {\em functions\/}.
In particular, in those cases where non-isomorphic vertex algebras have
equivalent representation categories (as it happens e.g.\ for WZW theories
based on $\mathfrak{so}(2n{+}1)$ at level 1 with $n$ taking different values 
congruent modulo 8), the TFT approach yields the
same link invariants in the description of the correlators. It is only after
substituting multivalued functions for the conformal blocks that two
such models have different correlators as actual functions of moduli and
insertion points.

\Itemize
\item[(ii)] An {\em algebra} $A$ in \calc.
\\
There are generically several distinct CFTs which share the same chiral algebra
$\CA$, the best known example being the A-D-E-classification of minimal models 
\cite{caiz2}. The additional datum we need to specify the CFT uniquely is an 
algebra $A$ in the tensor category $\calc$. 
\end{itemize}
More precisely, what is needed is
   \def\leftmargini{2.3em} \begin{itemize}
   \item
a {\em symmetric special Frobenius algebra\/} (see section I:3)
for obtaining a CFT defined on world sheets of type (3);
   \item
a {\em Jandl algebra\/}, i.e.\ a symmetric special Frobenius algebra
together with a {\em reversion\/}\\
(see section II:2), for obtaining a CFT defined on world sheets of type (4).
   \end{itemize}
At the level of the CFT, specifying a symmetric special Frobenius algebra
corresponds to providing
the structure constants for three boundary fields on the disk (denoted \bbb\
above) for a single boundary condition. To obtain a Jandl algebra, one 
must \cite{fuRs8} in addition provide a reversion (a braided analogue of an 
involution) on the boundary fields that preserve the given boundary condition.

\medskip

Suppose now we are given such a pair $(\calc , A)$ of data. The correlators
$C_\AA$ of the conformal field theory CFT($A$) associated to this pair
are obtained as follows. Given a world sheet $\X$, one constructs a particular
cobordism $\MX$, the {\em connecting manifold\/} \cite{fffs3}
together with a ribbon graph $R_\X$ embedded in $\MX$. The different
ingredients of $\X$, like field insertions, boundary conditions and
defect lines, are represented by specific parts of the ribbon
graph $R_\X$, and also the algebra $A$ enters in the definition of $R_\X$; 
this will be described in detail in section \ref{sec:fi}. 
Next one uses TFT($\calc$) to define the correlator of CFT($A$) on $\X$
as $C_\AA(\X) \,{:=}\, Z(\MX)$.

One can prove \cite{fjfrs} that the correlators supplied by the assignment
$\X \,{\mapsto}\, C_\AA(\X)$ are single-valued and consistent with 
factorisation or, equivalently, with its inverse operation, sewing. This 
implies in particular that the fundamental correlators
\BBB, \Bb, \bbb\ and \BX\ obtained in this manner solve the sewing constraints.

We conjecture that, conversely, {\em every\/} CFT build from conformal blocks
of the rational vertex algebra $\CA$ defined on world sheets of types (3) and 
(4) can be obtained from a suitable pair $(\calc,A)$ with $\calc\eq{\Rep}(\CA)$.
This converse assignment is not unique: several different algebras lead to one
and the same CFT. This corresponds to the freedom of choosing the particular 
boundary condition whose structure constants are used to define the algebra
structure. All such different algebras are, however, Morita equivalent.

\medskip

For theories defined on world sheets of type (3), it has been advocated in
\cite{bppz2,pezu6,pezu7,pezu8} that aspects of the
fundamental correlators are captured by the structure of a
weak Hopf algebra, respectively an Ocneanu double triangle algebra. Indeed,
a weak Hopf algebra furnishes (see \cite{ostr}, and e.g.\ \cite{ffrs4} for a 
review) a non-canonical description of a module category over a tensor category,
i.e.\ of the situation that in our TFT-based construction is
described by the algebra $A$ in the tensor category ${\Rep}(\CA)$.

\subsection{Plan of the paper}

In section \ref{sec:mb} we introduce the fusing matrices for the categories 
\calca\ of $A$-modules and \calcaa\ of $A$-bimodules. These are useful for
obtaining compact expressions for the structure constants. In section
\ref{sec:fi} the ribbon graph representation of field insertions
is described. We will consider three kinds of fields: boundary fields,
bulk fields, and defect fields. By a boundary field we mean either
a field that lives on a boundary with a given boundary condition or
a field that changes the boundary condition. Likewise, a defect field
can either live on a defect line of a given type, or change the defect type. 
A bulk field can thus be regarded as a special type of defect field,
namely one that connects the invisible defect to the invisible defect.

In section \ref{sec:rc} the ribbon invariants for the fundamental
correlators \BBB, \Bb, \bbb\ and \BX\ are computed, as well as the correlators
\DDD\ of three defect fields on the sphere, and \DX\ of one defect field on the
cross cap.  Up to this point the discussion proceeds entirely at the level of 
the tensor category \calc, i.e.\ of the monodromy data of the conformal blocks. 

We would also like to obtain the fundamental correlators as explicit functions 
of the field insertion points. The necessary calculations, which are presented 
in section \ref{sec:cf}, require additional input beyond the level of \calc, 
namely the vertex algebra $\CA$ with representation category ${\Rep}
(\CA)$ and the notions of intertwiners between $\CA$-\rep s and of conformal 
blocks, as well as the isomorphism between the state spaces of the TFT and 
the spaces of conformal blocks. These additional ingredients are the subject 
of section \ref{sec:cb}. Unlike the mathematical prerequisites employed in the 
earlier sections \ref{sec:mb}\,--\,\ref{sec:rc}, some of this machinery is not 
yet fully developed in the literature. However, in section \ref{sec:cf} we 
only need the two- and three-point conformal blocks, which are well understood.

\medskip

The formulas which give our results for correlators of boundary fields
$\varPsi$, bulk fields $\vhi$ and defect fields $\Thetha$
are listed in the following table.  

\medskip

{}~~~\begin{tabular}{ccccc}
           &   Ribbon  & Ribbon invariant & $\!\!$Ribbon invariant$\!\!$
           & Correlation \\[-2pt]
Correlator &~~invariant~~& in a basis & $\!\!$in Cardy case $A\eq\one\!\!$
           & function \\[3pt]
\hline \\[-3pt]
\bbb  & eq.~~\erf{eq:uhp-cobord}         & eq.~\erf{eq:3bnd-inv-basis} & 
        eq.~\erf{eq:3bnd-basis-cardy}    & eq.~\erf{eq:3bnd-corr} \\[4pt]
\Bb   & eq.~\erf{eq:1bulk1bnd-rib-2}     & eq.~\erf{eq:1B1b-basis} & 
        eq.~\erf{eq:1B1b-bas-Cardy}      & eq.~\erf{eq:1bulk-1bnd} \\[4pt]
\BBB  & eq.~\erf{eq:3bulk-rib}           & eq.~\erf{eq:3bulk-coeff} & 
        eq.~\erf{eq:3bulk-basis-Cardy}   & eq.~\erf{eq:3bulk}      \\[4pt]
\DDD  & eq.~\erf{eq:3defect-rib}         & eq.~\erf{eq:3def-basis} &
        eq.~\erf{eq:3def-basis-Cardy}    & eq.~\erf{eq:3defect}    \\[4pt]
\BX   & eq.~\erf{eq:xcap-ribbon-emb}     & eq.~\erf{chatphic} &
        eq.~\erf{eq:cardy-1bulk-xcap}    & eq.~\erf{eq:1B-xcap}    \\[4pt]
\DX   & eq.~\erf{eq:xcap-def-rib}        & eq.~\erf{eq:1D-Xcap-basis} &
        eq.~\erf{eq:1def-Xcap-bas-Cardy} & eq.~\erf{eq:1D-xcap} 
\\{}
\end{tabular}

\medskip

\noindent
{\bf Acknowledgements.}
We are indebted to N.~Potylitsina-Kube for her skillful help with the numerous 
illustrations. J.F.\ is supported by VR under project no.\ 621--2003--2385,
and C.S.\ is supported by the DFG project SCHW 1162/1-1.

\newpage

\sect{Fusing matrices for modules and bimodules}\label{sec:mb}

To exhibit the mathematical structure encoded by the bulk and  boundary
structure constants, we make use of the concept of module 
\cite{pare11,pare16,beRn,ostr} and bimodule categories. As the main focus
of this paper lies on the computation of correlators,  we do not develop
the mathematical formalism in all detail.

Given a tensor category $\calc$, a {\em right module category\/} over 
$\calc$ comes by definition with a prescription
for tensoring objects of $\calm$ with objects of $\calc$ from the right,
  \be
  \otimes_\calm :\quad \calm \times \calc \rightarrow \calm \,.  \ee
The tensor product $\otimes_\calm$ must be compatible with the tensor product 
$\otimes$ of $\calc$ in the sense that there are associativity isomorphisms
  \be
  (M\otimes_\calm U) \otimes_\calm V
  \,\cong\, M\otimes_\calm (U \oti   
  V)
  \ee
for $U,V \iN \objc$ and $M\iN\obj(\calm)$, such that the corresponding 
pentagon identity is satisfied, as well as unitality isomorphisms 
$M\,{\otimes_\calm}\,\one\,{\cong}\, M$ obeying appropriate triangle identities.

A {\em bimodule category\/} $\calb$ is a straightforward extension 
of this concept, allowing both for a left action of a tensor category $\calc$
and a right action of a possibly different tensor category $\cald$,
  \be
  \otimes_\calb^l :\quad \calc \times \calb \rightarrow \calb 
  \qquad {\rm and} \qquad
  \otimes_\calb^r :\quad \calb \times \cald \rightarrow \calb  \,.
  \ee
The compatibility conditions of the different tensor products are
given by associativity isomorphisms
  \be\begin{array}{ll}
  U \otimes_\calb^l ( V \otimes_\calb^l X) \,\cong\,
    ( U \,{\otimes_{\mathcal{C}}}\, V) \otimes_\calb^l X
    \qquad & {\rm (left~action)} \,,\\[5pt]
  (X\otimes_\calb^r R) \otimes_\calb^r S \,\cong\,
    X\otimes_\calb^r (R \,{\otimes_\cald}\, S)
    \qquad & {\rm (right~action)} \,,\\[5pt]
  U \otimes_\calb^l ( X \otimes_\calb^r R) \,\cong\,
    ( U \otimes_\calb^l X) \otimes_\calb^r R 
    \qquad & {\rm (left~and~right~action~commute)} 
  \eear\ee
for $U,V \iN \objc$, $R,S \iN \obj(\cald)$ and $X \iN \obj(\calb)$. Again the 
various associators must fulfill the corresponding pentagon identities,
and again there are also unitality isomorphisms satisfying triangle identities.
In our application it 
is natural to take the category $\calb$ to be a tensor category as well; then 
there are also the corresponding additional associators and pentagon identities.

In section \ref{sec:rc} we will see that the boundary structure constants 
encode the associator of a module category, while the structure constants of 
bulk and defect fields give the associator of a bimodule category. In the 
present section we define the expansion of these associators in a basis. 
This gives the fusing matrices for modules and bimodules.

\subsection{Fusing matrices for modules}

The right module category relevant for our purposes is the
category $\calca$ of left $A$-modules, where $A$ is an
algebra in a tensor category $\calc$ (see e.g.\ section I:4, or section 2.3 
of \cite{ffrs}; we take $\calc$ to be strict.) Given a left $A$-module
$M \eq (\dot M, \rho)$, we can tensor with any object $U$ of $\calc$ from 
the right so as to obtain again a (generically different) left $A$-module
$M\oti U \eq (\dot M \oti U, \rho \oti \id_U)$ (with all tensor
products taken in $\calc$). This defines a right action
$\otimes{:}\; \calca \,{\times}\, \calc \To \calca$.

We take $A$ to be a symmetric special Frobenius algebra and $\calc$ to be a
modular tensor category. Then also $\calca$ is semisimple and has a finite 
number of non-isomorphic simple objects \cite{fuSc16,kios}.
Let $\{ U_i \,|\, i\iN\II \}$ be a set of representatives of isomorphism
classes of simple objects of $\calc$ and $\{ M_\kappa \,|\, \kappa\iN\JJ \}$ 
a set of representatives of isomorphism classes of simple left $A$-modules. 
Thus lower case roman letters $i,j,k,...$ appear as labels for simple
objects $U_i, U_j, U_k, ... $ of $\calc$, while lower case greek letters 
$\mu,\nu,\kappa, ... $ from the upper range of the alphabet 
appear as labels for simple $A$-modules $M_\mu, M_\nu, M_\kappa, ... $ 
(later on, we will use such greek letters also for simple bimodules).

As a first step towards defining the fusing matrices for $\calca$
we choose bases in the spaces $\Hom_A(M_\mu \oti U_i, M_\nu)$,
  \bea \begin{picture}(190,60)(0,33)
  \put(60,0)     {\includeourbeautifulpicture 59 }
  \put(0,41)     {$ \psi_{(\mu i)\nu}^{\alpha} \;:= $}
  \put(112,41)   {$ \in \Hom_A(M_\mu \oti U_i, M_\nu)  \;. $}
  \put(59,-8.2)  {\scriptsize$ M_\mu $} 
  \put(59,91.2)  {\scriptsize$ M_\nu $} 
  \put(68,44.8)  {\scriptsize$ \alpha $} 
  \put(80.6,-8.2){\scriptsize$ U_i $} 
  \epicture25 \labl{eq:MUM-basis} \xlabp 59
We do not restrict this choice of basis in any way,
except for $i\eq0$, i.e.\ when $U_i\eq\one$. In this
case we demand the basis element to be $\psi_{(\mu 0)\mu}\eq\id_{M_\mu}$.
Both the basis morphisms \erf{eq:MUM-basis} and the basis morphisms 
$\lambda_{(ij)k}^{\alpha}\iN\Hom(U_i\oti U_j,U_k)$ that were chosen in 
(I:2.29) will be labelled by lower case greek letters 
$\alpha, \beta, \gamma, ... $, chosen from the beginning of the alphabet. 

We also need bases in the spaces $\Hom_A(M_\nu, M_\mu \oti U_i)$, which we 
denote by $\{ \bar\psi_{\alpha}^{(\mu i)\nu} \}$. We choose these bases 
such that they are dual to the $\psi_{(\mu i)\nu}^{\alpha}$ in the sense
that $\psi_{(\mu i)\nu}^{\alpha} \circ \bar\psi_{\beta}^{(\mu i)\nu}
{=}\, \delta_{\alpha,\beta}\, \id_{M_\nu}$. Just as for the morphisms
$\lambda_{(ij)k}^{\alpha}$ in (I:2.31), owing to the semisimplicity of $\calca$
the morphisms $\psi_{(\mu i)\nu}^{\alpha}$ obey the completeness relation
  \bea \begin{picture}(190,60)(0,40)
  \put(55,0)     {\Includeourbeautifulpicture 61a }
  \put(130,0)    {\Includeourbeautifulpicture 61b }
  \put(0,47)     {$ \dsty\sum_{\kappa\in\JJ}\;\sum_{\alpha=1}^{\Ann{i\mu}{\kappa}} $}
  \put(54,-8.2)  {\scriptsize$ M_\mu $} 
  \put(54,102.5) {\scriptsize$ M_\mu $} 
  \put(61,46.5)  {\scriptsize$ M_\kappa $} 
  \put(63,37.9)  {\scriptsize$ \alpha $} 
  \put(63,57.3)  {\scriptsize$ \bar\alpha $} 
  \put(75.5,-8.2){\scriptsize$ U_i $} 
  \put(75.5,102.5){\scriptsize$ U_i $} 
  \put(100,47)   {$ = $}
  \put(126,-8.2) {\scriptsize$ M_\mu $} 
  \put(126,102.5){\scriptsize$ M_\mu $} 
  \put(144.3,-8.2){\scriptsize$ U_i $} 
  \put(144.3,102.5){\scriptsize$ U_i $} 
  \epicture26 \labl{pic61} \xlabp 61
Here we abbreviate the dimension of the morphism space $\Hom_A(M_\mu \Oti U_i,
M_\kappa)$ by $\Ann{i\mu}{\kappa}$; according to proposition I:5.22, 
these non-negative integers give the annulus coefficients of the CFT.

The fusing matrices $\GG[A]$ relate two different bases of the space
$\Hom_A(M_\mu\Oti U_i \Oti U_j,M_\nu)$. We denote these bases
by $\{ b^{(1)}_{\alpha\kappa\beta} \}$ and $\{ b^{(2)}_{\gamma k\delta} \}$, 
with the individual vectors given by
  \bea \begin{picture}(330,60)(0,48)
                 \put(56,0){ \begin{picture}(0,0)
  \put(0,0)      {\Includeourbeautifulpicture 60a }
  \put(-10.4,46.5){\scriptsize$ M_{\kappa}^{} $} 
  \put(-1,-8.2)  {\scriptsize$ M_\mu $} 
  \put(-1,102.5) {\scriptsize$ M_\nu $} 
  \put(12,32.8)  {\scriptsize$ \alpha $} 
  \put(12,60)    {\scriptsize$ \beta $} 
  \put(20.5,-8.2){\scriptsize$ U_i $} 
  \put(40.8,-8.2){\scriptsize$ U_j $} 
                 \end{picture} }
                 \put(246,0){ \begin{picture}(0,0)
  \put(0,0)      {\Includeourbeautifulpicture 60b }
  \put(-1,-8.2)  {\scriptsize$ M_\mu $} 
  \put(-1,102.5) {\scriptsize$ M_\nu $} 
  \put(12,68.3)  {\scriptsize$ \gamma $} 
  \put(31.9,48.9){\scriptsize$ \delta $} 
  \put(16.5,-8.2){\scriptsize$ U_i $} 
  \put(19.2,57.5){\scriptsize$ U_k $} 
  \put(36.5,-8.2){\scriptsize$ U_j $} 
                 \end{picture} }
  \put(0,49)     {$ b^{(1)}_{\alpha\kappa\beta} \;:= $}
  \put(148,49)   {and $\qquad b^{(2)}_{\gamma k\delta} \;:= $}
  \epicture35 \labl{pic60} \xlabp 60
Let us verify that these are indeed bases. Linear independence can be seen by 
composing with the corresponding dual basis elements. Completeness of 
$\{ b^{(1)}_{\alpha\kappa\beta} \}$ is seen by applying \erf{pic61} twice.  
A similar reasoning establishes completeness of $\{b^{(2)}_{\gamma k\delta}\}$.
Also note that there are $\sum_\kappa \Ann{i \mu}{\kappa} \Ann{j\kappa}{\nu}$
basis elements of type $b^{(1)}$ and $\sum_k \N ijk \Ann{k \mu}{\nu}$ basis 
elements of type $b^{(2)}$. That these two numbers are indeed equal has
been established in theorem I:5.20(v). In CFT, equality of the two sums means 
that the annulus coefficients furnish a NIMrep of the fusion rules 
\cite{prss3,bppz2,gann17}.

The fusing matrix of the module category $\calca$ describes a change of basis
from $b^{(1)}$ to $b^{(2)}$. There exists a unique set of numbers 
$\GG[A]^{(\mu ij)\,\nu}_{\alpha\kappa\beta,\gamma k\delta}$ such that
  \bea \begin{picture}(280,59)(0,46)
                 \put(0,0){ \begin{picture}(0,0)
  \put(0,0)      {\Includeourbeautifulpicture 60a }
  \put(-10.4,46.5){\scriptsize$ M_{\kappa}^{} $}
  \put(-1,-8.2)  {\scriptsize$ M_\mu $}
  \put(-1,102.5) {\scriptsize$ M_\nu $}
  \put(12,32.8)  {\scriptsize$ \alpha $}
  \put(12,60)    {\scriptsize$ \beta $}
  \put(20.5,-8.2){\scriptsize$ U_i $}
  \put(40.8,-8.2){\scriptsize$ U_j $}
                 \end{picture} }
                 \put(220,0){ \begin{picture}(0,0)
  \put(0,0)      {\Includeourbeautifulpicture 60b }
  \put(-1,-8.2)  {\scriptsize$ M_\mu $}
  \put(-1,102.5) {\scriptsize$ M_\nu $}
  \put(12,68.3)  {\scriptsize$ \gamma $}
  \put(31.9,48.9){\scriptsize$ \delta $}
  \put(16.5,-8.2){\scriptsize$ U_i $}
  \put(19.2,57.5){\scriptsize$ U_k $}
  \put(36.5,-8.2){\scriptsize$ U_j $}
                 \end{picture} } 
  \put(65,46)    {$ =\; \dsty\sum_{k\in\II}\;
                 \sum_{\gamma=1}^{\Ann{k \mu}{\nu}}\; \sum_{\delta=1}^{\N ijk}\,
                 \GG[A]^{(\mu ij)\,\nu}_{\alpha\kappa\beta,\gamma k\delta} $}
  \epicture33 \labl{eq:GA-def} \xlabp 60
We denote the fusing matrix arising here by $\GG[A]$ rather than by
$\FF[A]$, because in the Cardy case $A\eq\one$ it reduces
to the inverse fusing matrix $\GG$ of $\calc$, see (I:2.40). By composing
\erf{eq:GA-def} with the basis of $\Hom_A(M_\nu,M_\mu\Oti U_i \Oti U_j)$ 
dual to $\{b^{(2)}_{\gamma k\delta}\}$,
one arrives at the following ribbon invariant for $\GG[A]$:
  \bea \begin{picture}(200,56)(0,32)
  \put(66,0)     {\Includeourbeautifulpicture 63a }
  \put(120,0)    {\Includeourbeautifulpicture 63b }
  \put(0,38)     {$ \GG[A]^{(\mu ij)\,\nu}_{\alpha\kappa\beta,\gamma k\delta} $}
  \put(62,-8.2)  {\scriptsize$ M_\nu $}
  \put(62,85.2)  {\scriptsize$ M_\nu $}
  \put(90,38)    {$ = $}
  \put(110,31.5) {\scriptsize$ M_\mu $}
  \put(110,59.1) {\scriptsize$ M_\kappa $}
  \put(113.5,50.9){\scriptsize$ \alpha $}
  \put(113.8,67.9){\scriptsize$ \beta $}
  \put(113.8,11.3){\scriptsize$ \bar\gamma $}
  \put(119,-8.2) {\scriptsize$ M_\nu $}
  \put(119,85.2) {\scriptsize$ M_\nu $}
  \put(131,36.2) {\scriptsize$ U_i $}
  \put(136,22.5) {\scriptsize$ U_k $}
  \put(146,57.3) {\scriptsize$ U_j $}
  \put(152.3,31.3){\tiny$ \bar\delta $}
  \epicture26 \labl{eq:GA-rib-inv} \xlabp 63
A possibility to compute these numbers is to first find
all simple $A$-modules $M_\mu$ by decomposing induced modules (see
section I:4.3), which is a linear (and finite) problem. 
Next one needs to identify the spaces $\Hom_A(M_\mu \Oti U_i, M_\nu)$ as 
subspaces of $\Hom(\dot M_\mu \Oti U_i , \dot M_\nu)$. This can be done
by finding the eigenspaces of the projector $f \,{\mapsto}\, \overline f$
that takes a morphism $f \iN \Hom(\dot M_\mu \Oti U_i , \dot M_\nu)$
to its $A$-averaged morphism $\overline f$, which is known to lie
in the subspace $\Hom_A(M_\mu \Oti U_i, M_\nu)$, see definition I:4.3
and lemma I:4.4. This is again a linear problem. 
These ingredients suffice to evaluate \erf{eq:GA-rib-inv}.

By construction, the matrices $\GG[A]$ solve the pentagon relation
  \bea \begin{picture}(420,150)(0,60)
  \put(0,55)     {\Includeourbeautifulpicture 64a }
  \put(0,55)     {\setlength{\unitlength}{.38pt}
                  \put(-5,12){\scriptsize$ \mu $} 
                  \put( 63, 5){\scriptsize$ i $} 
                  \put(152, 5){\scriptsize$ j $} 
                  \put(240, 5){\scriptsize$ k $} 
                  \put(120,203){\scriptsize$ \nu $} 
                  \put(55,71){\scriptsize$ \kappa $} 
                  \put(99,115){\scriptsize$ \lambda $} 
                  \setlength{\unitlength}{1pt}}
  \put(150,125)  {\Includeourbeautifulpicture 64b }
  \put(150,125)  {\setlength{\unitlength}{.38pt}
                  \put(-5,12){\scriptsize$ \mu $} 
                  \put( 80, 5){\scriptsize$ i $} 
                  \put(152, 5){\scriptsize$ j $} 
                  \put(240, 5){\scriptsize$ k $} 
                  \put(120,203){\scriptsize$ \nu $} 
                  \put(121,71){\scriptsize$ p $} 
                  \put(99,115){\scriptsize$ \lambda $} 
                  \setlength{\unitlength}{1pt}}
  \put(310,125)  {\Includeourbeautifulpicture 64c }
  \put(310,125)  {\setlength{\unitlength}{.38pt}
                  \put(-5,12){\scriptsize$ \mu $} 
                  \put( 80, 5){\scriptsize$ i $} 
                  \put(152, 5){\scriptsize$ j $} 
                  \put(240, 5){\scriptsize$ k $} 
                  \put(120,203){\scriptsize$ \nu $} 
                  \put(141,70){\scriptsize$ p $} 
                  \put(160,120){\scriptsize$ q $} 
                  \setlength{\unitlength}{1pt}}
  \put(150,0)    {\Includeourbeautifulpicture 64d }
  \put(150,0)    {\setlength{\unitlength}{.38pt}
                  \put(-7,12){\scriptsize$ \mu $} 
                  \put( 72, 5){\scriptsize$ i $} 
                  \put(157, 5){\scriptsize$ j $} 
                  \put(240, 5){\scriptsize$ k $} 
                  \put(120,203){\scriptsize$ \nu $} 
                  \put(81, 99){\scriptsize$ \kappa $} 
                  \put(181,99){\scriptsize$ r $} 
                  \setlength{\unitlength}{1pt}}
  \put(310,0)    {\Includeourbeautifulpicture 64e }
  \put(310,0)    {\setlength{\unitlength}{.38pt}
                  \put(-5,12){\scriptsize$ \mu $} 
                  \put( 80, 5){\scriptsize$ i $} 
                  \put(170, 5){\scriptsize$ j $} 
                  \put(240, 5){\scriptsize$ k $} 
                  \put(120,203){\scriptsize$ \nu $} 
                  \put(160,120){\scriptsize$ q $} 
                  \put(207,73){\scriptsize$ r $} 
                  \setlength{\unitlength}{1pt}}
  \put(90,110)   {\Includeourbeautifulpicture 64f }
    \put(92,128)   {\scriptsize$ \GG[A] $}
  \put(120,40)   {\Includeourbeautifulpicture 64g }
    \put(125,45)   {\scriptsize$ \GG[A] $}
  \put(256,40)   {\Includeourbeautifulpicture 64h }
    \put(270,33)   {\scriptsize$ \GG[A] $}
  \put(256,165)  {\Includeourbeautifulpicture 64h }
    \put(270,175)   {\scriptsize$ \GG[A] $}
  \put(341,81)   {\Includeourbeautifulpicture 64i }
    \put(390,100)   {\scriptsize$ \GG $}
    \put(334,100)   {\scriptsize$ \FF $}
  \epicture44 \labp 64
Explicitly, this pentagon reads
  \bea\displaystyle
  \sum_\eps 
  \GG[A]^{(\kappa j k)\,\nu}_{\alpha_2 \lambda \alpha_3 \, , \, \eps r \delta_2}\,
  \GG[A]^{(\mu i r)\,\nu}_{\alpha_1 \kappa \eps \, , \, \gamma_1 q \delta_1}
  \\[5pt] \displaystyle \hspace*{7em} = 
  \sum_p \sum_{\beta_1,\beta_2,\gamma_1}
  \GG[A]^{(\mu i j)\,\lambda}_{\alpha_1 \kappa \alpha_2 \, , \, \beta_1 p \beta_2}\,
  \GG[A]^{(\mu p k)\,\nu}_{\beta_1 \lambda \alpha_3 \, , \, \gamma_1 q \gamma_2}\,
  \GG^{(ijk)\,q}_{\beta_2 p \gamma_2 \, , \, \delta_1 r \delta_2} \,.
  \eear\labl{eq:module-pent-G}
If one prefers the fusing matrix $\FF$ of $\calc$ to appear directly,
rather than its inverse $\GG$, one may rewrite this in the form
  \bea\displaystyle
  \sum_{\beta_1} 
  \GG[A]^{(\mu i j)\,\lambda}_{\alpha_1 \kappa \alpha_2 \, , \, \beta_1 p \beta_2}\,
  \GG[A]^{(\mu p k)\,\nu}_{\beta_1 \lambda \alpha_3 \, , \, \gamma_1 q \gamma_2}
  \\[5pt] \displaystyle \hspace*{7em} = 
  \sum_r \sum_{\delta_1,\delta_2,\eps}
  \GG[A]^{(\kappa j k)\,\nu}_{\alpha_2 \lambda \alpha_3 \, , \, \eps r \delta_2}\,
  \GG[A]^{(\mu i r)\,\nu}_{\alpha_1 \kappa \eps \, , \, \gamma_1 q \delta_1}\,
  \FF^{(ijk)\,q}_{\delta_1 r \delta_2 \, , \, \beta_2 p \gamma_2} \,.
  \eear \hspace*{1.6em} \labl{eq:module-pent-F}
When all morphism spaces involved are one-dimensional, these relations
simplify considerably, e.g.\ \erf{eq:module-pent-G} reads
  \be
  \GG[A]^{(\kappa j k)\,\nu}_{\lambda \,,\, r}\,
  \GG[A]^{(\mu i r)\,\nu}_{\kappa \,,\, q}
  = \sum_p \GG[A]^{(\mu i j)\,\lambda}_{\kappa \,,\, p}\,
  \GG[A]^{(\mu p k)\,\nu}_{\lambda \,,\, q}\,
  \GG^{(ijk)\,q}_{p \,,\, r} \,.  \ee

\dtl{Remark}{rem:bnd-6j}
(i)~\,The fusing matrices $\GG[A]$ (or rather, their inverses $\FF[A]$)
also appear in an approach to CFT based on graphs and cells
\cite{bppz2,pezu6}, where the notation $^{(1)\!}F$ is used. The $^{(1)\!}F$ 
are required to solve a pentagon relation; the corresponding equation 
for $\GG[A]$ is \erf{eq:module-pent-G}. In fact, the associativity conditions
between the categories $\calc$, $\calca$ and $\calcaa$ give rise to a number of
pentagon relations, called the ``Big Pentagon'', for the corresponding 
associators. The associators and the pentagon relations can be visualised using
cells (known as Ocneanu cells) \cite{ocne10,bosz},
and can also be interpreted in terms a 2-category with two objects,
see e.g.\ \cite{ffrs4} for more details. 
\\[.3em]
(ii)~As observed for the Cardy case $A\eq\one$ in \cite{runk,fffs}, and for the
general case in \cite{bppz}, the pentagon for $\GG[A]$ is equivalent to the 
sewing (or factorisation) constraint \cite{lewe3} for four boundary fields 
on the upper half plane. If $\hatpsi^{\mu\nu}_a$ denotes a boundary field
which changes a boundary condition described by the $A$-module $M_\nu$
to $M_\mu$ (looking along the real axis towards $+\infty$) and has chiral 
representation label $a$, the boundary OPE symbolically takes the form
  \be
  \hatpsi^{\mu\nu}_a\, \hatpsi^{\nu\kappa}_a =~\sum_{c} c^{\mu\nu\kappa}_{abc} 
  \, \hatpsi^{\mu\kappa}_c \,,
  \ee
where the constants $c^{\mu\nu\kappa}_{abc}$ are the OPE coefficients\,%
  \footnote{~With all multiplicity labels and coordinate dependence
  in place, one obtains an equation of the form (I:3.11).}. 
Factorisation of the correlator of four boundary fields leads to
the constraint \cite{lewe3}
  \be
  c^{\mu\nu\kappa}_{abf} \, c^{\mu\kappa\rho}_{fcd} = 
  \sum_c c^{\nu\kappa\rho}_{bce} \, c^{\mu\nu\rho}_{aed}\, \Fs abcdef \,.
  \ee
By relation \erf{eq:module-pent-F}, this equation in the unknowns
$c$ is solved by
  \be
  c^{\mu\nu\kappa}_{abc} = \GG[A]^{(\mu a b)\,\kappa}_{\nu \,,\, c} \,,
  \labl{eq:bnd-GA-id}
where again all multiplicity labels are omitted.
\\
And indeed, when evaluating the
corresponding ribbon invariant in \erf{eq:bnd-3pt-rib} below, one finds
the explicit expression \erf{eq:3bnd-inv-basis} for the coefficient
occurring in the correlator of three boundary fields. This expression
implies\,\footnote{~%
  In \erf{eq:3bnd-inv-basis} the combination
  $c^{\kappa \nu \mu}_{ji \bar k} c^{\mu \kappa \mu}_{k \bar k 0}$
  appears because it describes a three-point correlator, rather than an
  OPE coefficient.}
\erf{eq:bnd-GA-id}. We thus arrive at the conclusion that the boundary OPE 
coefficients are given by the 6j-symbols of the module category $\calca$.

\subsection{Fusing matrices for bimodules}

The bimodule category that is needed for the description of the bulk
structure constants is of the form\,\footnote{%
  ~As mentioned in section \ref{sec:in-1}, we treat here only theories
  defined on surfaces of types (3) and (4). When one does not include
  surfaces with boundary, one can also consider heterotic theories,
  in which case more general bimodule categories
  $\calc \Times \calb \Times \overline{\cald} \To \calb$ occur.}
$\calc \Times \calb \Times \overline{\calc} \To \calb$. Here 
$\overline{\calc}$ denotes the tensor category dual\,\footnote{%
  ~The dual of a tensor category $(\calc,\otimes)$ is e.g.\ defined in section
  6.2 of \cite{ffrs}, where it was taken to be $(\calc^{\rm opp}, \otimes)$
  rather than $(\calc,\otimes^{\rm opp})$. These two categories
  are tensor equivalent via the functor $?^\vee$, 
  i.e.\ by taking the duals of objects and morphisms.}
to $\calc$ in the sense that if $\calc \eq (\calc , {\otimes})$ then 
$\overline{\calc} \eq (\calc,{\otimes}^{\rm opp})$. Denoting the quantities in 
$\overline{\calc}$ with a bar, this means that $\overline U \eq U$ and 
$\overline U \,\overline{\otimes}\,\overline V \eq V \Oti\, U$. The 
bimodule category $\calb$ of interest here
is the category $\calcaa$ of $A$-bimodules, see definition I:4.5. 

For the action of $\calc$ and $\overline{\calc}$ on $\calcaa$
there are, in each case, two possibilities. Let us start with 
$\otimes_{\calb}^l{:}\; \calc \Times \calcaa \To \calcaa$. Given an object $U$ 
of $\calc$ and an $A$-bimodule $X \eq (\dot X, \rho_l, \rho_r)$, we define 
the bimodules $U \,{\otimes^\pm} X$ as
  \bea
  U \,{\otimes^+}\, X := (U \Oti \dot X , 
    (\id_U\Oti\rho_l) \cir ( c_{U,A}^{~-1} \Oti \id_X), \id_U\Oti\rho_r )
  \qquad{\rm and} \\{}\\[-.7em]
  U \,{\otimes^-}\, X := (U \Oti \dot X , 
    (\id_U\Oti\rho_l) \cir ( c_{\!A,U}^{} \Oti \id_X), \id_U\Oti\rho_r ) \,,
  \eear\labl{UXUX}
respectively. (In graphical notation, for $\otimes^+$ the $U$-ribbon passes 
above the $A$-ribbon, while for $\otimes^-$ it passes below.)
We take the functor $\otimes_{\calb}^l$ to be given by $\otimes^+$.
Similarly we can define bimodules $X \,{\otimes^\pm} U$ when
tensoring with $U$ from the right as
  \bea
  X \,{\otimes^+}\, U := (\dot X \Oti U , \rho_l \Oti \id_U,
  (\rho_r \Oti \id_U) \cir (\id_X\Oti c_{U,A}^{}) ) 
  \qquad{\rm and} \\{}\\[-.8em]
  X \,{\otimes^-}\, U := (\dot X \Oti U , \rho_l \Oti \id_U,
  (\rho_r \Oti \id_U) \cir (\id_X\Oti c_{\!A,U}^{\,-1}) ) \,,
  \eear\labl{XUXU}
respectively. (In graphical notation, again for $\otimes^+$, $U$ passes
above $A$, while for $\otimes^-$, it passes below.)
We take the functor $\otimes_{\calb}^r$ to be given by $\otimes^-$.

The reason for choosing the actions of $\calc$ and $\overline{\calc}$
on $\calcaa$ in this particular way is that defect fields are to be labelled 
by elements of the space $\HomAA( U \,{\otimes^+} X \,{\otimes^-}\, V , Y)$
of bimodule morphisms between the $A$-bimodules
$U {\otimes^+} X {\otimes^-} V$ and $Y$. In \cite{fuRs4,fuRs8} instead
the space $\HomAA( X {\otimes^-} V, Y {\otimes^+} U^\vee )$
of bimodule morphisms was used. These two spaces are canonically
isomorphic; an isomorphism can be given as follows. 
\\[-2.3em]

\dtl{Lemma}{lem:defect-iso}
The mapping 
  \be
  \varphi :\quad \HomAA( X \,{\otimes^-}\, V, Y \,{\otimes^+}\, U^\vee )
  \rightarrow \HomAA( U \,{\otimes^+} X \,{\otimes^-}\, V , Y)
  \ee  
given by
  \be
  \varphi(f) := (\tilde d_U \oti\id_Y) \cir (\id_U \oti [c_{Y,U^\vee}^{}\cir f])
  \ee
is an isomorphism.

\medskip\noindent
Proof:\\
That $\varphi(f)$ commutes with the left and right action of $A$ follows by a 
straightforward use of the definitions. Further, one
readily checks that the map $\tilde\varphi$ acting as
  \be
  \tilde\varphi(g) := c_{Y,U^\vee}^{~-1}
  \cir (\id_{U^\vee} \oti g) \cir (\tilde b_U \oti \id_X \oti \id_V) 
  \ee
for $g\iN\HomAA( U{\otimes^+}X{\otimes^-}V , Y)$ is left and right
inverse to $\varphi$.
\qed

\medskip

It follows from proposition I:4.6 that the category of $A$-bimodules
is isomorphic to the category of left modules over the algebra
$A\oti A_{\rm op}$. Since $A$ is a symmetric special Frobenius algebra, so is
$A\oti A_{\rm op}$. It follows that the category of 
$A\oti A_{\rm op}$-modules is semisimple, and hence $\calcaa$ is semisimple,
too. Let us choose a set of representatives $\{ X_\kappa\,|\,\kappa\iN\KK\}$
of isomorphism classes of simple $A$-bimodules. 

The algebra $A$ is called {\em simple} iff it is a simple bimodule over itself 
or, equivalently, iff $Z(A)_{00}\eq1$, see definition 2.26 and remark 2.28(i) 
of \cite{ffrs} for details. If $A$ is simple, then we choose the 
representatives of simple modules such that $X_0 \eq A$.

Further, we choose bases $\{ \xi_{(i\mu j)\nu}^{\alpha} \}$ of the spaces
$\HomAA(U_i{\otimes^+}X_\mu{\otimes^-}U_j,X_\nu)$, with graphical representation
  \bea \begin{picture}(130,56)(0,38)
  \put(60,0)     {\includeourbeautifulpicture 65 }
  \put(0,41)     {$ \xi_{(i\mu j)\nu}^{\alpha} \;=: $}
  \put(57.2,-8.2){\scriptsize$ U_i $}
  \put(76.5,-8.2){\scriptsize$ X_\mu $}
  \put(76.5,90.8){\scriptsize$ X_\nu $}
  \put(89.8,40.2){\scriptsize$ \alpha $}
  \put(98.5,-8.2){\scriptsize$ U_j $}
  \epicture28 \labl{eq:xi-basis} \xlabp 65
For the special case $i\eq j\eq 0$ 
we choose $\xi_{(0\mu 0)\mu}\eq \id_{X_\mu}$.
The number of elements in the basis $\{ \xi_{(i\mu j)\nu}^{\alpha} \}$ of 
$\HomAA( U_i{\otimes^+}X_\mu{\otimes^-}U_j , X_\nu)$ is denoted by
$Z_{ij}^{X_\mu|X_\nu}$; more generally, for any two $A$-bimodules $X$ and $Y$ 
we set
  \be
  Z_{ij}^{X|Y} 
  := \dim \HomAA( U_i{\otimes^+}X{\otimes^-}U_j , Y)
  = \dim \HomAA( X{\otimes^-}U_j, Y{\otimes^+}U_{\bar\imath} ) \,;
  \labl{ZijXY}
the second equality follows from lemma \ref{lem:defect-iso} together with 
$U_i^\vee \,{\cong}\, U_{\bar\imath}$. We denote the basis elements 
dual to \erf{eq:xi-basis} by $\bar\xi_{\alpha}^{(i\mu j)\nu}{\in}\,
\HomAA(X_\nu,U_i{\otimes^+}X_\mu{\otimes^-}U_j)$; they obey
  \be
  \xi_{(i\mu j)\nu}^{\alpha} \cir
  \bar\xi_{\beta}^{(i\mu j)\nu} = \delta_{\alpha,\beta}\, \id_{X_\nu}
  \quad\ {\rm and} \quad\
  \sum_{\nu\in\KK} \sum_{\alpha=1}^{Z_{ij}^{X_\mu|X_\nu}}\!\!
  \bar\xi_{\alpha}^{(i\mu j)\nu} \cir
  \xi_{(i\mu j)\nu}^{\alpha} = \id_{U_i} \oti \id_{X_\mu} \oti \id_{U_j}
  \labl{eq:xi-bas-prop}
The definition \erf{ZijXY} of $Z_{ij}^{X|Y}$ is in accordance with the
definition via the ribbon invariant (I:5.151), as can be seen by
inserting the second identity of \erf{eq:xi-bas-prop} into
the ribbon graph (I:5.151) and using $Z_{00}^{X|Y} {=}\, \dim \HomAA(X,Y)$.
The numbers $Z_{ij}^{X|Y}$ also have a physical interpretation; they
describe the torus partition function with the insertion of two
parallel defect lines \cite{pezu5,pezu7}; see section I:5.10 for details.

To motivate the definition of the fusing matrices for $\calb\eq\calcaa$,
let us have a look at the left action of $\overline{\calc}$ on $\calb$. 
By semisimplicity,
the isomorphism $(X_\mu{\otimes_{\calb}^r}\overline{U}_i) \,{\otimes_{\calb}^r}\, 
\overline{U}_j \,{\cong}\, X_\mu \,{\otimes_{\calb}^r}\, (\overline{U}_i \, 
\overline{\otimes} \, \overline{U}_j)$ on objects gives rise to an isomorphism
  \bea\displaystyle
  \bigoplus_\kappa
  \Hom^{\calb}(X_\mu \,{\otimes_{\calb}^r}\, \overline{U}_i, X_\kappa) 
  \otimes
  \Hom^{\calb}(X_\kappa {\otimes_{\calb}^r}\, \overline{U}_j, X_\nu) 
  \\[5pt]\displaystyle
  \hspace*{6em} \stackrel{\cong}{\longrightarrow}\;
  \bigoplus_k \Hom^{\calb}(X_\mu \,{\otimes_{\calb}^r}\, \overline{U}_k, X_\nu) 
  \otimes \Hom^{\overline{\mathcal{C}}}(
  \overline{U}_i \, \overline{\otimes} \,\overline{U}_j , \overline{U}_k ) 
  \eear\labl{eq:right-action-cob}
of morphism spaces. By definition, the space $\Hom^{\overline{\mathcal{C}}}(
\overline{U}_i \, \overline{\otimes} \,\overline{U}_j , \overline{U}_k )$
of morphisms in $\overline{\calc}$ is equal to the
morphism space $\Hom^{\mathcal{C}}(U_j{\otimes}U_i,U_k)$ in $\calc$, while 
$\Hom^{\calb}(X_\mu{\otimes_{\calb}^r}\overline{U}_i, X_\kappa) 
\eq \HomAA(X_\mu{\otimes^{-}} U_i, X_\kappa)$ etc. 
We would still like to describe the isomorphism
\erf{eq:right-action-cob} with the help of the graphical rep\-resentation
of morphisms in the ribbon category $\calc$. To this end we need a 
visualisation for morphisms in $\overline{\calc}$. 
A natural possibility is to represent them graphically by
coupons that face the reader with their back side. Thus, using dashed lines
for ribbons with back side facing the reader (see the beginning of
section II:3 for conventions), we have
  \bea \begin{picture}(120,64)(0,43)
                 \put(0,0){ \begin{picture}(0,0)
  \put(0,0)      {\Includeourbeautifulpicture 66a }
  \put(-.6,-8.2) {\scriptsize$ U $}
  \put(14.8,106) {\scriptsize$ W $}
  \put(16.1,57.4){\tiny$ f $}
  \put(29.5,-8.2){\scriptsize$ V $}
                 \end{picture} }
  \put(56,50)    {$ = $}
                 \put(78,0){ \begin{picture}(0,0)
  \put(2,0)      {\Includeourbeautifulpicture 66b }
  \put(-.6,-8.2) {\scriptsize$ U $}
  \put(14.8,106) {\scriptsize$ W $}
  \put(16.1,57.4){\tiny$ f $}
  \put(29.5,-8.2){\scriptsize$ V $}
                 \end{picture} }
  \epicture26 \labp 66
for $f \iN \Hom^{\overline{\mathcal{C}}}(\overline{U}\,\overline{\otimes}\,
\overline{V},\overline{W}) \eq \Hom^{\mathcal{C}}(V{\otimes}U,W)$, 
where on the left hand side the coupon labelled by $f$ is showing the black
side to the reader while on the right hand side it is showing its white side.

\medskip

After these preliminaries we are ready to give the definition of the
fusing matrices $\FF[A|A]$ of the bimodule category $\calcaa$. We have
  \bea \begin{picture}(410,67)(0,46)
                 \put(0,0){ \begin{picture}(0,0)
  \put(0,0)      {\Includeourbeautifulpicture 67a }
  \put(-2.8,-8.2){\scriptsize$ U_i $}
  \put(17.5,-8.2){\scriptsize$ U_j $}
  \put(29.5,56.5){\scriptsize$ X_{\!\kappa} $}
  \put(37.5,-8.2){\scriptsize$ X_\mu $}
  \put(37.5,110) {\scriptsize$ X_\nu $}
  \put(49.8,49.6){\scriptsize$ \alpha $}
  \put(49.8,74.6){\scriptsize$ \beta $}
  \put(59.2,-8.2){\scriptsize$ U_k $}
  \put(79.5,-8.2){\scriptsize$ U_l $} 
                 \end{picture} }
                 \put(311,0){ \begin{picture}(0,0)
  \put(0,0)      {\Includeourbeautifulpicture 67b }
  \put(-2.8,-8.2){\scriptsize$ U_i $}
  \put(22.7,-8.2){\scriptsize$ U_j $}
  \put(24.7,66.2){\scriptsize$ U_{q_1} $}
  \put(41.8,-8.2){\scriptsize$ X_\mu $}
  \put(41.8,110) {\scriptsize$ X_\nu $}
  \put(54.8,75.9){\scriptsize$ \gamma $}
  \put(59.8,62.2){\scriptsize$ U_{q_2} $}
  \put(63.2,-8.2){\scriptsize$ U_k $}
  \put(17.2,46.3){\tiny$ \delta_1 $}
  \put(82.5,46.3){\tiny$ \delta_2 $}
  \put(89.1,-8.2){\scriptsize$ U_l $} 
                 \end{picture} }
  \put(97,50)    {$ = \dsty\sum_{q_1,q_2\in\II}\!
                 \sum_{\gamma=1}^{Z_{q_1 q_2}^{X_\mu|X_\nu}}
                 \sum_{\delta_1=1}^{\N ij{q_1}}\;\sum_{\delta_2=1}^{\N ij{q_2}}\,
     \FF[A|A]^{(ij\mu kl)\,\nu}_{\alpha\kappa\beta,\gamma q_1q_2\delta_1\delta_2} $}
  \epicture32 \labl{eq:FAA-def} \xlabp 67
Note that this encodes both the left action of $\calc$
and the right action of $\overline{\calc}$; the individual actions can be 
extracted by setting $i\eq j\eq 0$ or $k\eq l\eq 0$, respectively.

An expression for the bimodule fusing matrix $\FF[A|A]$ 
is obtained from \erf{eq:FAA-def} by composing with the dual basis. One finds
  \bea \begin{picture}(400,140)(0,51)
  \put(102,0)    {\Includeourbeautifulpicture 68a }
  \put(150,0)    {\Includeourbeautifulpicture 68b }
  \put(287,0)    {\Includeourbeautifulpicture 68c }
  \put(0,87)     {$ \FF[A|A]^{(ij\mu kl)\, \nu}_{\alpha \pi \beta, 
                 \gamma q_1 q_2 \delta_1 \delta_2} $}
  \put(98,-8)    {\scriptsize$ X_\nu $}
  \put(98,183)   {\scriptsize$ X_\nu $} 
  \put(121,87)   {$ = $}
  \put(192,-8)   {\scriptsize$ X_\nu $}
  \put(192,183)  {\scriptsize$ X_\nu $} 
  \put(182,146)  {\scriptsize$ \beta $}
  \put(184,129)  {\scriptsize$ X_\pi $}
  \put(182,122)  {\scriptsize$ \alpha $}
  \put(184,85)   {\scriptsize$ X_\mu $}
  \put(182,29)   {\scriptsize$ \overline{\gamma} $}
  \put(166.5,102){\scriptsize$ U_j $}
  \put(219,102)  {\scriptsize$ U_k $} 
  \put(148,122)  {\scriptsize$ U_i $}
  \put(237,120)  {\scriptsize$ U_l $} 
  \put(153,52)   {\scriptsize$ \overline{\delta_1} $}
  \put(233,52)   {\scriptsize$ \overline{\delta_2} $} 
  \put(170,42.1) {\scriptsize$ U_{q_1} $}
  \put(216,42.1) {\scriptsize$ U_{q_2} $} 
  \put(260,87)   {$ = $} 
  \put(329,-8)   {\scriptsize$ X_\nu $}
  \put(329,183)  {\scriptsize$ X_\nu $} 
  \put(320,146)  {\scriptsize$ \beta $} 
  \put(320,122)  {\scriptsize$ \alpha $}
  \put(320,85)   {\scriptsize$ X_\mu $}
  \put(320,29)   {\scriptsize$ \overline\gamma $} 
  \put(321.5,129){\scriptsize$ X_\pi $}
  \put(285,120)  {\scriptsize$ U_i $}
  \put(375,120)  {\scriptsize$ U_l $} 
  \put(290,52)   {\scriptsize$ \overline{\delta_1} $}
  \put(368,51.6) {\scriptsize$ \overline{\delta_2} $} 
  \put(303,102)  {\scriptsize$ U_j $}
  \put(354,102)  {\scriptsize$ U_k $} 
  \put(303,42)   {\scriptsize$ U_{q_1} $}
  \put(354,42)   {\scriptsize$ U_{q_2} $} 
  \epicture34 \labl{eq:FAA-rib-inv} \xlabp 68
Evaluating this expression is again a linear
problem, but it is more complicated than the procedure for evaluating 
\erf{eq:GA-rib-inv}. First one must find all simple $A$-bimodules. 
By proposition I:4.6, this is the same as finding all
simple $A{\otimes}A_{\rm op}$-modules, which (via the method
of induced modules) is a finite, linear problem. Next, the subspaces
$\HomAA( U_i\,{\otimes^+}\,X\,{\otimes^- U_j}, Y)$ of
$\Hom( U_i\oti\dot X\oti U_j, \dot Y)$
can be determined as images of the projection operators 
  \bea \begin{picture}(180,93)(0,51)
  \put(62,0)    {\includeourbeautifulpicture 69 }
  \put(0,63)      {$ P(f) \;:= $}
  \put(59,-9.5)   {\scriptsize$ U_i $}
  \put(154,-9.5)  {\scriptsize$ U_j $}
  \put(106,-9.5)  {\scriptsize$ X $}
  \put(59,78)     {\scriptsize$ A $}
  \put(95,58)     {\scriptsize$ A $}
  \put(154.8,60)  {\scriptsize$ A $}
  \put(108,87)    {\scriptsize $f$}
  \put(108,140)   {\scriptsize$ Y $} 
  \epicture35 \labp 69 
for $f\iN\Hom( U_i\oti\dot X\oti U_j, \dot Y)$.
This is a linear problem, after the solution of which  
one has all the ingredients at hand that are needed to evaluate 
\erf{eq:FAA-rib-inv}.

By construction, the $\FF[A|A]$ satisfy a pentagon relation; schematically, this
relation looks like
  \bea \begin{picture}(420,280)(10,85)
  \put(0,110)    {\Includeourbeautifulpicture 70a }
  \put(0,110)     {\setlength{\unitlength}{.38pt}
                  \put(144.3,23){\scriptsize$ \mu $} 
                  \put(144.3,150){\scriptsize$ \kappa $} 
                  \put(144.3,270){\scriptsize$ \lambda $} 
                  \put(144.3,375){\scriptsize$ \nu $} 
                  \put(  1,23){\scriptsize$ i $} 
                  \put( 52,23){\scriptsize$ j $} 
                  \put(103,23){\scriptsize$ k $} 
                  \put(213,23){\scriptsize$ l $} 
                  \put(266,23){\scriptsize$ m $} 
                  \put(319,23){\scriptsize$ n $} 
                  \setlength{\unitlength}{1pt}}
  \put(150,220)  {\Includeourbeautifulpicture 70b }
  \put(310,220)  {\Includeourbeautifulpicture 70c }
  \put(150,0)    {\Includeourbeautifulpicture 70d }
  \put(310,0)    {\Includeourbeautifulpicture 70e }
  \put(310,0)     {\setlength{\unitlength}{.38pt}
                  \put(144.3,23){\scriptsize$ \mu $} 
                  \put(144.3,375){\scriptsize$ \nu $} 
                  \put(  1,23){\scriptsize$ i $} 
                  \put( 52,23){\scriptsize$ j $} 
                  \put(102,23){\scriptsize$ k $} 
                  \put(213,23){\scriptsize$ l $} 
                  \put(264,23){\scriptsize$ m $} 
                  \put(319,23){\scriptsize$ n $} 
                  \put(104,251){\scriptsize$ r $} 
                  \put(47,126){\scriptsize$ t $} 
                  \put(217,251){\scriptsize$ s $} 
                  \put(271,126){\scriptsize$ u $} 
                  \setlength{\unitlength}{1pt}}
  \put(92,228)   {\scriptsize$ \FF[A|A] $}
  \put(122,80)   {\scriptsize$ \FF[A|A] $}
  \put(276,300)  {\scriptsize$ \FF[A|A] $}
  \put(276,80)   {\scriptsize$ \FF[A|A] $}
  \put(376,184)  {\scriptsize$ \FF \cdot \FF $}
  \put(100,210)  {\Includeourbeautifulpicture 64f }
  \put(130,72)   {\Includeourbeautifulpicture 64g }
  \put(265,70)   {\Includeourbeautifulpicture 64h }
  \put(265,290)  {\Includeourbeautifulpicture 64h }
  \put(368,158)  {\Includeourbeautifulpicture 64j }
  \epicture61 \labp 70
and explicitly it reads
  \bea\displaystyle \hsp{-.7}
  \sum_\delta
  \FF[A|A]^{(ij \kappa mn)\, \nu}
          _{\alpha_2 \lambda \alpha_3 \,,\, \delta t u \eps_1 \varphi_1} \,
  \FF[A|A]^{(tk \mu lu)\,\nu}
          _{\alpha_1 \kappa \delta \,,\,  \gamma_1 r s \eps_2 \varphi_2}
  \\[5pt] \hsp{2.2} \displaystyle 
  = \sum_{p,q} \sum_{\beta_1,\beta_2,\beta_3,\gamma_2,\gamma_3} \hspace*{-1em}
  \FF[A|A]^{(jk \mu lm)\,\lambda}
          _{\alpha_1 \kappa \alpha_2 \,,\, \beta_1 p q \beta_2 \beta_3} \,
  \FF[A|A]^{(ip \mu qn)\,\nu}
          _{\beta_1 \lambda \alpha_3 \,,\, \gamma_1 r s \gamma_2 \gamma_3} \,
  \FF^{(ijk)\,r}_{\gamma_2 p \beta_2 \,,\, \eps_1 t \eps_2} \,
  \FF^{(nml)\,s}_{\gamma_3 q \beta_3 \,,\, \varphi_1 u \varphi_2} .
  \eear\labl{eq:FAA-pentagon}

\dtl{Remark}{rem:bulk-6j}
(i)~\,In remark \ref{rem:bnd-6j}(ii) we have seen that the boundary OPE is
given by the 6j-symbols of the module category $\calca$. In fact, similar 
considerations apply to bulk and defect fields.\\
Let $\vPhi_\alpha$ be a bulk field with chiral/anti-chiral representation
labels $\alpha_l$, $\alpha_r$. Omitting all position dependence and
multiplicity indices, the OPE of two bulk fields can symbolically we written as
  \be
  \vPhi_\alpha \, \vPhi_\beta = \sum_{\gamma} C_{\alpha\beta}^{~\;\gamma} \,
  \vPhi_\gamma \,.
  \ee
The OPE coefficients $C_{\alpha\beta}^{~~\gamma}$ must solve a factorisation
constraint coming from the four-point correlator on the sphere 
\cite{sono2,lewe3,dofa},
  \be
  C_{\alpha\beta}^{~\;\varphi}\, C_{\varphi\gamma}^{~\;\delta} 
  = \sum_{\eps}
  C_{\alpha\eps}^{~\;\delta} \, C_{\beta\gamma}^{~\;\eps} \,
  \Fs{\alpha_l}{\beta_l}{\gamma_l}{\delta_l}{\eps_l}{\varphi_l} \,
  \Fs{\alpha_r}{\beta_r}{\gamma_r}{\delta_r}{\eps_r}{\varphi_r} \,,
  \labl{eq:4bulk-fact}
where again all multiplicity indices and summations are omitted. If we identify
  \be
  C_{\alpha\beta}^{~~\gamma} =
  \FF[A|A]^{(\alpha_l \beta_l 0 \beta_r \alpha_r) 0}_{\cdot 0 \cdot , 
  \cdot \gamma_l \gamma_r \cdot \cdot } \,,
  \labl{eq:bulk-OPE-FAA}
then \erf{eq:4bulk-fact} is equivalent to the pentagon relation 
\erf{eq:FAA-pentagon} for the $\FF[A|A]$-matrices. Recall that
the bimodule index $0$ refers to the algebra $A$ itself (we take
$A$ to be simple).
\\
In the TFT approach, the factorisation constraints hold by construction,
and indeed the ribbon invariant \erf{eq:3bulk-rib} for the
correlator of three bulk fields on the sphere (to be constructed
in section \ref{sec:rc-3bulk} below) leads to the coefficient
\erf{eq:3bulk-coeff}, which implies \erf{eq:bulk-OPE-FAA}. Note that
in \erf{eq:3bulk-coeff} two factors $\FF[A|A]$ appear because this
equation describes the coefficient for a three-point correlator, which
is a product of two OPE coefficients.
\\
Let us now turn to defect fields. Denote by $\Thetha^{\mu\nu}_\alpha$
a defect field which changes a defect described by a bimodule $X_\nu$
into a defect described by $X_\mu$. 
One can think of a bulk field $\vPhi_\alpha$ as a special defect field
$\Thetha^{\mu\nu}_\alpha$ which changes the invisible defect 
(labelled by $A$ itself)
to the invisible defect, i.e.\ with $\mu$ and $\nu$ set to zero. 
One can convince oneself that the OPE-coefficients of defect fields
have to fulfill a factorisation condition similar to 
\erf{eq:4bulk-fact} and accordingly we would expect that these OPE-coefficients
are also given by matrix elements of $\FF[A|A]$. As before, this does indeed
result from the TFT-computation, see \erf{eq:3def-basis} below.
Thus, just like is the case of boundary fields, the OPE coefficients of
defect fields can be seen as the coefficients of a suitable associator
morphism.
\\[5pt]
(ii)~It is also worth recalling a well-known geometric interpretation of the 
pentagon identity for the $\FF$-matrices (see e.g.\ \cite{dujn}).
This uses a description of the 6j-symbols $\FF$ of a semisimple tensor category
as tetrahedra, with edges labelled by simple objects $U_i$ and faces by 
morphisms $\Hom(U_i\oti U_j,U_k)$, i.e.\ schematically
  \bea \begin{picture}(140,40)(0,18)
  \put(40,0)     {\includeourbeautifulpicture 41 }
  \put(0,23)     {$ \FF \;= $}
  \epicture06 \labp 41
In this description, the pentagon identity corresponds to the equality between
two ways to obtain the body of a `double tetrahedron': first, by gluing two 
tetrahedra along a common face, and second, by gluing three tetrahedra along 
an edge common to all three of them (namely the edge that, in the first 
description, connects the two vertices not belonging to the distinguished face)
and along three faces each of which is common to two of the tetrahedra. In the 
resulting equation, the labels of the common faces and the internal edge are 
to be summed over. 

To obtain an analogous interpretation of the pentagon relation
\erf{eq:FAA-pentagon} for bimodules, one can represent the $\FF[A|A]$ as prisms
  \bea \begin{picture}(200,41)(0,21)
  \put(70,0)     {\includeourbeautifulpicture 42 }
  \put(0,24)     {$ \FF[A|A] \;= $}
  \epicture11 \labp 42
in which the edges of the two triangular faces are labelled by simple
objects of \calc\ and the remaining edges by simple $A$-bimodules.
The pentagon \erf{eq:FAA-pentagon} then describes again the equality between
two ways to obtain a certain geometric body, this time a `double prism':
on one side (two $\FF[A|A]$s) two prisms are glued along a common
quadrangle; on the other side, they are instead glued along a quadrangle of 
which two edges are those edges of the original outer quadrangles that were 
not involved in the previous gluing; at the remaining two edges the quadrangle 
must touch triangles, and this is achieved by gluing a tetrahedron to each of 
the original pairs of triangles. Again, the labels of common faces and internal 
edges are to be summed over. Pictorially,
  \bea \begin{picture}(335,50)(0,46)
  \put(0,0)      {\Includeourbeautifulpicture 44a }
  \put(194,0)    {\Includeourbeautifulpicture 44b }
  \put(151,48)   {$ = $}
  \epicture19 \labp 44

\bigskip{}

\subsection{Expressions in a basis} \label{sec:mb-basis}

As already pointed out, the fusing matrices $\GG[A]$ and $\FF[A|A]$
are completely determined by the modular tensor category $\calc$, the
symmetric special Frobenius algebra $A$, and a choice of bases in the 
relevant morphism spaces. While the explicit expressions
are not very illuminating, we still quickly go through the calculations,
because later on we will need some of the quantities entering these computations.

We can describe simple subobjects of an object $V$ of $\calc$ by
bases $\{b_{(i,\alpha)}^V\}$ of the morphism spaces $\Hom(U_i,V)$
and the dual bases $\{b^{(i,\alpha)}_V\}$ of $\Hom(V,U_i)$, satisfying 
$b_{V}^{(j,\beta)} \cir b^V_{(i,\alpha)} \eq\delta_i^j\, \delta_\alpha^\beta 
\,\id_{U_i}^{}$ and the completeness property $\sum_{i\in\II}\sum_\alpha 
b^V_{(i,\alpha)} \cir b_{V}^{(i,\alpha)}\eq\id_V$. Their graphical notation is
  \bea \begin{picture}(180,34)(0,21)
  \put(45,0) {\Includeourbeautifulpicture 77a }
  \put(-5,20)     {$ b^V_{(i,\alpha)}=: $}
  \put(46.1,-7.8) {\scriptsize$ U_i $}
  \put(46.1,46.9) {\scriptsize$ V $}
  \put(53.1,21)   {\tiny$ \alpha $}
  \put(173,0){\Includeourbeautifulpicture 77b }
  \put(77,20)     {$ , $}
  \put(122,20)    {$ b_V^{(j,\beta)}=: $}
  \put(172.9,-8.4){\scriptsize$ V $}
  \put(174.2,48.4){\scriptsize$ U_j $}
  \put(181.3,21.2){\tiny$\bar\beta $}
  \epicture11 \labl{pic77} \xlabp 77
For modules, this amounts to the choice of morphisms introduced in (I:4.21).
For bimodules we take analogously
$b_{(i,\alpha)}^X\iN\Hom(U_i,\dot X)$ and $b^{(i,\alpha)}_X\iN\Hom(\dot X,U_i)$.
In terms of these embedding and restriction morphisms we can now expand the 
morphisms $\psi_{(\mu k)\nu}^{\alpha}$ and $\xi_{(i\mu j)\nu}^{\alpha}$ 
described before, as well as their duals. We choose the conventions
  \bea \begin{picture}(360,169)(0,46)
                 \put(0,115){\begin{picture}(0,0)
  \put(60,0)     {\Includeourbeautifulpicture 78a }
  \put(312,0)    {\Includeourbeautifulpicture 78b }
  \put(0,42)     {$ \psi^\alpha_{(\mu k)\nu} \;\equiv $}
  \put(105,42)   {$\dsty := \sum_{m,n\in\II}
                 \sum_{\gamma_1=1}^{\langle U_m, \dot M_\mu\rangle}
                 \sum_{\gamma_2=1}^{\langle U_n, \dot M_\nu\rangle}
                 \sum_{\delta=1}^{\N mkn}\,
                 [\psi^\alpha_{(\mu k)\nu}]^{mn}_{\gamma_1 \gamma_2 \delta} $}
  \put(58,-7)    {\scriptsize$ M_\mu $}
  \put(58,91)    {\scriptsize$ M_\nu $}
  \put(81,-7)    {\scriptsize$ U_k $}
  \put(73,44.5)  {\scriptsize$ \alpha $} 
  \put(310,-7)   {\scriptsize$ M_\mu $}
  \put(310,91)   {\scriptsize$ M_\nu $}
  \put(304,20)   {\scriptsize$ \overline\gamma_1 $}
  \put(304,67)   {\scriptsize$ \gamma_2 $}
  \put(331,-7)   {\scriptsize$ U_k $}
  \put(328,45)   {\scriptsize$ \delta $}
  \put(307,30)   {\scriptsize$ m $}
  \put(307,56)   {\scriptsize$ n $} 
                 \end{picture}}
                 \put(0,0){\begin{picture}(0,0)
  \put(60,0)     {\Includeourbeautifulpicture 79a }
  \put(312,0)    {\Includeourbeautifulpicture 79b }
  \put(0,42)     {$ \bar\psi_\alpha^{(\mu k)\nu} \;\equiv $}
  \put(105,42)   {$\dsty := \sum_{m,n\in\II} 
                 \sum_{\gamma_1=1}^{\langle U_m, \dot M_\mu\rangle}
                 \sum_{\gamma_2=1}^{\langle U_n, \dot M_\nu\rangle}
                 \sum_{\delta=1}^{\N mkn}\,
                 [\bar\psi_\alpha^{(\mu k)\nu}]^{mn}_{\gamma_1 \gamma_2 \delta} $}
                 \end{picture}}
  \put(58,91)    {\scriptsize$ M_\mu $}
  \put(58,-7)    {\scriptsize$ M_\nu $}
  \put(80,91)    {\scriptsize$ U_k $}
  \put(73,41)    {\scriptsize$ \overline{\alpha} $}
  \put(310,91)   {\scriptsize$ M_\mu $}
  \put(310,-7)   {\scriptsize$ M_\nu $}
  \put(304,67)   {\scriptsize$ \gamma_1 $}
  \put(304,20)   {\scriptsize$ \overline\gamma_2 $}
  \put(331,91)   {\scriptsize$ U_k $}
  \put(324.5,49.5) {\scriptsize$ \overline\delta $}
  \put(308.7,28) {\scriptsize$ n $}
  \put(307,55)   {\scriptsize$ m $} 
  \epicture33 \labl{eq:psi-expansion} \xlabp 78 \xlabp 79
for the module basis; the expansion coefficients
$[\psi^\alpha_{(\mu k)\nu}]^{mn}_{\gamma_1 \gamma_2 \delta}$ are
complex numbers. Analogously, for bimodules we choose
  \bea \begin{picture}(370,246)(0,39)
                 \put(0,155){\begin{picture}(0,0)
  \put(64,0)     {\Includeourbeautifulpicture 80a }
  \put(300,0)    {\Includeourbeautifulpicture 80b }
  \put(0,59)     {$ \xi^\alpha_{(i\mu j)\nu} \;\equiv $}
  \put(127,59)   {$\dsty := \sum_{l,m,n\in\II}\,
                 \sum_{\gamma_1,\gamma_2,\delta_1,\delta_2}
                 [\xi^\alpha_{(i\mu j)\nu}]^{lmn}_{\gamma_1 \gamma_2 \delta_1 \delta_2} $}
  \put(32,33)  {} 
  \put(81,-7)    {\scriptsize$ X_\mu $}
  \put(82,122)   {\scriptsize$ X_\nu $}
  \put(61,-7)    {\scriptsize$ U_i $}
  \put(103,-7)   {\scriptsize$ U_j $}
  \put(95,59)    {\scriptsize$ \alpha $}
  \put(315,-7)   {\scriptsize$ X_\mu $}
  \put(315,122)  {\scriptsize$ X_\nu $}
  \put(307,20)   {\scriptsize$ \overline\gamma_1 $}
  \put(307,95)   {\scriptsize$ \gamma_2 $}
  \put(330,47)   {\scriptsize$ \delta_1 $}
  \put(300,77)   {\scriptsize$ \delta_2 $}
  \put(321,84)   {\scriptsize$ n $}
  \put(321,63)   {\scriptsize$ m $}
  \put(314.5,32) {\scriptsize$ l $}
  \put(297,-7)   {\scriptsize$ U_i $}
  \put(334,-7)   {\scriptsize$ U_j $}
                 \end{picture}}
                 \put(0,0){\begin{picture}(0,0)
  \put(64,0)     {\Includeourbeautifulpicture 81a }
  \put(300,0)    {\Includeourbeautifulpicture 81b }
  \put(0,59)     {$ \bar\xi_\alpha^{(i\mu j)\nu} \;\equiv $}
  \put(127,59)   {$\dsty := \sum_{l,m,n\in\II}\,
                 \sum_{\gamma_1,\gamma_2,\delta_1,\delta_2}
                 [\bar\xi_\alpha^{(i\mu j)\nu}]^{lmn}_{\gamma_1 \gamma_2 \delta_1 \delta_2} $}
                 \end{picture}}
  \put(82,122.8) {\scriptsize$ X_\mu $}
  \put(80,-7)    {\scriptsize$ X_\nu $}
  \put(61,122.8) {\scriptsize$ U_i $}
  \put(103,122.8){\scriptsize$ U_j $}
  \put(95,57)    {\scriptsize$ \overline{\alpha} $}
  \put(315,122)  {\scriptsize$ X_\mu $}
  \put(313.5,-7) {\scriptsize$ X_\nu $}
  \put(307,20)   {\scriptsize$ \overline\gamma_2 $}
  \put(307,95)   {\scriptsize$ \gamma_1 $}
  \put(307,47)   {\scriptsize$ \overline{\delta_2} $}
  \put(325,79)   {\scriptsize$ \overline{\delta_1} $}
  \put(313,84)   {\scriptsize$ l $}
  \put(321,54)   {\scriptsize$ m $}
  \put(312,28)   {\scriptsize$ n $}
  \put(297,122)  {\scriptsize$ U_i $}
  \put(334,122)  {\scriptsize$ U_j $}
  \epicture33 \labl{eq:xi-expansion} \xlabp 80 \xlabp 81
Substituting the expansion \erf{eq:psi-expansion}
into the ribbon invariant \erf{eq:GA-rib-inv} for $\GG[A]$ and
taking the trace on both sides of \erf{eq:GA-rib-inv} immediately leads to
  \be
  \GG[A]^{(\mu i j)\, \nu}_{\alpha\kappa\beta , \gamma k \delta}
  = \sum_{m,n,r}
  \Frac{\dim(U_n)}{\dim(\dot M_\nu)}
  \sum_{\stackrel{\gamma_1, \gamma_2, \gamma_3, \gamma_4,}{\eps_1,\eps_2,\eps_3}}
  [\psi^\alpha_{(\mu i)\kappa}]^{mr}_{\gamma_2 \gamma_3 \eps_2}
  [\psi^\beta_{(\kappa j)\nu}]^{rn}_{\gamma_3 \gamma_1 \eps_3}
  [\bar\psi_\gamma^{(\mu k)\nu}]^{mn}_{\gamma_2 \gamma_1 \eps_1}
  \GG^{(mij)\,n}_{\eps_2 r \eps_3 , \eps_1 k \delta} \,.
  \labl{eq:GA-basis}
For the fusion matrices of the bimodule category one must substitute 
\erf{eq:xi-expansion} into \erf{eq:FAA-rib-inv}. 
Using also the braiding relation (I:2.41), this results in 
  \bea \begin{picture}(420,191)(0,23)
  \put(305,-2)   {\includeourbeautifulpicture 82 }
  \put(-14,130)  {$\dsty \FF[A|A]^{(ij\mu kl)\,\nu}_{\alpha\pi\beta,
                 \gamma q_1 q_2 \delta_1 \delta_2} 
                 \,=\, \Frac{1}{\dim(\dot X_\nu^{\phantom|})} 
                 \sum_{x_1,\dots,x_6} \sum_{\varphi_1,\varphi_2,\varphi_3}
                 \sum_{\eps_1,\dots,\eps_6} $}
  \put(17,97)    {$\dsty [\xi^\alpha_{(j\mu k)\pi}]^{x_5 x_4 x_3}
                 _{\varphi_3\varphi_2\eps_4\eps_3}\,
                 [\xi^\beta_{(i\pi l)\nu}]^{x_3 x_2 x_1}
                 _{\varphi_2\varphi_1\eps_2\eps_1}\,
                 [\bar\xi_\gamma^{(q_1 \mu q_2)\nu}]^{x_5 x_6 x_1}
                 _{\varphi_3\varphi_1\eps_5\eps_6}\!
                 \sum_{\rho_1} \RR^{-(lk)\,q_2}_{\rho_1 \delta_2} $}
  \put(323.5,210){\scriptsize$ x_1 $}
  \put(334.5,202){\scriptsize$ \epsilon_1 $}
  \put(332,183)  {\scriptsize$ x_2 $}
  \put(353,172)  {\scriptsize$ \epsilon_2 $}
  \put(300,160)  {\scriptsize$ i $}
  \put(312,118.5){\scriptsize$ \delta_1 $}
  \put(316,70)   {\scriptsize$ q_1 $}
  \put(329,25.5) {\scriptsize$ \epsilon_6 $}
  \put(323.5,9)  {\scriptsize$ x_1 $}
  \put(343,33)   {\scriptsize$ x_6 $}
  \put(348,56)   {\scriptsize$ \epsilon_5 $}
  \put(366,88.5) {\scriptsize$ \rho_1 $}
  \put(362,64)   {\scriptsize$ q_2 $}
  \put(381,120)  {\scriptsize$ l $}
  \put(343,80)   {\scriptsize$ x_5 $}
  \put(355,112)  {\scriptsize$ \epsilon_4 $}
  \put(362,98)   {\scriptsize$ k $}
  \put(336,145)  {\scriptsize$ \epsilon_3 $}
  \put(318,130)  {\scriptsize$ j $}
   \put(332,160) {\scriptsize$ x_3 $}
  \put(344,130)  {\scriptsize$ x_4 $} 
  \epicture13 \labl{eq:FAA-calc-rib} \xlabp 82
Applying the fusing relation (I:2.40) first to the vertex labelled $\epsilon_3$
and then to the vertex labelled $\epsilon_4$ finally yields
  \be  \bearll \displaystyle
  \FF[A|A]^{(ij\mu kl)\,\nu}_{\alpha\pi\beta, \gamma q_1 q_2 \delta_1 \delta_2}
  = \!\!& \displaystyle
  \sum_{n_1,\dots,n_6} 
  \Frac{\dim(U_{n_1})}{\dim(\dot X_\nu)} \sum_{\varphi_1,\varphi_2,\varphi_3}
  \sum_{\eps_1,\dots,\eps_6}
  \\{}\\[-.6em]& \displaystyle \hspace*{2.5em}
  [\xi^\alpha_{(j\mu k)\pi}]^{n_5 n_4 n_3}_{\varphi_3\varphi_2\eps_4\eps_3}
  [\xi^\beta_{(i\pi l)\nu}]^{n_3 n_2 n_1}_{\varphi_2\varphi_1\eps_2\eps_1}
  [\bar\xi_\gamma^{(q_1 \mu q_2)\nu}]^{
     n_5 n_6 n_1}_{\varphi_3\varphi_1\eps_5\eps_6}
  \\[10pt]& \displaystyle \hspace*{2.5em}
  \sum_{\rho_1,\rho_2,\rho_3} \RR^{-(lk)\,q_2}_{\rho_1 \delta_2} \,
  \GG^{(j n_4 l)\, n_2}_{\eps_3 n_3 \eps_2\,,\,\rho_2 n_6 \rho_3}
  \GG^{(n_5 k l)\,n_6}_{\eps_4 n_4 \rho_3 \,,\,\eps_5 q_2 \rho_1}
  \FF^{(ij n_6)\,n_1}_{\eps_1 n_2 \rho_2\,,\,\delta_1 q_1 \eps_6} \,.
  \eear\labl{eq:FAA-basis}
In writing the expansions \erf{eq:psi-expansion} and \erf{eq:xi-expansion}, as 
well as the expressions \erf{eq:GA-basis} and \erf{eq:FAA-basis} for the fusing 
matrices, we have included all multiplicity indices. In the Cardy case 
$A\eq\one$ treated below, as well as later on in section \ref{sec:rc-Cardy}, 
we make the simplifying assumption that the fusion rules in $\calc$ obey 
$\N ijk \iN \{0,1\}$ so as to somewhat reduce the notational burden.

\subsection{Example: The Cardy case}

Let us now work out the fusion matrices for the module and bimodule categories
in the case $A\eq\one$. This choice for the algebra $A$ corresponds to having 
the charge conjugation modular invariant as the partition function on the
torus. As already mentioned, for further simplification we assume in the
sequel that the fusion rules of the modular tensor category $\calc$ satisfy 
$\N ijk\iN\{0,1\}$; this allows us to suppress many of the multiplicity labels.

The simple $\one$-modules are just the simple objects of $\calc$,
so that $\calca \,{\cong}\, \calc$ as categories. We choose the 
representatives of simple modules as $M_\mu\eq U_\mu$ with $\mu\iN\II$. 
Similarly, for the bimodules we have $\calcaa \,{\cong}\,\calc$, this time
even as tensor categories, and we choose the representatives of simple bimodules
as $X_\mu\eq U_\mu$, $\mu\iN\II$. Note that we continue to use greek letters to 
label simple modules and bimodules to avoid confusion, even though now the 
three index sets $\II$, $\JJ$ and $\KK$ are identical.

Next we must specify the basis elements $\psi^\alpha_{(\mu k)\nu}$ and 
$\xi^\gamma_{(i\mu j)\nu}$ that we will use. We make the choices
  \bea \begin{picture}(365,72)(0,28)
  \put(80,8)     {\includeourbeautifulpicture 72 }
  \put(300,0)    {\includeourbeautifulpicture 73 }
  \put(0,40)     {$ \psi_{(\mu k)\nu} \,:=\, \beta^{\mu,\nu}_k $}
  \put(150,40)   {and $\qquad \xi^\gamma_{(i \mu j) \nu} \,:=\,
                 t_j^{\,-1}\, {\alpha^{\mu|\nu}_{ij,\gamma}} $}
  \put(77,-1)    {\scriptsize$ U_\mu $}
  \put(98,-1)    {\scriptsize$ U_k $}
  \put(86.5,78)  {\scriptsize$ U_\nu $}
  \put(316,-7)   {\scriptsize$ U_\mu $}
  \put(297,-7)   {\scriptsize$ U_i $}
  \put(335,-7)   {\scriptsize$ U_j $}
  \put(320,43.5) {\scriptsize$ U_\gamma $}
  \put(316,94)   {\scriptsize$ U_\nu $} 
  \epicture15 \labl{eq:cardy-bases} \xlabp 72 \xlabp 73
where the (non-zero) constants $\beta^{\mu,\nu}_k$ and $\alpha^{\mu|\nu}_{ij,
\gamma}$ can be chosen at will; later on they will serve as normalisations of 
the bulk, boundary and defect fields. The phases $t_j$ appearing in 
\erf{eq:cardy-bases} have been defined in (II:3.43). If $\calc$ is the \rep\
category of a conformal vertex algebra we take $t_j\eq\exp(-\pi\ii\Delta_j)$
with $\Delta_j$ the conformal highest weight of the \rep\ labelled by $j$.
The phases $t_j$ are convenient because they will lead to simpler expressions 
for the bulk two-point function on the sphere and for the bulk one-point 
function on the disk.

Because of $\psi_{(\mu k)\nu} \iN \HomA(M_\mu \Oti U_k, M_\nu) \,{\cong}\, 
\Hom(U_\mu\Oti U_k, U_\nu)$, there is no need for a multiplicity label in 
the first formula in \erf{eq:cardy-bases}. In contrast, the dimension of the
morphism space $\HomAA(U_i{\otimes}^+ X_\mu {\otimes}^- U_j,X_\nu)
\eq \Hom(U_i \oti U_\mu \oti U_j, U_\nu)$ can be larger than one, and
accordingly the basis elements $\xi^\gamma_{(i \mu j) \nu}$ are labelled by 
those simple objects $U_\gamma$ that occur in the fusion 
$U_\mu \oti U_j$ and for which $\N i\gamma\nu \,{\ne}\, 0$.
In the notation introduced in \erf{eq:psi-expansion}
and \erf{eq:xi-expansion} the basis choice \erf{eq:cardy-bases} reads
  \be
  [\psi_{(\mu k)\nu}]^{mn}_{\cdot\cdot\cdot}
  = \beta^{\mu,\nu}_k \, \delta_{m\mu} \,\delta_{n\nu}
  \qquad{\rm and}\qquad
  [\xi^\gamma_{(i\mu j)\nu}]^{xyz}_{\cdot\cdot\cdot\cdot}
  = t_j^{\,-1} \, \alpha^{\mu|\nu}_{ij,\gamma}\, \delta_{\mu x}\, \delta_{\gamma y}\,
  \delta_{\nu z} \,.
  \labl{eq:cardy-bas-rel}
The dot `$\,\cdot\,$' stands for a multiplicity label that can take only a
single value, the corresponding morphism space being one-dimensional.

We also need the duals of the bases \erf{eq:cardy-bases}. One
easily checks that these are given by
  \bea \begin{picture}(350,72)(0,28)
  \put(83,8)     {\includeourbeautifulpicture 74 }
  \put(280,0)    {\includeourbeautifulpicture 75 }
  \put(0,40)     {$\dsty \bar\psi^{(\mu k)\nu}\,=\,\frac{1}{\beta^{\mu,\nu}_k} $}
  \put(150,40)   {and $\qquad\dsty \bar\xi_\gamma^{(i \mu j) \nu} \,=\, 
                 \frac{t_j}{\alpha^{\mu|\nu}_{ij,\gamma}} $}
  \put(79,78.5)  {\scriptsize$ U_\mu $}
  \put(100,78.5) {\scriptsize$ U_k $}
  \put(90.2,0)   {\scriptsize$ U_\nu $} 
  \put(296,93)   {\scriptsize$ U_\mu $}
  \put(277,93)   {\scriptsize$ U_i $}
  \put(314,93)   {\scriptsize$ U_j $}
  \put(300.8,40) {\scriptsize$ U_\gamma $} 
  \put(295,-8)   {\scriptsize$ U_\nu $} 
  \epicture15 \labp 74 \xlabp 75 
Substituting the choice of basis \erf{eq:cardy-bases} into 
\erf{eq:GA-rib-inv} immediately yields 
  \be
  \GG[\one]^{(\mu i j)\,\nu}_{\kappa \,,\, k} =
  \frac{\beta^{\kappa,\nu}_j \beta^{\mu,\kappa}_i}{\beta^{\mu,\nu}_k}
  \,\GG^{(\mu i j)\,\nu}_{\kappa\,k} 
  \labl{eq:GA-bas-Cardy}
for the fusion matrices of $\calca$. 

For the fusion matrices of the 
bimodule category we deduce from \erf{eq:FAA-rib-inv} that
  \bea \begin{picture}(330,178)(0,50)
  \put(75,0)     {\Includeourbeautifulpicture 76a }
  \put(239,0)    {\Includeourbeautifulpicture 76b }
  \put(0,106)    {$ \FF[\one|\one]^{(ij\mu kl)\,\nu}_{\alpha\pi\beta,\gamma pq} $}
  \put(99,106)   {$\dsty =\ \frac{
                   \alpha^{\pi|\nu}_{il,\beta}\, \alpha^{\mu|\pi}_{jk,\alpha}}
                   {\alpha^{\mu|\nu}_{pq,\gamma}}\,
                   \Rs{-}lk{\,q}\,\frac{t_q}{t_l\,t_k} $}
  \put(72,-7.7)  {\scriptsize$ U_\nu $}
  \put(72,224)   {\scriptsize$ U_\nu $}
  \put(264,224)  {\scriptsize$ U_\nu $}
  \put(264,-7.7) {\scriptsize$ U_\nu $}
  \put(278,35)   {\scriptsize$ \gamma $}
  \put(297,65)   {\scriptsize$ q $}
  \put(251,60)   {\scriptsize$ p $}
  \put(269,80)   {\scriptsize$ \mu $}
  \put(235,160)  {\scriptsize$ i $}
  \put(270,185)  {\scriptsize$ \beta $}
  \put(270,160)  {\scriptsize$ \pi $}
  \put(253,130)  {\scriptsize$ j $}
  \put(316,130)  {\scriptsize$ l $}
  \put(279,130)  {\scriptsize$ \alpha $}
  \put(297,98)   {\scriptsize$ k $} 
  \epicture36 \labp 76
Similarly to the calculation in \erf{eq:FAA-calc-rib}, applying the fusion 
relation (I:2.40) first to the vertex for $\Hom(U_j\oti U_\alpha,U_\pi)$ and
then to the one for $\Hom(U_\mu\oti U_k,U_\alpha)$ results in the expression
  \be
  \FF[\one|\one]^{(ij\mu kl)\, \nu}_{\alpha \pi \beta, \gamma p q} \,=\, 
  \frac{ \alpha^{\pi|\nu}_{il,\beta} \, \alpha^{\mu|\pi}_{jk,\alpha}}
  {\alpha^{\mu|\nu}_{pq,\gamma}} \,
  \Rs{-}lk{\,q} \frac{t_q}{t_l\,t_k} \, \Gs j\alpha l\beta\pi\gamma \,
  \Gs \mu kl\gamma \alpha q \, \Fs ij\gamma\nu\beta p .
  \labl{eq:FAA-bas-Cardy}


\sect{Ribbon graph representation of field insertions}\label{sec:fi}

In this section we present the ribbon graph representation
for field insertions on boundaries, in the bulk and on defect lines.
This will be used in section \ref{sec:rc} to express structure 
constants as invariants of ribbon graphs in three-manifolds.

\medskip

The ribbon graph representations will be given explicitly
only for the case of unoriented world sheets. Recall from \cite{fuRs8} that
in this case the relevant algebraic structure to describe
the full CFT is a Jandl algebra. 

It is then straightforward to obtain the oriented case as well. Of course, 
given an oriented world sheet, one can just forget the orientation to obtain 
an unoriented world sheet, for which one can apply the construction below. 
However, this is not the correct procedure to use, since in the two cases 
different algebraic structures are relevant:

\begin{center}\begin{tabular}{lcl}
  oriented world sheets & $\rightarrow$ & 
    symmetric special Frobenius algebra \\[3pt]
  unoriented world sheets  & $\rightarrow$ &  Jandl algebra
\end{tabular}\end{center}

\noindent
Thus for oriented world sheets one must formulate 
the construction of the ribbon graphs in a way that is applicable for 
{\em any\/} symmetric special Frobenius algebra, not just for those
admitting a reversion. This can be done as follows.

First, as already emphasised, instead of a Jandl algebra, just a symmetric 
special Frobenius  algebra is to be used.  Second, whenever in the unoriented 
case an orientation $\orr_1$ of a boundary component, respectively $\orr_2$ 
of a surface, enters the construction, then in the corresponding oriented
case a canonical choice is provided by the orientation of the world sheet.
In the un\-ori\-en\-ted case there are equivalence relations
linking the two possible choices for $\orr_{1,2}$, and this
is in fact the only place where the reversion of the Jandl algebra 
enters. In the oriented case these equivalence relations
are not needed and hence only the symmetric special Frobenius
structure is used.

For example, recall from section II:3 that in the unoriented case
boundary conditions are labelled by equivalence classes
$[M,\orr_1]$, where $M$ is a left $A$-module and $\orr_1$ is
an orientation of the boundary component. The equivalence relation (II:3.1) 
is that $(M,\orr_1) \,{\sim}\, (M',\orr_1')$
if either $\orr_1'\eq\orr_1$ and $M'\,{\cong}\, M$
or if $\orr_1'\eq{-}\orr_1$ and $M'\,{\cong}\, M^\sigma$, where
$M^\sigma$ is the module conjugate to $M$ (see section II:2.3).
To obtain the prescription in the oriented case, we use the
orientation $\orr_{\partial \X}$ of the boundary that is induced by the world
sheet orientation $\orr_\X$ to select the pair $(M,\orr_{\partial \X})$. 
The second alternative in the equivalence relation is then obsolete.

\subsection{A note on conventions}\label{sec:fi-conv}

The construction of the ribbon graph described in this section departs slightly
from the one used in \cite{fuRs4} and \cite{fuRs8}, resulting from a number 
of choices of conventions that have to be made. However, in the absence of 
field insertions the final result, i.e.\ the ribbon 
graph embedded in the connecting manifold, is still the
same as in \cite{fuRs4,fuRs8}. To see this let us summarise the conventions
we choose and point out how they differ from those in \cite{fuRs4,fuRs8}.

\subsubsection*{Category of three-dimensional cobordisms}

Recall from the summary in sections I:2.3 and I:2.4 that a three-dimensional
topological field theory furnishes a functor $(Z,\calh)$ from the cobordism 
category $\coborc$ to the category of finite-dimensional
complex vector spaces; such a functor can be constructed from any modular 
tensor category $\calc$. It turns out to be convenient to give a
definition of the objects and morphisms of $\coborc$ slightly
different from the one used in section I:2.4.
  
Objects of $\cobord$ are extended surfaces and morphisms 
of $\cobord$ are cobordisms between extended surfaces.
An {\em extended surface\/} $E$ consists of the following data:
\ITemize
\item[\Nxt] 
A compact oriented two-dimensional manifold without
boundary, also denoted by $E$.
\item[\Nxt]
A finite set of marked points -- that is, of quadruples
$(p_i,[\gamma_i],V_i,\eps_i)$, where
the $p_i\iN E$ are mutually distinct points of the surface $E$
and $[\gamma_i]$ is a germ of arcs\,%
  \footnote{~By a {\em germ of arcs\/} we mean an equivalence class $[\gamma]$
  of continuous embeddings $\gamma$ of intervals 
  $[-\delta,\delta] \,{\subset}\, \reals$ into the extended surface $E$. 
  Two embeddings $\gamma{:}~[-\delta,\delta] \To E$ and
  $\gamma'{:}~[-\delta',\delta'] \To E$ are equivalent if there is a 
  positive $\eps\,{<}\,\delta,\delta'$ such that $\gamma$ and $\gamma'$
  are equal when restricted to the interval $[-\eps,\eps]$.
  }
$\gamma_i{:}\; [-\delta,\delta] \To E$ with $\gamma_i(0)\eq p_i$.
Furthermore, $V_i \iN \obj(\calc)$, and $\eps_i\iN\{\pm 1\}$ is a sign.
\item[\Nxt]
A Lagrangian subspace $\lambda \,{\subset}\, H_1(E,\reals)$.
\end{itemize}
A morphism $E \To E'$ is a {\em cobordism\/} $\M$, consisting of
the following data:
\begin{itemize}
\item[\Nxt]
A compact oriented three-manifold, also denoted by $\M$, such that
$\partial \M \eq(-E){\sqcup}E'$.
\\
Here $-E$ is obtained from $E$ by reversing its $2$-orientation
and replacing any marked point $(p,[\gamma],U,\eps)$ by
$(p,[\tilde\gamma],U,-\eps)$ with $\tilde\gamma(t)\eq\gamma(-t)$,
see section I:2.4. The boundary $\partial \M$ of a cobordism is 
oriented according to the inward pointing normal.
\item[\Nxt]
A ribbon graph $R$ inside $\M$ such that for each marked point
$(p,[\gamma],U,\eps)$ of $(-E){\sqcup}E'$ there is a ribbon ending on
$(-E){\sqcup}E'$ in the following way.
Let $\varphi{:}\; [0,1]\Times[-\frac1{10},\frac1{10}]\To \M$ be the
parametrisation\,\footnote{%
  ~By a {\em parametrisation\/} of a ribbon $S$ we mean the following.
  $S$ has a $2$-orientation of its surface and a $1$-orientation of its core.
  The rectangle $[0,1]\Times[-\frac1{10},\frac1{10}]$ carries a natural
  $2$-orientation as a subset of $\reals^2$, and the interval
  $[0,1]\Times\{0\}$ carries a natural $1$-orientation as a subset
  of the $x$-axis. A parametrisation $\varphi$ of $S$ is
  required to preserve both of these orientations.
  }
of a ribbon $S$ embedded in $\M$ and labelled by $U\iN\objc$. If $\eps\eq{+}1$,
then the core of $S$ must point away from $\partial \M$ and the end of $S$
must induce the germ $[\gamma]$, i.e.\ $\varphi(0,0)\eq p$ and
$[\varphi(0,t)]\eq [\gamma(t)]$. If $\eps\eq{-}1$, then the core of $S$
must point towards $\partial \M$, and $\varphi(1,0)\eq p$ as well as
$[\varphi(1,-t)] \eq [\gamma(t)]$, the presence of the minus sign implying
that the orientation of the arc is reversed.
\end{itemize}
The difference to the definition used in section I:2.4 is that we do not
take a surface with embedded arcs, but rather with germs of arcs.\,%
  \footnote{~This convention should be compared to those of \cite{TUra}, which
  uses the arcs themselves, and of \cite{BAki}, which uses tangential vectors
  at the insertion points $p$.
  The present convention is a compromise between the two. Only the
  behaviour in an arbitrarily small neighbourhood of $p$ is relevant, but it
  does not need a differential structure on $E$.
  }
The reason for using this definition of extended surface will become clear in
section \ref{sec:cb-link} where we assign a germ of arcs to a germ of local
coordinates.

Note also that the allowed ways for a ribbon 
to end on an arc-germ of an extended surface depend on the orientations
of arc, ribbon-surface and ribbon-core, but not on the 
orientation of the three-manifold $\M$ or of the extended surface $E$.
The two allowed possibilities are -- looked at from `above' and `below'
$E$, as well as from `in front of' and `behind' the ribbon -- the following:
  \bea 
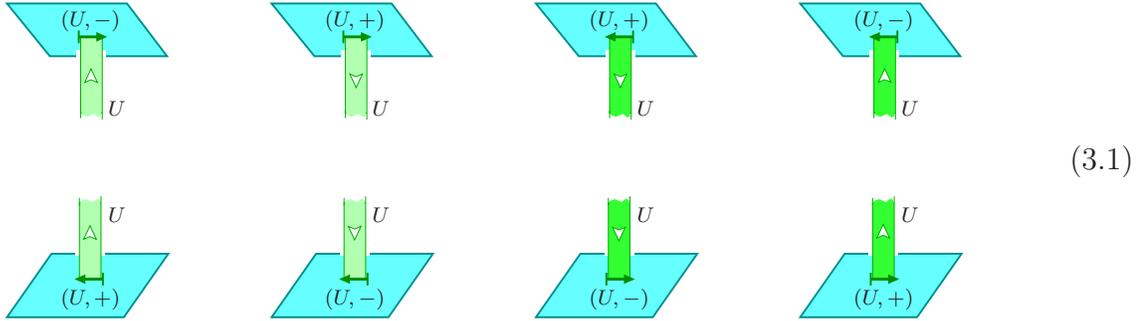
\begin{picture}(370,85)(0,41)
  \put(0,75)    {\Includeourbeautifulpicture 08a }
  \put(100,75)  {\Includeourbeautifulpicture 08b }
  \put(200,75)  {\Includeourbeautifulpicture 08c }
  \put(300,75)  {\Includeourbeautifulpicture 08d } 
  \put(0,0)     {\Includeourbeautifulpicture 08e }
  \put(100,0)   {\Includeourbeautifulpicture 08f }
  \put(200,0)   {\Includeourbeautifulpicture 08g }
  \put(300,0)   {\Includeourbeautifulpicture 08h }
  \put(21,5.6)    {\scriptsize$(U,+) $}
  \put(121,5.6)   {\scriptsize$(U,-) $}
  \put(221,5.6)   {\scriptsize$(U,-) $}
  \put(321,5.6)   {\scriptsize$(U,+) $}
  \put(21,111)    {\scriptsize$(U,-) $}
  \put(121,111)   {\scriptsize$(U,+) $}
  \put(221,111)   {\scriptsize$(U,+) $}
  \put(321,111)   {\scriptsize$(U,-) $}
  \put(39,37)     {\scriptsize$ U $}
  \put(139,37)    {\scriptsize$ U $}
  \put(239,37)    {\scriptsize$ U $}
  \put(339,37)    {\scriptsize$ U $} 
  \put(39,77)     {\scriptsize$ U $}
  \put(139,77)    {\scriptsize$ U $}
  \put(239,77)    {\scriptsize$ U $}
  \put(339,77)    {\scriptsize$ U $}
  \epicture24 \labl{eq:ribbon-arc-end} \xlabp 08 
(For instance, the first picture in the first row is the same as the fourth 
picture in the first row, as well as the second and third picture in the 
second row.)

\subsubsection*{Ribbon graph embedded in the connecting manifold} 

In the TFT description of
correlators for oriented world sheets, we have the following conventions:
\begin{itemize} 
\item[\Nxt] 
  The correlator for a world sheet $X$ is described by a cobordism $\emptyset 
  \,{\overset{\MX\,}{\longrightarrow}}\,\hat\X$, where $\hat\X$ is the double of
  $\X$. In particular, $\hat\X$ is the {\em outgoing\/} part of the boundary
  $\partial\MX$.
\item[\Nxt] 
  Field insertions on the world sheet $\X$ lead to marked points on the
  double $\hat\X$. For the ribbon graph inside the cobordism $\emptyset 
  \,{\overset{\MX\,}{\longrightarrow}}\, \hat\X$ we take all ribbons ending on 
  $\partial\MX$ to have cores pointing {\em away\/} from the boundary, 
  i.e.\ $\eps\eq{+}1$ for all marked points on $\hat\X$.
\item[\Nxt] 
The non-negative integers (compare formulas (I:5.30) and (I:5.65), and lemma 
\ref{lem:defect-iso}) 
  \be
  Z(A)_{ij} := \dim\,\HomAA(\alpha^-_\AA(U_j),\alpha^+_\AA(U_{\bar\imath}))
  = \dim\,\HomAA(U_i\oT+A\ot-U_j,A)
  \,, \ee
with $\alpha^\pm_\AA(U)$ denoting $\alpha$-induced\,%
  \footnote{~For the notion of $\alpha$-induction, which is due to \cite{lore},
  see e.g.\ definition 2.21 of \cite{ffrs}. The definition, as well as a list
  of references, is also given in section I:5.4 where, however, the notation
  $\alpha^{(\pm)}$ is used in place of $\alpha^\pm_\AA$. \label{alphafootnote} }
bimodules, count the number of linearly independent primary bulk fields 
$\vPhi_{ij}(z,\bar z)$ that have chiral representation label $i$ and
anti-chiral representation label $j$.
\item[\Nxt] 
Boundary conditions are labelled by left $A$-modules.
\end{itemize}
As a consequence (compare (I:5.30) or \erf{eq:bulk-insert} below),
when constructing the ribbon graph in $\MX$ for an oriented world sheet $\X$,
the ribbons embedded in $\X$ must be inserted in such a manner that their 
`white' side faces the boundary of $\MX$ which has the same orientation as $\X$. 
In particular, the orientation of the ribbon-surfaces is {\em opposite\/} to 
that of the world sheet $\X$.

This clashes with the prescription given in section I:5.1, 
according to which orientation of the world sheet and the ribbons agree. The
reason is that in \cite{fuRs4} implicitly, the boundary of a cobordism was taken to be
oriented according to the {\em outward\/} pointing normal. However due to
the conventions above we should use the {\em inward\/} pointing normal. 

It is not difficult to convince oneself that {\em all pictures and formulas
in\/} \cite{fuRs4} {\em and\/} \cite{fuRs8} {\em remain unchanged\/} provided that 
\begin{itemize}
\item[--] one uses the
{\em inward} pointing normal to orient the boundary of a cobordism,
and that
\item[--] the prescription in sections I:5 and II:3.1 is modified 
such that ribbons embedded in $\X\,{\subset}\, \MX$ have surface orientation
{\em opposite} to the (local) orientation of $\X$, and similarly
their core has {\em opposite} orientation to $\partial \X$ if they
lie on the world sheet boundary.
\end{itemize}
{}From hereon we 
use this modified prescription both in the oriented and in the unoriented case.

\subsection{Boundary fields}\label{sec:fi-bnd}

A connected component $\bndb$ of the boundary $\partial \X$
of the world sheet has the topology of a circle. On $\bndb$
there may be several insertions of boundary fields, and accordingly the
intervals between the field insertions may be labelled by
different boundary conditions. Below we will see that the
configurations of fields and boundary conditions on such a 
boundary component are again labelled
by equivalence classes, and that this labelling generalises
the one for boundaries without field insertions.

In the sequel we say that two curves $\varpi$ and $\mu$ are {\em aligned} 
in a point $p$ of the world sheet iff there exist parametrisations 
$\varpi'(t)$ of $\varpi$ and 
$\mu'(t)$ of $\mu$ such that $\varpi'(0) \eq p \eq \mu'(0)$ and 
$[\varpi'] \eq [\mu']$.

\subsubsection*{Labelling of boundary insertions}

A {\em boundary field} $\hatpsi$ is a collection 
  \be
  \hatpsi = ( M , N , V , \psi , p , [\gamma] )
  \ee
of the following data:
$M, N$ are left $A$-modules, $V$ is an object of $\calc$,
the morphism $\psi$ is an element of $\HomA(M\oti V,N)$,
$p \iN \partial \X$ is a point on the boundary of the world sheet
$\X$, and finally $[\gamma]$ is an arc germ such that $\gamma(0)\eq p$ and
that $\gamma$ is aligned to $\partial \X$ in $p$. 

Denote by $\bndb([\gamma_1],...\,,[\gamma_n])$ a connected
component of $\partial \X$ together with $n$ arc germs 
$[\gamma_k]$ such that $\gamma_k$ is aligned to $\partial \X$ in
$\gamma_k(0)$. The points $\gamma_k(0)$ mark the insertion points of the
boundary fields. A boundary component 
$\bndb([\gamma_1],...\,,[\gamma_n])$ with $n$ boundary field 
insertions is labelled by equivalence classes of tuples
  \be
  (M_1, ...\, , M_{n'} ; \hatpsi_1, ...\, , \hatpsi_n ; \orr_1) 
  \labl{eq:bnd-tupel}
subject to the following conditions.
\\
\nxt $\orr_1$ is an orientation of the boundary circle $\bndb$.
\\
\nxt The points $p_k \eq \gamma_k(0)$ are ordered
such that when passing along the boundary circle $\bndb$ 
opposite to the direction 
given by $\orr_1$, one passes from $p_k$ to $p_{k+1}$. (Here
and below it is understood that $p_0 \eq p_n$ and 
$p_{n+1} \eq p_1$, and analogously for $M_k$ and $\hatpsi_k$.)
\\
\nxt The $M_1,...\,,M_{n'}$ are left $A$-modules. $M_k$ 
labels the boundary condition on the stretch of
boundary between $p_{k-1}$ and $p_k$. Thus if there
are no field insertions, then $n'\eq1$, and otherwise $n'\eq n$.
\\
\nxt The $\hatpsi_1,...\,,\hatpsi_n$ are boundary fields with
defining data
  \be
  \hatpsi_k = ( M_k, M_{k+1} , V_k, \psi_k, p_k, [\gamma_k] ) \,.
  \labl{eq:bnd-psik}

To define the equivalence relation we need 
\\[-2.3em]

\dtl{Definition}{def:psi-sig}
Let $A$ be a Jandl algebra in $\calc$.
Given an object $V$ of $\calc$ and two left $A$-modules $M$ and $N$, 
$s\,{\equiv}\,s_{M,N,V}$ is the map $s{:}\ \Hom(\dot M\oti V, \dot N) \To
\Hom(\dot N^\vee \oti V , \dot M^\vee)$ defined by
  \bea \begin{picture}(150,89)(0,50)
  \put(56,0)     {\includeourbeautifulpicture 07 }
  \put(0,56)     {$ s(\psi) \;:= $}
  \put(87,130)   {\scriptsize$ \dot{M}^\vee $}
  \put(92.5,-9)  {\scriptsize$ \dot{N}^\vee $}
  \put(122.8,-7) {\scriptsize$ V $}
  \put(93.4,72.5)  {\scriptsize$ \psi $}
  \epicture27 \labl{eq:bnd-s-def} \xlabp 07
for $\psi\iN\Hom(\dot M\oti V, \dot N)$.

\medskip

\dtl{Lemma}{lem:psi-s}
Let $A$ and $s$ be as in definition \ref{def:psi-sig}. If
$\psi\iN\HomA(M\oti V,N)$, then $s(\psi) \iN \HomA(N^\sigma\oti V,M^\sigma)$.

\medskip\noindent
Proof:\\
The assertion follows by direct computation from the defining properties
of conjugate left modules, as given in definition II:2.6.
\qed

We can now formulate the equivalence relation on labellings
\erf{eq:bnd-tupel} of a boundary component $\bndb$ with
(or without) field insertions. Consider two tuples
  \be
  (M_1, ...\, , M_{n'} ; \hatpsi_1, ...\, , \hatpsi_n ; \orr_1) 
  \quad\ {\rm and} \quad\
  (M'_1, ...\, , M'_{n'} ; \hatpsi'_1, ...\, , \hatpsi'_n ; \orr'_1) \,,
  \labl{eq:bnd-equiv-def}
where $\hatpsi_k\eq( M_k, M_{k+1} , V_k, \psi_k, p_k, [\gamma_k] )$ and
$\hatpsi_k'\eq( M_k', M_{k+1}' , V'_k, \psi_k', p_k', [\gamma_k'] )$.
We assume that the numbering of boundary insertions is done such that 
either $p_k \eq p_k'$ or $p_k \eq p_{n-k}'$ (otherwise the numbering 
must be shifted appropriately). 

\dtl{Definition}{bnd-equiv-def}
The set of labels for the boundary component $\bndb([\gamma_1],...\,,[\gamma_n])$
is the set
  \be
  \mathcal{B}\big(\bndb([\gamma_1],...\,,[\gamma_n]) \big)
  := \big\{ (M_1, ...\, , M_{n'} ; \hatpsi_1, ...\, , 
  \hatpsi_n ; \orr_1) \big\} /{\sim}
  \ee
of equivalence classes with respect to the following
equivalence relation $\sim$\,: The two tuples \erf{eq:bnd-equiv-def} 
are equivalent iff one of the following two conditions is satisfied:
\\[3pt]
(i) We have $\orr_1'\eq\orr_1$, $V_k'\eq V_k$, $p_k'\eq p_{k}$, 
$[\gamma_k'] \eq [\gamma_{k}]$, and for $k\eq1,...\,,n'$ there exist isomorphisms 
$\varphi_k \iN \HomA(M_k', M_k)$ such that 
  \be
  \psi_k = \varphi_{k+1} \cir \psi'_k \cir (\varphi_k^{-1} \oti \id_{V_k})\,.
  \labl{eq:bnd-equiv-def-i}
(ii) $\orr_1' \eq {-}\orr_1$, $V_k' \eq V_{n-k}$, $p_k' \eq p_{n-k}$, 
$[\gamma_k'] \eq [\gamma_{n-k}]$, and for $k\eq1,...\,,n'$ there exist isomorphisms
$\varphi_k \iN \HomA(M_k', M_{n'-k+1}^\sigma)$ such that 
\be
  s(\psi_{n-k}) = \left\{ \begin{array}{ll}
  \varphi_{k+1} \cir \psi'_k \cir 
    (\varphi_{k}^{-1} \otimes \id_{V_k'})
  ~ & {\rm if~~} \orr_1' \eq \orr(\gamma_k') \,, \\{}\\[-.8em]
  \varphi_{k+1} \cir \psi'_k \cir 
    (\varphi_{k}^{-1} \otimes \theta_{V_k'})
  ~ & {\rm if~~} \orr_1' \eq {-}\orr(\gamma_k') \,,  \eear\right.
  \labl{eq:bnd-equiv-def-ii}
where $\orr(\gamma_k')$ denotes the local orientation of $\bndb$
induced by the arc-germ $[\gamma_k']$.

\subsubsection*{Marked points on the double}

Every boundary field insertion $\hatpsi\eq(M,N,V,\psi,p,[\gamma])$ on $\X$ gives
rise to one marked point on the double $\hat\X$. This marked point is given by
  \be
  (\tilde p, [\tilde \gamma], V, +)
  \ee
with $\tilde p\eq [p,\pm \orr_2(p)]$. Recall from \erf{eq:double-def}
that $\hat\X$ is the orientation bundle over $\X$ with the two
orientations identified over the boundary $\partial \X$. To obtain the germ 
$[\tilde\gamma]$, first choose a representative $\gamma$ of $[\gamma]$. Since 
$\gamma$ is aligned with $\partial\X$ at $p$, by definition there exists a
$\delta\,{>}\,0$ such that $\gamma {:}\; [-\delta,\delta] \To \partial \X$.
A representative $\tilde\gamma{:}\; [-\delta,\delta] \To \hat \X$ of
$[\tilde\gamma]$ is then given by $\tilde\gamma(t)\,{:=}\, [\gamma(t),\pm\orr_2]$.

Given a labelling $B\iN\mathcal{B}\big(\bndb([\gamma_1],...\,,[\gamma_n]) \big)$,
one can verify that the set of marked points on $\hat \X$ obtained in the
way described above is indeed independent of the choice of representative of $B$.

\subsubsection*{Ribbon graph embedded in the connecting manifold}

Note that in a small neighbourhood of a point $p$ on the boundary
of the world sheet $\X$, the connecting manifold $\MX$ from (II:3.8)
looks as follows:
  \bea  
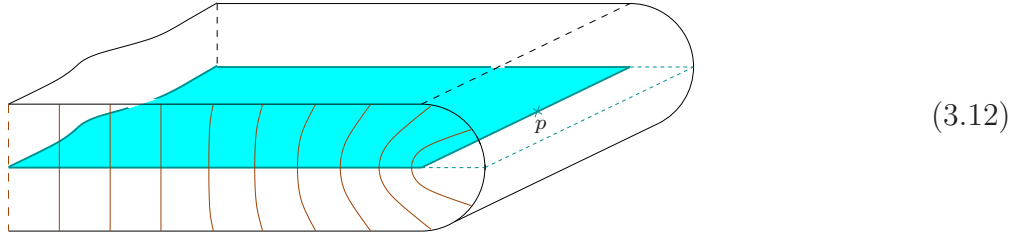
\begin{picture}(280,66)(0,32)
  \put(2,0)         {\includeourbeautifulpicture 09 }
  \put(202,39)         {\scriptsize $p$}
  \epicture20 \labl{eq:con-mf-bnd} \xlabp 09
Indicated are also the intervals $[-1,1]$, which according to the identification
rule of the general prescription degenerate into an interval $[0,1]$ above each
boundary point. The ribbon graph representation of the boundary component
$\bndb([\gamma_1],...\,,[\gamma_n])$ labelled by an equivalence class 
$B \eq [M_1, ...\, , M_{n'} ; \hatpsi_1, ...\, , \hatpsi_n ; \orr_1]$ in 
$\mathcal{B}\big(\bndb([\gamma_1],...\,,[\gamma_n]) \big)$
is now constructed as follows.

\ITemize
\item[\Nxt] 
Choose  a representative 
$(M_1, ...\, , M_{n'} ; \hatpsi_1, ...\, , \hatpsi_n ; \orr_1)$ of 
the equivalence class $B$ -- \underline{Choice \#1}.

\item[\Nxt] 
Choose a dual triangulation of $\X$ such that the insertion points of $\bndb$ 
lie on edges of the triangulation -- \underline{Choice \#2}.

\item[\Nxt] 
The orientation $\orr_1$ of the boundary circle $\bndb$ together with the 
inward pointing normal induces an orientation of its neighbourhood in the 
world sheet $\X$. This fixes a local orientation of $\X$ close to $\bndb$, which
agrees with both the bulk and the boundary orientation of the upper half plane.

\item[\Nxt] At each vertex of the triangulation which lies
on the boundary $\bndb$ place the element\,\footnote{%
  ~Note that the orientations in this picture are opposite to those in
  (II:3.12). This is due to the change in convention
  explained in section \ref{sec:fi-conv}.} 
  \bea  
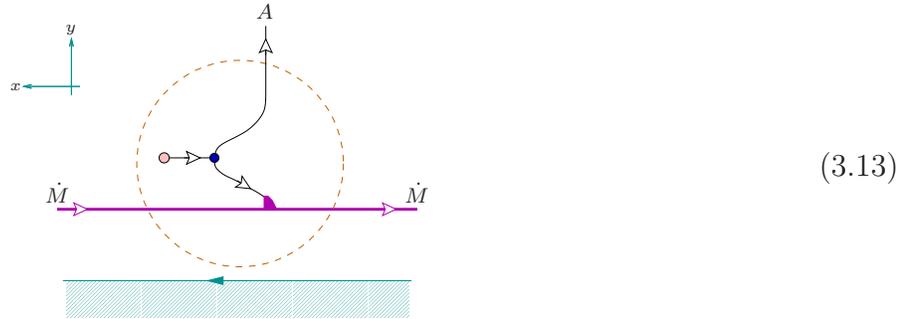
\begin{picture}(195,87)(0,37)
  \put(2,0)         {\includeourbeautifulpicture 43 }
                \put(-1.5,19){
  \put(20,89.5)     {\tiny$y$}
  \put(-1,67.3)     {\tiny$x$}
  \put(12.3,24.8)   {\scriptsize$ \dot M $}
  \put(92,94.3)     {\scriptsize$ A $}
  \put(148.3,24.8)  {\scriptsize$ \dot M $}
                           } 
  \epicture24 \labl{eq:bnd-vertex} \xlabp 43
in such a way that the orientation of bulk and boundary agree with that
around the vertex.

\item[\Nxt] 
At each insertion point $\hatpsi_k \eq (M_k, M_{k+1}, V_k, \psi_k, p_k , 
[\gamma_k])$ place one of the two elements
  \bea \begin{picture}(360,57)(0,45)
  \put(0,0)       {\Includeourbeautifulpicture 10a }
  \put(200,0)     {\Includeourbeautifulpicture 10b }
  \put(37,64)     {\scriptsize$ M_k $}
  \put(129,65)    {\scriptsize$ M_{k+1} $}
  \put(88,54.5)   {\scriptsize$ \psi_k $}
  \put(67,23)     {\scriptsize$ V_k $} 
  \put(244,64)    {\scriptsize$ M_k $}
  \put(328,65)    {\scriptsize$ M_{k+1} $}
  \put(288.1,54.5){\scriptsize$ \psi_k $}
  \put(267,38)    {\scriptsize$ V_k $} 
  \put(17,96.5)   {\tiny$y$}
  \put(-5,74.3)   {\tiny$x$}
  \put(217,96.5)  {\tiny$y$}
  \put(195,74.3)  {\tiny$x$}
  \epicture26 \labl{eq:bnd-ribbon} \xlabp 10
depending on the orientation of the arc-germ $[\gamma_k]$ relative to that of 
the boundary (recall also the convention in \erf{eq:ribbon-arc-end}). Shown in 
\erf{eq:bnd-ribbon} is a horizontal section of the connecting manifold 
\erf{eq:con-mf-bnd}. Correspondingly the lower boundary in \erf{eq:bnd-ribbon} 
is that of $\MX$ while the ribbons $M_k$ and $M_{k+1}$ are placed on the 
boundary of $\X$ as embedded in $\MX$. The arrow on the boundary in 
\erf{eq:bnd-ribbon} indicates the orientation of $\partial\X$ (transported 
to $\partial\MX$ along the preferred intervals). The ribbon graph must be 
placed in the plane of the world sheet $\X$ (embedded in $\MX$) in such a 
way that the bulk and boundary orientation of \erf{eq:bnd-ribbon} agree 
with the local orientations at the insertion point $p_k\iN\X$. 
\end{itemize}

\noindent
The prescription involves two choices, and we proceed to
show that different choices lead to equivalent ribbon graphs.
\\[5pt]
\nxt\underline{Choice \#2}:\\[2pt]
As in sections I:5.1 and II:3.1, independence of the triangulation follows from 
the identities (I:5.11) and (I:5.12). However, if there are field insertions on 
the boundary we need an additional move to relate any one dual triangulation to 
any other dual triangulation. It consists of taking a vertex of the 
triangulation past a boundary insertion,
  \bea \begin{picture}(352,57)(0,44)
  \put(0,0)     {\Includeourbeautifulpicture 11a }
  \put(184,0)   {\Includeourbeautifulpicture 11b }
  \put(164,33)  {$ = $}
  \put(10,63)   {\scriptsize$ M $}
  \put(67,63)   {\scriptsize$ M $}
  \put(140,63)  {\scriptsize$ N $}
  \put(195,63)  {\scriptsize$ M $}
  \put(329,63)  {\scriptsize$ N $}
  \put(51,84)   {\scriptsize$ A $}
  \put(295,98)  {\scriptsize$ A $}
  \put(87,54)   {\scriptsize$ \psi $}
  \put(246,54)  {\scriptsize$ \psi $}
  \put(65,34)   {\scriptsize$ V $}
  \put(225,34)  {\scriptsize$ V $}
  \put(267,46)  {\scriptsize$ N $}
  \epicture24 \labl{eq:bnd-ch2} \xlabp 11
together with a similar identity for an insertion of (\ref{eq:bnd-ribbon}\,b).
The dashed circle in \erf{eq:bnd-ch2} represents the piece of ribbon graph 
inside the dashed circle in \erf{eq:bnd-vertex}. This identity is a direct 
consequence of the fact that the morphism $\psi$ in $\hatpsi \eq (M,N,V,\psi,
p,[\gamma])$ is an intertwiner of $A$-modules, $\psi \iN \HomA(M \Oti V, N)$.
\\[5pt]
\nxt\underline{Choice \#1}:
\\[2pt]
Denote by 
$R \eq (M_1, ...\, , M_{n'} ; \hatpsi_1, ...\, , \hatpsi_n ; \orr_1)$ and
$R' \eq (M'_1, ...\, , M'_{n'} ; \hatpsi'_1, ...\, , \hatpsi'_n ; \orr'_1)$
two representatives of $B$. 
We will apply the above construction to the representative $R'$ and
show that the resulting ribbon graph can be turned into the one obtained from
$R$. According to definition \ref{bnd-equiv-def} we must consider two cases.
\\[3pt]
(i)~\,$\orr_1' \eq \orr_1$. Then also $V_k' \eq V_k$, $p_k' \eq p_{k}$ and
$[\gamma_k'] \eq [\gamma_{k}]$, and there exists a choice of isomorphisms 
$\varphi_k \iN \HomA(M_k', M_k)$ fulfilling \erf{eq:bnd-equiv-def-i}.
Starting with the ribbon graph constructed from $R'$, we insert on
each ribbon $M_k'$ the identity morphism in the form
$\id_{M_k'} \eq \varphi_k^{-1} \cir \varphi_k$, in such a way that
$\varphi_k^{-1}$ comes next to the field insertion $\hatpsi_k'$.
Then we move the morphism $\varphi_k$ to the other end of the $M_k'$-ribbon. 
As in (II:3.17), $\varphi_k$ can be taken past vertices of the
triangulation, which may lie on the boundary interval labelled by
$M_k'$. After these manipulations, every field insertion
$\hatpsi_k'$ comes with the morphism $\varphi_k^{-1}$ to one side and 
with $\varphi_{k+1}$ to the other. In other words, for each $k$ the morphism
$\psi_k'$ appearing in $\hatpsi_k'$ is replaced by
$\varphi_{k+1} \cir \psi_k' \cir (\varphi_k^{-1} \oti \id_{V_k})
\,{\equiv}\,\psi_k$. We thus remain with the ribbon graph constructed from $R$.
\\[4pt]
(ii)~$\orr_1' \eq {-}\orr_1$. Then also $V_k' \eq V_{n-k}$, $p_k' \eq p_{n-k}$
and $[\gamma_k'] \eq  [\gamma_{n-k}]$, and there exist isomorphisms 
$\varphi_k \iN \HomA(M_k', M_{n'-k+1}^\sigma)$, for $k \eq 1,...\,,n'$,
fulfilling \erf{eq:bnd-equiv-def-ii}. We shall demonstrate the equivalence of 
ribbon graphs in some detail for the second case in \erf{eq:bnd-equiv-def-ii}; 
the first case can be seen similarly. Around the field insertion $\hatpsi_k'$ 
the ribbon graph obtained from $R'$, which we denote by $[R']$, looks as follows:
  \bea \begin{picture}(220,39)(0,30)
  \put(55,0)     {\includeourbeautifulpicture 12 }
  \put(0,34)     {$ [R'] \ = $} 
  \put(165,66)   {\scriptsize$ M'_k $}
  \put(70,66)    {\scriptsize$ M'_{k+1} $}
  \put(142,40)   {\scriptsize$ V'_k $}
  \put(136.8,4)  {\scriptsize$ p'_k $}
  \put(74,-2)    {\scriptsize$ \orr'_1 $}
  \put(76,23)    {\scriptsize$ \orr_1 $}
  \put(108.4,54) {\scriptsize$ \psi'_k $}
  \epicture15 \labl{eq:bnd-equiv-def-iii} \xlabp 12
The two opposite orientations of the boundary are indicated. We are facing the 
black side of the ribbons because in order to make the local orientations of 
\erf{eq:bnd-equiv-def-iii} induced by $\orr_1'$ agree with those in 
(\ref{eq:bnd-ribbon}\,b), the element (\ref{eq:bnd-ribbon}\,b) must be turned 
`upside down' before it is inserted. As in case (i) we insert 
$\id_{M_l'} \eq \varphi_l^{-1}{\circ}\,\varphi_l$
on each $M_l'$, close to $\hatpsi_l'$. After applying in addition
a half-twist and using that, by definition, $\dot M^\sigma \eq \dot M^\vee$ as
an object, this leads to
  \bea \begin{picture}(310,44)(0,34)
  \put(55,0)     {\includeourbeautifulpicture 13 }
  \put(0,34)     {$ [R'] \ = $}    
  \put(140,66)   {\scriptsize$ M'_k $}
  \put(70,66)    {\scriptsize$ M'_{k+1} $}
  \put(197,67)   {\scriptsize$ M_{n-k+1} $}
  \put(272,66)   {\scriptsize$ M'_k $}  
  \put(142,40)   {\scriptsize$ V'_k $}
  \put(136.2,4)  {\scriptsize$ p'_k $}
  \put(73,-2)    {\scriptsize$ \orr'_1 $}
  \put(77,23)    {\scriptsize$ \orr_1 $}
  \put(108.6,55) {\scriptsize$ \psi'_k $}
  \put(166,58.6) {\scriptsize$ \varphi^{-1}_k $}
  \put(247.8,59) {\scriptsize$ \varphi_k $} 
  \epicture21 \labp 13
By a reasoning similar to that of (II:3.18) this specific combination
of half-twist and $\varphi_k$ can be taken past vertices of the
triangulation that lie on the boundary, i.e.\ past locations at which
$A$-ribbons arriving from the bulk are attached to the module-ribbons
on the boundary. Taking all of the $\varphi_l$ to the neighbouring
field insertion leads to a ribbon graph which close to $\hatpsi_k'$ looks as 
  \bea \begin{picture}(300,48)(0,39)
  \put(55,0)     {\includeourbeautifulpicture 14 }
  \put(0,39)     {$ [R'] \ = $}    
  \put(125,64)   {\scriptsize$ M'_{k+1} $}
  \put(61,65)    {\scriptsize$ M_{n-k} $}
  \put(258,65)   {\scriptsize$ M_{n-k+1} $}
  \put(187,65.7) {\scriptsize$ M'_k $} 
  \put(188,40)   {\scriptsize$ V'_k $}
  \put(182.4,4)  {\scriptsize$ p'_k $}
  \put(73,-2)    {\scriptsize$ \orr'_1 $}
  \put(77,23)    {\scriptsize$ \orr_1 $}
  \put(153.5,54) {\scriptsize$ \psi'_k $}
  \put(212,59)   {\tiny$ \varphi^{-1}_k $}
  \put(105,59)   {\tiny$ \varphi_{k+1} $}
  \epicture29 \labp 14
The morphism $\varphi_{k+1}$ has arrived from the insertion
$\hatpsi_{k+1}'$ which is not visible in the section of the
full ribbon graph $[R']$ that we display. Next rotate
the coupons labelled by $\varphi_{k+1}$, $\psi_k'$ and
$\varphi_{k}^{-1}$ by $180^{\circ}$ in the manner already indicated in
figure \erp 14, and deform the resulting ribbon graph slightly,
  \bea \begin{picture}(370,62)(0,47)
  \put(55,0)     {\includeourbeautifulpicture 57 }
  \put(0,53)     {$ [R'] \ = $}    
  \put(57,79)    {\scriptsize$ M_{n-k} $}
  \put(327,80)   {\scriptsize$ M_{n-k+1} $}
  \put(217,72)   {\scriptsize$ M'_{k+1} $}
  \put(176,78)   {\scriptsize$ M'_k $} 
  \put(108.3,34) {\scriptsize$ V'_k\eq V_{n-k} $}
  \put(102,5)    {\scriptsize$ p'_k=p_{n-k} $}
  \put(71,-2)    {\scriptsize$ \orr'_1 $}
  \put(76,23)    {\scriptsize$ \orr_1 $} 
  \put(202,63.8) {\tiny{$ \psi'_k $}}
  \put(160,73)   {\tiny{$ \varphi^{-1}_k $}}
  \put(240.9,65) {\tiny{$ \varphi_{k+1} $}}
  \epicture33 \labp 57
By the second alternative in \erf{eq:bnd-equiv-def-ii}, the morphism in the 
dashed box is equal to $s(\psi_{n-k})$.  After replacing the dashed box by 
the explicit form of $s(\psi_{n-k})$ in \erf{eq:bnd-s-def},
the ribbon graph can be deformed so as to obtain the 
ribbon graph constructed from the representative $R$ of $B$.

\medskip
\noindent
We have thus established that the ribbon graph that is obtained
by our prescription for a boundary component 
$\bndb([\gamma_1],...\,,[\gamma_n])$ labelled by an equivalence class in 
$\mathcal{B}\big(\bndb([\gamma_1],...\,,[\gamma_n])\big)$ 
is independent of the choices involved.

\subsection{Bulk fields} \label{sec:fi-bulk}

A bulk field insertion is described by a point $p$ in the
interior of $\X$, together with an arc-germ $[\gamma]$ at $p$
and a label $\PHI$. The set of labels for bulk insertions
at $[\gamma]$ will be denoted by $\mathcalD([\gamma])$. Similarly as for 
boundary fields, this set consists of equivalence classes of certain tuples.

\subsubsection*{Labelling of bulk fields}

The equivalence relation will be formulated in terms of a linear
map $\omega^A_{UV}$ that is defined as follows.
\\[-2.3em]

\dtl{Definition}{def:bulk-omega}
For $A$ a Jandl algebra in $\calc$, the morphism $\omega^A_{UV}{:}\; 
\Hom(U\oti A\oti V, A) \To \Hom(V\oti A\oti U, A)$ is defined by
  \bea \begin{picture}(170,90)(0,43)
  \put(75,0)     {\includeourbeautifulpicture 92 }
  \put(0,60)     {$ \omega^A_{UV}(\phi) \ := $}    
  \put(99.3,46.4) {\scriptsize$ \sigma^{\!-\!1} $}
  \put(102.4,87.9){\scriptsize$ \sigma $}
  \put(101,125.7){\scriptsize$ A $}
  \put(87,-8)    {\scriptsize$ V $}
  \put(100.5,-8) {\scriptsize$ A $}
  \put(115,-8)   {\scriptsize$ U $}
  \put(102,68.3) {\scriptsize$ \phi $}
  \epicture27 \labl{eq:bulk-om-def} \xlabp 92
for $\phi\iN\Hom(U\oti A\oti V, A)$.

\medskip

Some properties of the map $\omega^A_{UV}$ that will be needed below 
are listed in the following lemma. 
\\[-2.2em]

\dtl{Lemma}{lem:omega-prop}
With $A$ and $\omega^A_{UV}$ as in definition \ref{def:bulk-omega}, we have:
\\[.2em]
(i)~\,For any $\phi\iN\Hom(U\oti A\oti V, A)$ one
has $\omega^A_{VU}\big(\omega^A_{UV}(\phi)\big) \eq \phi$.
\\[.3em]
(ii)~If $\phi\iN\HomAA(U\oT+A\ot-V,A)$,
then $\omega^A_{UV}(\phi) \iN \HomAA(V\oT+A\ot-U,A)$.

\medskip\noindent
Proof:\\
(i)~\,Writing out the ribbon graph for the morphism $\omega^A_{VU}\big(
\omega^A_{UV}(\phi)\big)$ and using $\sigma\cir\sigma\eq\theta_A$,
one can deform the resulting ribbon graph (which amounts to using that
$\calc$ is sovereign) so as to be left with the graph for $\phi$.
\\[.3em]
(ii)~One must verify that the morphism $\omega^A_{UV}(\phi)$ commutes
with the left and right action of $A$ if $\phi$ itself does.
This can be done by a straightforward computation using the
definitions \erf{UXUX} and \erf{XUXU} of the left and right action of $A$
as well as the defining properties (II:2.6) of the reversion $\sigma$ on $A$.
\qed

\dtl{Definition}{def:bulk-PHI}
Let $[\gamma]$ be an arc-germ around a point $p\eq\gamma(0)$ in
the interior of $\X$. The possible bulk fields that can be
inserted at $p$ are labelled by equivalence classes 
  \be
  \mathcalD([\gamma]) = \big\{
  (i,j,\phi,p,[\gamma],\orr_2(p))\big\} /{\sim} \,.
  \ee
Here $i,j \iN \II$ label simple objects,
$\phi$ is a morphism in $\HomAA(U_i\,{\otimes^+}A\,{\otimes^-}U_j,A)$,
the insertion point $p$ is at $\gamma(0)$, and $\orr_2(p)$ is a local
orientation of the world sheet $\X$ around $p$. The
equivalence relation is defined as
  \be
  \big(i,j,\phi,p,[\gamma],\orr_2(p)\big) \sim 
  \big(j,i,\omega^A_{U_iU_j}(\phi),p,[\gamma],-\orr_2(p)\big)\,.
  \ee

\subsubsection*{Marked points on the double}

Each bulk field insertion $\PHI$ on the world sheet $\X$ gives
rise to {\em two\/} marked points on the double $\hat \X$. If
$\big(i,j,\phi,p,[\gamma],\orr_2(p)\big)$ is a representative
of $\PHI$, then these two points are
  \be
  (\tilde p_i, [\tilde\gamma_i] , U_i , +) \qquad {\rm and} \qquad
  (\tilde p_j, [\tilde\gamma_j] , U_j , +) \,;
  \labl{eq:bulk-marked-Xh}
the terms appearing in these tuples are given by $\tilde p_i \eq [p,\orr_2(p)]$, 
$\tilde p_j \eq [p,-\orr_2(p)]$,
as well as $\tilde\gamma_i(t) \eq [\gamma(t),\orr_2(\gamma(t))]$ and
$\tilde\gamma_j(t) \eq [\gamma(t),-\orr_2(\gamma(t))]$. Here $\orr_2(\gamma(t))$ 
is obtained by the extension of $\orr_2(p)$ to a neighbourhood of $p$.

Just like in the case of boundary fields, one verifies that the marked points 
\erf{eq:bulk-marked-Xh} on $\hat \X$ are independent of the choice of
representative of $\PHI$.

\subsubsection*{Ribbon graph embedded in the connecting manifold}

The ribbon graph representation of a bulk field $\PHI$ is
constructed as follows.

\ITemize
\item[\Nxt] 
Choose  a representative $(i,j,\phi,p,[\gamma],\orr_2(p))$
of $\PHI$ -- \underline{Choice \#1}.

\item[\Nxt] Choose a dual triangulation of $\X$ such that the insertion point 
$p$ lies on an edge and such that the arc $\gamma$ is aligned with this edge 
at $p\eq\gamma(0)$ -- \underline{Choice \#2}.

\item[\Nxt] The bulk insertion is treated as a two-valent vertex that is to
be included in the dual triangulation. Locally around the insertion
point $p$, the world sheet has the orientation $\orr_2$. In
$\MX$ place the following ribbon graph:
  
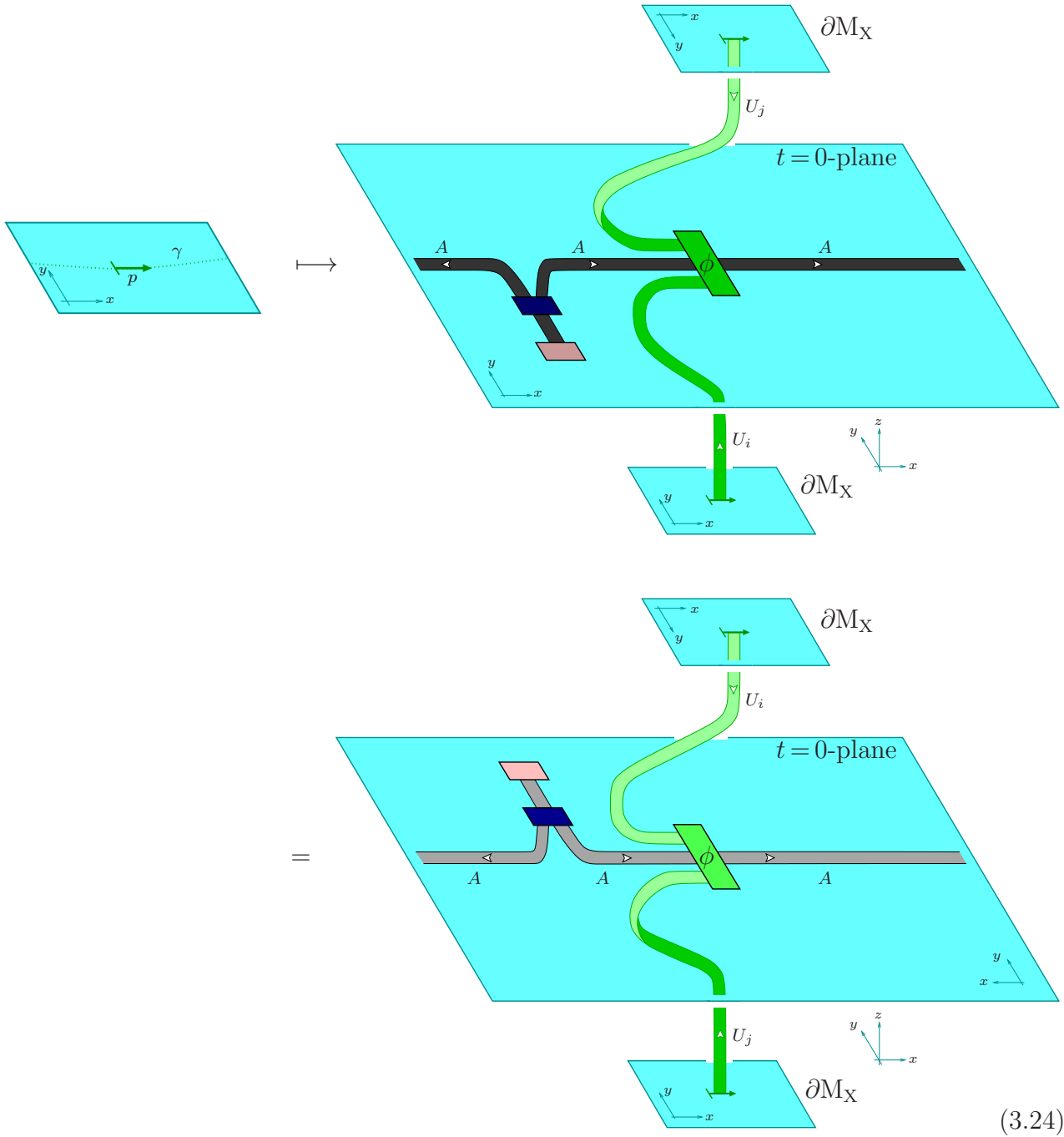
\begin{eqnarray}
     \begin{picture}(420,240)(5,0)
  \put(-20,100)  {\Includeourbeautifulpicture 94a }
  \put(130,0)    {\Includeournicelargepicture 94b }
  \put(350,227)    {$\partial\MX$}
  \put(341,19)     {$\partial\MX$}
  \put(291,234)    {\tiny $x$}
  \put(284,221)    {\tiny $y$}
  \put(26,105.5)   {\tiny $x$}
  \put(-4.4,119)   {\tiny $y$}
  \put(56,127)     {\scriptsize $\gamma$}
  \put(36,115)     {\scriptsize $p$}
  \put(316,192)    {\scriptsize$ U_j$}
  \put(310,41)     {\scriptsize$ U_i$}
  \put(175,128)    {\scriptsize$A$}
  \put(237,128)    {\scriptsize$A$}
  \put(349,128)    {\scriptsize$A$}
  \put(295,120)    {$\phi$}
  \put(330,168.1)     {$t\eq0$-plane}
  \put(220,63)     {\tiny $x$}
  \put(199,77)     {\tiny $y$}
  \put(390,30)     {\tiny $x$}
  \put(374.3,50)   {\tiny $z$}
  \put(363,46)     {\tiny $y$}
  \put(279,16)     {\tiny $y$}
  \put(297.8,4)    {\tiny $x$}
  \put(112,120)  {$ \longmapsto $}
  \end{picture} \nonumber
  \\                 \begin{picture}(420,265)(5,0)
  \put(130,0)    {\includeournicelargepicture 93 }
  \put(110,120)  {$ = $}
  \put(350,227)    {$\partial\MX$}
  \put(343,14)     {$\partial\MX$}
  \put(291,234)    {\tiny $x$}
  \put(284,221)    {\tiny $y$}
  \put(316,192)    {\scriptsize$ U_i$}
  \put(310,40)     {\scriptsize$ U_j$}
  \put(190,111)    {\scriptsize$A$}
  \put(248,111)    {\scriptsize$A$}
  \put(349,111)    {\scriptsize$A$}
  \put(295,120)    {$\phi$}
  \put(330,168.1)     {$t\eq0$-plane}
  \put(422,65)     {\tiny $x$}
  \put(428,77)     {\tiny $y$}
  \put(390,30)     {\tiny $x$}
  \put(374.3,50)   {\tiny $z$}
  \put(363,46)     {\tiny $y$}
  \put(279,16)     {\tiny $y$}
  \put(297.8,4)    {\tiny $x$}
  \end{picture} \nonumber 
  \\[-1.5em]{} \label{eq:bulk-insert}
  \end{eqnarray} \xlabp 93 \xlabp 94
In this figure we display the ribbon graph twice, once looked upon from 
`above' the connecting manifold and once looked at from `below'. In any case, 
the orientation of the $A$-ribbons is opposite to that of the world sheet and 
their white side faces the insertion point of the chiral field label $U_i$.
The ribbon graph \erf{eq:bulk-insert} must be placed in $\MX$ in such a manner 
that the embedding respects the orientation of the three-manifold $\MX$, the 
local orientation $\orr_2$ of the world sheet $\X$, as well as the orientation 
of the arcs on the boundary of $\MX$. There is precisely one way to do this; 
no further choice needs to be made.
\end{itemize}

\noindent
We proceed to establish independence of the two choices.
\\[5pt]
\nxt\underline{Choice \#2}:\\[2pt]
Independence of the dual triangulation follows from the identities (I:5.11)
together with a move that allows to take three-valent vertices of the 
triangulation past the two-valent vertex formed by the bulk field insertion
  \bea \begin{picture}(420,48)(0,4) 
  \put(0,0)     {\Includeourbeautifulpicture 18a 
                 \put(105,8){ $\phi$ }
                }
  \put(230,0)   {\Includeourbeautifulpicture 18b 
                 \put(62,8){ $\phi$ }
                }
  \put(202,11)  {$ = $}
  \epicture-4 \labl{eq:bulk-ch2} \xlabp 18
and together with similar identities for different local orientations and 
for situations where the $A$-ribbon arrives from the other side. The empty 
dashed circle in picture \erf{eq:bulk-ch2} stands for the element (II:3.11), 
the dashed rectangular box for (II:3.13), and the dashed circle with
inscribed $\phi$ for the element \erf{eq:bulk-insert}.
Analogously as for (I:5.11) and (I:5.12),\,%
    \footnote{~For detailed derivations, see \cite{fjfrs}.}
the equality follows in a straightforward way upon substituting the definitions
and using that the morphism $\phi$ is an intertwiner of $A$-bimodules.
\\[5pt]
\nxt\underline{Choice \#1}:\\[2pt]
Let $(i,j,\phi,p,[\gamma],\orr_2)$ and $(j,i,\phi',p,[\gamma],-\orr_2)$ 
be the two representatives of $\PHI$, i.e.\ we have 
$\phi'\eq\omega^A_{U_iU_j}(\phi)$. We must show that
  \bea \begin{picture}(350,15)(0,12) 
  \put(0,0)     {\Includeourbeautifulpicture 19a 
                 \put(62,8){ $\phi$ }
                }
  \put(199,0)   {\Includeourbeautifulpicture 19b 
                 \put(60,8){ $\phi'$ }
                }
  \put(166,9)   {$ = $}
  \epicture01 \labl{eq:bulk-ch1} \xlabp 19
The dashed boxes stand for (II:3.13) and (II:3.14), respectively, 
while the dashed circle with $\phi$ or $\phi'$ inscribed stands for the 
element \erf{eq:bulk-insert}, inserted with the appropriate orientation.

To facilitate drawing, let us rotate the middle plane of \erf{eq:bulk-insert} 
into the paper plane and use the blackboard framing convention. Equation 
\erf{eq:bulk-ch1} then looks as follows.
  \bea \begin{picture}(420,75)(8,43)
  \put(0,0)      {\Includeourbeautifulpicture 95a }
  \put(215,0)    {\Includeourbeautifulpicture 95b }
  \put(190,57)   {$ = $}
  \put(30,62.8)  {\scriptsize$ A $}
  \put(115,63.5) {\scriptsize$ A $}
  \put(150,63.5) {\scriptsize$ A $}
  \put(215,63.5) {\scriptsize$ A $}
  \put(390,63.5) {\scriptsize$ A $}
  \put(80,59)    {\scriptsize$ \phi $}
  \put(294,59)   {\scriptsize$ \phi' $}
  \put(47.7,94)  {\scriptsize$ i $}
  \put(269.5,95) {\scriptsize$ i $}
  \put(270,35)   {\scriptsize$ j $}
  \put(46.7,38)  {\scriptsize$ j $}
  \put(235,58.8) {\scriptsize$ \sigma $}
  \put(344.5,58.8){\scriptsize$ \sigma $} 
  \epicture22 \labl{eq:bulk-ch1-b} \xlabp 95
In the right figure one must insert the element \erf{eq:bulk-insert} `upside 
down' because the local orientation of the world sheet is inverted with 
respect to the left figure. As compared to \erf{eq:bulk-ch1}, on both sides 
of the equation one coproduct has been removed against one product using
the Frobenius property. In the right figure of \erf{eq:bulk-ch1-b} the central 
part of the ribbon graph can be rotated by $180^\circ$ as indicated. Using 
also $\theta_A^{-1}\cir\sigma\eq\sigma^{-1}$ this leads to
  \bea \begin{picture}(320,77)(8,46)
  \put(65,0)     {\includeourbeautifulpicture 96 }
  \put(0,57)     {$ {\rm \erf{eq:bulk-ch1-b}}\ = $}
  \put(64,63.2)  {\scriptsize$ A $}
  \put(239.4,66) {\scriptsize$ A $}
  \put(271,66)   {\scriptsize$ A $} 
  \put(175.5,59) {\scriptsize$ \phi' $} 
  \put(108,93)   {\scriptsize$ i $}
  \put(112,38)   {\scriptsize$ j $}
  \put(207,80)   {\scriptsize$ j $} 
  \put(155.5,59) {\scriptsize$ \sigma^{-1} $}
  \put(196.5,59) {\scriptsize$ \sigma $}
  \epicture26 \labl{eq:bulk-ch1-c} \xlabp 96
Comparing with \erf{eq:bulk-om-def} we see that
the part of the ribbon graph inside the dashed box amounts to the morphism
  \be
  { \raisebox{-3pt}{\dashbox{3.5}(63,14){{\rm dashed~box}}} }_{\phantom|} \;=
  \omega^A_{U_{j_{\phantom|}^{}}\!U_i}(\phi') = \phi \,,
  \ee
where the last equality follows from the equivalence relation for bulk labels 
which forces $\phi'\eq\omega^A_{U_iU_j}(\phi)$, together with
lemma \ref{lem:omega-prop}. Thus \erf{eq:bulk-ch1-c} is equal to the left hand 
side of \erf{eq:bulk-ch1-b}, establishing the claimed equality between the 
left hand side and the right hand side of \erf{eq:bulk-ch1-b}.

\medskip\noindent
This finishes the demonstration that the ribbon graph for a bulk field 
insertion is independent of the choices involved in its construction.

\subsection{Defect fields}\label{sec:fi-defect}

Together with the defect fields treated in this
section, the three types of fields we consider are
  \bea \begin{picture}(340,65)(0,8)
  \put(22,18)    {\Includeourbeautifulpicture 22a }
  \put(125,0)    {\Includeourbeautifulpicture 22b }
  \put(260,14)   {\Includeourbeautifulpicture 22c }
  \put(0,55)     { {\em bulk field} }
  \put(113,55)   { {\em boundary field} }
  \put(252,55)   { {\em defect field} }
  \epicture-5 \labl{eq:fields-consd} \xlabp 22
Note, however, that this is not the most general way to treat defects. 
In fact, upon forming operator products, the first two kinds of fields in
\erf{eq:fields-consd} close amongst each other. But once we include also 
the defect field, we can generate the following more general
class of fields by taking operator products,
  \bea \begin{picture}(220,49)(0,7)
  \put(9,4)      {\Includeourbeautifulpicture 23a }
  \put(155,0)    {\Includeourbeautifulpicture 23b }
  \epicture-1 \labl{eq:fields-not-consd} \xlabp 23
i.e.\ a boundary field with several defect lines attached and a defect field 
with more than two defect lines ending on it. For example, a field of the 
first kind can be obtained by taking the operator product of a boundary 
field and a defect field of the types listed in \erf{eq:fields-consd}. 
While fields of the type \erf{eq:fields-not-consd} and their operator 
products can be treated in the TFT-formalism in a natural way, in this paper 
we will nonetheless only consider fields of the types \erf{eq:fields-consd}.

\subsubsection*{Labelling of defect fields}

Recall from section II:3.8 that a circular defect $\dftd$ without
field insertions and with orientable neighbourhood is labelled
by equivalence classes of triples $[X,\orr_1,\orr_2]$, where
$X$ is an $A$-bimodule, $\orr_1$ is an orientation of the
defect circle and $\orr_2$ is an orientation of its neighbourhood.
The equivalence relation is given in (II:3.150).

In order to describe the labelling of a defect circle with field insertions 
$\dftd([\gamma_1],...\,,[\gamma_n])$ we first need the following definition.
\\[-2.3em]

\dtl{Definition}{def:defect-equiv}
The linear maps
  \beaa\displaystyle
  & u :\quad \Hom(U_i \oti X\oti U_j, Y) \to \Hom(U_j \oti X\oti U_i, Y) 
  \\{}\\[-.8em]  {\rm and} \quad& \displaystyle
  \tilde u :\quad \Hom(U_i \oti X\oti U_j, Y) \to
      \Hom(U_i \oti Y^\vee{\otimes}\, U_j, X^\vee) 
  \eear\ee
are defined by
  \bea \begin{picture}(360,93)(0,50)
  \put(290,0)    {\Includeourbeautifulpicture 97a }
  \put(60,0)     {\Includeourbeautifulpicture 97b }
  \put(230,0){
   \put(67,-9.5)   {\scriptsize$U_i$}
   \put(114,-9.5)  {\scriptsize$U_j$}
   \put(90,-9.5)   {\scriptsize$ Y^\vee $}
   \put(91,71)     {\small $\vartheta$}
   \put(90,144)    {\scriptsize$ X^\vee $}
  }
  \put(-230,0){
   \put(334,-9.5)  {\scriptsize$U_i$} 
   \put(295,-9.5)  {\scriptsize$U_j$}
   \put(315,-9.5)  {\scriptsize$X$}
   \put(316.5,71)  {\small $\vartheta$}
   \put(316,144.5) {\scriptsize$Y$}
  }
  \put(0,68)       {$ u(\vartheta) \;:= $}
  \put(170,64)     { and }
  \put(230,64)     {$ \tilde u(\vartheta) \;:= $}
  \epicture39 \labl{eq:uv-def} \xlabp 97
for $\vartheta\iN\Hom(U_i \oti X\oti U_j, Y)$.

\medskip

The maps $u$ and $\tilde u$ will enter into the formulation of the
equivalence relation for defect labels. This also requires 
the following lemma, which can be verified by direct calculation.
\\[-2.3em]

\dtl{Lemma}{lem:defect-equiv}
Let $A$ be an algebra and let $X,Y$ be $A$-bimodules. 
Then $\tilde u$ restricts to an isomorphism
  \be
  \tilde u :\quad \HomAA(U_i\,{\otimes^+}\,X\,{\otimes^-}\,U_j,Y) \to
  \HomAA(U_i\,{\otimes^+}\,Y^v\,{\otimes^-}\,U_j,X^v) \,.
  \ee
If $A$ is a Jandl algebra, then in addition $u$ restricts to an isomorphism
  \be
  u :\quad \HomAA(U_i\,{\otimes^+}\,X\,{\otimes^-}\,U_j,Y) \to
  \HomAA(U_j\,{\otimes^+}\,X^s\,{\otimes^-}\,U_i,Y^s) \,.
  \ee
Here $X^v,Y^v$ and $X^s,Y^s$ denote two of the three 
duals one has for the bimodules $X,Y$, as defined in (II:2.40).

\medskip

After these preliminaries we turn to the description of defect labels.
Let $A$ be a Jandl algebra. A {\em defect field\/} $\Thetha$ 
is a collection of data
  \be
  \Thetha = \big(\,X,\orr_2(X),Y,\orr_2(Y),i,j,\vartheta,p,[\gamma],
  \orr_2(p)\,\big) \,.
  \ee
Here $i,j\iN\II$, $X,Y$ are $A$-bimodules, and $\orr_2(X)$ and $\orr_2(Y)$ 
are local orientations of the world sheet on a neighbourhood of the
defect line labelled by $X$ and $Y$, respectively; $\vartheta$ is a morphism 
  \be
  \vartheta \in \HomAA(U_i\,{\otimes^+}\,Z\,{\otimes^-}\,U_j,Z') \,,
  \ee
where
  \be
  Z =   \left\{ \bearll
    X~&  {\rm if}~~ \orr_2(X)=\orr_2(p) \,, \\[.2em]
    X^s~&  {\rm if}~~ \orr_2(X)=-\orr_2(p) 
  \eear\right.
  \quad\ {\rm and} \quad\
  Z' =   
  \left\{ \bearll
    Y~& {\rm if}~~ \orr_2(Y)=\orr_2(p) \,, \\[.2em]
    Y^s~& {\rm if}~~ \orr_2(Y)=-\orr_2(p) \,.  
  \eear\right.
  \labl{eq:defect-ZZ-XY}
Finally $p\eq\gamma(0)$ is the insertion point of the defect field,
and $[\gamma]$ is a germ of arcs aligned with the defect line at $p$.  

\medskip

A defect circle $\dftd([\gamma_1],...\,,[\gamma_n])$ on the world
sheet is labelled by equivalence classes of tuples
  \be
  \big(\, X_1,\orr_2(X_1),...\,,X_{n'},\orr_2(X_{n'}) \,;\,
  \Thetha_1, ...\, , \Thetha_n \,;~ \orr_1(\dftd) \,\big)  \,,
  \labl{eq:defect-tupel}
subject to the following conditions.
\ITemize
\item[\Nxt] 
$\orr_1(\dftd)$ is an orientation of the defect circle $\dftd$.
\item[\Nxt]
The points $p_k\eq\gamma_k(0)$ are ordered
such that when passing along the defect circle $\dftd$ in the direction 
opposite to $\orr_1(\dftd)$, one passes from $p_k$ to $p_{k+1}$. Here
and below it is understood that $p_0 \,{\equiv}\, p_n$ and 
$p_{n+1} \,{\equiv}\, p_1$, and similarly for $X_k$ and $\Thetha_k$.
\item[\Nxt]
The $X_1,...\,,X_{n'}$ are $A$-bimodules. $X_k$ 
labels the defect interval between $p_{k-1}$ and $p_k$. Thus if there
are no field insertions, then $n'\eq 1$, and otherwise $n'\eq n$.
\item[\Nxt]
$\orr_2(X_k)$ is a local orientation\,\footnote{%
  ~If the neighbourhood of the whole defect circle is non-orientable
  we require that there is at least one field insertion.  
  This field insertion can transform like an identity field, though, in
  which case correlators are independent of the insertion
  point of this field (recall the corresponding discussion in the
  second half of section II:3.8).}
of the world sheet on a
neighbourhood of the defect interval between $p_{k-1}$ and $p_k$.
\item[\Nxt]
$\Thetha_k$ is a defect field with defining data
  \be
  \Thetha_k = \big(\, X_{k},\orr_2(X_{k}),X_{k+1},\orr_2(X_{k+1}),i_k,j_k,
  \vartheta_k,p_k,[\gamma_k], \orr_2(p_k)\,\big)  \,.
  \ee
\end{itemize}

\medskip

Because of the large number of orientations entering the labelling of a defect 
circle, the equivalence relation on the tuples \erf{eq:defect-tupel} becomes 
somewhat technical; we include it merely for completeness of the presentation.

\medskip

The relation is conveniently formulated in terms of five basic relations. 
The full equivalence class is obtained by completing the relations generated 
by the basic relations with respect to transitivity and reflexivity. Consider 
the two tuples
  \bea\displaystyle
  T = \big(\, X_1,\orr_2(X_1),...\,,X_{n'},\orr_2(X_{n'}) \,;\,
  \Thetha_1, ...\, , \Thetha_n \,;\, \orr_1(\dftd) \,\big) \,,
      \\[7pt]\displaystyle
  T' = \big(\, X'_1,\orr_2'(X'_1),...\,,X'_{n'},\orr_2'(X'_{n'}) \,;
  \Thetha_1', ...\, , \Thetha_n' \,;\, \orr_1'(\dftd) \,\big) \,.
  \eear\ee
The basic relations are:
\\[4pt]
\nxt \underline{Relation 1}: Shift of labels.
\\[.2em]
The tuples $T$ and $T'$ are equivalent if their data are related by
$\orr_1'(\dftd) \eq \orr_1^{}(\dftd)$ and, for every 
$k$, $X'_k \eq X_{k+1}^{}$, $\orr_2'(X_k') \eq \orr_2(X_{k+1}^{})$ 
and $\Thetha_k'\eq \Thetha_{k+1}^{}$.
\\[4pt]
\nxt \underline{Relation 2}: 
Reverse orientation of the defect circle, $\orr_1'(\dftd)\eq{-}\orr_1(\dftd)^{}$. 
\\[.2em]
Suppose $\orr_1'(\dftd) \eq {-} \orr_1(\dftd)$. Then $T$ and $T'$ are
equivalent if the following conditions hold.
First, the ordering of the points is reversed in both labellings,
$[\gamma_k'] \eq [\gamma_{n-k}]$. Second, the bimodules labelling the defect
intervals are related by $X_{k} \eq (X_{n-k+1}')^v$.
Finally, the defect fields $\Thetha_{k}$ are related to the fields
  \be
  \Thetha_{\ell}' = 
  \big(\, X_{\ell}',\orr_2'(X_{\ell}'),X_{\ell+1}',\orr_2'(X_{\ell+1}'),
  i_{\ell}',j_{\ell}',\vartheta_{\ell}',p_{\ell}',[\gamma_{\ell}'],
        \orr_2'(p_{\ell}')\, \big) 
  \ee
via
  \bea\displaystyle
  \Thetha_{k} = 
  \big(\, (X_{n-k+1}')^v,\orr_2'(X_{n-k+1}'), (X_{n-k}')^v,\orr_2'(X_{n-k}'),
  \\[5pt]\displaystyle\hspace{8em}
  i_{n-k}',j_{n-k}', \tilde \beta \cir \tilde u(\vartheta_{n-k}') 
  \cir (\alpha \oti \beta \oti \tilde \alpha) , p_{n-k}',[\gamma_{n-k}'],
  \orr_2'(p_{n-k}') \,\big) \,, 
  \eear\labl{eq:muk-rel5}
where $\tilde u$ is the map defined in \erf{eq:uv-def}, and
  \beaa
  \displaystyle
  \alpha = \begin{cases}
    \id_{U_{i_{n-k}'}}  & {\rm if}~ \orr_1(\gamma_k) \eq \orr_1(\dftd) \,, \\
    \theta_{U_{i_{n-k}'}}  & {\rm if}~ \orr_1(\gamma_k) \eq {-}\orr_1(\dftd)\,,    
  \end{cases}
  &
  \beta = \begin{cases}
    \id_{(X_{n-k+1}')^\vee} & \!\!{\rm if}~ \orr_2(p_k) \eq \orr_2(X_k) \,, \\
    \theta_{(X_{n-k+1}')^\vee}^{-1} &\!\!{\rm if}~\orr_2(p_k) \eq {-}\orr_2(X_k)
    \,, \end{cases}
  \\[2em] \displaystyle
  \tilde\alpha = \begin{cases}
    \id_{U_{j_{n-k}'}} & {\rm if}~ \orr_1(\gamma_k) \eq \orr_1(\dftd) \,, \\
    \theta_{U_{j_{n-k}'}}^{-1} & {\rm if}~ \orr_1(\gamma_k) \eq {-}\orr_1(\dftd) \,,   
  \end{cases}
  ~~ &
  \tilde\beta = \begin{cases}
    \id_{(X_{n-k}')^\vee} & \!\!{\rm if}~ \orr_2(p_k) \eq \orr_2(X_{k+1}) \,, \\
    \theta_{(X_{n-k}')^\vee} & \!\!{\rm if}~ \orr_2(p_k) \eq {-}\orr_2(X_{k+1}) \,. 
  \end{cases}
  \eear\ee
\\[4pt]
For the next relations, select any $k_0\iN\{1,2,...\,,n'\}$.
\\[4pt]
\nxt \underline{Relation 3}: Isomorphism of bimodules,
$X_{k_0}' \,{\cong}\, X_{k_0}$.
\\[.2em]
The tuples $T$ and $T'$ are equivalent if all data of $T'$ coincide with 
the corresponding data of $T$, possibly with the exception of
$X'_{k_0}$, $\Thetha'_{k_0-1}$ and $\Thetha'_{k_0}$. 
In the latter case we require the existence of an isomorphism
$\varphi\iN\HomAA(X_{k_0}',X_{k_0})$ such that 
  \be
  \vartheta'_{k_0-1} = \varphi^{-1} \cir \vartheta_{k_0-1}
  \quad\ {\rm and} \quad\
  \vartheta'_{k_0} = \vartheta_{k_0} \cir
  (\id_{U_{i_{k_0}}}\oti\varphi\oti\id_{U_{j_{k_0}}}) \,,
  \ee
where $\vartheta'$ and $\vartheta$ are the morphisms in the defining data
of $\Thetha'$ and $\Thetha$.
\\[4pt]
\nxt \underline{Relation 4}: Reverse orientation at a field insertion,
$\orr_2'(p_{k_0}) \eq {-} \orr_2(p_{k_0})$. 
\\[.2em]
The tuples $T$ and $T'$ are equivalent if
all data of $T'$ and $T$ agree, with the exception of
$\Thetha'_{k_0}$, for which it is instead required that if
  \be
  \Thetha_{k_0} = 
  \big(\, X_{k_0},\orr_2(X_{k_0}),X_{k_0+1},\orr_2(X_{k_0+1}),
  i_{k_0},j_{k_0},\vartheta_{k_0},p_{k_0},[\gamma_{k_0}],\orr_2(p_{k_0})\,\big) \,, 
  \ee
then $\Thetha'_{k_0}$ is given by
  \bea\displaystyle
  \Thetha_{k_0}' = 
  \big(\, X_{k_0},\orr_2(X_{k_0}),X_{k_0+1},\orr_2(X_{k_0+1}),
  \\[5pt]\displaystyle\hspace{9em}
  j_{k_0},i_{k_0}, \tilde \beta \cir u(\vartheta_{k_0})
  \cir (\alpha \oti \beta \oti \tilde \alpha)
  ,p_{k_0},[\gamma_{k_0}], -\orr_2(p_{k_0})\, \big) \,,
  \eear\ee
where $u$ is the map defined in \erf{eq:uv-def}, and
  \beaa
  \displaystyle
  \alpha = \begin{cases}
    \theta_{U_{j_{k_0}}}^{-1}  & {\rm if}~ \orr_1(\gamma_{k_0}) \eq \orr_1(\dftd) \,, \\
    \id_{U_{j_{k_0}}}  & {\rm if}~ \orr_1(\gamma_{k_0}) \eq {-}\orr_1(\dftd)  \,,  
  \end{cases}
  \quad &
  \beta = \begin{cases}
    \id_{X_{k_0}} & {\rm if}~ \orr_2(p_{k_0}) \eq \orr_2(X_{k_0}) \,, \\
    \theta_{X_{k_0}} & {\rm if}~ \orr_2(p_{k_0}) \eq {-}\orr_2(X_{k_0})\,,  
  \end{cases}
  \\[2em] \displaystyle
  \tilde\alpha = \begin{cases}
    \theta_{U_{i_{k_0}}} & {\rm if}~ \orr_1(\gamma_{k_0}) \eq \orr_1(\dftd) \,, \\
    \id_{U_{i_{k_0}}} & {\rm if}~ \orr_1(\gamma_{k_0}) \eq {-}\orr_1(\dftd) \,,   
  \end{cases}
  \quad &
  \tilde\beta = \begin{cases}
    \id_{X_{k_0+1}} & {\rm if}~ \orr_2(p_{k_0}) \eq \orr_2(X_{k_0+1})\,,  \\
    \theta_{X_{k_0+1}}^{-1} & {\rm if}~ \orr_2(p_{k_0}) \eq {-}\orr_2(X_{k_0+1})\,.  
  \end{cases}
  \eear\ee
\\[3pt]
\nxt \underline{Relation 5}: Reverse orientation at a defect segment,
$\orr_2'(X_{k_0}') \eq {-} \orr_2(X_{k_0})$. 
\\[.2em]
The tuples $T$ and $T'$ are equivalent if all data of $T'$ and $T$ agree, with 
the exception of $X_{k_0}$ and $\orr_2(X_{k_0})$, as well as 
$\Thetha_{k_0-1}$ and $\Thetha_{k_0}$. For the latter we demand
  \be
  (X_{k_0}')^s = X_{k_0}
  \qquad {\rm and} \qquad \orr_2'(X_{k_0}') = - \orr_2(X_{k_0}) \,,
  \ee
as well as
  \bea\displaystyle
  \Thetha_{k_0-1} = 
  \big(\, X_{k_0-1}',\orr_2'(X_{k_0-1}'),(X_{k_0}')^s,-\orr_2'(X_{k_0}'),
  \\[5pt]\displaystyle\hspace{9em}
    i_{k_0-1}',j_{k_0-1}', \tilde\beta \cir \vartheta_{k_0-1}',
    p_{k_0-1}',[\gamma_{k_0-1}'], \orr_2'(p_{k_0-1}')\, \big) \,,  
  \\[10pt]\displaystyle
  \Thetha_{k_0} = 
  \big(\, (X_{k_0}')^s,-\orr_2'(X_{k_0}'),X_{k_0+1}',\orr_2'(X_{k_0+1}'),
  \\[5pt]\displaystyle\hspace{9em}
    i_{k_0}',j_{k_0}', \vartheta_{k_0}' \cir (\id_{U_{i_{k_0}'}} {\otimes}\, 
    \beta \oti \id_{U_{j_{k_0}'}}) , p_{k_0}',[\gamma_{k_0}'],
    \orr_2'(p_{k_0}')\, \big)  \,,  
  \eear\ee
where
  \be
  \beta = \begin{cases}
    \id_{X_{k_0}'} & {\rm if}~ \orr_2(p_{k_0}) \eq \orr_2(X_{k_0}) \,, \\
    \theta_{X_{k_0}'}^{-1} & {\rm if}~ \orr_2(p_{k_0}) \eq {-}\orr_2(X_{k_0})\,,  
  \end{cases}
  \qquad
  \tilde\beta = \begin{cases}
    \id_{X_{k_0}'} & {\rm if}~ \orr_2(p_{k_0-1}) \eq \orr_2(X_{k_0}) \,, \\
    \theta_{X_{k_0}'} & {\rm if}~ \orr_2(p_{k_0-1}) \eq {-}\orr_2(X_{k_0})\,.  
  \end{cases}
\ee

\medskip

The set of labels for the defect circle $\dftd([\gamma_1],...\,,[\gamma_n])$ is 
given by the equivalence classes
  \be
  \mathcalD(\dftd([\gamma_1],...\,,[\gamma_n]))
  = \big\{\,\big(\, X_1,\orr_2(X_1),...\,,X_{n'},\orr_2(X_{n'}) \,;\,
  \Thetha_1, ...\, , \Thetha_n \,;\,\orr_1(\dftd) \,\big) \,
  \big\} / {\sim} \,,
  \ee
where $\sim$ is the equivalence relation generated
by the five basic relations given above.

\subsubsection*{Marked points on the double}

For a defect insertion on the world sheet $\X$
the marked points on the double $\hat \X$ are the same as 
for bulk fields. Each insertion
  \be
  \Thetha = \big(\, X,\orr_2(X),Y,\orr_2(Y),i,j,\vartheta,p,[\gamma],
  \orr_2(p)\, \big) 
  \ee
on the world sheet $\X$ gives rise to two marked points
  \be
  (\tilde p_i, [\tilde\gamma_i] , U_i , +) 
  \qquad {\rm and} \qquad
  (\tilde p_j, [\tilde\gamma_j] , U_j , +) 
\ee
on the double $\hat \X$. Here again
$\tilde p_i \eq [p,\orr_2(p)]$, $\tilde p_j \eq [p,-\orr_2(p)]$,
as well as $\tilde\gamma_i(t) \eq [\gamma(t),\orr_2(\gamma(t))]$ and
$\tilde\gamma_j(t) \eq [\gamma(t),-\orr_2(\gamma(t))]$.

\subsubsection*{Ribbon graph embedded in the connecting manifold}

The ribbon graph representation of a 
defect circle $\dftd([\gamma_1],...\,,[\gamma_n])$ labelled
by an equivalence class 
$D \iN \mathcalD(\dftd([\gamma_1],...\,,[\gamma_n]))$ is 
constructed as follows.
\ITemize
\item[\Nxt] 
Choose  a representative $\big( X_1,\orr_2(X_1),...\,,X_{n'},\orr_2(X_{n'}) ;
\Thetha_1, ...\, , \Thetha_n ; \orr_1(\dftd) \big)$
of $D$ -- \underline{Choice \#1}.
\item[\Nxt] Choose a dual triangulation of $\X$ which contains
the defect circle as a subset, such that each defect insertion lies on an
edge of the dual triangulation -- \underline{Choice \#2}.
\item[\Nxt] At each insertion of a defect field
$\Thetha \eq \big(X,\orr_2(X),Y,\orr_2(Y),i,j,\vartheta,p,[\gamma],
\orr_2(p)\big)$ place -- depending on the various orientations -- one of the 
following eight ribbon graphs:
  \begin{eqnarray}  \begin{picture}(400,250)(0,0)
  \put(65,0)     {\Includeournicelargepicture 98a }
  \put(325,30)     {\tiny $x$}
  \put(310,50)     {\tiny $z$}
  \put(302,47)     {\tiny $y$}
  \put(234,4)      {\tiny $x$}
  \put(214,17)     {\tiny $y$}
  \put(93,170)     {\tiny $x$}
  \put(86,160)     {\tiny $y$}
  \put(226,235)    {\tiny $x$}
  \put(220,225)    {\tiny $y$}
  \put(250,192)    {\scriptsize $U_i$}
  \put(244,41)     {\scriptsize $U_j$}
  \put(172,129)    {\scriptsize $Z$}
  \put(277,129)    {\scriptsize $Z'$}
  \put(186,106)    {\small $\rm d $}
  \put(266,106)    {\small $\rm d $}
  \put(112,105)    {\scriptsize $\orr_1(\rm d)$}
  \put(316,105)    {\scriptsize $\orr_1(\rm d)$}
  \put(229.6,126)  {\begin{turn}{280} $\vartheta$\end{turn}}
  \put(-29,228)  { Cases (1)\,--\,(4)\,: }
  \end{picture} \nonumber
  \\                 
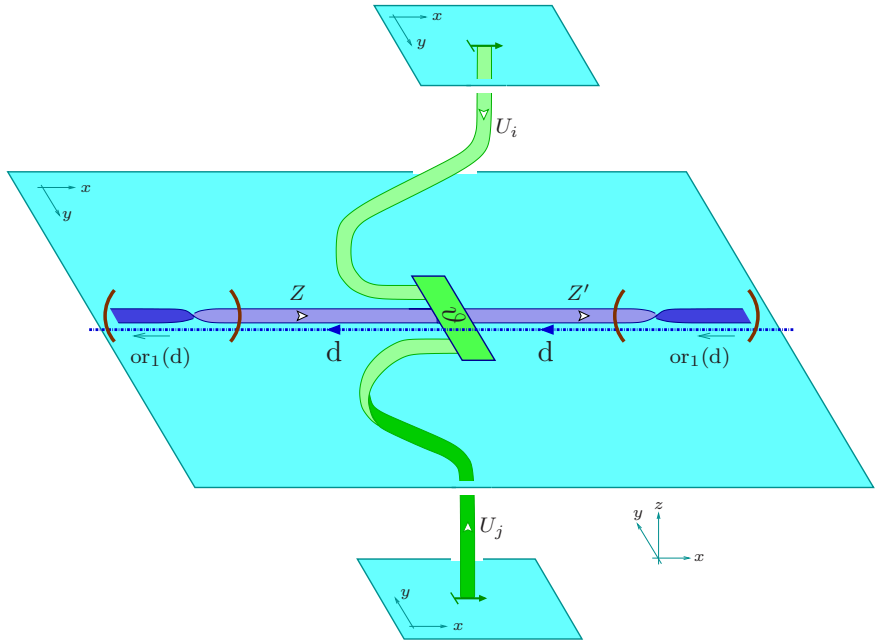
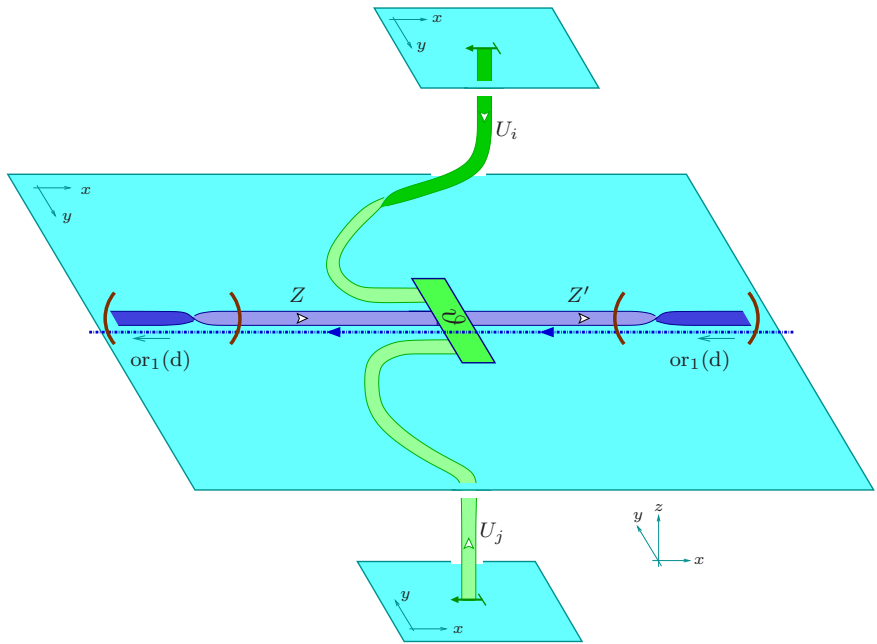
\begin{picture}(400,260)(0,0)
  \put(65,0)     {\Includeournicelargepicture 98b }
  \put(325,30)     {\tiny $x$}
  \put(310,50)     {\tiny $z$}
  \put(302,47)     {\tiny $y$}
  \put(234,4)      {\tiny $x$}
  \put(214,17)     {\tiny $y$}
  \put(93,170)     {\tiny $x$}
  \put(86,160)     {\tiny $y$}
  \put(226,235)    {\tiny $x$}
  \put(220,225)    {\tiny $y$}
  \put(250,192)    {\scriptsize $U_i$}
  \put(244,40)     {\scriptsize $U_j$}
  \put(172,129)    {\scriptsize $Z$}
  \put(277,129)    {\scriptsize $Z'$}
  \put(112,105)    {\scriptsize $\orr_1(\rm d)$}
  \put(316,105)    {\scriptsize $\orr_1(\rm d)$}
  \put(229.6,126)  {\begin{turn}{280} $\vartheta$\end{turn}}
  \put(-29,228)  { Cases (5)\,--\,(8)\,: } 
  \end{picture} \nonumber
  \\[-1.6em]{} \label{eq:pic41}
  \end{eqnarray} \xlabp 98
Here in all cases the coupon labelled by $\vartheta$ is placed in such a way that 
its orientation is opposite to $\orr_2(p)$. The bimodule ribbons are always 
placed such that the orientation of their cores is opposite to $\orr_1(\dftd)$.
The bimodules $Z$ and $Z'$ are related to $X$ and $Y$ as in 
\erf{eq:defect-ZZ-XY}. In cases (1)\,--\,(4) the orientation of
the arc-germ $[\gamma]$ is equal to $-\orr_1(\dftd)$. Those parts 
of the ribbon graphs that we have put in brackets are to be inserted only if
$\orr_2(X) \eq {-}\orr_2(p)$ and $\orr_2(Y) \eq {-}\orr_2(p)$, respectively.
In cases (5)\,--\,(8) the orientation of the arc-germ $[\gamma]$ is equal to 
$+\orr_1(\dftd)$; the parts of the ribbon graph in brackets are to be 
understood in the same way as in cases (1)\,--\,(4).
\end{itemize}

\noindent
We only briefly sketch how to establish independence of these
choices.
\\[5pt]
\nxt\underline{Choice \#2}:\\[2pt]
As for bulk fields the fact that the morphism $\vartheta$ in a
defect field $\Thetha$ is an intertwiner of $A$-bimodules
allows one to move vertices of the dual triangulation past
insertion points of defect fields.
\\[5pt]
\nxt\underline{Choice \#1}:\\[2pt]
Suppose $T$ and $T'$ are two representatives of the
equivalence class $D$ labelling the defect circle. We have
to establish that if $T$ and $T'$ are linked by one of
the five basic relations above, then the construction leads
to equivalent ribbon graphs. This can be done by straightforward, though
tedious, computations. We refrain from presenting them in any detail, except
for the example that $T$ and $T'$ are linked by relation~2.

If $T$ and $T'$ are linked by relation~2, then in particular
$p_k \eq p'_{n-k}$. Select a point $p\eq p_{k_0}\eq p'_{n-k_0}$ and
consider only the fragment of the ribbon graph for $T$ and $T'$ close to $p$. 
Suppose further that $\orr_1(\gamma_{k_0}) \eq \orr_1(\dftd)$, $\orr_2(p_{k_0})
\eq\orr_2(X_{k_0})$ and $\orr_2(p_{k_0})\eq{-}\orr_2(X_{k_0+1})$. The other 
instances of relation~2 follow by similar considerations. Denote the fragments 
close to $p$ by $[T]$ and $[T']$. Applying the construction above gives 
  \begin{eqnarray} \begin{picture}(420,130)(21,0)
  \put(37,0)    {\INcludeourbeautifulpicture 110a }
  \put(293,0)   {\INcludeourbeautifulpicture 110b }
  \put(50,65)     {\scriptsize$X_{k_0}$}
  \put(188,65)    {\scriptsize$X_{k_0+1}$}
  \put(103,35)    {\scriptsize$j_{k_0}$}
  \put(103,79)    {\scriptsize$i_{k_0}$}
  \put(61,42)     {\scriptsize$\orr_1(\rm d)$}
  \put(125.5,66)  {\begin{turn}{270}\small$\vartheta_{k_0}$\end{turn}}
  \put(301,65)    {\scriptsize$X'_{n-k_0+1}$}
  \put(451,65)    {\scriptsize$X'_{n-k_0}$}
  \put(392,32)    {\scriptsize$j'_{n-k_0}$}
  \put(392,82)    {\scriptsize$i'_{n-k_0}$}
  \put(307,42)    {\scriptsize$\orr_1(\rm d)'$}
  \put(356.9,46.9){\begin{turn}{90}\small$\vartheta'_{\!n-k_0}$\end{turn}}
  \put(0,56)    {$ [T] \;= $}
  \put(252,56)  {$ [T'] \;= $}
  \end{picture} \nonumber
  \\[.1em]{} \label{eq:dft-choice-TT}
  \end{eqnarray} \Xlabp 110
Here we displayed the graphs \erf{eq:pic41} using blackboard framing. 
The upper and lower boundary in these figures represent the upper
and lower boundary of the connecting manifold $\M_X$.
According to \erf{eq:muk-rel5} in the graph $[T]$ we should substitute
$X_{k_0} \eq (X_{n-k_0+1}')^v$, $X_{k_0+1} \eq (X_{n-k_0}')^v$,
$i_{k_0} \eq i_{n-k_0}'$, $j_{k_0} \eq j_{n-k_0}'$ and
$\vartheta_{k_0} \eq \theta_{(X_{n-k_0}')^\vee} \cir \tilde u(\vartheta_{n-k_0}')$.
This results in
  \bea \begin{picture}(340,84)(0,40)
  \put(50,0)    {\Includeourbeautifulpicture 111 }
  \put(0,56)    {$ [T] \;= $}
  \put(48,65)     {\scriptsize$X'_{n-k_0+1}$}
  \put(273,65)    {\scriptsize$X'_{n-k_0}$}
  \put(112,22)    {\scriptsize$j'_{k_0}$}
  \put(113,95)    {\scriptsize$i'_{k_0}$}
  \put(132.8,71.8){\begin{turn}{270}\small$\vartheta'_{\!n-k_0}$\end{turn}}
  \epicture21 \labl{eq:dft-choice-or1} \Xlabp 111
One can now verify that this ribbon graph fragment can be
deformed into $[T']$ in \erf{eq:dft-choice-TT}.


\sect{Ribbon graphs for structure constants}\label{sec:rc}

The correlation function for an arbitrary world sheet $\X$, which may have 
boundaries and field insertions, and which may or may not be orientable,
can be computed by cutting $\X$ into
smaller pieces and summing over intermediate states. This procedure
is known as `sewing'. In principle, every correlator can be obtained in this 
way from a small number of basic building blocks, namely the correlators of
\\[.5em]
\hspace*{3.5em}--- three boundary fields on the disk;
\\[.4em]
\hspace*{3.5em}--- one bulk field and one boundary field on the disk;
\\[.4em]
\hspace*{3.5em}--- three bulk fields on the sphere;
\\[.4em]
\hspace*{3.5em}--- one bulk field on the cross cap.
\\[.5em]
(The last of these is only needed if non-orientable world sheets are admitted.)
Some correlators involving defect fields will be presented as well, namely
\\[.5em]
\hspace*{3.5em}--- three defect fields on the sphere;
\\[.4em]
\hspace*{3.5em}--- one defect field on the cross cap.
\\[.5em]
However, recall from the discussion in section \ref{sec:fi-defect}
that this does not correspond to the most general way to treat defects. When 
cutting a world sheet involving defect lines into smaller pieces, 
one is forced to also consider fields of
the type \erf{eq:fields-not-consd}, which we will not do in this paper.

\medskip

The aim of this and the next two sections is to derive expressions for the 
building blocks listed above in the form 
  \be
  \mbox{(\,correlator\,)}
  ~=~ \mbox{(\,const\,)} ~\times~ \mbox{(\,conformal block\,)} \,. 
  \labl{corr = c x block}
The constants appearing here, to be referred to as {\em structure constants\/},
will be obtained in sections \ref{sec:rc-bnd3pt}\,--\,\ref{sec:rc-1defect-xcap} 
as ribbon invariants in the 3-d TFT. The conformal blocks are then introduced 
in section \ref{sec:cb}, and finally the connection to correlation functions
is made in section \ref{sec:cf}.

Since a given world sheet $\X$ can be reduced to the fundamental building 
blocks in several different ways, the latter must fulfill a set of consistency 
conditions, the so-called sewing or factorisation constraints 
\cite{bepz,sono2,sono3,cale,lewe3,fips,bppz2}. The correlators computed in 
the TFT approach satisfy these constraints by construction \cite{fffs3,fjfrs}.

\medskip

When giving the correlators on the disk and on the sphere in sections 
\ref{sec:rc-bnd3pt}\,--\,\ref{sec:rc-3defect}, we will fix a (bulk and boundary)
orientation for the world sheet from the start. In this way the expressions
we obtain are valid for oriented CFTs as well. In the unoriented
case we can change the orientation without affecting the result for the 
correlator, by using the equivalence relations of section \ref{sec:fi}.

\subsection{Standard bases of conformal blocks on the sphere}
\label{sec:standbas}

Let the world sheet $\X$ be one of the building blocks listed above. To compute 
the constants in \erf{corr = c x block} we first construct the ribbon graph
representation in $\MX$ of the correlator on $\X$ according to the rules
given in section \ref{sec:fi}. The TFT assigns to $\MX$ an element $C(\X)$
in the space $\calh(\hat \X)$ of blocks on the double $\hat \X$. In the cases
considered here, $\hat\X$ is either the sphere $S^2$ or the disjoint
union $S^2{\sqcup}(-S^2)$.
The constants are the coefficients occurring in the expansion of $C(\X)$ 
in a standard basis of blocks in $\calh(\hat\X)$. In the present section we 
describe our convention for the standard basis in $\calh(S^2)$. In section 
\ref{sec:cb} this basis will be related to products of chiral vertex operators.

\medskip

Take the sphere $S^2$ to be parametrised as $\reals^2 \,{\cup}\, \{\infty\}$.
Define the standard extended surface $E_{0,n}[i_1,\cdots i_n]$ to be $\reals^2 
\,{\cup}\, \{\infty\}$ with the standard orientation, together with the marked 
points $\{((k,0),[\gamma_k],U_{i_k},+) \,|\, k\eq1,...\,,n \}$ and 
$\gamma_k(t)\eq (k{+}t,0)$. If it is clear from the context what 
representations occur at the marked points, we use the shorthand
notation $E_{0,n}$ for $E_{0,n}[i_1,\cdots i_n]$. Let the cobordism
$B_{0,n} \,{\equiv}\, B_{0,n}[i_1,...\,,i_n;
p_1,...\,, p_{n-3};\delta_1 ,...\,,\delta_{n-2}]$ be given by the
three-manifold $\{ (x,y,z)\iN\reals^3\,|\,z\,{\ge}\,0 \}\,{\cup}\,\{\infty\}$
together with the ribbon graph
  \bea \begin{picture}(360,200)(0,44)
  \put(40,0)     {\includeourbeautifulpicture 02 }
  \put(0,124)    {$ B_{0,n}^{} \ = $}
  \put(73.4,68)  {\scriptsize$ z $}
  \put(92,43.7)  {\scriptsize$ y $}
  \put(335,21)   {\scriptsize$ x $}
  \put(108,13)   {\scriptsize$ 1 $}
  \put(170,13)   {\scriptsize$ 2 $}
  \put(216,13)   {\scriptsize$ 3 $}
  \put(306,14)   {\scriptsize$ n $}
  \put(111,62)   {$ U_{\!i_{1}}^{} $}
  \put(158.2,62) {$ U_{\!i_{2}}^{} $}
  \put(205.2,62) {$ U_{\!i_{3}}^{} $}
  \put(295.2,62) {$ U_{\!i_{n}}^{} $}
  \put(139,77)   {\scriptsize$ \delta_{1}^{} $}
  \put(162,123)  {\scriptsize$ \delta_{2}^{} $}
  \put(208.8,213){\scriptsize$ . $}
  \put(131,106)  {$ U_{\!p_{1}^{}}^{} $}
  \put(150,144)  {$ U_{\!p_{2}^{}}^{} $}
  \put(177,197)  {$ U_{\!\overline{\imath_{n}^{}}}^{} $}
  \epicture32 \labl{eq:standard-cobord} \xlabp 02
where the $\delta_m$ label the basis elements (I:2.29) in the spaces of
three-point couplings. For example $\delta_1\eq1,...\,,\N{i_2}{i_1}{p_1}$
labels the basis of $\Hom(U_{i_{2}} \oti U_{i_{1}} , U_{p_{1}} )$.
Note that in \erf{eq:standard-cobord} all ribbons are oriented in such a way 
that their black side is facing the reader. The linear map $Z(B_{0,n},
\emptyset,E_{0,n}) {:}\; \complex \To\calh(E_{0,n})$, applied to $1\iN\complex$,
and with all possible choices of the labels $p_1,...\,,p_{n-3}$ and 
$\delta_1,...\,,\delta_{n-2}$, gives the standard basis in the space 
$\calh(E_{0,n})$. For a proof that this is indeed a basis, see e.g.\ 
lemma 2.1.3 in chapter IV of \cite{TUra} or section 4.4 of \cite{BAki}. 
Situations in which the insertion points are at different positions
or in which the arcs are oriented in a different way can be obtained
from \erf{eq:standard-cobord} by continuous deformation.

\subsection{Boundary three-point function on the disk}\label{sec:rc-bnd3pt}

Consider the correlator of three boundary fields on a disk.  
For future convenience we regard the disk as the upper half plane plus a 
point. What is shown in the picture below is a piece of the upper half plane:
  \bea 
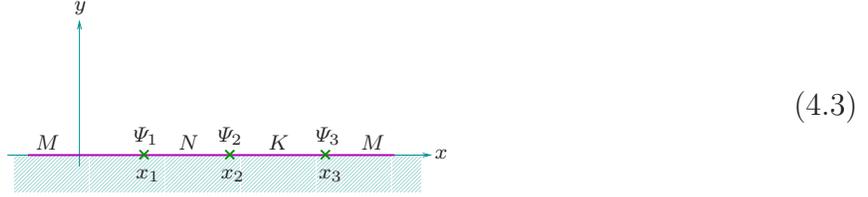
\begin{picture}(160,43)(0,33)
  \put(0,0)      {\Includeourbeautifulpicture 24b }
  \put(11,16)    {\scriptsize$ M $}
  \put(25.6,69)  {\scriptsize$ y $}
  \put(162,12.6) {\scriptsize$ x $}
  \put(49,5)     {\scriptsize$ x_1 $}
  \put(81,5)     {\scriptsize$ x_2 $}
  \put(118,5)    {\scriptsize$ x_3 $}
  \put(48,19)    {\scriptsize$ \hatpsi_1 $}
  \put(80,19)    {\scriptsize$ \hatpsi_2 $}
  \put(117,19)   {\scriptsize$ \hatpsi_3 $}
  \put(134,16)   {\scriptsize$ M $}
  \put(99,16)    {\scriptsize$ K $}
  \put(65,16)    {\scriptsize$ N $}
  \epicture22 \labp 24
On the world sheet we choose the standard orientation for bulk and
boundary, as indicated. The three boundary fields are given by
  \be \bearl
  \hatpsi_1 = (N,M,U_i,\psi_1,x_1,[\gamma_1])\,, \qquad
  \hatpsi_2 = (K,N,U_j,\psi_2,x_2,[\gamma_2])\,, \\{}\\[-.7em]
  \hatpsi_3 = (M,K,U_k,\psi_3,x_3,[\gamma_3])\,.  \eear \ee
Here the coordinates of the
insertion points satisfy $0\,{<}\,x_1\,{<}\,x_2\,{<}\,x_3$, and the arc germs 
$[\gamma_\ell]$ at these points are given by $\gamma_\ell(t) \eq (x_\ell{+}t,0)$
for $\ell\iN\{1,2,3\}$. Further, $U_i$, $U_j$, $U_k$ are simple objects, while
$M, N, K$ are left $A$-modules.

Following the procedure for boundary fields described in section 
\ref{sec:fi-bnd}, one finds that the correlator is determined by the cobordism
  \bea 
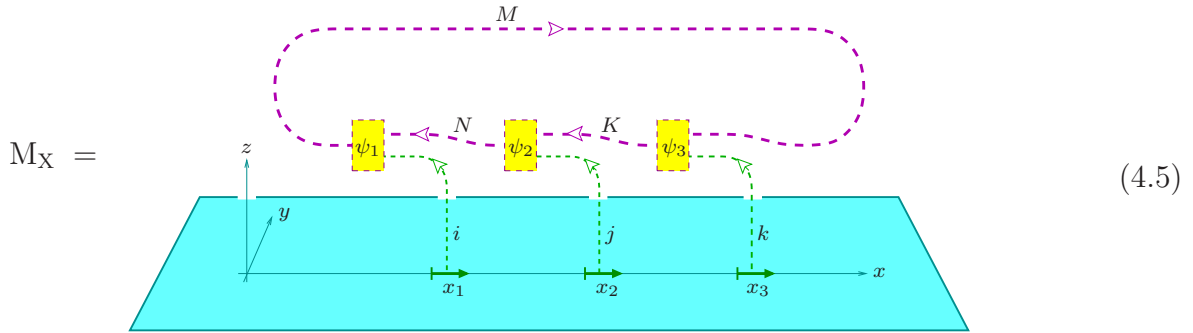
\begin{picture}(380,95)(0,31)
  \put(45,0)     {\includeourbeautifulpicture 25 }
  \put(0,64)     {$ \MX \;= $}
  \put(87.8,68)  {\scriptsize$ z $}
  \put(102,43.7){\scriptsize$ y $}
  \put(327,21)   {\scriptsize$ x $}
  \put(168,35)   {\scriptsize$ i $}
  \put(225.6,35) {\scriptsize$ j $}
  \put(283,35)   {\scriptsize$ k $}
  \put(164,15)   {\scriptsize$ x_1 $}
  \put(222,15)   {\scriptsize$ x_2 $}
  \put(279,15)   {\scriptsize$ x_3 $}
  \put(168,76)   {\scriptsize$ N $}
  \put(184,118)  {\scriptsize$ M $}
  \put(224,76)   {\scriptsize$ K $}
  \put(131,68)   {\scriptsize$ \psi_1 $}
  \put(189,68)   {\scriptsize$ \psi_2 $}
  \put(247,68)   {\scriptsize$ \psi_3 $}
  \epicture19 \labl{eq:uhp-cobord} \xlabp 25
In more detail, this cobordism is obtained as follows. First note that in
the situation at hand any $A$-ribbons resulting from a dual triangulation 
of the world sheet $\X$ can be eliminated by repeatedly using the moves 
(I:5.11) and (I:5.12). Thus only the module ribbons with the field 
insertions as in figure \erf{eq:bnd-ribbon} remain. As explained in section 
\ref{sec:fi-conv}, all ribbons are inserted with orientation opposite to 
that of the world sheet. Also note that the three-manifold displayed in 
\erf{eq:uhp-cobord} is topologically equivalent to the
one indicated in \erf{eq:con-mf-bnd}.

According to section \ref{sec:standbas}, the structure
constants $c(M \hatpsi_1 N \hatpsi_2 K \hatpsi_3 M)_\delta$
are defined to be the coefficients occurring when expanding
\erf{eq:uhp-cobord} in terms of a standard basis for the space
of blocks $\calh(\hat X)$. In the case under consideration, $\hat X$ is the 
sphere with field insertions at $x_1,\, x_2,\, x_3$. The standard basis
is given by the cobordisms
  \bea \begin{picture}(400,77)(0,46)
  \put(90,0)      {\includeourbeautifulpicture 26 }
  \put(0,69)      {$ B(x_1,x_2,x_3)_\delta^{} \;= $}
  \put(133.4,68)  {\scriptsize$ z $}
  \put(148,43)    {\scriptsize$ y $} 
  \put(372,21)    {\scriptsize$ x $}
  \put(213,56)    {\scriptsize$ i $}
  \put(262.7,56)  {\scriptsize$ j $}
  \put(328.6,56)  {\scriptsize$ k $}
  \put(208,15)    {\scriptsize$ x_1 $}
  \put(263,15)    {\scriptsize$ x_2 $}
  \put(323,15)    {\scriptsize$ x_3 $}
  \put(260,109)   {\scriptsize$ \overline{k} $}
  \put(231,89)    {\scriptsize$ \delta $}
  \epicture27 \labp 26
which are obtained from \erf{eq:standard-cobord} by shifting the insertion 
points appropriately. The defining equation for the structure constants thus 
reads
  \be
  Z(\MX,\emptyset,\hat \X) = \sum_{\delta=1}^{\N ij{\bar k}} 
  c(M \hatpsi_1 N \hatpsi_2 K \hatpsi_3 M)_\delta\,
  Z(B(x_1,x_2,x_3)_\delta , \emptyset , \hat \X) \,.
  \labl{eq:uhp-3pt-cobord}
To determine the coefficients $c_\delta$ we compose both sides of the equality 
with $Z(\overline{B}(x_1,x_2,x_3)_\delta,\hat \X,\emptyset)$ from the left,
where $\overline{B}_\delta$ is the cobordism dual to $B_\delta$. This removes 
the summation and results in the desired ribbon invariant expressing $c_\delta$:
  \bea \begin{picture}(420,288)(0,0)
  \put(140,150)     {\begin{picture}(0,0)
  \put(0,0)       {\Includeourbeautifulpicture 27a }
                    \end{picture} }
  \put(140,0)       {\begin{picture}(0,0)
  \put(0,0)       {\Includeourbeautifulpicture 27b }
                    \end{picture} }
  \put(0,219)     {$ c(M \hatpsi_1 N \hatpsi_2 K \hatpsi_3 M)_\delta \;= $}
  \put(110,69)    {$ = $}
  \put(290,87)    {\scriptsize$ N $}
  \put(230,129)   {\scriptsize$ M $}
  \put(230,86)    {\scriptsize$ K $} 
  \put(294,64)    {\scriptsize$ i $}
  \put(245,64)    {\scriptsize$ j $}
  \put(185,44)    {\scriptsize$ k $} 
  \put(250,23)    {\scriptsize$ \overline{k} $} 
  \put(270,38)    {\scriptsize$ \overline{\delta} $} 
  \put(324.3,80)  {\scriptsize$ \psi_1 $}
  \put(266.6,80)  {\scriptsize$ \psi_2 $}
  \put(208.6,80)  {\scriptsize$ \psi_3 $}
  \put(393.7,123.5) {$ \Stn $} 
  \put(206,238)   {\scriptsize$ N $}
  \put(222,280)   {\scriptsize$ M $}
  \put(265,237)   {\scriptsize$ K $} 
  \put(209,215)   {\scriptsize$ i $}
  \put(257,214)   {\scriptsize$ j $}
  \put(324,208)   {\scriptsize$ k $}
  \put(250,160)   {\scriptsize$ \overline{k} $} 
  \put(231.7,188) {\scriptsize$ \overline{\delta} $} 
  \put(172.2,230) {\scriptsize$ \psi_1 $}
  \put(229.7,230) {\scriptsize$ \psi_2 $}
  \put(286.9,230) {\scriptsize$ \psi_3 $}
  \put(394.5,273.7) {$ \Stn $} 
  \epicture{-1}1 \labl{eq:bnd-3pt-rib} \xlabp 27
Here the second equality amounts to rotating the ribbon graph in $S^3$ by 
$180^\circ$ degrees, so that one is looking at the white sides of the ribbons, 
rather than their black sides. 
Also, we introduced the notation $\Stn$ to indicate the graphical representation
of the {\em normalised\/} invariant of a ribbon graph in $S^3$. Recall from 
the end of section I:5.1 that when the picture of a ribbon graph in a 
three-manifold $\M$ is drawn, it represents the invariant $Z(\M)$ 
{\em except\/} when $\M\eq S^3$, in which case it rather represents 
$Z(\M)/S_{0,0}$. Since $S_{0,0}\eq Z(S^3)$, this amounts to assigning the 
value $1$ to the empty graph in $S^3$. To stress this difference,
we label figures representing normalised invariants in $S^3$ by $\Stn$,
rather than by $S^3$. (This notational distinction was not made in 
\cite{fuRs4,fuRs8,fuRs9}, where the normalisation convention for $S^3$ is used 
implicitly.) Recall also that, by definition, we have
  \be 
  S_{0,0}^{~\,-2} = \sum_{i \in \II} \dim(U_i)^2_{} \,.
  \ee  
       
\medskip

Let us express the constant $c(M \hatpsi_1 N \hatpsi_2 K \hatpsi_3 M)_\delta$
in terms of the basis introduced in \erf{eq:MUM-basis}. To this end we take 
the modules $M, N, K$ to be simple. A correlator labelled by reducible
modules can be expressed as a sum of correlators labelled by
simple modules. We set
  \be
  M=M_\mu \,, \quad N=M_\nu \,, \quad K=M_\kappa \ee
and
  \be
  \psi_1 = \psi_{(\nu i)\mu}^{\alpha_1} \,, \quad
  \psi_2 = \psi_{(\kappa j)\nu}^{\alpha_2} \,, \quad
  \psi_3 = \psi_{(\mu k)\kappa}^{\alpha_3} \,.  \ee
It is then straightforward to compute the invariant \erf{eq:bnd-3pt-rib} 
from relation \erf{eq:GA-def} for the fusing matrix of the module category 
$\calca$. The result is
  \be
  c(M \hatpsi_1 N \hatpsi_2 K \hatpsi_3 M)_\delta
  = \dim(\dot M_\mu) \!\! \sum_{\beta=1}^{\Ann{\bar k M_\kappa}{M_\mu}} \!\!\!\!
  \GG[A]^{(\kappa ji)\mu}_{\alpha_2 \nu \alpha_1 \,,\, \beta \bar k \delta}\,
  \GG[A]^{(\mu k \bar k)\mu}_{\alpha_3 \kappa \beta \,,\, \cdot 0 \cdot} \,.
  \labl{eq:3bnd-inv-basis}
The dot `$\,\cdot\,$' stands again for a multiplicity label that can only 
take a single value, the corresponding morphism space being one-dimensional.

Formula \erf{eq:3bnd-inv-basis} is in agreement with equation (4.16) of
\cite{bppz2}. Note that in \cite{bppz2}, the quantities
$^{(1)\!}F$ are taken as an input, whereas the corresponding 
quantities $\GG[A]$ can be computed (with some effort) in terms of 
the multiplication of $A$ and the fusing matrices $\FF$ of $\calc$,
see section \ref{sec:mb-basis}.

Expressions for the structure constants of three boundary fields in rational 
conformal field theory have been obtained previously in various cases
for models with charge conjugation invariant \cite{runk,bppz2,fffs2,fffs3} 
and non-diagonal models \cite{runk2,brSc2}.

\subsection{One bulk and one boundary field on the disk}
\label{sec:rc-1bulk1bnd}

The next correlator we consider is that of a bulk field $\vhi$ and a boundary 
field $\vsi$ on the disk, which we again describe as the upper half plane:
  \bea \begin{picture}(160,32)(0,43)
  \put(0,0)       {\includeourbeautifulpicture 29 }
  \put(150,12)    {\scriptsize$ x $}
  \put(26,68.6)   {\scriptsize$ y $}
  \put(9,17)      {\scriptsize$ M $}
  \put(126,17)    {\scriptsize$ M $}
  \put(105,17)    {\scriptsize$ \vsi $}
  \put(59,49)     {\scriptsize$ \vhi $}
  \put(104,6)     {\scriptsize$ s $}
  \put(60,38.5)   {\scriptsize$ p $}
  \epicture24 \labl{eq:1bulk-1bnd-geom} \xlabp 29
The dotted line indicates the dual triangulation that we will
use later on in the construction of the ribbon graph.
The bulk and boundary fields are the tuples
  \be
  \vhi = (i,j,\phi,p,[\gamma_\phi],\orr_2) \qquad {\rm and} \qquad
  \vsi = (M,M,U_k,\psi,s,[\gamma_\psi]) \,,
  \ee
where $i,j\iN\II$ label simple objects, $\phi\iN\HomAA(U_i\otiP A\otim U_j,A)$ 
and $\psi\iN\HomA(M\oti U_k,M)$. Further, the insertion points are 
$p \,{\equiv}\, (x,y)$ and $s \,{\equiv}\, (s,0)$ with
$y\,{>}\,0$ and $s\,{>}\,x\,{>}\,0$, and the two arc germs
are given by $\gamma_\phi(t)\eq(x{+}t,y)$ and $\gamma_\psi(t)\eq(s{+}t,0)$,
respectively. The orientation $\orr_2$ is the standard
orientation of the upper half plane at the insertion point of the bulk field.

\medskip

The construction of section \ref{sec:fi} yields the following
ribbon graph inscribed in the connecting manifold:
  
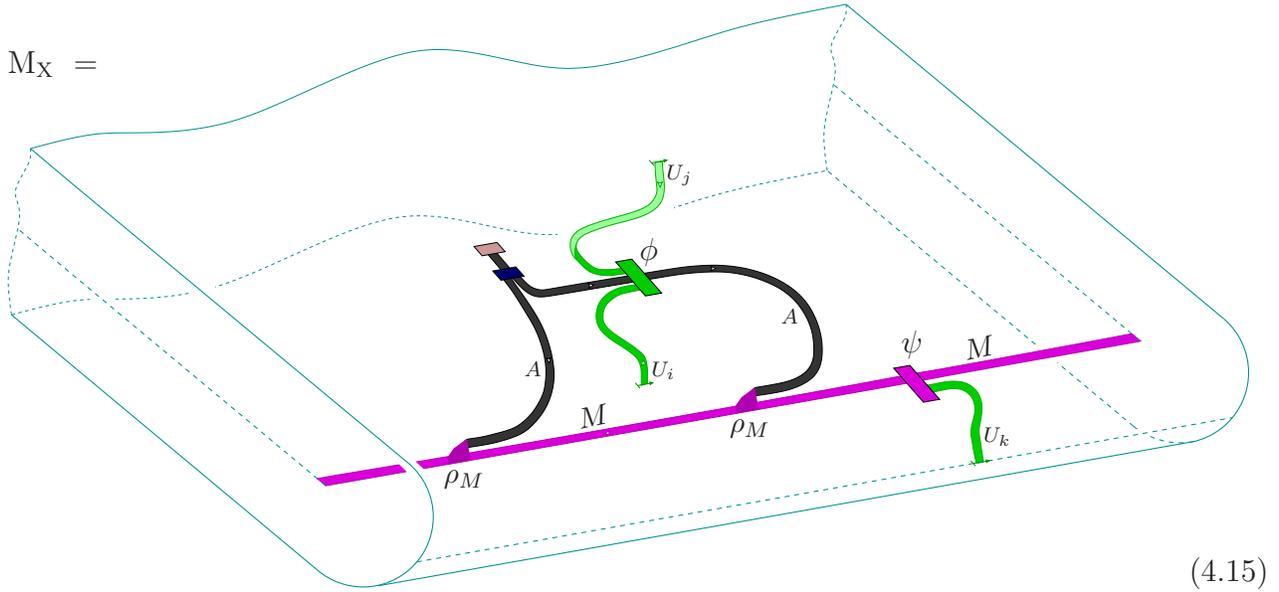
\begin{eqnarray} \begin{picture}(420,225)(15,0)
  \put(0,0)      {\includeourniceLargepicture 99 }
  \put(369,55)     {\scriptsize $U_k$}
  \put(243.5,80)   {\scriptsize $U_i$}
  \put(249.2,154.5){\scriptsize $U_j$}
  \put(165,39.8)   {\small ${\rho_M}$}
  \put(273,58.8)   {\small ${\rho_M}$}
  \put(215,60)     {\begin{turn}{12}\small $M$\end{turn}}
  \put(361,86)     {\begin{turn}{12}\small $M$\end{turn}}
  \put(239,124)    {$\phi$}
  \put(338,89)     {$\psi$}
  \put(195.5,80)   {\scriptsize $A$}
  \put(292.5,100)  {\scriptsize $A$}
  \put(0,195)      {$ \MX~ = $}
  \end{picture} \nonumber
  \\[-1.7em]{} \label{eq:1bulk1bnd-rib}
  \end{eqnarray} \xlabp 99
Here the three-manifold has been drawn as in \erf{eq:con-mf-bnd}.
Also, in \erf{eq:1bulk1bnd-rib} it is understood that the $M$-ribbon
that leaves the picture to the left and to the right is closed to a large 
loop. One may now `unfold' the boundary $\partial\MX$ so as to arrive at 
the equivalent representation of the three-manifold used also in 
\erf{eq:uhp-cobord}. This yields
  \bea 
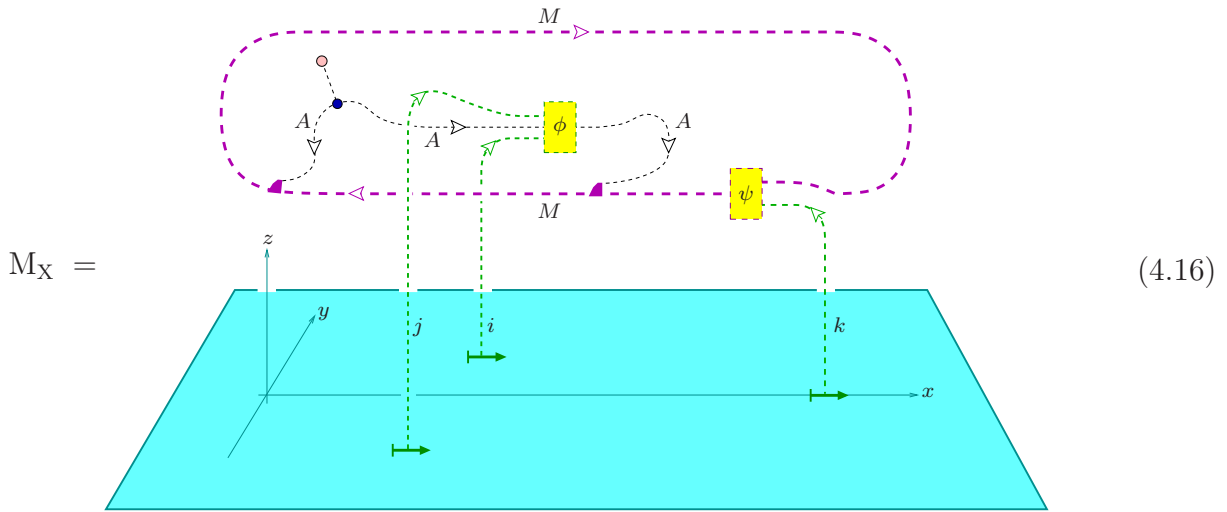
\begin{picture}(400,139)(0,50)
  \put(36,0)     {\Includeourbeautifulpicture 100 }
  \put(96,101)     {\scriptsize$ z $}
  \put(345.4,43.2) {\scriptsize$ x $}
  \put(117,74)     {\scriptsize$ y $} 
  \put(181,68)     {\scriptsize$ i $}
  \put(153,68)     {\scriptsize$ j $}
  \put(312,68)     {\scriptsize$ k $} 
  \put(276,118)    {\scriptsize$ \psi $}
  \put(200,111)    {\scriptsize$ M $}
  \put(108,145)    {\scriptsize$ A $}
  \put(157,138)    {\scriptsize$ A $}
  \put(252,145)    {\scriptsize$ A $} 
  \put(200,185)    {\scriptsize$ M $}
  \put(206,144)    {\scriptsize$ \phi $} 
  \put(0,90)       {$ \MX \;= $}
  \epicture33 \labl{eq:1bulk1bnd-rib-2} \Xlabp 100
To verify that this cobordism is indeed the same as \erf{eq:1bulk1bnd-rib},
one best works in the full ribbon representation and translates
the picture to blackboard framing only at the very end.

The basis of blocks in $\calh(\hat\X)$ in which we expand
$Z(\MX,\emptyset,\hat\X)$ is given by
  \bea \begin{picture}(410,130)(0,43)
  \put(55,0)     {\includeourbeautifulpicture 32 }
  \put(329,34)     {\small $s$}
  \put(249.5,31.4) {\small $3$}
  \put(210,31.4)   {\small $2$}
  \put(171,31.4)   {\small $1$}
  \put(165,14)     {\scriptsize $\bar{p}$}
  \put(207.5,70)   {\scriptsize $p$}
  \put(150,73)     {\scriptsize $y$}
  \put(132,102)    {\scriptsize $z$}
  \put(363,38)     {\scriptsize $x$}
  \put(196,122)    {\scriptsize $U_i$}
  \put(165,122)    {\scriptsize $U_j$}
  \put(180,141)    {\scriptsize $\delta$}
  \put(208,156)    {\scriptsize $U_{\bar k}$}
  \put(334,122)    {\scriptsize $U_k$}
  \put(0,83)       {$ B(x,y,s)_\delta \;= $}
  \epicture29 \labl{eq:Bxys} \xlabp 32
Also indicated in the figure is the continuation path that one has to
choose starting from the standard situation \erf{eq:standard-cobord}.
The structure constants $c(\vhi; M \vsi)_\delta$ are the expansion 
coefficients of \erf{eq:1bulk1bnd-rib-2} in terms of \erf{eq:Bxys},
  \be
  Z(\MX , \emptyset , \hat \X) = 
  \sum_{\delta} c(\vhi; M \vsi)_\delta\,
  Z(B(x,y,s)_\delta , \emptyset , \hat \X) \,.  \labl{eq:1B1b-TFT}
Gluing the cobordism dual to $B(x,y,s)_\delta$ selects one term in the 
sum on the \rhs\ and thus yields, after a rotation by $180^\circ$,
the ribbon graph 
  \bea \begin{picture}(370,133)(0,53)
  \put(86,0)     {\Includeourbeautifulpicture 101 } 
  \put(234,88)   {\scriptsize$ i $}
  \put(270,88)   {\scriptsize$ j $} 
  \put(149,10)   {\scriptsize$ k $}
  \put(205,8)    {\scriptsize$ \overline{k} $} 
  \put(147,108)  {\scriptsize$ \psi $} 
  \put(295,103)  {\scriptsize$ M $}
  \put(305,124)  {\scriptsize$ A $}
  \put(253,129)  {\scriptsize$ A $}
  \put(161,134)  {\scriptsize$ A $} 
  \put(217,175)  {\scriptsize$ M $}
  \put(207,134)  {\scriptsize$ \phi $}
  \put(368.5,172.2){$ \Stn $}
  \put(0,88)     {$ c(\vhi; M \vsi)_\delta \;= $}
  \epicture20 \labl{eq:1bulk-1bnd-sc} \Xlabp 101

\dtl{Remark}{rem:1bulk-uhp}
An interesting special case is the correlator of one bulk
field on the upper half plane, without any boundary field insertion.
The corresponding structure constant $c(\vhi;M)$ can be obtained 
from \erf{eq:1bulk-1bnd-sc} by taking the boundary field to be the 
identity, $\vsi \eq (M,M,\one,\id_{\dot M},[\gamma])$. Doing so, one gets
  \bea \begin{picture}(290,152)(0,46)
  \put(76,0)     {\Includeourbeautifulpicture 102 }
  \put(0,92)     {$ c(\vhi; M) \;= $}
  \put(157,58)   {\scriptsize$ i $}
  \put(178,58)   {\scriptsize$ \overline{i} $}
  \put(175,108)  {\scriptsize$ i $}
  \put(195,108)  {\scriptsize$ \overline{i} $} 
  \put(73,108)   {\scriptsize$ i $} 
  \put(167,80)   {\scriptsize$ \phi $} 
  \put(170.3,39.6) {\scriptsize$ A $}
  \put(160,101)  {\scriptsize$ A $}
  \put(148,130)  {\scriptsize$ A $} 
  \put(190,193)  {\scriptsize$ M $}
  \put(240,66)   {\scriptsize$ M $}
  \put(279,164.5){$ \Stn $}
  \epicture32 \labl{eq:1bulk-sc} \Xlabp 102
To obtain \erf{eq:1bulk-sc} from \erf{eq:1bulk-1bnd-sc}, one needs to get 
rid of one of the $A$-ribbons that are attached to the 
$M$-ribbon, by first moving the two representation morphisms close to each 
other, and then using the representation property (I:4.2), taking the 
resulting multiplication morphism past $\phi$ (which is allowed because 
$\phi$ commutes in particular with the left action of $A$ on itself). The 
resulting $A$-loop can be omitted owing to specialness, so that one
is left with the unit $\eta$. To arrive at \erf{eq:1bulk-sc} it then only 
remains to deform the resulting ribbon graph a bit.
\\
A ribbon graph of the form \erf{eq:1bulk-sc} was already encountered in
the definition of the quantity $\tilde S^A$ in (II:3.89). It follows that
  \be
  c(\vhi; M ) = \frac{\tilde S^A(M;U_i,\tilde\phi)}{S_{0,0}}
  \,, \ee
where $\tilde\phi$ is the morphism enclosed in a dashed box
in \erf{eq:1bulk-sc}. 

\medskip

Next we express the invariant \erf{eq:1bulk-1bnd-sc} in a basis, making the 
simplifying assumption that $\N ijk \iN \{0,1\}$ and that $\langle U_i,A 
\rangle \iN \{0,1\}$. This implies in particular that $A$ is haploid and thus 
simple. By convention, the bimodules are then labelled in such a way that 
$X_0\eq A$. Let us choose the boundary condition and the morphisms for 
the bulk and boundary field in $c(\vhi;M \hatpsi)$ as
  \be
  M = M_\mu \,, \qquad
  \psi = \psi^\alpha_{(\mu k)\mu} \,, \qquad
  \phi = \xi^\beta_{(i0j)0} \,.
  \labl{eq:1bulk1bnd-basis-def}
Note that \erf{eq:1bulk-1bnd-sc} can be slightly simplified by 
removing the $A$-ribbon that arrives at $\phi$ in the same way as explained 
in remark \ref{rem:1bulk-uhp}. Substituting \erf{eq:1bulk1bnd-basis-def} 
into \erf{eq:1bulk-1bnd-sc} leads to the ribbon graph
  \bea \begin{picture}(340,260)(0,0)
  \put(90,30)   {\includeourbeautifulpicture 83}
  \put(206,90)     {\scriptsize $U_i$}
  \put(233,90)     {\scriptsize $U_j$}
  \put(232.4,168.3){\scriptsize $\beta$}
  \put(141.4,78)   {\scriptsize $\alpha$}
  \put(187.4,46)   {\scriptsize $U_{\bar k}$}
  \put(163,46)     {\scriptsize $U_k$}
  \put(224.7,134)  {\scriptsize $A$}
  \put(230,204.3)  {\scriptsize $A$}
  \put(279.6,170)  {\scriptsize $M_{\mu}$}
  \put(94.7,140)   {\scriptsize $M_{\mu}$}
  \put(323,230.3)  {$ \Stn $}
  \put(0,140)      {$ c(\phi;M \hatpsi) ~= $}
  \put(58,0)       {$ \dsty 
            =~ \sum_{m,n,a} \sum_{\rho,\sigma} [\xi^\beta_{(i0j)0}]^{0ja}
            \rho^{{M_\mu}\,(m\rho)}_{a\,(n\sigma)} 
            [\psi^\alpha_{(\mu k)\mu}]^{mn}_{\rho\sigma}\,I(i,j,k,a,m,n) \,,$}
  \epicture-5 \labl{eq:1B1b-basis} \xlabp 83
where in the second step we have substituted the expansions
of $\psi^\alpha_{(\mu k)\mu}$ and $\xi^\beta_{(i0j)0}$
given in \erf{eq:psi-expansion} and \erf{eq:xi-expansion}.
The coefficients $\rho^M$, introduced in (I:4.56), express a representation 
morphism in a basis. The function $I(i,j,k,a,m,n)$ is the ribbon invariant
  \bea \begin{picture}(365,220)(0,0)
  \put(120,0)   {\includeourbeautifulpicture 84}
  \put(0,100)     {$ \dsty I(i,j,k,a,m,n) ~= $}
  \put(220,90)    {\scriptsize $i$}
  \put(282,82)    {\scriptsize $j$}
  \put(196,11.5)  {\scriptsize $k$}
  \put(223,11.5)  {\scriptsize $\bar{k}$}
  \put(251,160)   {\scriptsize $a$}
  \put(254,86.5)  {\scriptsize $n$}
  \put(125,107)   {\scriptsize $m$}
  \put(134,0)     {\small \dashbox{2.2}(11,11){1} }
  \put(201,84)    {\small \dashbox{2.2}(11,11){2} }
  \put(265,121.2) {\small \dashbox{2.2}(11,11){3} }
  \put(320.7,192.5){$ \Stn $}
  \epicture-7 \labl{eq:1B1b-I-def} \xlabp 84
To evaluate this invariant one first applies the appropriate relation from 
appendix II:A.1 to the dashed box marked {\small\dashbox{2.2}(11,11){1}}\,.
Next one inserts a complete basis (I:2.31) at {\small\dashbox{2.2}(11,11){2}}
and removes all braidings with the help of (I:2.41). Finally one applies an 
$\FF$-move (II:2.36) to {\small\dashbox{2.2}(11,11){3}}\,, so as to obtain
  \be
  I(i,j,k,a,m,n) = 
  \GG^{(mk\bar k)m}_{n0}\, \Rs{}nam \dim(U_m)
  \sum_p \frac{\theta_n \theta_i}{\theta_p}\, \Fs nijmap\, \Gs nijmp{\bar k} \,.
  \ee

The correlators for one bulk field and one boundary field on the upper half 
plane, i.e.\ the structure 
constants describing the coupling of bulk to boundary fields, have first been 
considered in \cite{cale}. There, a constraint is derived which these 
coefficients have to solve and an explicit expression for their square is found 
for the case of the charge conjugation invariant. Since then, solutions for the 
coefficients themselves have been found, both for models with charge conjugation
invariant and for more general cases \cite{runk,bppz2,runk2,fgrs,fffs3}.
Also, in equation (5.6) of \cite{pezu6} the bulk-boundary coefficients of 
spinless bulk fields are expressed through $^{(1)\!}F$.  

The coupling of a bulk field to the identity on the boundary, i.e.\ the 
one-point function of a bulk field on the disk as in remark \ref{rem:1bulk-uhp},
is easier to obtain than the general bulk-boundary structure constants, as it 
can already be recovered from the NIM-rep data (multiplicities in the tensor 
products of $\calc$ and $\calca$) and does not require knowledge of the 
6j-symbols \cite{card9,cale}. These coefficients, which determine the boundary 
state, have been calculated for many models.

\subsection{Three bulk fields on the sphere} \label{sec:rc-3bulk}

We now turn to the correlator of three bulk fields 
on the sphere. Analogously as in the discussion of the disk, we describe the 
sphere as $\reals^2$ plus a point. The correlator then takes the form
  \bea \begin{picture}(150,76)(0,27)
  \put(0,0)     {\includeourbeautifulpicture 35 }
  \put(2,100)    {\scriptsize $y$}
  \put(151,27)   {\small $x$}
  \put(39,38)    {\scriptsize $p_1$}
  \put(75,3)     {\scriptsize $p_2$}
  \put(111,55)   {\scriptsize $p_3$}
  \put(38,51)    {\small $\vhi_1$}
  \put(74,17)    {\small $\vhi_2$}
  \put(109,68)   {\small $\vhi_3$}
  \epicture14 \labl{eq:3bulk-geom} \xlabp 35
The dotted line again indicates the dual triangulation that will be used 
in the construction of the ribbon graph. The three bulk fields are tuples
  \be
  \vhi_1 = (i,j,\phi_1,p_1,[\gamma_1],\orr_2) \,, \quad
  \vhi_2 = (k,l,\phi_2,p_2,[\gamma_2],\orr_2) \,, \quad
  \vhi_3 = (m,n,\phi_3,p_3,[\gamma_3],\orr_2) \,. \ee
Here $i,j,k,l,m,n \iN \II$ label simple objects. $\phi_1$ is a morphism in
$\HomAA(U_i\,{\otimes^+}\,A\,{\otimes^-}\,U_j,A)$, and similarly one has
$\phi_2\iN\HomAA(U_k\,{\otimes^+}\,A\,{\otimes^-}\,U_l,A)$ and
$\phi_3\iN\HomAA(U_m\,{\otimes^+}\,A\,{\otimes^-}\,U_n,A)$.  
The three fields are inserted at $p_1$, $p_2$ and $p_3$,
respectively, with arc germs $[\gamma_1]$, $[\gamma_2]$, $[\gamma_3]$ given by
$\gamma_{1,2,3}(t)\eq p_{1,2,3} \,{+}\, (t,0)$. The orientations
$\orr_2$ around the insertion points of the three bulk fields
are given by the standard orientation of $\reals^2$.

    \newpage

The construction of section \ref{sec:fi} gives the following
ribbon graph inscribed in the connecting manifold:
  
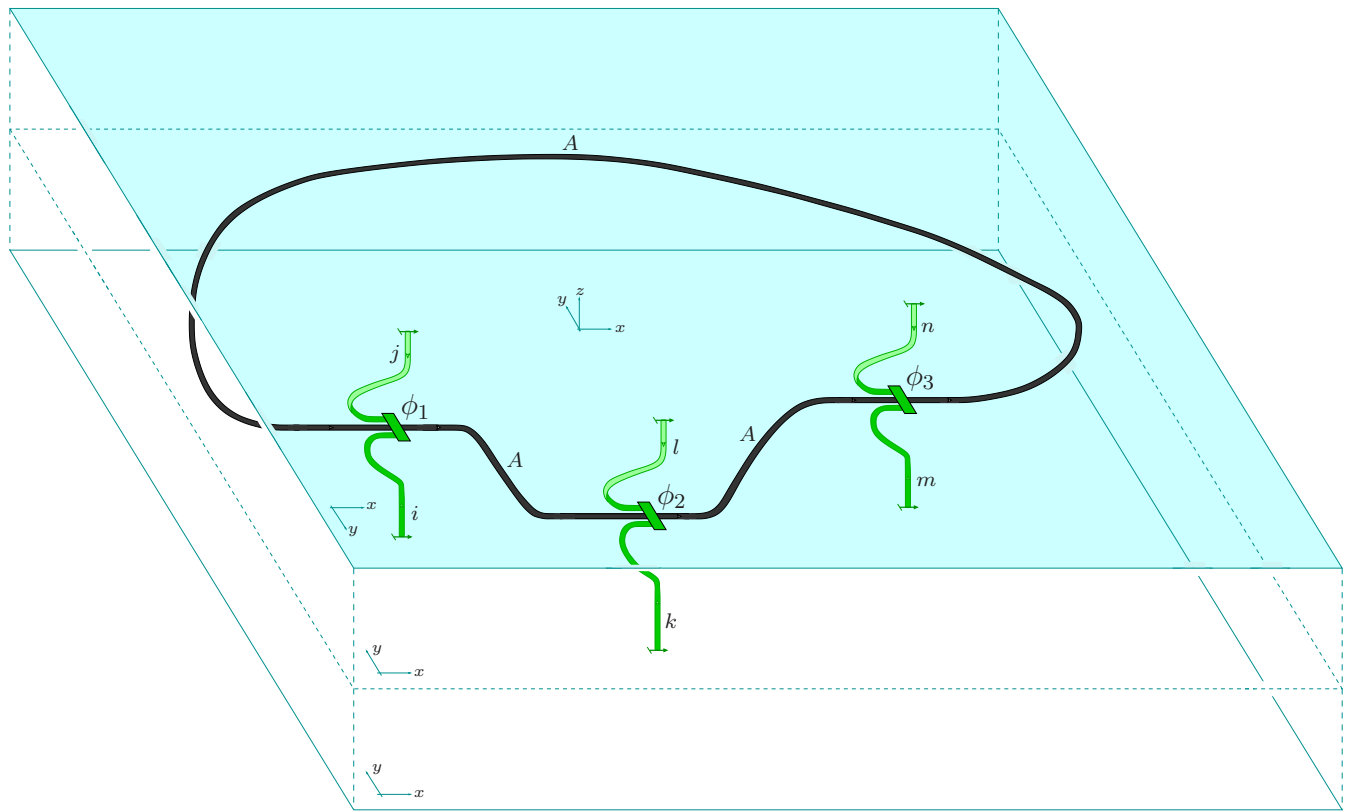
\begin{eqnarray} \begin{picture}(420,333)(25,0)
  \put(0,0)      {\Includeournicehugepicture 103 }
  \put(153,5)       {\tiny $x$}
  \put(137,15)      {\tiny $y$}
  \put(153,51)      {\tiny $x$}
  \put(137,60)      {\tiny $y$}
  \put(135,114)     {\tiny $x$}
  \put(128,105)     {\tiny $y$}
  \put(229,181)     {\tiny $x$}
  \put(207,192)     {\tiny $y$}
  \put(214,195)     {\tiny $z$}
  \put(152,110)     {\scriptsize $i$}
  \put(144,171)     {\scriptsize $j$}
  \put(251,135)     {\scriptsize $l$}
  \put(248,69)      {\scriptsize $k$}
  \put(345,181)     {\scriptsize $n$}
  \put(343.2,123)   {\scriptsize $m$}
  \put(276,140)     {\scriptsize $A$}
  \put(188,130)     {\scriptsize $A$}
  \put(209,250)     {\scriptsize $A$}
  \put(148,150)     {\small $\phi_1$}
  \put(245,116)     {\small $\phi_2$}
  \put(339,161)     {\small $\phi_3$}
  \put(5,320)       {$ \MX~ = $}
  \end{picture} \nonumber
  \\[.5em]{} \label{eq:3bulk-rib}
  \end{eqnarray} \Xlabp 103
As a basis for the space $\calh(\hat\X)$ we take the cobordisms
  \begin{eqnarray} \begin{picture}(420,434)(0,0)
  \put(102,220)  {\Includeourbeautifulpicture 37a }
  \put(175,0)    {\Includeourbeautifulpicture 37b }
  \put(202,371)     {\scriptsize $p_1$}
  \put(200,335)     {\scriptsize $j$}
  \put(217,286)     {\scriptsize $\nu$}
  \put(240,407.5)   {\scriptsize $p_2$}
  \put(246,335)     {\scriptsize $l$}
  \put(307.8,388.5) {\scriptsize $p_3$}
  \put(313.5,322)   {\scriptsize $n$}
  \put(243,267)     {\scriptsize $\bar{n}$}
  \put(360,423)     { $B^3$}
  \put(132,359)     {\tiny $y$}
  \put(143,341)     {\tiny $x$}
  \put(119,364)     {\tiny $z$}
  \put(275,58.5)    {\scriptsize $p_1$}
  \put(273,89)      {\scriptsize $i$}
  \put(288.7,140)   {\scriptsize $\mu$}
  \put(312.5,17.3)  {\scriptsize $p_2$}
  \put(319,90)      {\scriptsize $k$}
  \put(381,38.8)    {\scriptsize $p_3$}
  \put(387,129)     {\scriptsize $m$}
  \put(307,155)     {\scriptsize $\bar{m}$}
  \put(426,198.5)   { $B^3$}
  \put(233,143)     {\tiny $y$}
  \put(242,127)     {\tiny $x$}
  \put(219,151)     {\tiny $z$}
  \put(0,330)      {$ B(p_1,p_2,p_3)_{\mu\nu} \;= $}
  \put(148,90)     {$ \sqcup $}
  \end{picture} \nonumber
  \\[-1.4em]{} \label{eq:3bulk-bas}
  \end{eqnarray} \xlabp 37
where the symbol `$B^3$' indicates that the manifolds are solid three-balls,
with orientations as indicated, while `$\sqcup$'
denotes the disjoint union of the two three-manifolds. 
Expanding \erf{eq:3bulk-rib} in terms of \erf{eq:3bulk-bas} defines the bulk 
structure constants $c(\vhi_1 \vhi_2 \vhi_3)_{\mu\nu}$ via
  \be
  Z(\MX , \emptyset , \hat \X) = 
  \sum_{\mu\nu} c(\vhi_1 \vhi_2 \vhi_3)_{\mu\nu} \,
  Z(B(p_1,p_2,p_3)_{\mu\nu},\emptyset,\hat \X) \,. 
  \labl{Z3bul}
By gluing the cobordism dual to $B(p_1,p_2,p_3)_{\mu\nu}$ we get rid of the 
summation on the \rhs, thus obtaining the ribbon graph 
  \bea \begin{picture}(360,139)(0,40)
  \put(105,0)     {\Includeourbeautifulpicture 104 }
  \put(-65,0){
  \put(218,61)    {\scriptsize $i$}
  \put(262,61)    {\scriptsize $k$}
  \put(267,8)     {\scriptsize $\bar{m}$}
  \put(300,8)     {\scriptsize ${m}$}
  \put(249,32)    {\footnotesize $\bar{\mu}$}
  \put(237.9,80)  {\footnotesize $\phi_1$}
  \put(295.3,80)  {\footnotesize $\phi_2$}
  \put(353.1,80)  {\footnotesize $\phi_3$}
  \put(253.4,85)  {\scriptsize $A$}
  \put(312,85)    {\scriptsize $A$}
  \put(330,173)   {\scriptsize $A$}
  \put(235,130)   {\footnotesize $\bar{\nu}$}
  \put(420.6,165.5){ \small $\Stn$}
  \put(217,102)   {\scriptsize $j$}
  \put(264,101)   {\scriptsize $l$}
  \put(261,150)   {\scriptsize $\bar{n}$}
  \put(310,150)   {\scriptsize ${n}$}
  }
  \put(-10,86)      {$ \dsty c(\vhi_1\vhi_2\vhi_3)_{\mu\nu} \;=\;\frac{1}{S_{0,0}} $}
  \epicture20 \labl{eq:3bulk-sc-aux} \Xlabp 104
The additional factor $1/S_{0,0}$ comes from the normalisation of the cobordism
dual to $B(p_1,p_2,p_3)_{\mu\nu}$, which consists of two copies of $B^3$.

\medskip

Let us assume that the algebra $A$ is simple, and thus by convention 
$X_0\eq A$. Also, we choose the morphisms entering the three bulk fields 
in $c(\vhi_1 \vhi_2 \vhi_3)_{\eps\varphi}$ as
  \be
  \phi_1 = \xi^{\alpha_1}_{(i0j)0} \, , \qquad
  \phi_2 = \xi^{\alpha_2}_{(k0l)0} \, , \qquad
  \phi_3 = \xi^{\alpha_3}_{(m0n)0} \,.
  \ee
Inserting these expressions in \erf{eq:3bulk-sc-aux} and rotating the
resulting graph in such a way that the white side of the $A$-ribbons
faces the reader results in
  \bea \begin{picture}(315,147)(0,52)
  \put(99,0)     {\includeourbeautifulpicture 85 }
  \put(144,75)    {\scriptsize $i$}
  \put(129,75)    {\scriptsize $k$}
  \put(91.5,60)   {\scriptsize $m$} 
  \put(140,32)    {\scriptsize $\bar{m}$}
  \put(153,115)   {\footnotesize $\alpha_1$}
  \put(153,140)   {\footnotesize $\alpha_2$}
  \put(153,163)   {\footnotesize $\alpha_3$}
  \put(173,21)    {\scriptsize $X_0$}
  \put(173.6,121) {\scriptsize $X_0$}
  \put(173.6,148) {\scriptsize $X_0$}
  \put(132,46)    {\footnotesize $\bar{\varepsilon}$}
  \put(301,185)   { \small $\Stn$}
  \put(195,74)    {\scriptsize $j$}
  \put(210.9,74)  {\scriptsize $l$}
  \put(197.7,32)  {\scriptsize $\bar{n}$}
  \put(242,60)    {\scriptsize ${n}$}
  \put(207,49)    {\footnotesize $\bar{\varphi}$}
  \put(-19,94) {$ \dsty c(\vhi_1\vhi_2\vhi_3)_{\eps\varphi}~=~\frac{1}{S_{0,0}} $}
  \epicture36 \labl{eq:3bulk-basis-aux} \xlabp 85
Applying now the definition \erf{eq:FAA-def} of the $\FF[A|A]$-matrices
twice, first to the pair $\alpha_1$, $\alpha_2$ of morphisms, and then to the 
remaining two bimodule morphisms, the invariant \erf{eq:3bulk-basis-aux} becomes
  \be
  c(\vhi_1 \vhi_2 \vhi_3)_{\eps\varphi} = \frac{\dim(A)}{S_{0,0}}
  \sum_{\beta} \FF[A|A]^{(ki0jl)0}_{\alpha_1 0 \alpha_2\,,\,
    \beta \bar m \bar n \eps \varphi} \,
  \FF[A|A]^{(m \bar m 0 \bar n n)0}_{\beta 0 \alpha_3 \,,\,
  \cdot 0 0 \cdot \cdot} \,.
  \labl{eq:3bulk-coeff}
Multiplying this with the inverse of the two-point correlator, one finds that 
the bulk OPE coefficients are given precisely by entries of the matrix 
$\FF[A|A]$, as announced in remark \ref{rem:bulk-6j}.

Of the various types of structure constants considered in this paper, the 
coefficient for the correlator of three bulk fields has the longest history. 
Explicit expressions have first been obtained for the A-series Virasoro minimal 
models \cite{dofa2,dofa3}, and subsequently for many models with charge 
conjugation invariant \cite{knza,zafa2,chfl,musS2,kikkmo} and more general 
theories \cite{dmfs,jf12,fukl,petk2,dotr,kaki,fuks,rest,pezu,bppz2}.

\subsection{Three defect fields on the sphere} \label{sec:rc-3defect}

The calculation for three defect fields is almost the same as for three bulk 
fields. This is not surprising, as a bulk field is just a special kind
of defect field, namely the one connecting the invisible defect to the 
invisible defect. We consider three defect fields on a defect circle,
  \bea 
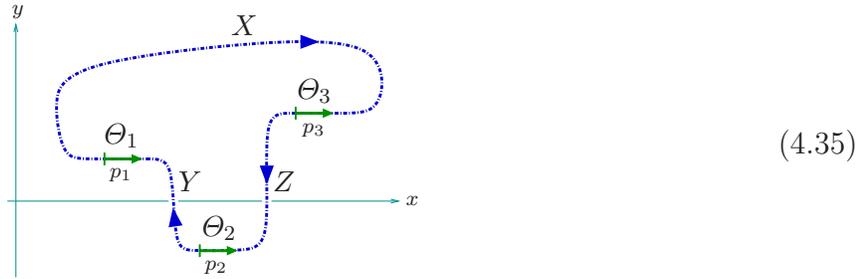
\begin{picture}(160,76)(0,30)
  \put(0,0)     {\includeourbeautifulpicture 46 }
  \put(2,100)     {\scriptsize $y$}
  \put(151,27.7)  {\scriptsize $x$}
  \put(39,38)     {\scriptsize $p_1$}
  \put(75,3)      {\scriptsize $p_2$}
  \put(112,55)    {\scriptsize $p_3$}
  \put(37,51)     {\small $\Thetha_1$}
  \put(74,16)     {\small $\Thetha_2$}
  \put(110,68)    {\small $\Thetha_3$}
  \put(65,33)     {\small $Y$}
  \put(100.8,33)  {\small $Z$}
  \put(85,92)     {\small $X$} 
  \epicture14 \labl{eq:3defect-geom} \xlabp 46
The three defect fields are tuples
  \bea
  \Thetha_1=(X,\orr_2,Y,\orr_2,i,j,\vartheta_1,p_1,[\gamma_1],\orr_2)\,,\\[.6em]
  \Thetha_2=(Y,\orr_2,Z,\orr_2,k,l,\vartheta_2,p_2,[\gamma_2],\orr_2)\,,\\[.6em]
  \Thetha_3=(Z,\orr_2,X,\orr_2,m,n,\vartheta_3,p_3,[\gamma_3],\orr_2)\,.
  \eear\ee
Here $\orr_2$ is the standard orientation of $\reals^2$, $i,j,k,l,m,n\in\II$ 
label simple objects, and $\vartheta_{1,2,3}$ are morphisms
  \bea
  \vartheta_1 \in \HomAA(U_i\,{\otimes^+}\,X\,{\otimes^-}\,U_j,Y) \,,\\[.6em]
  \vartheta_2 \in \HomAA(U_k\,{\otimes^+}\,Y\,{\otimes^-}\,U_l,Z) \,,\\[.6em]
  \vartheta_3 \in \HomAA(U_m\,{\otimes^+}\,Z\,{\otimes^-}\,U_n,X) \,.
  \eear\ee
The three fields are inserted at $p_1$, $p_2$, $p_3$, with arc germs 
$[\gamma_{1,2,3}]$ given by $\gamma_{1,2,3}(t)\eq p_{1,2,3} \,{+}\, (t,0)$. 

\medskip

The ribbon graph differs from the one for three bulk fields only in that the 
annular $A$-ribbon is replaced by the appropriate bimodule ribbons. Thus we 
obtain
  
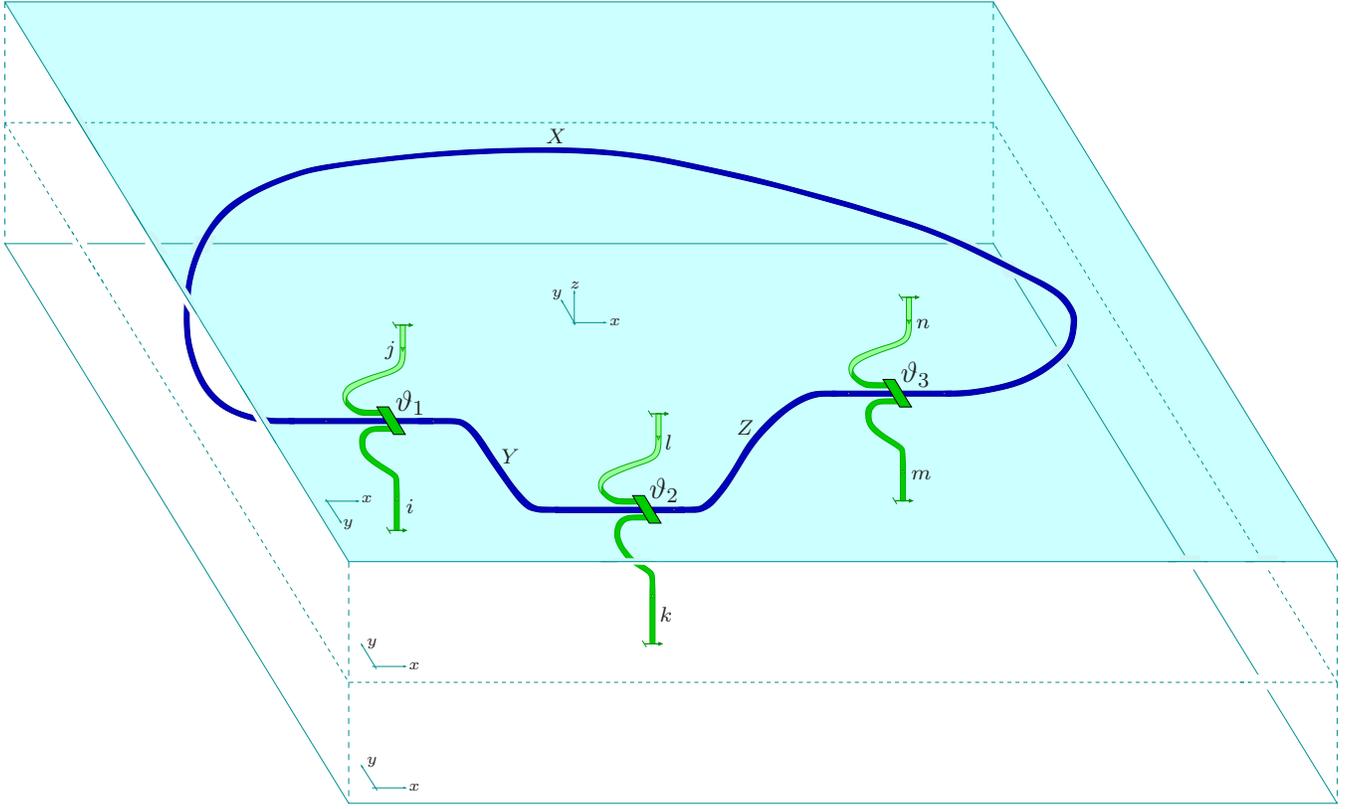
\begin{eqnarray} \begin{picture}(420,333)(25,0)
  \put(0,0)      {\includeournicehugepicture 47 }
  \put(153,5)       {\tiny $x$}
  \put(137,15)      {\tiny $y$}
  \put(153,51)      {\tiny $x$}
  \put(137,60)      {\tiny $y$}
  \put(135,114)     {\tiny $x$}
  \put(128,105)     {\tiny $y$}
  \put(229,181)     {\tiny $x$}
  \put(207,192)     {\tiny $y$}
  \put(214,195)     {\tiny $z$}
  \put(152,110)     {\scriptsize $i$}
  \put(144,170)     {\scriptsize $j$}
  \put(250,134)     {\scriptsize $l$}
  \put(248,69)      {\scriptsize $k$}
  \put(345,180)     {\scriptsize $n$}
  \put(343,123)     {\scriptsize $m$}
  \put(277,140)     {\scriptsize $Z$}
  \put(188,129)     {\scriptsize $Y$}
  \put(205,250)     {\scriptsize $X$}
  \put(148,149)     {\small $\vartheta_1$}
  \put(244,115.7)   {\small $\vartheta_2$}
  \put(339,159.7)   {\small $\vartheta_3$}
  \put(5,320)       {$ \MX~ = $}
  \end{picture} \nonumber
  \\[.5em]{} \label{eq:3defect-rib}
  \end{eqnarray} \xlabp 47
Expanding this in the basis \erf{eq:3bulk-bas} defines
the structure constants of three defect fields via
  \be
  Z(\MX , \emptyset , \hat \X) = 
  \sum_{\mu\nu} c(X,\Thetha_1, Y ,\Thetha_2 ,Z, \Thetha_3, X)_{\mu\nu} \,
  Z(B(p_1,p_2,p_3)_{\mu\nu},\emptyset,\hat \X) \,. 
  \labl{Z3def}
Gluing the cobordism dual to $B(p_1,p_2,p_3)_{\mu\nu}$ yields a ribbon 
invariant very similar to \erf{eq:3bulk-sc-aux}:
  \bea \begin{picture}(400,185)(0,0)
  \put(170,0)   {\includeourbeautifulpicture 86}
  \put(218,61)    {\scriptsize $i$}
  \put(262,61)    {\scriptsize $k$}
  \put(267,8)     {\scriptsize $\bar{m}$}
  \put(300,8)     {\scriptsize ${m}$}
  \put(249,32)    {\footnotesize $\bar{\mu}$}
  \put(237.9,80)  {\footnotesize $\vartheta_1$}
  \put(295.3,80)  {\footnotesize $\vartheta_2$}
  \put(353.1,80)  {\footnotesize $\vartheta_3$}
  \put(256,85)    {\scriptsize $Y$}
  \put(312,85)    {\scriptsize $Z$}
  \put(330,176)   {\scriptsize $X$}
  \put(235,130)   {\footnotesize $\bar{\nu}$}
  \put(421,165.4) { \small $\Stn$}
  \put(218,101)   {\scriptsize $j$}
  \put(264,101)   {\scriptsize $l$}
  \put(261,150)   {\scriptsize $\bar{n}$}
  \put(310,150)   {\scriptsize ${n}$}
  \put(-20,80)    {$ \dsty c(X,\Thetha_1, Y ,\Thetha_2 ,Z, \Thetha_3, X)
                     _{\mu\nu}^{} \;=~ \frac{1}{S_{0,0}} $}
  \epicture{-1}4 \labl{eq:3defect-sc} \xlabp 86
The calculation of the invariant \erf{eq:3defect-sc} runs along the 
same line as in the case of three bulk fields. Let us choose
  \be
  X = X_\mu \,, ~~ Y = X_\nu \,, ~~ Z = X_\kappa \,, \quad
  \vartheta_1 = \xi^{\alpha_1}_{(i \mu j)\nu} \,, ~~
  \vartheta_2 = \xi^{\alpha_2}_{(k \nu l)\kappa} \,, ~~
  \vartheta_3 = \xi^{\alpha_3}_{(m \kappa n)\mu} \,.
  \ee
Inserting this in \erf{eq:3defect-sc} leads to a ribbon invariant similar
to \erf{eq:3bulk-basis-aux}, which can again be evaluated with the help
of the definition \erf{eq:FAA-def}. We find
  \be
  c(X_\mu,\Thetha_1, X_\nu ,\Thetha_2 ,X_\kappa, \Thetha_3, X_\mu)_{\eps\varphi} 
  = \frac{\dim(\dot X_\mu)}{S_{0,0}}
  \sum_{\beta} \FF[A|A]^{(ki\mu jl)\kappa}_{\alpha_1 \nu \alpha_2\,,\,
    \beta \bar m \bar n \eps \varphi} \,
  \FF[A|A]^{(m \bar m \mu \bar n n)\mu}_{\beta \kappa \alpha_3 \,,\,
  \cdot 0 0 \cdot \cdot} \,.  \labl{eq:3def-basis}
Just like for three bulk fields, multiplying this expression with the inverse 
of the two-point correlator one finds that the OPE coefficients for defect 
fields are entries of the matrix $\FF[A|A]$.

\subsection{One bulk field on the cross cap} \label{sec:rc-1bulk-xcap}

After dealing with the fundamental correlators allowing to obtain all
orientable world sheets by sewing, we finally consider the additional
world sheet that is needed to construct all non-orientable surfaces as well,
i.e.\ the cross cap. Correspondingly, the algebra $A$ is now a Jandl algebra.

We present the cross cap as $\reals^2{/}{\sim}$, with the equivalence relation 
$\sim$ identifying points according to the rule $u \,{\sim}\,\sigrp(u)$,
where in standard coordinates $x,y$ on $\reals^2$, $\sigrp$ acts as
  \be
  \sigrp(x,y) := \Frac{-1}{x^2+y^2}\, (x,y) \,.
  \ee 
Consider now the correlator of one bulk field on the cross cap,
with the bulk field given by $\vhi\eq (i,j,\phi,p,[\gamma],\orr_2)$:
  \bea 
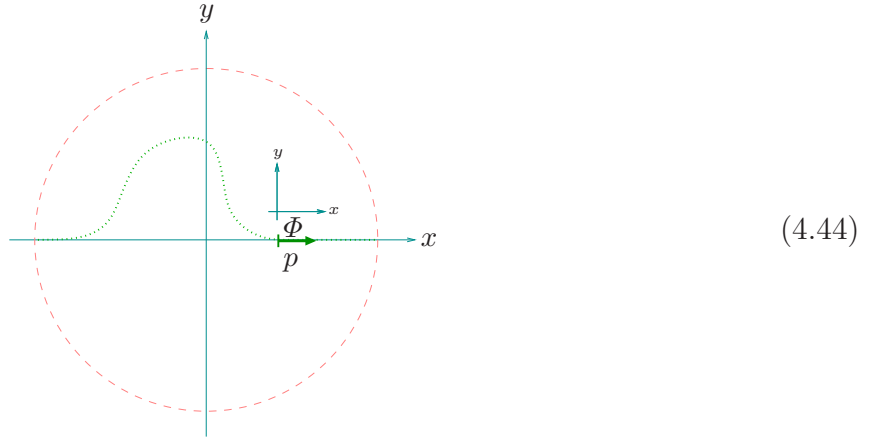
\begin{picture}(160,170)(0,0)
  \put(0,0)   {\Includeourbeautifulpicture 105}
  \put(72,159)     {\small $y$}
  \put(156,72.5)   {\small $x$}
  \put(121,84.5)   {\tiny $x$}
  \put(100,106.4)  {\tiny $y$} 
  \put(104,65)   {\small $p$}
  \put(104,76)   {\small $\vhi$}
  \epicture-4 \labl{eq:1bulk-xcap-geom} \Xlabp 105
As before, $i,j\iN\II$, $\phi\iN\HomAA(U_i\otiP A\otim U_j,A)$
and $\gamma(t)\eq p \,{+}\, (t,0)$, where $p$ is the insertion point
of the bulk field. The local orientation $\orr_2$ around $p$
is chosen as indicated in figure \erf{eq:1bulk-xcap-geom}.
The interior of the dashed circle is a fundamental domain of the equivalence
relation $u \,{\sim}\,\sigrp(u)$. On its boundary, i.e.\ for $|u|\eq 1$,
points are identified according to $u \,{\sim}\,{-}u$.

\medskip

Similarly as in (II:3.10), one can write the connecting manifold as
  \be
  \MX = \hat \X \times [-1,0] \,{/} \sim
  \qquad {\rm with} \quad (u,0) \,{\sim}\, (\sigrp(u),0) 
  {\rm ~~for~all~} u \iN \hat\X \,.  \ee
For the cross cap we obtain in this way the three-manifold
  \bea 
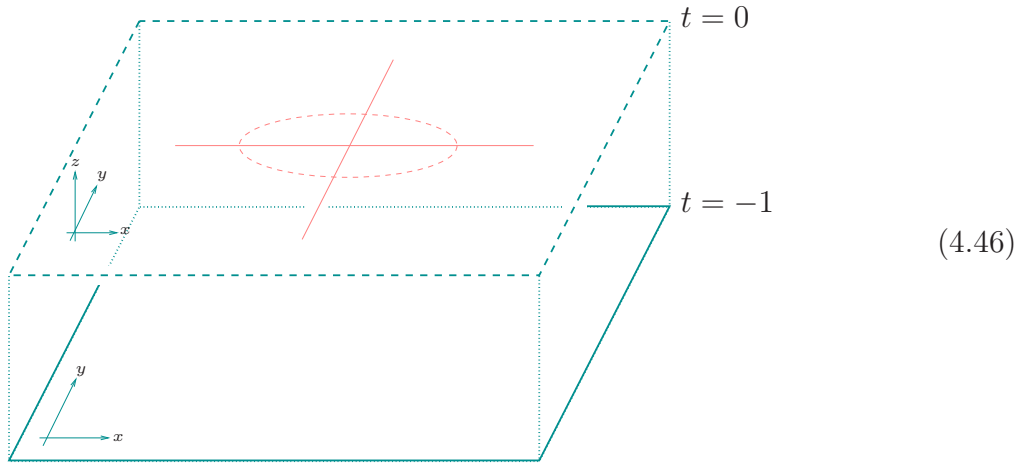
\begin{picture}(280,180)(0,0)
  \put(0,0)   {\Includeourbeautifulpicture 106}
  \put(26,34)      {\tiny $y$}
  \put(40,8.5)     {\tiny $x$}
  \put(42.6,86.7)  {\tiny $x$}
  \put(34,108)     {\tiny $y$}
  \put(24,112.3)   {\tiny $z$}
  \put(251,95)     {\ $t=-1$}
  \put(251,165)    {\ $t=0$}
  \epicture-4 \labl{eq:xcap-3mfld}  \Xlabp 106
with orientations as indicated, and with points in the $t{=}0$\,-plane
to be identified according to $u \,{\sim}\, \sigrp(u)$. 
Next we draw the ribbon graph embedded in the $t{=}0$\,-plane:
  \bea 
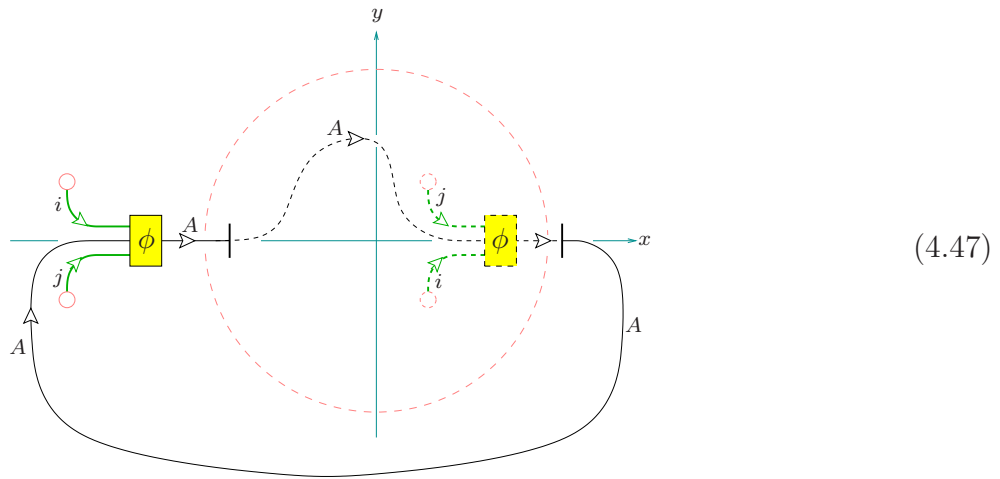
\begin{picture}(260,160)(0,20)
  \put(0,0)   {\Includeourbeautifulpicture 107}
  \put(17,101)     {\scriptsize $i$}
  \put(16.6,74)    {\scriptsize $j$}
  \put(161.5,104.5){\scriptsize $j$}
  \put(160.9,71.9) {\scriptsize $i$}
  \put(182.4,87)   {\small $\phi$}
  \put(48.5,87)    {\small $\phi$}
  \put(0,47)       {\scriptsize $A$}
  \put(65,93)      {\scriptsize $A$}
  \put(120,130)    {\scriptsize $A$}
  \put(233,55)     {\scriptsize $A$}
  \put(137,174)    {\scriptsize $y$}
  \put(238,88.2)   {\scriptsize $x$}
  \epicture07 \labl{eq:xcap-ribgr}  \Xlabp 107
Note that there is only a single field insertion, but as a consequence of the 
identification $u \,{\sim}\,\sigrp(u)$ the ribbon graph is duplicated in the 
picture. In \erf{eq:xcap-ribgr} the $U_i$- and $U_j$-ribbons end in dashed 
circles; this way we indicate the points at which they leave the $t{=}0$-plane 
so as to connect to the marked points on the boundary of $\MX$. 
Recall also the notation (II:3.4) for the element connecting
two $A$-ribbons with opposite orientation.

The complete ribbon graph, lifted out of the $t{=}0$-plane to $t\,{<}\,0$ and 
rotated into blackboard framing, looks as follows
(again one better uses full ribbon notation to verify this move):
  \bea \begin{picture}(360,226)(0,41)
  \put(58,0)   {\Includeourbeautifulpicture 108}
  \put(79.6,34)    {\tiny $y$}
  \put(94.6,6)     {\tiny $x$}
  \put(80,102.2)   {\tiny $x$}
  \put(71,124)     {\tiny $y$}
  \put(61.6,128)   {\tiny $z$}
  \put(310,93)     {\ $t\eq{-}1$}
  \put(310,254)    {\ $t\eq0$}
  \put(204,106)    {\scriptsize $i$}
  \put(126,108)    {\scriptsize $j$}
  \put(204,150)    {\scriptsize $j$}
  \put(227.5,126)  {\small $\phi$}
  \put(183,131)    {\scriptsize $A$}
  \put(282,131)    {\scriptsize $A$}
  \put(0,123)  {$ \MX \;= $}
  \epicture26 \labl{eq:xcap-ribbon-emb}  \Xlabp 108 
The space $\calh(\hat X)$ in this case is one-dimensional if 
$j\eq\bar\imath$ and zero-dimensional otherwise. For $j\eq\bar \imath$ we select
  \begin{eqnarray} \begin{picture}(420,163)(10,0)
  \put(44,0)  {\Includeourbeautifulpicture 52a}
  \put(81,25)      {\tiny $y$}
  \put(87,4)       {\tiny $x$}
  \put(116,70)     {\scriptsize $\bar\imath$}
  \put(211,70)     {\scriptsize $i$}
  \put(242,-70)    {\scriptsize $\bar\imath$}
  \put(338,-70)    {\scriptsize $i$}
  \put(274,159)    {\footnotesize $B^3$}
  \put(452,6)      {\footnotesize $B^3$}
  \put(0,80)       {$ B_2 \;= $}
              \end{picture}
  \nonumber\\  \begin{picture}(420,150)(0,0)
  \put(190,0)  {\Includeourbeautifulpicture 52b}
  \put(221.5,132)  {\tiny $y$}
  \put(260,152.5)  {\tiny $x$}
  \put(160,72)     {$ = $}
  \end{picture}
  \nonumber\\[-2.2em] \label{eq:xcap-aux-1}
  \end{eqnarray} \xlabp 52
as the cobordism describing the basis element in the space of blocks.
The structure constant appearing in
  \be
  Z(\MX , \emptyset , \hat \X) = c(\vhi) \, Z(B_2,\emptyset,\hat \X) 
  \labl{eq:xcap-1bulk-expand}
is again found by composing $Z(\MX,\emptyset,\hat \X)$ with the cobordism dual 
to \erf{eq:xcap-aux-1}. After a suitable rotation of the 
three-manifold, this results in
  \bea \begin{picture}(420,210)(10,40)
  \put(50,0)    {\INcludeourbeautifulpicture 109a}
  \put(282,0)   {\INcludeourbeautifulpicture 109b}
  \put(78,154)     {\scriptsize $\bar\imath$}
  \put(164,159)    {\scriptsize $i$}
  \put(187,118)    {\begin{turn}{90}\small $\phi$\end{turn}}
  \put(146,123)    {\scriptsize $A$}
  \put(165,85)     {\scriptsize $\bar\imath$}
  \put(202.3,73.3) {\tiny $\sigma$}
  \put(206,46)     {\scriptsize $A$}
  \put(238.1,194)  {\footnotesize $\mathbb{RP}^3$}
  \put(317,70)     {\scriptsize $\bar\imath$}
  \put(367,165)    {\scriptsize $i$}
  \put(398,133.4)  {\small $\phi$}
  \put(339,49)     {\scriptsize $A$}
  \put(389,109)    {\scriptsize $A$}
  \put(403,146)    {\scriptsize $A$}
  \put(417,90)     {\scriptsize $\bar\imath$}
  \put(417,222)    {\scriptsize $\bar\imath$}
  \put(367,96.5)   {\tiny $\sigma$}
  \put(437,45)     {\scriptsize $A$}
  \put(454.4,246.5){\footnotesize $\mathbb{RP}^3$} 
  \put(-3,127)     {$\dsty c(\vhi) \;=\; \frac{1}{S_{0,0}} $}
  \put(262,127)    {$ =~ \dsty \frac{1}{S_{0,0}} $}
  \epicture18 \labl{eq:xcap-cobord}  \Xlabp 109

In the second equality, after deforming the ribbon graph, we use that $\phi$
is a morphism of bimodules, $\phi\iN\HomAA(U_i\otiP A\otim U_j,A)$.
Comparing to (II:3.91), we conclude that the resulting ribbon graph is just
  \be
  c(\vhi) = \Gamma^\sigma(U_{\bar\imath},\phi') \,/\, S_{0,0} \,,
  \labl{chatphic}
where $\phi'$ is the morphism inside the dashed box on the \rhs\ of 
\erf{eq:xcap-cobord}. $\Gamma^\sigma$ is the ribbon invariant that
was introduced in (II:3.90), which was computed in a basis in (II:3.110).  

\medskip

One-point functions on a cross cap have been first obtained for free
bosons, fermions and ghosts in \cite{clny2,poca} and for the Ising model in
\cite{fips}. The cross cap Ishibashi states were discussed in \cite{ishi}. 
A set of cross cap one-point functions for the Cardy case in rational CFTs 
was given in \cite{prss,fffs3}, and for the D-series of su(2) in \cite{prss2}. 
Simple current techniques allow one to find more solutions in the Cardy case, as well 
as for more general simple current invariants \cite{huss,huss2,husc,fhssw,brho}. 
In particular, Gepner (and related) models have recently received a lot of 
attention, see e.g.\ \cite{bhhw,blwe,dihs2}.

\subsection{One defect field on the cross cap} \label{sec:rc-1defect-xcap}

As already noted in section \ref{sec:rc-3defect}, replacing bulk fields by 
defect fields changes very little in the calculation of a correlator. Thus we
discuss the correlator of one defect field on the cross cap only very briefly.
The geometric setup for the correlator is 
  \bea 
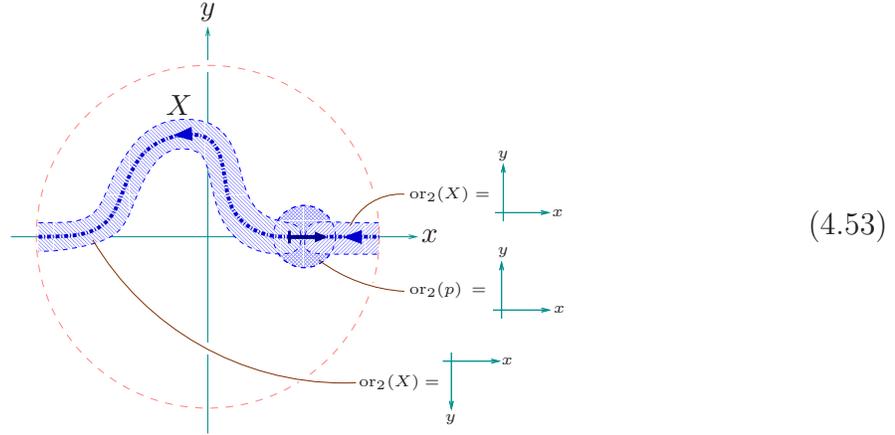
\begin{picture}(180,166)(0,0)
  \put(0,0)   {\Includeourbeautifulpicture 112}
  \put(205,82.4)   {\tiny $x$}
  \put(184.4,104.5){\tiny $y$} 
  \put(205.6,46)   {\tiny $x$}
  \put(184.4,68)   {\tiny $y$} 
  \put(186,26.4)   {\tiny $x$} 
  \put(164.6,5)    {\tiny $y$}
  \put(155.6,72.7) {\small $x$}
  \put(71.9,158.6) {\small $y$}
  \put(59,121)     {\small $X$} 
  \put(151,89.3)   {\tiny $\orr_2(X)=$}
  \put(151,53)     {\tiny $\orr_2(p)\,=$}
  \put(132,18)     {\tiny $\orr_2(X)=$} 
  \epicture-9 \Labp 112
where the defect field is given by 
$\Thetha \eq (X,-\orr_2,X,\orr_2,i,j,\vartheta,p,[\gamma],\orr_2)$ 
with $\gamma(t) \eq p{+}(t,0)$ and $\vartheta\iN\HomAA(U_i\otip X^s\otim U_j,X)$.
The choice of $\gamma$ implies that at the insertion point of the defect field
the defect runs parallel to the real axis. The orientations $\orr_2(p)$ around 
the insertion point and $\orr_2(X)$ around the defect line are chosen as 
indicated.  When giving $\orr_2(X)$ one must take the orientation-reversing 
identification of points on $\reals^2$ into account.

The connecting manifold $\M_X$ can be found by the same method that led to 
\erf{eq:xcap-ribbon-emb}. The only difference is that the $A$-ribbon gets 
replaced by an $X$-ribbon, and that there is a half-twist in accordance with 
\erf{eq:pic41}. The resulting ribbon graph is, in blackboard framing,
  \bea 
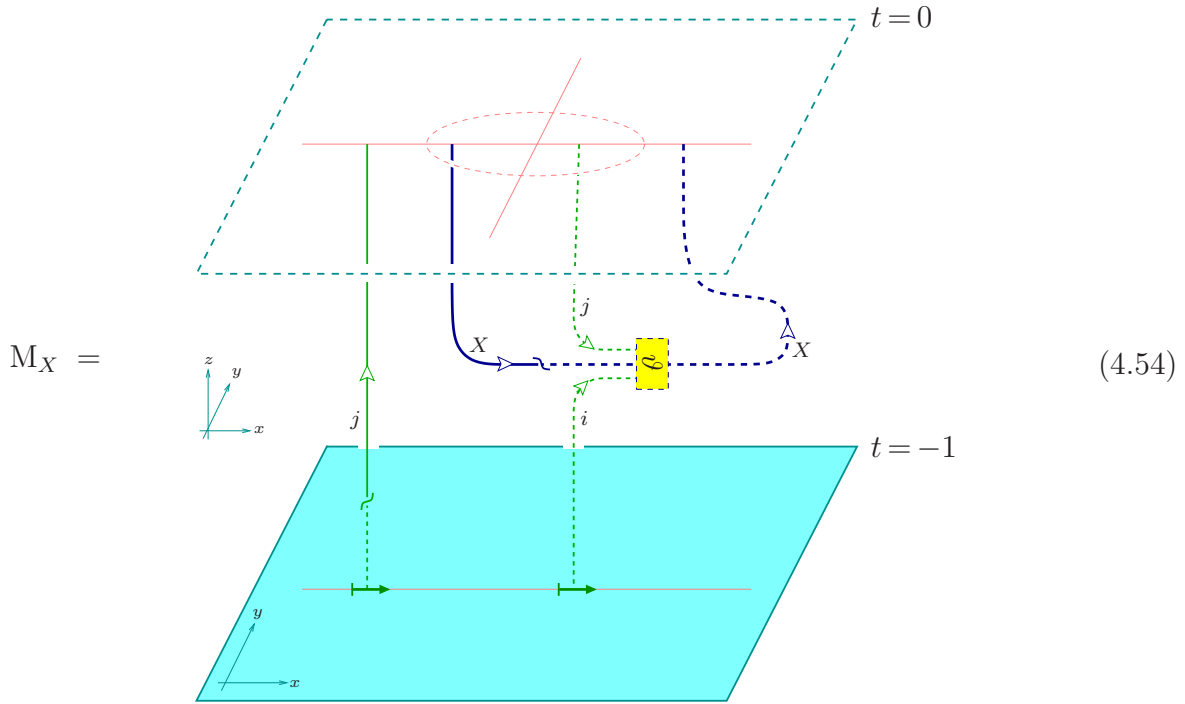
\begin{picture}(370,224)(0,40)
  \put(70,0)    {\Includeourbeautifulpicture 113}
  \put(92.6,102)   {\tiny $x$}
  \put(84,124)     {\tiny $y$}
  \put(73.6,128.2) {\tiny $z$}
  \put(106,6)      {\tiny $x$}
  \put(92,33)      {\tiny $y$}
  \put(322,93)     {\ $t\eq{-}1$}
  \put(322,256)    {\ $t\eq0$}
  \put(216,105)    {\scriptsize $i$}
  \put(129,105)    {\scriptsize $j$}
  \put(216,148)    {\scriptsize $j$}
  \put(174,133)    {\scriptsize $X$}
  \put(296,131)    {\scriptsize $X$}
  \put(239.5,132)  {\begin{turn}{270}\small $\vartheta$\end{turn}}
  \put(0,127)   {$ \M_X \;= $}
  \epicture26 \labl{eq:xcap-def-rib}  \Xlabp 113
Composing with the dual of the cobordism \erf{eq:xcap-aux-1}
one determines the constant $c(X,\Thetha)$ appearing in the relation 
$Z(\MX,\emptyset,\hat\X)\eq c(X,\Thetha)\,Z(B_2,\emptyset,\hat \X)$ to be
  \bea \begin{picture}(300,170)(0,36)
  \put(70,0)    {\Includeourbeautifulpicture 114}
  \put(147,182)    {\scriptsize $i$}
  \put(256,115)    {\scriptsize $X$}
  \put(150,105)    {\scriptsize $X^s_{}$}
  \put(115,182)    {\scriptsize $\bar\imath$}
  \put(204,105)    {\scriptsize $\bar\imath$}
  \put(193,139)    {\small $\vartheta$}
  \put(277.7,193.4){\footnotesize $\mathbb{RP}^3$}
  \put(-15,112)    {$\dsty c(X,\Thetha) ~=~ \frac{1}{S_{0,0}} $}
  \epicture11 \labl{eq:xcap-def-inv}  \Xlabp 114

\medskip

To work out the invariant \erf{eq:xcap-def-inv} we need to introduce 
a bit more notation. Note that the morphism $\vartheta$ entering the data for 
the defect field $\Thetha$ is an element of 
$\HomAA(U_i\,{\otimes^+}\,X^s\,{\otimes^-}\,U_{\bar\imath},X)$.
If we choose $X$ to be the simple bimodule $X_\nu$, then
also $X^s \,{\equiv}\, (X_\nu)^s$ is a simple bimodule. Define a map
  \be
  s:\quad \KK \longrightarrow \KK
  \ee
via $(X_\nu)^s \,{\cong}\, X_{s(\nu)}$. We explicitly choose, once
and for all, a set of isomorphisms
  \be
  \Pi^s_\nu \in \HomAA(X_\nu , (X_{s(\nu)})^s) \,.  \ee
This is similar to the choice of $\pi_i \iN \Hom(U_i,(U_{\bar\imath})^\vee)$ 
made in (I:2.23). For our purposes we need to expand the inverse of 
$\Pi^s_\nu$ in a basis,
  \bea \begin{picture}(200,55)(0,38)
  \put(0,0)    {\Includeourbeautifulpicture 87a}
  \put(190,0)  {\Includeourbeautifulpicture 87b}
  \put(7,91)   {\scriptsize $\dot{X}_{\nu}$}
  \put(7,-9)   {\scriptsize $\dot{X}_{\nu}$}
  \put(0.3,43.4){\tiny $\Bigl(\Pi_{\nu}^s\Bigr)^{\hspace{-0.1cm}-1\!}$}
  \put(189,91) {\scriptsize $\dot X_{\nu}$}
  \put(189,-9) {\scriptsize $\dot X_{\nu}$}
  \put(199,60) {\scriptsize $\beta$}
  \put(198,23) {\scriptsize $\alpha$}
  \put(196.4,41) {\scriptsize $x$}
  \put(50,43)  {$ =~\dsty\sum_{x\in\II}
                   \sum_{\alpha,\beta=1}^{\langle U_x,\dot X_\nu \rangle}
                   \big[ (\Pi^s_\nu)^{-1} \big]^x_{\alpha\beta} $}
  \epicture24 \labl{eq:Pis-expansion} \xlabp 87
In writing this relation we used that as objects in $\calc$, $X_\nu$ and
$(X_{s(\nu)})^s$ are both equal to $\dot X_\nu$. They only differ
in the left and right actions of $A$.

After these preliminaries, let us evaluate the invariant \erf{eq:xcap-def-inv}.
We make again the simplifying assumption that $\N ijk\iN\{0,1\}$.  Let us choose
  \be
  X = X_\nu
  \qquad {\rm and} \qquad
  \vartheta = \xi^\alpha_{(i s(\nu) \bar\imath) \nu}
  \cir \big(\id_{U_i}\Oti\,(\Pi^s_{s(\nu)})^{-1}\Oti\, \id_{U_{\bar\imath}}\big)
  \,. \ee
Substituting this into \erf{eq:xcap-def-inv} results in
  \begin{eqnarray} \begin{picture}(410,220)(0,0)
  \put(75,0)    {\includeourbeautifulpicture 88}
  \put(-5,115)      {$\dsty c(X,\Thetha) ~=~ \frac{1}{S_{0,0}} $}
  \put(152,205)    {\scriptsize $i$}
  \put(200,137)    {\scriptsize $X_{\!s(\nu)}$}
  \put(262,169)    {\scriptsize $X_{\nu}$}
  \put(147,107)    {\scriptsize $(X_{\nu})^s_{}$}
  \put(121,205)    {\scriptsize $\bar\imath$}
  \put(221,115)    {\scriptsize $\bar\imath$}
  \put(210.5,162.9){\scriptsize $\alpha$}
  \put(190.4,124)  {\tiny $\Pi^{s~-1}_{\!s(\nu)}$}
  \put(282.7,217)  {\footnotesize $\mathbb{RP}^3$}
  \end{picture} \nonumber \\[.4em] 
  =\; \dsty \frac{1}{S_{0,0}} \sum_{x,y} \sum_{\gamma_1,\gamma_2}
           [\xi^\alpha_{(i s(\nu) \bar\imath) \nu}]^{xyx}_{\gamma_1 \gamma_2}\,
           [(\Pi^s_{s(\nu)})^{-1}]^x_{\gamma_2 \gamma_1} I(i,x,y) \,. ~~~
  \label{eq:1D-Xcap-basis}
  \end{eqnarray}  \xlabp 88
In the second equality we have replaced $\xi^\alpha_{(i s(\nu) 
\bar\imath) \nu}$ and $(\Pi^s_{s(\nu)})^{-1}$ by their expansions 
\erf{eq:xi-expansion} and \erf{eq:Pis-expansion}, respectively. If one
in addition drags the points where the $U_{\bar\imath}$-ribbon touches the
identification plane along the paths indicated\,\footnote{%
  ~When moving the $U_{\bar\imath}$-ribbon one must
  carefully take into account the identification. As discussed
  at the end of section II:3.5 this introduces additional twists.}
one obtains the following representation of the invariant $I(i,x,y)$:
  \bea \begin{picture}(385,150)(0,75)
  \put(85,0)    {\includeourbeautifulpicture 89}
  \put(94,138)     {\scriptsize $\bar\imath$}
  \put(148,133)    {\scriptsize $i$}
  \put(227,58)     {\scriptsize $\bar\imath$}
  \put(177,54)     {\scriptsize $x$}
  \put(203.8,134)  {\scriptsize $y$}
  \put(281.8,152)  {\scriptsize $\bar\imath$}
  \put(326,152)    {\scriptsize $x$}
  \put(362.7,217)  {\footnotesize $\mathbb{RP}^3$}
  \put(0,123)      {$ I(i,x,y) \;= $}
  \epicture44 \labp 89
To compute $I(i,x,y)$, we apply the relation (II:3.73) to the morphism that is
indicated by the dashed box. Thereby $I(i,x,y)$, which according to \erp 89 is
the invariant of a graph in $\RP$, is expressed as the invariant of a graph in 
$S^3$. In the latter form, $I(i,x,y)$ is straightforward to evaluate; we find
  \be
  I(i,x,y) = S_{0,0}\, \kappa^{-1} \sum_{j\in\II} S_{0,j}\, \theta_j^{\,-2}
  \,\frac{\theta_x}{\theta_y} \, \frac{1}{\Rs{}iyx} \, 
  \Gs yi{\bar\imath}yx0  s_{j,y}
  = P_{0,y} \, \frac{ \theta_x\, \Gs yi{\bar\imath}yx0}{t_y\, \Rs{}iyx} \,.
  \ee
In the second step we made use of the definition of the $P$-matrix 
\cite{bedu2,bisa3}, see (II:3.122), and used the notations
  \be
  P_{i,j} = \kappa^{-1} \hat P_{i,j}\,,\quad
  \hat P_{i,j} = (\hat T^{1/2}  S \hat T^2 S \hat T^{1/2})_{ij}\,,\quad
  \hat T_{ij} = \theta_i^{\,-1} \delta_{ij}\,,\quad
  \hat T_{ij}^{\,1/2} = t_i^{\,-1} \delta_{ij}\,;
  \labl{eq:P-def}
finally, $\kappa \eq S_{0,0} \sum_k\! \theta_k^{\,-1} (\dim\,U_k)^2$
is the charge of the modular category $\calc$. If $\calc$ is the
representation category of a RVOA, it is given by
$\kappa \eq \eE^{\pi \ii c/4}$, with $c$ the central charge of the chiral CFT.

\subsection{Example: The Cardy case} \label{sec:rc-Cardy}

In this section we apply the TFT formalism to the simplest situation,
i.e.\ to the algebra $A\eq\one$. This case has already been treated\,\footnote{%
  ~In \cite{fffs2,fffs3} the structure constants were derived for a specific 
  normalisation of the bulk and boundary fields. We find it helpful to 
  reproduce the results with normalisations kept arbitrary. Also, structure 
  constants of defect fields where not considered in \cite{fffs2,fffs3}.
  }
in \cite{fffs2,fffs3}. Furthermore we make again the
simplifying assumption that $\N ijk\iN\{0,1\}$.

With the exception of the structure constants for three defect
fields on the sphere \erf{eq:3def-basis-Cardy}
and one defect field on the cross cap
\erf{eq:1def-Xcap-bas-Cardy}, the
results of this section are not new and merely serve to illustrate
the general formalism developed above. References to previous
results in the literature are listed at the end of sections
\ref{sec:rc-bnd3pt}\,--\,\ref{sec:rc-1bulk-xcap} above.

\subsubsection*{Field content}

As is immediate from (I:5.30) and (I:5.19), for $A\eq\one$ the torus
partition function is given by the charge conjugation modular invariant, 
$Z(\one)_{ij} \eq \delta_{i,\bar\jmath}$. Thus there is a bulk field $\vhi_i$ 
for each simple object. In the notation of section \ref{sec:fi-bulk}, we choose 
  \be
  \vhi_i := \big(i,\bar\imath, \phi_i, p, [\gamma], \orr_2(p) \big)\,,
  \ee
where the morphism $\phi_i\eq\xi^{\bar\imath}_{(i0\bar\imath)0}$ 
is given by the basis element introduced in \erf{eq:cardy-bases}.

The boundary conditions are in one-to-one correspondence with the simple 
objects $U_i$, too, and the annulus coefficients are given by the fusion rules 
\cite{card9}. Thus we have $\JJ\eq\II$ and $\Ann{i\mu}{\nu} \eq \N \mu i\nu$, 
compare remark I:5.21(i). Nonetheless we continue use greek letters to label 
boundary conditions. Let us denote the field that changes a boundary condition 
$\nu$ to $\mu$ and is labelled by the simple object $U_k$ by $\vsi^{\mu\nu}_k$. 
Thus in terms of the notation of section \ref{sec:fi-bnd}, we choose
  \be
  \vsi^{\mu\nu}_k := (M_\mu, M_\nu, U_k, \psi, p, [\gamma] ) \,,
  \ee
where the morphism $\psi \equiv \psi_{(\mu k)\nu}$ is the basis element
defined in \erf{eq:cardy-bases}.

The defects are in one-to-one correspondence with the simple objects 
$U_i$ as well, so that $\KK\eq\II$. Moreover, the twisted partition 
functions are again expressed in terms of the fusion rules \cite{pezu5};
as follows from the ribbon representation (I:5.151), we have 
$Z^{\mu|\nu}_{kl} \eq \sum_{k\in\II} \N klm \N m\mu\nu$. The number
$Z^{\mu|\nu}_{kl}$ gives the multiplicity of defect fields carrying 
chiral and antichiral labels $U_k$ and $U_l$, respectively, that change a 
$\mu$-defect into
a $\nu$-defect. Let us denote such fields by $\Thetha^{\mu|\nu}_{kl,\alpha}$.
To be specific, in the notation of \ref{sec:fi-defect} we choose
  \be
  \Thetha^{\mu|\nu}_{kl,\alpha} := (X_\mu, \orr_2(p), X_\nu, \orr_2(p), 
  k,l,\xi, p, [\gamma], \orr_2(p))\,,  \ee
where $\xi \equiv \xi^\alpha_{(k\mu l)\nu}$. The multiplicity label $\alpha$ 
takes all values in $\II$ for which $\N\mu l\gamma \N k\gamma\nu \,{\ne}\, 0$.

\subsubsection*{Three boundary fields}

Let us first consider the case of three boundary fields
$\vsi^{\nu\mu}_i$, $\vsi^{\kappa\nu}_j$ and 
$\vsi^{\mu\kappa}_k$ on the disk, changing boundary conditions from $\mu$ 
to $\nu$ to $\kappa$ to $\mu$ as we move along the real axis. Substituting 
the expression \erf{eq:GA-bas-Cardy} for the fusion matrices of $\calca$
into the general expression \erf{eq:3bnd-inv-basis} immediately results in
  \be
  c(M_\mu \vsi^{\nu\mu}_i M_\nu \vsi^{\kappa\nu}_j M_\kappa 
  \vsi^{\mu\kappa}_k M_\mu)
  = \beta^{\nu\mu}_i \beta^{\kappa\nu}_j \beta^{\mu\kappa}_k \,
  \Gs\kappa ji\mu\nu{\bar k}\,\Gs\mu k{\bar k}\mu\kappa 0\, \dim(U_\mu)
  \,.  \labl{eq:3bnd-basis-cardy}
Recall that the constants $\beta^{\nu\mu}_i$ etc.\ give the normalisation
of the boundary fields. 

\dt{Remark}
In view of the results in \cite{runk,bppz2,fffs2,fffs3,pezu6}
one might have expected that the $\FF$-matrices appear in
\erf{eq:3bnd-basis-cardy}, rather than their inverses $\GG$.
This is, however, merely an issue of normalisation and of labelling of 
the boundary conditions, and illustrates the fact that the labelling
of boundary conditions by simple objects is not canonical. Indeed,
at the end of this section we will see that by suitably changing
the normalisation of the boundary fields, in addition
to relabelling the boundary conditions as $\mu \,{\mapsto}\, \bar\mu$,
the structure constants \erf{eq:3bnd-basis-cardy} can equivalently
be expressed in terms of $\FF$-matrices.

\subsubsection*{Three defect fields}

Next we consider three defect fields. The structure constant of three bulk 
fields can then be obtained as a special case thereof. We consider a circular 
defect consisting of three segments of defect types $\mu$, $\nu$ and $\kappa$, 
respectively, see \erf{eq:3defect-geom}. The change of defect type is effected
by three defect fields $\Thetha^{\mu|\nu}_{ij,\alpha_1}$, 
$\Thetha^{\nu|\kappa}_{kl,\alpha_2}$ and $\Thetha^{\kappa|\mu}_{mn,\alpha_3}$.
The corresponding structure constant is obtained by substituting
\erf{eq:FAA-bas-Cardy} into \erf{eq:3def-basis}, resulting in
  \bea\displaystyle
  c(X_\mu, \Thetha^{\mu|\nu}_{ij,\alpha_1}, 
  X_\nu , \Thetha^{\nu|\kappa}_{kl,\alpha_2} ,
  X_\kappa, \Thetha^{\kappa|\mu}_{mn,\alpha_3} , X_\mu)
  \\{}\\[-.7em] \displaystyle \hspace*{7em}
  = 
  \alpha^{\mu|\nu}_{ij,\alpha_1} \alpha^{\nu|\kappa}_{kl,\alpha_2} 
  \alpha^{\kappa|\mu}_{mn,\alpha_3}
  \frac{ \Rs-lj{\bar n}\Rs-n{\bar n}0 }{ t_l\,t_j\,t_n }\, 
  \frac{ \dim(U_\mu) }{S_{0,0}}
  \\{}\\[-.7em] \displaystyle \hspace*{7em}\quad \sum_\beta
  \Gs i{\alpha_1}l{\alpha_2}\nu\beta
  \Gs \mu jl\beta{\alpha_1}{\bar n}
  \Fs ki\beta\kappa{\alpha_2}{\bar m}
  \Gs{\bar m}\beta n{\alpha_3}\kappa\mu
  \Gs\mu{\bar n}n\mu\beta 0
  \Fs m{\bar m}\mu\mu{\alpha_3}0 \,.
  \eear\labl{eq:3def-basis-Cardy}
The constants $\alpha^{\mu|\nu}_{ij,\alpha_1}$ etc.\ determine
the normalisation of the defect fields.

\subsubsection*{Three bulk fields}

The structure constants of three bulk fields $\vhi_i$, $\vhi_j$ and $\vhi_k$
on the sphere are obtained by restricting formula \erf{eq:3def-basis-Cardy} 
to the case $\mu\eq\nu\eq\kappa\eq0$.
The expression then simplifies a lot, leading to
  \be
  c(\vhi_i,\vhi_j,\vhi_k)
  = \alpha_i \, \alpha_j \, \alpha_k \,
  \Rs-{\bar\jmath}{\bar\imath}k \, \frac{t_k}{t_i\,t_j} \,
  \frac{\Gs i{\bar\imath}{\bar\jmath}{\bar\jmath}0k \,
  \Fs jik0{\bar\jmath}{\bar k}} {S_{0,0} \, \dim(U_k)} \,.
  \labl{eq:3bulk-basis-Cardy}
Here $\alpha_i \,{\equiv}\, \alpha^{0|0}_{i \bar\imath, \bar\imath}$
etc.\ give the normalisation of the bulk fields. In deriving
this formula we made use of the relation
  \be
  \Gs p{\bar p}pp00 = \Rs{}{\bar p}p0\, \theta_p\, \dim(U_p)^{-1}\,,
  \labl{eq:G=Rthdim}
which can be verified analogously as the corresponding equation
for $\FF$ in (I:2.45). 

The normalisation of the bulk fields can be read off from the two-point 
structure constant, divided by the squared norm of the vacuum (expressed 
as the two-point function of the identity field).
Setting $m\eq 0$ and $j\eq \bar\imath$ in 
\erf{eq:3bulk-basis-Cardy} and using once more \erf{eq:G=Rthdim} gives
  \be
  \frac{c(\vhi_i,\vhi_{\bar\imath})}{c(\one,\one)}
  = \frac{\alpha_i \, \alpha_{\bar\imath}}{\dim(U_i)} \,.  \ee

\subsubsection*{One bulk and one boundary field}

Consider now the correlator for one bulk field $\vhi_i$ and one boundary
field $\vsi^{\mu\mu}_k$ on the disk, with boundary condition labelled by 
$\mu$. To evaluate the general expression \erf{eq:1B1b-basis}, first note 
that the expansion coefficients of the representation morphism are now 
trivial, $\rho^{M \,(m)}_{0,(m)}\eq 1$. Also recall relation
\erf{eq:cardy-bas-rel} for the expansion coefficients of $\xi$
and $\psi$. We then find
  \be
  c(\vhi_i,M_\mu,\vsi^{\mu\mu}_k) = t_i^{\,-1}\,
  \alpha_i\,\beta^{\mu\mu}_k\,
  \GG^{(\mu k\bar k)\mu}_{\mu 0} \, \dim(U_\mu)
  \sum_p \frac{\theta_\mu \theta_i}{\theta_p} \,
  \Fs \mu i{\bar\imath}\mu 0p \Gs \mu i{\bar\imath}\mu p{\bar k} .
  \labl{eq:1B1b-bas-Cardy}
Of particular interest is the correlator of just one bulk field on the disk,
which is obtained from
\erf{eq:1B1b-bas-Cardy} by setting $k\eq1$. Using the identities
  \be
  \Fs\mu i{\bar\imath}\mu 0p\, \Gs\mu i{\bar\imath}\mu p0
  = \frac{\dim(U_p)}{\dim(U_i)\, \dim(U_\mu)}
  \qquad {\rm and} \qquad
  s_{i,\bar\jmath}
  = \sum_{k\in\II} \frac{\theta_i \, \theta_j}{\theta_k}\, \N ijk \dim(U_k)
  \,, \ee
one finds
  \be
  c(\vhi_i,M_\mu) = 
  \frac{1}{t_i}\, \frac{\alpha_i}{\dim(U_i)}\,s_{i,\bar\mu}\,.
  \labl{eq:Cardy-1disc}
The phase $t_i^{\,-1}$ will cancel a corresponding phase in the correlator, 
see equation \erf{eq:1bulk} below. Also note that here one might have 
expected $s_{i,\mu}$ instead of $s_{i,\bar\mu}$. This difference amounts to 
the freedom to choose the labelling of the boundary conditions at will. 
Indeed, our labelling is related to the one leading to $s_{i,\mu}$ 
by $\mu\,{\leftrightarrow}\,\bar\mu$;
we will return to this point at the end of the section.

\subsubsection*{One defect field on the cross cap}

Consider a defect of type $\nu$ on a cross cap, together with
a defect field insertion $\theta^{s(\nu)|\nu}_{i\bar\imath,\alpha}$.
To evaluate \erf{eq:1D-Xcap-basis} we need to make a definite choice for
the isomorphisms $\Pi^s_\nu$. First note that for $A\eq\one$ one has
$s(\nu)\eq\nu$ for all $\nu\iN\KK$. It is convenient to 
choose $\Pi^s_\nu \eq t_\nu \id_{U_\nu}$. One then finds
  \be
  c(X_\nu,\Theta^{\nu|\nu}_{i\bar\imath,\alpha}) = 
  \alpha^{\nu|\nu}_{i\bar\imath,\alpha}\,
  \Rs-\alpha i\nu \frac{t_\nu}{t_i\,t_\alpha}\,
  \Gs\alpha i{\bar\imath}\alpha\nu 0
  \frac{P_{0,\alpha}}{S_{0,0}} \,.
  \labl{eq:1def-Xcap-bas-Cardy}

\subsubsection*{One bulk field on the cross cap}

The structure constants of one bulk field $\vhi_i$ on the cross cap
are obtained by setting $\nu\eq 0$ in \erf{eq:1def-Xcap-bas-Cardy}:
  \be
  c(\vhi_i) = \frac{\alpha_i}{\dim(U_i)}\, 
  \frac{P_{0,\bar\imath}}{S_{0,0}} \,.
  \labl{eq:cardy-1bulk-xcap}

\subsubsection*{Change of normalisation for boundary fields}

Let us now return to the question of expressing the structure constant 
\erf{eq:3bnd-basis-cardy} of three boundary fields in terms of $\FF$-matrices 
instead of their inverses $\GG$. We would like to demonstrate that there 
is a normalisation of the boundary fields in which \erf{eq:3bnd-basis-cardy}
is expressed in terms of $\FF$, but with all boundary labels
$\mu$, $\nu$, $\kappa$ replaced by their conjugates.

Introduce numbers $\gamma_{\mu i}^\nu$ by relating two different basis 
vectors of $\Hom(U_\mu \oti U_i, U_\nu)$, according to
  \bea \begin{picture}(265,140)(0,7)
  \put(0,0)   {\Includeourbeautifulpicture 90a}
  \put(125,0)   {\Includeourbeautifulpicture 90b}
  \put(-2.3,-9.2)  {\scriptsize $\mu$}
  \put(51,-9.2)    {\scriptsize $i$}
  \put(25,140)     {\scriptsize $\nu$}
  \put(131,-9.2)   {\scriptsize $\mu$}
  \put(165,-9.2)   {\scriptsize $i$}
  \put(126,86)     {\scriptsize $\bar{\mu}$}
  \put(128,37)     {\small $\pi_{\mu}$}
  \put(253,44)     {\scriptsize $\bar{\nu}$}
  \put(242.5,88)   {\small $\pi_{\nu}^{\!-1}$}
  \put(248.5,143)  {\scriptsize $\nu$}
  \put(80,66)      {$ \dsty  =~ \gamma_{\mu i}^\nu $}
  \epicture07 \labl{eq:G-F-aux1} \xlabp 90
A short computation shows that this implies the relation 
  \bea \begin{picture}(325,140)(0,7)
  \put(0,0)   {\Includeourbeautifulpicture 91a}
  \put(170,0)   {\Includeourbeautifulpicture 91b}
  \put(-1.6,141)   {\scriptsize $\mu$}
  \put(51,140)     {\scriptsize $i$}
  \put(24.5,-8.6)  {\scriptsize $\nu$}
  \put(177,145)    {\scriptsize $\mu$}
  \put(210,144)    {\scriptsize $i$}
  \put(171,52)     {\scriptsize $\bar{\mu}$}
  \put(172,100.3)  {\footnotesize $\pi_{\mu}^{\!-1}$}
  \put(298,97)     {\scriptsize $\bar{\nu}$}
  \put(290,47)     {\small $\pi_{\nu}$}
  \put(293,-8.6)   {\scriptsize $\nu$}
  \put(75,66)      {$ \dsty  =~ \frac{\dim(U_\nu)}{\dim(U_\mu)} \,
                   \frac{1}{\gamma_{\mu i}^\nu}  $}
  \epicture00 \labl{eq:G-F-aux2} \xlabp 91
between the dual basis vectors. Consider now the ribbon representation of
$\Gs\kappa ji\mu\nu{\bar k}$ as in (I:2.40). Take the trace of that 
relation and replace the three basis vectors $\lambda_{(\nu i)\mu}$, 
$\lambda_{(\kappa j)\nu}$ and $\Upsilon^{(\kappa \bar k) \mu}$ by the 
corresponding right hand sides of \erf{eq:G-F-aux1} and \erf{eq:G-F-aux2}.
It is then not difficult to see that the resulting ribbon graph can be 
deformed to equal the trace of the left hand side of
(I:2.37). Altogether we obtain
  \be
  \Gs\kappa ji\mu\nu{\bar k}
  = \frac{ \gamma_{\nu i}^\mu \, \gamma_{\kappa j}^\nu}{
    \gamma_{\kappa \bar k}^\mu} \, 
  \Fs ji{\bar \mu}{\bar\kappa}{\bar\nu}{\bar k}\,.  \ee
Substituting this relation into \erf{eq:3bnd-basis-cardy}
results in the equality
  \be
  c(M_\mu \vsi^{\nu\mu}_i M_\nu \vsi^{\kappa\nu}_j M_\kappa 
  \vsi^{\mu\kappa}_k M_\mu)
  = 
  (\gamma_{\nu i}^\mu \, \beta^{\nu\mu}_i) \,
  (\gamma_{\kappa j}^\nu \, \beta^{\kappa\nu}_j) \,
  (\gamma_{\mu k}^\kappa \, \beta^{\mu\kappa}_k) \,
  \Fs ji{\bar\mu}{\bar\kappa}{\bar\nu}{\bar k}\,
  \Fs k{\bar k}{\bar\mu}{\bar\mu}{\bar\kappa}0\, \dim(U_\mu) \,.
  \ee
We conclude that by changing the normalisation of the boundary fields and 
renaming the boundary conditions as $\mu \,{\to}\, \bar\mu$, one 
arrives at the usual expression of the boundary structure constants in 
terms of $\FF$-matrix elements.


\sect{Conformal blocks on the complex plane}\label{sec:cb}

In the study of CFT correlators there are two different aspects,
a complex-analytic and a purely topological one.
The construction of the correlators, as presented in detail in sections 
\ref{sec:mb}\,--\,\ref{sec:rc} for the fundamental correlators \BBB, \Bb, 
\bbb\ and \BX, as well as for the defect correlators \DDD\ and \DX, 
proceeds entirely at the topological level.
Given this construction, one would next like to address the complex-analytic
aspect of the correlators, thereby obtaining them as actual functions of
the insertion points. This aspect is treated in the present section and
in section \ref{sec:cf}. We start in sections \ref{sec:cb-cvo} and 
\ref{sec:cb-spaces} by recalling the concepts of
vertex algebras and bundles of conformal blocks. 

The relation between the 3-d TFT formulation of the correlators
and the complex-analytic description relies on a number of assumptions,
thus restricting the class of vertex algebras $\CA$ to
which our construction can be applied. First, we require the category 
${\Rep}(\CA)$ of representations of $\CA$ to be a modular tensor category. 
Sufficient conditions on $\CA$ which 
ensure this property of ${\Rep}(\CA)$ have been given in \cite{huan21}.

Given $\CA$ we can define bundles of conformal blocks for Riemann surfaces 
with marked points.
Our second condition is to demand that these spaces are finite-dimensional, and 
that for surfaces of arbitrary genus the conformal blocks give rise to the same 
monodromy and factorisation data as the 3-d TFT constructed from ${\Rep}(\CA)$.
So far there is no complete treatment of this issue in the literature. However, 
many results have been obtained, see e.g.~\cite{Mose,HUan,FRbe,BAki,LEli,huan14}.  
For the purposes of this paper we do not need a complex-analytic construction
of correlators in its full generality, though, as we only consider particularly 
simple cases, namely the fundamental correlators listed above. For these, at 
most conformal three-point blocks on the Riemann sphere are needed.

For the purposes of this paper, we will call the class of vertex 
algebras satisfying these conditions rational.
{}From now on we take $\CA$ to be rational in this sense.

Another important issue is the metric dependence of the correlation functions 
through the Weyl anomaly, which arises whenever the Virasoro central charge of 
$\CA$ is nonzero. Here again we do not need a general treatment, because we will
be interested in the ratio of a correlator and of the corresponding correlator
with all field insertions removed (see section \ref{sec:ws-db-cmf} below).
The presence of field insertions does not affect the metric-dependence
of a correlator, and hence the Weyl anomaly cancels in such ratios.

\subsection{Chiral vertex operators}\label{sec:cb-cvo}

In this section we describe the space of conformal
blocks on the Riemann sphere $\mathbb{P}^1$, realised as
$\mathbb{P}^1\eq\complex\,{\cup}\,\{\infty\}$. To this end we collect a 
few notions from the theory of vertex algebras, in particular that
of an intertwiner between representations of vertex algebras.
However, this section neither is intended as a review of vertex algebras, nor 
is it as general as it could be in a vertex algebra setting. Rather, the main
purpose is to specify some notations and conventions needed in the description 
of the CFT correlation functions. For a much more detailed exposition of
vertex algebras the reader may consult e.g.\ \cite{FRbe,LEli,hule3.5}.

\subsubsection*{Vertex algebras and intertwiners}

One formalisation of the physical concept of a chiral algebra for a \cft\ is 
the structure of a conformal vertex algebra $\CA$. This is a vector space 
$\RCA_\Omega$, the space of states, which is $\zet_{\ge 0}$-graded with  
finite-dimensional homogeneous subspaces, equipped 
  with some additional structure. In particular, there is a state-field
  correspondence, assigning to every $W{\in}\, \RCA_\Omega$ a 
field $W(z)\eq Y(W;z) \iN \End(\RCA_\Omega)\fose{z,z^{-1}}$.
Here $\End({\rm V})\fose{z}$, for $\rm V$ a vector space, denotes the space 
of formal power series in the indeterminate $z$ with coefficients in 
$\End({\rm V})$. The state space $\RCA_{\Omega}$, also called the vacuum 
module, contains a distinguished vector $v_\Omega$, the vacuum state, 
for which the associated field is the identity, $Y(v_\Omega;z) \eq 
\id_{\RCA_{\Omega}}$, and for which $Y(u;z)v_\Omega \eq u + O(z)$ for all 
$u\iN\RCA_\Omega$. The grading of the state space $\RCA_{\Omega}$ is given by 
the conformal weight. The vacuum $v_\Omega$ is a non-zero vector in the
component of weight zero, which we assume to be one-di\-men\-si\-o\-nal.
The component of weight two contains the Virasoro vector $v_{\rm vir}$,
whose associated field $Y(v_{\rm vir};z) \eq T(z)$ is the
(holomorphic component of the) stress tensor.

We will be interested in representations of the vertex algebra $\CA$ and 
intertwiners between them. A representation of $\CA$ of lowest conformal weight
$\Delta$ consists of a $(\Delta{+}\zet_{\ge 0})$-graded vector space $\RCA$ and 
a map $\rho{:}~\RCA_{\Omega} \To \End(\RCA)\fose{z^{\pm1}}$, subject to some 
additional consistency conditions. An intertwining operator between
three modules $\RCA_i$, $\RCA_j$, $\RCA_k$ with 
conformal highest weights $\Delta_{i},\, \Delta_{j},\, \Delta_{k}$ is a map
  \be
  \VO ijk{\cdot\,}z :\quad \RCA_j \longrightarrow
  z^{\Delta_i-\Delta_j-\Delta_k} \Hom(\RCA_k,\RCA_i)\fose{z^{\pm1}} \,,
  \ee
satisfying a condition to be described in a moment.

In the sequel we make the simplifying assumption that certain power series 
of our interest possess a non-empty domain of convergence when the indeterminate
$z$ is replaced by a complex variable, again denoted by $z$; 
they can then be regarded as complex-valued functions analytic in that domain.
For instance, suppose we are given some fields $\varphi_i(z) \iN z^{\alpha_i}
\Hom(\RCA_i,\RCA_{i+1})\fose{z^{\pm1 }}$, which may be either fields $W(z)$ in 
the chiral algebra (in which case $\RCA_{i+1}\eq\RCA_i$) or intertwining 
operators $\VO ip{,i{+}1}vz$ for some $\CA$-\rep\ $\RCA_p$ and some 
$v \iN \RCA_p$.
For any two homogeneous vectors $a \iN \RCA_1$ and\,\footnote{~%
  Here $\RCA_i^*$ is the restricted dual of 
  the vector space $\RCA_i$, which by definition is the direct sum 
  of the duals of the (finite-dimensional) homogeneous subspaces of $\RCA_i$,
  i.e.\ $\RCA_i^* \eq \bigoplus_{n=0\!}^{\infty} \RCA_{(n)}^*$ for
  $\RCA_i\eq\bigoplus_{n=0\!}^{\infty} \RCA_{(n)}$.}
$b^* \iN \RCA_{n+1}^*$ we have
  \be
  (b^* , \varphi_n(z_n) \cdots \varphi_1(z_1)\, a)
  \in z_1^{\alpha_1} \cdots z_n^{\alpha_n}\,
  \complex\fose{z_1^{\pm 1},...\,, z_n^{\pm 1}} 
  \labl{bphia}
with suitable numbers $\alpha_i$. It is then assumed that, when taking the 
$z_j$ in the power series \erf{bphia} to be complex variables (and thereby 
interpreting them as field insertion points in the complex plane),
this series converges for all $(z_1,z_2,...\,, z_n)\iN\complex^n$ 
satisfying $|z_n| \,{>}\, \dots \,{>}\, |z_1| \,{>}\, 0$.

As another example where we assume convergence, let us formulate the condition 
for $\VO ijk{\cdot\,}z$ to be an intertwining operator.
For $a^* \iN \RCA_i^*$, $b \iN \RCA_j$ and $c\iN\RCA_k$,
consider the three formal power series
  \bea
  f_1(z) = (\,a^* ,\, \rho_i(W(z))\, \VO ijkbw \,c\,)  \, ,\\{}\\[-.8em]
  f_2(z) = (\,a^* ,\, \VO ijk{\rho_j(W(z{-}w))\,b}w \,c\,) \, ,\\{}\\[-.8em]
  f_3(z) = (\,a^* ,\, \VO ijkbw \, \rho_k(W(z)) \,c\,) \,,
  \eear\ee
which are elements of $w^{\Delta_i-\Delta_j-\Delta_k} \complex
\fose{z^{\pm 1}{,} w^{\pm 1}{,} (z{-}w)^{-1}}$. What we assume is that, with 
$z$ and $w$ interpreted as complex variables, each of these three series has a 
non-zero domain of convergence. Usually $f_1(z)$ converges
for $|z|\,{>}\,|w|$, $f_2(z)$ for $|z{-}w|\,{<}\,|w|$, and $f_3(z)$
for $|w|\,{>}\,|z|\,{>}\,0$. Then the condition for
$\VO ijk{\cdot\,}w$ to be an intertwining operator is that the
three functions $f_{1,2,3}(z)$ coincide upon analytic continuation and 
can have poles only at $z\eq0$, $w$ and $\infty$.

\medskip

As already mentioned, we demand that ${\Rep}(\CA)$ is a modular tensor category.
In particular, $\CA$ has only finitely many inequivalent irreducible 
representations,
and every $\CA$-\rep\ is fully reducible. We choose, once and for all, a set
  \be
  \{ \SCA_i \,|\, i \iN \II \}
  \labl{eq:CA-irreps}
of representatives of isomorphism classes of irreducible $\CA$-modules. By 
$\SCA_{\bar k}$ we denote the element in this list that is isomorphic to the 
dual module ${(\SCA_k)}^\vee_{}$. We also take $\SCA_0\eq\RCA_\Omega$.

The morphisms of the tensor category $\CA$ are intertwiners of $\CA$-modules. 
In particular, $\Hom_\CA(\SCA_j \oti \SCA_k, \SCA_i)$ is the vector space of 
intertwiners $\VO ijk{\cdot\,}z$ between $\SCA_j$, $\SCA_k$ and $\SCA_i$.
As $\CA$ is rational, the dimension $\N jki$ of this space is finite.

\subsubsection*{Choice of basis for intertwining operators}

We choose bases
  \be
  \{ \VOb ijk{\delta}{\cdot\,}z \,|\, \delta\eq1,...\,,\N jki \}
  \subset \Hom_\CA(\SCA_j \oti \SCA_k, \SCA_i) \ee
in the spaces of intertwiners $\VO ijk{\cdot\,}z$. We do not restrict this 
choice, except when $\SCA_j$ or $\SCA_k$ is the vacuum module. Take first 
$j\eq0$; then $\N jki\eq\delta_{ki}$. Thus $\VO i0i{\,\cdot\,}z{:}~
\SCA_0 \,{\to}\,\End(\SCA_i)\fose{z^{\pm1}}$ is unique up to multiplication 
by a scalar. As a basis, we take the \rep\ morphism itself, ${\mathrm V}^i_{0i}
\eq\rho_i$. Then in particular the field corresponding to the vacuum $v_\Omega$ 
is the identity,
  \be
  \VO i0i{v_\Omega}z = \id_{\SCA_i} \,.  \ee
Similarly, for $k\eq0$ we normalise 
$\VO ii0{\,\cdot\,}z{:}~\SCA_i \,{\to}\,\Hom(\SCA_0,\SCA_i)\fose{z^{\pm1}}$
via the condition that 
  \be
  \VO ii0uz v_\Omega = u + O(z)  \ee
for all $u\iN\SCA_i$.

\subsubsection*{Two- and three-point conformal blocks}

Let $v_\Omega^*$ be the unique vector of weight zero in $\RCA_\Omega^*$
satisfying $v_\Omega^*(v_\Omega) \eq 1$. For $X$ a product of
intertwiners, we write
$\langle 0|\, X \,|0\rangle$ for the number $v_\Omega^*(X v_\Omega)$.

For products of at most three intertwiners, the functional dependence 
of $\langle 0|\, X \,|0\rangle$ on the insertion points 
is fixed by conformal invariance.
For the case of two intertwiners one finds explicitly
  \bea\displaystyle
  \langle 0|\, \VO 0{\bar k}kv{z_2}\, \VO kk0u{z_1}\, |0\rangle
  = (z_2{-}z_1)^{-2\Delta_k(u)} B_{\bar k k}(v,u)\,,
  \\{}\\[-.7em]\displaystyle  {\rm with} \qquad
  B_{\bar k k}(v,u) := \langle 0|\,\VO 0{\bar k}kv{2}\,\VO kk0u{1}\,|0\rangle
  \,. \eear\labl{eq:2pt-block}
Here we take $u \iN \SCA_k$ and $v \iN \SCA_{\bar k}$ to be 
homogeneous of conformal weight $\Delta_k(u)$ and $\Delta_{\bar k}(v)$,
respectively. The two-point conformal block vanishes unless 
$\Delta_k(u)\eq\Delta_{\bar k}(v)$. 

For three intertwiners one gets
  $$ \begin{array}{l} 
  \langle 0|\, \VO 0k{\bar k}w{z_3}
            \, \VO {\bar k,\delta}jiv{z_2}
	    \, \VO ii0u{z_1} \,|0\rangle
  \\{}\\[-.7em] \quad\; =
  (z_3{-}z_2)^{\Delta_i(u)-\Delta_j(v)-\Delta_k(w)} 
  (z_3{-}z_1)^{\Delta_j(v)-\Delta_i(u)-\Delta_k(w)} 
  (z_2{-}z_1)^{\Delta_k(w)-\Delta_i(u)-\Delta_j(v)} B^\delta_{kji}(w,v,u)\,,
  \eear $$   \be  {\rm with} \qquad 
  B^\delta_{kji}(w,v,u) := 2^{-\Delta_j(v)+\Delta_i(u)+\Delta_k(w)} \,
  \langle 0|\, \VO 0k{\bar k}w{3}\, \VO {\bar k,\delta}jiv{2}\,
	     \VO ii0u{1} \,|0\rangle \,, \qquad \qquad
  \labl{eq:3pt-block}
where $u\iN\SCA_i$, $v\iN\SCA_j$ and $w\iN\SCA_k$ and, as before,
$\Delta_i(u)$, $\Delta_j(v)$ and $\Delta_k(w)$ are the conformal weights of 
the respective vectors.

\subsubsection*{Local coordinate changes}

For discussing the relationship to three-dimensional topological field theory 
in the next section, one needs to introduce the group $\AutO$ of local 
coordinate changes or, equivalently, locally invertible function germs. We 
describe $\AutO$ briefly; for details see e.g.\ section 6.3.1 of \cite{FRbe}.

The group $\AutO$ can be identified with a subset of the formal power 
series $\complex\fose{z}$ consisting of those $f \iN \complex\fose{z}$ 
that are of the form $f(z)\eq a_1 z + a_2 z^2 + \dots$ with $a_1\,{\ne}\,0$.
Since a $\CA$-module $\RCA$ is in particular a $\Vir$-module,
there is an action $R$ of $(\AutO)_{\rm opp}$ on the module $\RCA$,
  \be
  R\,{:}\quad \AutO \to \End(\RCA) \qquad {\rm such~that} \qquad
  R(\id) = \id \quad{\rm and}\quad R(f\cir g) = R(g) R(f) \,.
  \ee
This action $R$ is constructed as follows. For the coordinate change
$f\iN \AutO$ that is given by $f(z)\eq\sum_{k=1}^\infty a_k z^k$,
set $D \,{:=}\, \sum_{j=1}^\infty\! v_j t^{j+1} \frac\partial{\partial t}$,
with the coefficients $v_j$ determined by solving
$v_0\exp(D)t\eq f(t)$ order by order in $t$. The first few coefficients are
  \be
  v_0 = a_1 \, , \qquad
  v_1 = \frac{a_2}{a_1} \, ,\qquad
  v_2 = \frac{a_3}{a_1} - \Big(\frac{a_2}{a_1}\Big)^2 .
  \ee
The operator $R$ is then given by
$R(f) \,{:=}\, \exp(-\sum_{j=1}^\infty v_j L_j) {v_0}^{\!-L_0}$.
Actually we will later rather need the inverse of $R(f)$, which takes the form
  \be
  R(f)^{-1} = {v_0}^{\!L_0}\, \exp\!\Big(\sum_{j=1}^\infty v_j L_j\Big) \,.
  \labl{eq:Rinv-op}
For example, if $|\Delta\rangle$ is a $\Vir$-highest weight vector of
$L_0$-eigenvalue $\Delta$, then for the first two levels one finds
  \be\bearll
  R(f)^{-1} |\Delta\rangle & = (f')^\Delta\, |\Delta\rangle \qquad{\rm and}
  \\{}\\[-.5em]
  R(f)^{-1} L_{-1} |\Delta\rangle \!\!\!&=  (f')^\Delta\, \big(\,
  f' L_{-1} \, |\Delta\rangle + \Delta f''{/}f'\, |\Delta\rangle \,\big) \,,
  \eear\labl{eq:Rinv-action}
with $f'$ and $f''$ denoting derivatives of $f$ at $z\eq0$,
i.e.\ $f'\eq a_1$ and $f''\eq 2 a_2$.  Define a functional $\kappa_t$ via
  \be
  \kappa_t(f)(z) := f(z{+}t)-f(t) \,.
  \ee
Then for vertex operators we have the identity \cite[section~5.4]{HUan} 
  \be
  Y(v;f(z)) = R(f)^{-1}\, Y(R(\kappa_z(f))v;z) R(f) \,,
  \labl{eq:Y-xfer}
where again $f$ satisfies $f(0)\eq0$ and $f'(0) \neq 0$.


\subsection{Spaces of conformal blocks}\label{sec:cb-spaces}

\subsubsection*{Extended Riemann surfaces}

Recall from section \ref{sec:fi-conv} that the 3-d TFT assigns to every extended
surface $E$ a vector space $\calh(E)$. On the other hand, for the (chiral) 
2-d CFT we need surfaces with a complex structure. The appropriate objects
here are {\em extended Riemann surfaces\/}. An extended Riemann surface
$E^c$ is given by the following data:
\def\leftmargini{1em}
\begin{itemize}
    \item[\Nxt]  
A compact Riemann surface without boundary, also denoted by $E^c$.
    \item[\Nxt]  
A finite ordered set of {\em marked points\/}, i.e.\ triples 
$(p_i,[\varphi_i],\RCA_i)$,
where the $p_i \iN E^c$ are mutually distinct points, $[\varphi_i]$ is
a germ of injective holomorphic functions from a small disk
$D_\delta \,{\subset}\, \complex$ around 0 to $E^c$ such that 
$\varphi_i(0)\eq p_i$, and $\RCA_i$ is a module of the vertex algebra $\CA$.
    \item[\Nxt] 
A Lagrangian submodule $\lambda^c \,{\subset}\, H_1(E^c,\zet)$ --
that is, a $\zet$-submodule $\lambda^c$ of $H_1(E^c,\zet)$ such
that the intersection form $\langle \,\cdot\, , \cdot\, \rangle$
vanishes on $\lambda^c$ and that any $x \iN H_1(E^c,\zet)$ obeying
$\langle x , y \rangle \eq 0$ for all $y\iN\lambda^c$ lies in $\lambda^c$.
\end{itemize}
Given extended Riemann surfaces $E^c$ and $\tilde E^c$ with marked
points $(p_i,[\varphi_i],\RCA_i)$ and $(\tilde p_i,[\tilde \varphi_i],\RCA_i)$, 
respectively, an {\em isomorphism\/} $f{:}~ E^c \To \tilde E^c$ of extended 
Riemann surfaces is a holomorphic bijection of the underlying Riemann surfaces 
that is compatible with the (ordered) marked points in the sense that
  \be
  (\tilde p_i,[\tilde \varphi_i],\RCA_i) = (f(p_i),[f{\circ}\varphi_i],\RCA_i)
  \,. \ee

\subsubsection*{Bundles of conformal blocks}

Let $E^c$ be an extended Riemann surface with marked points
$(p_i,[\varphi_i],\RCA_i)$, $i\eq1,2,...\,,n$. Denote by
${(\RCA_1 \oti \cdots \oti \RCA_n)}^*_{}$ the space of linear functions\,%
  \footnote{~Here ${\otimes}\equiv\otimes_\complex$ denotes the tensor
  product of complex vector spaces, not the tensor product in $\Rep(\CA)$.}
$\RCA_1 \oti \cdots \oti \RCA_n \To \complex$. To an extended Riemann surface 
$E^c$ the chiral CFT assigns the space of conformal blocks on $E^c$, denoted 
by $\calh^c(E^c)$. It is defined as a subspace\,%
  \footnote{~The subspace $\calh^c(E^c)\,{\subset}\,{(\RCA_1\oti\cdots\oti 
  \RCA_n)}^*_{}$ is described via a compatibility condition with the action of 
  $\CA$. For details we refer e.g.\ to
  \cite[section 4]{Scfw} and \cite[chapter 9]{FRbe}.}
of $(\RCA_1 \oti \cdots \oti \RCA_n)^*_{}$. As mentioned at the beginning 
of section \ref{sec:cb} we assume all $\calh^c(E^c)$ to be finite-dimensional.
One can show that if
$E^c$ and $\tilde E^c$ are isomorphic as extended Riemann surfaces, then 
$\calh^c(E^c)\eq\calh^c(\tilde E^c)$. Note that this is an {\em equality\/}
of subspaces of ${(\RCA_1 \oti \cdots \oti \RCA_n)}^*_{}$, not just an 
isomorphism. In other words, there exists a well-defined assignment 
$[E^c] \,{\mapsto}\,\calh^c(E^c)$
on isomorphism classes $[E^c]$ of extended Riemann surfaces.

Denote by $\hat\calm_{g,n}(\RCA_1,\dots,\RCA_n)$ the moduli space of extended 
Riemann surfaces of genus $g$ with $n$ marked points, such that the $k$th 
point is labelled by the $\CA$-module $\RCA_k$. 
If the choice of representations is obvious from the context, or not important
to the argument, we will write $\hat\calm_{g,n}$ in place of
$\hat\calm_{g,n}(\RCA_1,\dots,\RCA_n)$. The same convention will be applied to 
related objects defined below.
Consider the trivial vector bundle
  \be
  \hat\calm_{g,n}(\RCA_1,\dots,\RCA_n)\times {(\RCA_1\oti\cdots\oti\RCA_n)}^*_{}
  \rightarrow \hat\calm_{g,n}(\RCA_1,\dots,\RCA_n) \,.
  \labl{eq:cb-triv-bundle}
Each point in $\hat\calm_{g,n}$ is an isomorphism class
$[E^c]$ of extended Riemann surfaces, to which we can assign the subspace
$\calh^c(E^c) \,{\subset}\,{(\RCA_1 \oti \cdots \oti \RCA_n)}^*_{}$. This 
defines a subbundle of \erf{eq:cb-triv-bundle}, the {\em bundle of conformal 
blocks\/} on $\hat\calm_{g,n}(\RCA_1,\dots,\RCA_n)$; we denote it by
  \be
  \cbs_{g,n}(\RCA_1,\dots,\RCA_n)
  \rightarrow \hat\calm_{g,n}(\RCA_1,\dots,\RCA_n) \,.
  \ee

\medskip

The bundle $\cbs_{g,n}$ comes equipped with a projectively flat connection, 
the \KZ connection, see \cite{frsh2} and e.g.\ \cite[chapter~17]{FRbe}.
Via parallel transport, this connection allows us to assign an
isomorphism $U_\gamma{:}~\calh^c(E^c) \To \calh^c(\tilde E^c)$ to 
every curve $\gamma{:}~[0,1] \To \hat\calm_{g,n}$
with $\gamma(0) \eq [E^c]$ and $\gamma(1) \eq [\tilde E^c]$.

\subsubsection*{Two examples of the map $U_\gamma$}

We will later use two instances of the map $U_\gamma$, the case that $\gamma$ 
corresponds to a change of local coordinates and the case that $\gamma$ amounts 
to moving the insertion points. 
Restricting to these two directions in $\hat\calm_{g,n}$ -- moving
insertion points and changing local coordinates while the complex
structure of the underlying Riemann surface remains unaltered -- 
results in a flat (not only projectively flat) connection.
$U_\gamma$ then depends only on the homotopy class of $\gamma$.

In both examples considered below, $E^c$ denotes an extended 
Riemann surface with marked points $(p_i,[\varphi_i],\RCA_i)$, 
$i\eq1,2,...\,,n$. We will consider curves with $\gamma(0)\eq[E^c]$.

\medskip

For the first example, define a family $E^c(t)$ of extended Riemann surfaces 
by taking $E^c(t)$ to be given by $E^c$, with the local coordinate at the
$k$th marked point replaced by $[\varphi_k \cir \eta_t]$ 
with $\eta_t{:}~ D_\eps \To D_\eps$ such that 
$\eta_t(0)\eq0$ and $\eta_0 \eq \id$. This defines a curve in $\hat\calm_{g,n}$
by setting $\gamma(t)\,{:=}\,[E^c(t)]$. For this curve the map $U_\gamma$ 
takes the form, for $\beta\iN\calh^c(E^c)$,
  \be
  (U_\gamma \beta)(v_1,...\,,v_n) =
  \beta(v_1,...\,,\hat R(\eta)^{-1}v_k,...\,,v_n) \in \calh^c(\tilde E^c)
  \,, \labl{eq:U-coord-change}
where $\tilde E^c \eq E^c(1)$. Note that we cannot directly take the map
$R(\eta_1)^{-1}$ as defined in \erf{eq:Rinv-op} because if $\RCA_k$ 
has non-integral conformal weight, then the expression $(\eta_1')^{\Delta_k}$ 
in \erf{eq:Rinv-action} is only defined up to a phase. Instead, we set
  \be
  \hat R(\eta)^{-1} := \hat r(1) \,,
  \ee
where
  \be \begin{array}{rcl}
  \hat r :\quad [0,1] &\longrightarrow& \End(\RCA_k)\\[2pt]
  t &\longmapsto& R(\eta_t)^{-1} \,,
  \eear\labl{eq:U-coord-Rhat-def}
and where now $R(\eta_t)^{-1}$ is the map defined in \erf{eq:Rinv-op}, with
the phase ambiguity resolved by demanding $\hat r$ to be continuous and to obey
$\hat r(0) \eq \id$. 

\medskip

For the second example consider the curves $\gamma_\delta(t) \,{:=}\, [E^c(t)]$
with $\delta\iN\complex$. Here $E^c(t)$ is again equal to $E^c$, but with the 
$k$th marked point replaced this time by $(p^t_k,[\varphi_k\cir\eta_t],\RCA_k)$, 
where $\eta_t(\zeta)\eq\zeta\,{+}\,t\delta$ and $p_k^t\eq\varphi_k(\eta_t(0))$. 
If $\varphi_k$ is defined on $D_\eps$, then $\varphi_k \cir \eta_t$ is a
well-defined local coordinate around $p^t_k$ as long as $\delta\,{<}\,\eps$.
With these definitions, in $\tilde E^c \eq E^c(1)$ the point $p_k\eq\varphi_k(0)$
has been moved to $\tilde p_k\eq\varphi_k(\delta)$. The map $U_{\gamma_\delta}$
must obey, for each $\beta \iN \calh^c(E^c)$,
  \be
  \frac{\rm d}{{\rm d}\delta} (U_{\gamma_\delta} \beta)(v_1,...\,,v_n)\big|
  _{\delta=0} = \beta(v_1,...\,,L_{-1}v_k,...\,,v_n) \,.
  \labl{eq:KZ-shift-coord}

\subsubsection*{Renumbering marked points}

Given an extended Riemann surface $E^c$ with marked points
$(p_i,[\varphi_i],\RCA_i)$, $i\eq1,2,...\,,n$, and a permutation
$\pi$ of $\{1,2,...\,,n\}$, one can define a new extended Riemann
surface $\tilde E^c \eq P_\pi(E^c)$ by taking the same underlying
Riemann surface, but now ordering the marked points as
$(\tilde p_i,[\tilde \varphi_i],\tilde \RCA_i) \eq 
(p_{\pi^{-1}(i)},[\varphi_{\pi^{-1}(i)}],\RCA_{\pi^{-1}(i)})$.

In the same spirit, one can define an isomorphism
$\calp_\pi {:}~ \calh^c(E^c) \To \calh^c(P_\pi(E^c))$ of conformal block
spaces via $(\calp_\pi \beta)(\tilde v_1,...\,,\tilde v_n)
\,{:=}\, \beta(\tilde v_{\pi(1)},...\,,\tilde v_{\pi(n)})$,
where $\tilde v_k \iN \tilde\RCA_k \eq \RCA_{\pi^{-1}(k)}$. This extends
to a bundle isomorphism, which we will also denote by $\calp_\pi$,
  \be
  \calp_\pi :\quad \cbs_{g,n}(\RCA_1,...\,,\RCA_n)
  \longrightarrow \cbs_{g,n}(\RCA_{\pi^{-1}(1)},...\,,\RCA_{\pi^{-1}(n)}) \,.
  \ee

\subsubsection*{Conformal blocks on the Riemann sphere}

Let us describe the bundle $\cbs_{0,n}$ more explicitly. Define the
standard extended Riemann surface $E^c_{0,n}[i_1,i_2,...\,,i_n]$
to be the Riemann sphere, realised as $\complex \,{\cup}\, \{\infty\}$, 
with marked points $\{(k,[\varphi_k],\SCA_{i_k}) \,|\, k\eq1,2,...\,,n \}$,
with $\varphi_k$ given by
$\varphi_k(z)\eq z{+}k$. When it is clear from the context what 
representations occur at the marked points, we use again the shorthand
notation $E^c_{0,n}$ for $E^c_{0,n}[i_1,i_2,...\,,i_n]$.
A basis for $\calh^c(E^c_{0,n})$ is provided by the conformal blocks
  \be
  \beta_{0,n}[i_1,...\,,i_n \,;\, p_1,...\,,p_{n-3} \,;\,
  \delta_1,...\,,\delta_{n-2}] \,\equiv\, \beta_{0,n} \,,
  \ee
given by a product of intertwiners,
  \be
  \beta_{0,n} :\quad
  v_1{\otimes}\cdots{\otimes}v_n \, \mapsto\,
  \langle 0 |\, \VOs^0_{i_n,\overline{i_n}}\big(v_n,n\big)  \cdots
  \VOs^{p_1,\delta_1}_{i_2,i_1}\big(v_2,2\big)\,
  \VOs^{i_1}_{i_1,0}\big(v_1,1\big) \,|0\rangle \,,
  \labl{eq:standard-block}
where $i_1,...\,,p_{n-3} \iN \II$ and 
$\delta_1,...\,,\delta_{n-2}$ are multiplicity labels.

The moduli space $\hat\calm_{0,n}$ is connected,
and the \KZ connection on $\hat\calm_{0,n}$ is flat. 
To define the bundle $\cbs_{0,n}$, pick a curve $\gamma$ from 
$[E^c_{0,n}]$ to some $[\tilde E^c] \iN \hat\calm_{0,n}$. Then we have
$\calh^c(\tilde E^c) \eq U_\gamma(\calh^c(E^c_{0,n}))$, as a subspace
of ${(\SCA_{i_1} \oti \cdots \oti \SCA_{i_n})}^*_{}$. 
(Note that while this defines the subspace, it does not provide a canonical
basis, since $U_\gamma$ itself depends on the homotopy class of $\gamma$,
whereas the image of $U_\gamma$ does not.) 
In fact, if $E^c$ is given by $\complex \,{\cup}\, \{\infty\}$ with
marked points $(z_k,[\varphi_k],\SCA_{i_k})$, then a basis of $\calh^c(E^c)$
can be given in terms of products of intertwiners,
  \bea\displaystyle
  \!\!\!\!\beta_{p_1,...\,,p_{n-3} \,;\, \delta_1 \dots \delta_{n-2}} : 
  \quad v_1{\otimes}\cdots{\otimes}v_n 
    \\[3pt]\displaystyle
  \qquad\ \longmapsto\; \langle 0 |\, 
  \VOs^0_{i_n,\overline{i_n}}
    \!\Big(R\big(\kappa_0(\varphi_n)\big)^{-1} v_n \,,\, z_n\Big)\,
  \VOs^{\overline{i_n},\delta_{n-2}}_{i_{n-1},p_{n-3}}
    \!\Big(R\big(\kappa_0(\varphi_{n-1})\big)^{-1} v_{n-1}\,,\,z_{n-1}\Big)\,
  \cdots
  \\{}\\[-.7em]\displaystyle
  \hsp{9.9} \cdots\,
  \VOs^{p_1,\delta_1}_{i_2,i_1}
    \!\Big(R\big(\kappa_0(\varphi_2)\big)^{-1} v_2\,,\,z_2\Big)\,
  \VOs^{i_1}_{i_1,0}
    \!\Big(R\big(\kappa_0(\varphi_1)\big)^{-1} v_1\,,\,z_1\Big)\,
  |0\rangle \,.
  \eear\labl{eq:H0n-conf-block}
The choice of branch cuts that are required when evaluating 
\erf{eq:H0n-conf-block} is related to the path in $\hat\calm_{0,n}(i_1,...\,,
i_n)$ chosen to obtain \erf{eq:H0n-conf-block} from \erf{eq:standard-block}.


\subsection{Relation to TFT state spaces}\label{sec:cb-link}

In the three-dimensional topological field theory
given by the Chern\hy Simons path integral, the space of states on a
component of the boundary of a three-manifold
can be thought of as the space of conformal blocks of the corresponding 
WZW model on that same surface \cite{witt27}. One can in fact 
extract the \KZ equation for field insertions
on the complex plane from the Chern\hy Simons path integral \cite{frki2}.

In our approach to topological and conformal
field theory the path integral does not play any role. Instead, the space of
states for the 3-d TFT and the space of conformal blocks for the 2-d CFT 
are constructed independently, and it is a separate task to find an isomorphism
between the two spaces. In this section we make such an isomorphism explicit
in the case that the surface is the Riemann sphere.
The TFT that is used in the present and the following section
is the one obtained from the modular tensor category $\calc \eq {\rm Rep}(\CA)$
of representations of the chiral algebra $\CA$.

\subsubsection*{Extended surfaces from extended Riemann surfaces}

To an extended Riemann surface $E^c$ one can assign an extended surface  
  \be
  E = \ff(E^c) 
  \ee
as follows. As a two-manifold, $E$ coincides with $E^c$. Further, a marked point
$(p,[\varphi],\RCA)$ on $E^c$ becomes the marked point $(p,[\gamma],\RCA,+)$ on 
$E$, where the arc-germ $[\gamma]$ is defined by $\gamma(t)\,{:=}\,\varphi(t)$, 
that is, by the real axis of the local coordinate system around $p$. 
The Lagrangian subspace of $H_1(E,\reals)$ is obtained as $\lambda\eq \lambda^c 
\,{\otimes}_{\zet}\, \reals$ from $\lambda^c \,{\subset}\, H_1(E^c,\zet)$.
The assignment $\ff$ thus acts on extended Riemann surfaces by forgetting the 
complex structure and by replacing germs of local coordinates by germs of arcs.%
  \,\footnote{~This also constitutes the main reason for using a slightly
  different definition of extended surface as compared to section I:2.4. While
  there is no canonical way to associate an {\em arc\/} to a germ of local
  coordinates, there
  is no problem in assigning a {\em germ of arcs\/} in the way described.}
For instance, recalling the definition of the standard extended surfaces 
$E_{0,n}[i_1,...\,,i_n]$ in section \ref{sec:standbas} and of the standard 
extended Riemann surfaces $E^c_{0,n}[i_1,...\,,i_n]$ above, we see that
  \be
  \ff(E^c_{0,n}) = E_{0,n} \,.  \ee

\subsubsection*{The topological counterparts of $P_\pi$ and $U_\gamma$} 

Recall that for an extended Riemann surface $E^c$,
$P_\pi(E^c)$ describes a permutation of the order of the 
marked points on $E^c$. Since the set of marked points on an 
extended surface $E$ is not ordered, we simply have 
  \be
  \ff \cir P_\pi = \ff \,.  \ee

To be able to formulate the analogue of $U_\gamma$ in the two examples 
discussed in section \ref{sec:cb-spaces} above, we need the following concepts.
By a family $E^c(t)$ of extended Riemann surfaces over
a Riemann surface $\Sigma^c$ we mean that each extended surface 
$E^c(t)$ is given by $\Sigma^c$ together with a set of marked points
$(p_k^t, [\varphi_k^t], \RCA_k)$ such that $p_k^t$ and $\varphi_k^t$ depend
smoothly on $t$. A family $E(t)$ of extended surfaces over
a two-manifold $\Sigma$ is defined in the same way. 
If $E^c(t)$ is a family of extended Riemann surfaces over
$\Sigma^c$, then $\ff(E^c(t))$ is a family of extended surfaces
over $\ff(\Sigma^c)$.

To a family $E^c(t)$ with $t \iN [0,1]$ we can assign a curve in the
moduli space $\hat\calm_{g,n}$ by setting $\gamma_{E^c}(t) 
\,{:=}\, [E^c(t)]$.  We then also get the isomorphism 
$U_{\gamma_{E^c}^{}}{:}~ \calh^c(E^c(0)) \To \calh^c(E^c(1))$.  

On the topological side, from a family $E(t)$ over $\Sigma$, with
$t \iN [0,1]$, we can construct a cobordism $\M_E {:}~ E(0) \To
E(1)$. As a three-manifold, $\M_E$ is given by the cylinder 
$\Sigma \Times [0,1]$. The ribbon graph in $\M_E$ is fixed by demanding
that the ribbons intersect the slice of $\M_E$ at `time' $t$ 
as prescribed by the marked points on $E(t)$. The topological
counterpart of $U_{\gamma_{E^c}^{}}$ is then the linear map
$Z(\M_E) {:}~ \calh(E(0)) \To \calh(E(1))$ that is supplied by the TFT.

\medskip

In this context the Riemann sphere with marked points is special:
any curve in $\hat\calm_{0,n}$ can be obtained as
$\gamma_{E^c}^{}$ for a family of extended Riemann surfaces over
a single fixed base Riemann surface, which we take to be
$\complex \,{\cup}\, \{ \infty \}$. For surfaces of higher genus one 
needs to vary the complex structure moduli of the underlying Riemann
surfaces as well. {}From here on we restrict our attention to the 
Riemann sphere with marked points.

\subsubsection*{The isomorphism in the case of the Riemann sphere}

Let $[E^c] \iN \hat\calm_{0,n}$. We would like to describe an isomorphism 
  \be
  \Alpha :\quad \calh^c(E^c) \to \calh(\ff(E^c)) 
  \ee
between the space of
conformal blocks associated to the extended Riemann surface $E^c$ and the
space of states of the TFT on the extended surface $\ff(E^c)$.

The TFT encodes the monodromy data of the conformal blocks. This must
be reflected in the isomorphism $\Alpha$. More precisely, we require
$\Alpha$ to fulfill 
  \be
  \Alpha \cir \calp_\pi = \Alpha \qquad {\rm and} \qquad
  \Alpha \cir U_{\gamma_{E^c}^{}} = Z(\M_{\ff(E^c)}) \cir \Alpha 
  \labl{eq:Alpha-prop}
for any permutation $\pi$ and any family $E^c(t)$ over $\complex \,{\cup}\, 
\{ \infty \}$. In addition to \erf{eq:Alpha-prop}, the isomorphism $\Alpha$ 
must be consistent with cutting and gluing of extended Riemann surfaces. 
For extended surfaces of higher genus, there are further conditions arising
from varying the complex structure moduli of the underlying Riemann surface. 
We will not analyse the latter two types of conditions in this paper. Instead 
we assume that $\Alpha$ does satisfy these conditions, and
content ourselves with describing $\Alpha$ for
points in $\hat\calm_{0,n}$ and illustrating its consistency with
\erf{eq:Alpha-prop} in some examples.

\medskip

For the Riemann sphere we fix the isomorphism $\Alpha$ by demanding that the 
standard conformal block $\beta_{0,n} \iN \calh(E^c_{0,n})$ from 
\erf{eq:standard-block} gets mapped to the state described by the cobordism 
$B_{0,n}$ from \erf{eq:standard-cobord},
  \bea
  \Alpha( \beta_{0,n}[i_1,...\,,i_n \,;\, p_1,...\,,p_{n-3} \,;\,
  \delta_1,...\,,\delta_{n-2}])
  \\[7pt] \hsp6
  = Z( B_{0,n}[i_1,...\,,i_n \,;\, p_1,...\,,p_{n-3} \,;\,
  \delta_1,...\,,\delta_{n-2}] , \emptyset, E_{0,n}[i_1,...\,,i_n]
  )\,1 \,. 
  \eear\labl{eq:Alpha-on-seRS}
Since $\hat\calm_{0,n}$ is connected, relation \erf{eq:Alpha-prop} can be used 
to determine $\Alpha$ for all points in $\hat\calm_{0,n}$, starting from its 
value at the point $[E^c_{0,n}]$ as given by \erf{eq:Alpha-on-seRS}.
This way of defining $\Alpha$ is consistent iff \erf{eq:Alpha-prop} holds for
every closed loop in $\hat\calm_{0,n}$. The first property in 
\erf{eq:Alpha-prop} just means that $\Alpha$ is independent of the ordering of 
the marked points of the extended Riemann surface.
The second property in \erf{eq:Alpha-prop} can be reformulated as follows. Two 
representations $\rho^c_{0,n}$ and $\rho_{0,n}^{}$ of $\pi_1(\hat\calm_{0,n})$
on $\calh^c(E_{0,n}^c)$ and on $\calh(E_{0,n})$, respectively, are given by
  \be
  \rho^c_{0,n}([\gamma^c]) :=  U_{\gamma^c}  \qquad {\rm and} \qquad 
  \rho_{0,n}^{}([\gamma^c]) := Z(\M_{\ff(E^c)}) \,,
  \ee
were the $\gamma^c$
are paths starting and ending on $[E^c_{0,n}]$ and $E^c$ is a family over
$\complex \,{\cup}\, \{ \infty \}$ such that $\gamma_{E^c}^{}\eq\gamma^c$.
The isomorphism $\Alpha$ must intertwine these two representations,
  \be
  \Alpha \cir \rho^c_{0,n} = \rho_{0,n}^{} \cir \Alpha \,.
  \labl{eq:Alpha-inter}
Below we will verify the intertwining property in two examples, and sketch 
an argument for it to hold in general at the end of section \ref{sec:cb-braid}.

\subsubsection*{Example: Closed path resulting in a twist}

Define a family of extended Riemann surfaces $E^c(t)$ over 
$\complex \,{\cup}\, \{ \infty \}$ by taking $E^c(t)$ to be equal to 
$E^c_{0,n}[i_1,...\,,i_n]$, except that the $k$th marked point gets replaced by 
$(k,\varphi_k \cir \eta_t, \SCA_{i_k})$ with $\eta_t(z) \eq \eE^{2\pi \ii t} z$.
Using $U$ to compute the transport of a block $\beta_{0,n}$ along the closed 
path $\gamma_{E^c}^{}$ gives
  \be
  (U_{\gamma_{E^c}^{}} \beta_{0,n})(v_1,...\,,v_n)
  = \eE^{2 \pi \ii \Delta_{i_k}} \beta_{0,n}(v_1,...\,,v_n) \,.
  \ee
To see this, start from formula \erf{eq:U-coord-change} and use
\erf{eq:Rinv-action} and \erf{eq:U-coord-Rhat-def} so as to deduce
first that $R(\eta_t)^{-1}v_k \eq \eE^{2\pi \ii t \Delta(v_k)} v_k$ and
then, also using that $\Delta_{i_k} \eq \Delta(v_k) \bmod \zet$,
that $\hat R(\eta)v_k \eq \eE^{2 \pi \ii \Delta_{i_k}} v_k$. We thus find 
  \be
  \rho^c_{0,n}([\gamma_{E^c}]) = \theta_{i_k}^{-1}\, \id_{\calh^c(E^c_{0,n})}\,.
  \ee
Further, let $E(t) \,{:=}\, \ff(E^c(t))$; then the cobordism
$\M_E{:}~ E(0) \To E(1)$ is given by
  \bea \begin{picture}(330,155)(10,40)
  \put(60,0)    {\Includeourbeautifulpicture 115 }
  \put(85,66)     {\tiny $x$}
  \put(78,90)     {\tiny $y$}
  \put(65.3,99.3) {\tiny $z$}
  \put(100,8)     {\tiny $x$}
  \put(87,33)     {\tiny $y$}
  \put(138,86)    {\small $S_{i_{k-1}}$}
  \put(191.4,86)  {\small $S_{i_{k}}$}
  \put(244,86)    {\small $S_{i_{k+1}}$}
  \put(312,183)   { $t\eq0$}
  \put(311,44)    { $t\eq1$}
  \put(0,89)      {$ \M_E \;= $}
  \epicture25 \labl{eq:twist-ex-MU} \Xlabp 115
with the orientations of three-manifold and ribbons as indicated.
Once oriented by the inward pointing normal, the boundary at
$t\eq0$ is given by $-E(0)\eq{-}E_{0,n}$ and the boundary at $t\eq1$
by $E(1) \eq E_{0,n}$, as it should be. By definition of the twist
(see (I:2.9)), the middle ribbon in \erf{eq:twist-ex-MU}
can be replaced by $\theta_{i_k}^{-1}$ times a straight ribbon, so that
  \be
  Z(\M_E) = \theta_{i_k}^{-1} Z( E_{0,n} \Times [0,1] ) \,.
  \ee
Since $Z$ applied to the cylinder over $E_{0,n}$ gives the
identity on $\calh(E_{0,n})$ we obtain
  \be
  \rho_{0,n}([\gamma_{E^c}]) = \theta_{i_k}^{-1} \, \id_{\calh(E_{0,n})} \,.
  \ee
For the closed path considered in this example, the desired identity 
\erf{eq:Alpha-inter} thus even holds independent of $\Alpha$ (and hence provides
a consistency check of the conventions described in section \ref{sec:fi-conv}).

\subsubsection*{Example: Closed path resulting in a double braiding}

In this example we deal again with a family $E^c(t)$ over
$\complex \,{\cup}\, \{ \infty \}$. We consider the situation that each of the 
extended Riemann surfaces $E^c(t)$ has three marked points, 
$(1,\varphi_1(z),\SCA_{i})$, $(p_2^t,\varphi_2^t(z),\SCA_{j})$ and
$(3,\varphi_3(z),\SCA_{k})$, with the functions $\varphi_i$ given by
$\varphi_1(z) \eq z{+}1$, $\varphi_2^t(z) \eq z{+}1{+}\eE^{2 \pi \ii t}$, 
$\varphi_3(z) \eq z{+}3$ and $p_2^t\eq \varphi_2^t(0)$.
Note that $E^c(0)\eq E^c(1)\eq E^c_{0,3}[i,j,k]$. 
A basis $\beta^\delta$ for the conformal blocks in $\calh^c(E^c_{0,3})$ 
is given by \erf{eq:3pt-block} with $z_1\eq1$, $z_2\eq2$, $z_3\eq3$. 
Transporting $\beta^\delta$ along $\gamma_{E^c}^{}$ from $0$ to $t$
results in the conformal block $\beta^\delta_t \iN\calh(E^c(t))$ given by
  \bea\displaystyle
  \beta^\delta_t(u,v,w) = 
  \langle 0 |\, \VO 0k{\bar k}w3\, \VO{\bar k}jiv{1{+}\eE^{2\pi \ii t}}\,
  \VO ii0u1 \,|0\rangle
  \\[8pt]\displaystyle
  \hspace{2.5em} = B^\delta_{kji}(w,v,u) \,
  (2{-}\eE^{2\pi \ii t})^{\Delta(u)-\Delta(v)-\Delta(w)} 
  2^{\Delta(v)-\Delta(u)-\Delta(w)} 
  \eE^{2\pi \ii t(\Delta(w)-\Delta(u)-\Delta(v))} \,.
  \eear\labl{eq:dbraid-ex1}
This block is obtained by setting $z_1\eq1$, $z_2\eq1{+}\eE^{2\pi \ii t}$
and $z_3\eq3$ in \erf{eq:3pt-block}, as follows from \erf{eq:KZ-shift-coord}
together with the property $\VO ijk{L_{-1}v}{z} \eq \frac{\rm d}{{\rm d}z}
\VO ijk{v}{z}$ of the intertwiners. Setting $t\eq1$ we obtain
  \be
  \beta^\delta_1(u,v,w) = 
  \eE^{2\pi \ii (\Delta_k-\Delta_i-\Delta_j)} \, \beta^\delta(u,v,w) \,,
  \ee
where we also used $\Delta(u) \eq \Delta_i \bmod \zet$ etc. Thus
  \be
  \rho^c_{0,3}([\gamma_{E^c}^{}]) = \frac{\theta_i \theta_j}{\theta_k} \,
  \id_{\calh^c(E^c_{0,3})} \,.
  \ee
Regarding the topological side, the cobordism $\M_E$ is easily found to be
  \bea \begin{picture}(375,130)(10,50)
  \put(60,0)    {\Includeourbeautifulpicture 116 }
  \put(90.3,70.5) {\tiny $x$}
  \put(77,96)     {\tiny $y$}
  \put(64,105)    {\tiny $z$}
  \put(325,21)    {\tiny $x$}
  \put(118.5,46)  {\tiny $y$}
  \put(325,148)   {\tiny $x$}
  \put(114,164.5) {\tiny $y$}
  \put(336,164)   { $t\eq0$}
  \put(336,42)    { $t\eq1$}
  \put(157,12)    {\small $1$}
  \put(214,12)    {\small $2$}
  \put(273,12)    {\small $3$}
  \put(157,152)   {\small $1$}
  \put(214,152)   {\small $2$}
  \put(273,152)   {\small $3$}
  \put(110,84)    {\scriptsize $j$}
  \put(162,83)    {\scriptsize $i$}
  \put(277.2,83)  {\scriptsize $k$}
  \put(0,81)      {$ \M_E \;= $}
  \epicture36 \Labp 116
Composing this cobordism with $B_{0,3}$ results in
  \begin{eqnarray} \begin{picture}(420,170)(10,0)
  \put(80,0)    {\Includeourbeautifulpicture 117 }
  \put(0,86)    {$ Z(\M_E) \cir Z(B_{0,3}) ~= $}
  \put(375,86)  {$ =~ \dsty\frac{\theta_i\theta_j}{\theta_k}\,Z(B_{0,3}) \;, $}
  \put(256,81)  {\scriptsize$ i $}
  \put(199.5,81){\scriptsize$ j $} 
  \put(318.4,81){\scriptsize$ k $}
  \put(251,158) {\scriptsize$ \overline{k} $}
  \put(198,13)  {\scriptsize$ 1 $}
  \put(256,13)  {\scriptsize$ 2 $}
  \put(314,13)  {\scriptsize$ 3 $}
  \put(221,138) {\scriptsize$ \delta $} 
  \put(123.2,67){\tiny$ z $}
  \put(138.4,45){\tiny$ y $}
  \put(361.2,21){\tiny$ x $}
  \end{picture} \nonumber \\[-1.6em]{}
  \label{pic4-117}  
  \end{eqnarray} \Xlabp 117
where we applied (I:2.41) twice and then substituted (I:2.43) (or rather the 
corresponding relations for $\RR^{-}$ instead of $\RR$). Thus for $\rho_{0,3}$ 
we get
  \be
  \rho_{0,3}([\gamma_{E^c}^{}]) = \frac{\theta_i \theta_j}{\theta_k} \,
  \id_{\calh(E_{0,3})} \,.
  \ee
This shows that also in this example the identity 
\erf{eq:Alpha-inter} holds independently of $\Alpha$, giving
another consistency check of the conventions in section  \ref{sec:fi-conv}.

\subsection{The braid matrices}\label{sec:cb-braid}

It is common to express various quantities of interest in terms of
matrices $\Bsmat ijkl{}{} \,{\equiv}\, (\Bsmat ijklpq)$ which are certain
specific combinations of fusing matrices $\FF$ and braiding matrices $\RR$
\cite{frki,Mose}. Conversely, one can recover $\FF$ and $\RR$ from $\BBmat$.
In terms of conformal blocks, the braid matrices $\BBmat$ can be deduced 
from the exchange of two insertion points in a four-point block on the 
sphere. Using the map $\Alpha$ one then verifies that this definition of 
$\BBmat$ is equivalent to the one used in the 3-d TFT context.

\subsubsection*{Braid matrices in modular tensor categories}

Similarly to the definition of the fusing matrices in (I:2.36), for a modular 
tensor category $\calc$ one defines the {\em braid matrices\/} $\Bsmat ijkl
{}{}$ to be the matrices whose entries appear as coefficients in the relation
  \bea \begin{picture}(275,89)(0,32) 
  \put(0,0)     {\Includeourbeautifulpicture 05a }
  \put(200,0)   {\Includeourbeautifulpicture 05b }
  \put(85,50)   {$ \dsty =~\sum_q \sum_{\gamma,\delta}
                 \Bmat{i\,j}klpq\alpha\beta\gamma\delta $}
  \put(-1,-8)   {\scriptsize$ i $}
  \put(28,-8)   {\scriptsize$ j $}
  \put(58,-8)   {\scriptsize$ k $} 
  \put(199,-8)  {\scriptsize$ i $}
  \put(228,-8)  {\scriptsize$ j $}
  \put(258,-8)  {\scriptsize$ k $} 
  \put(248,55)  {\scriptsize$ \delta $}
  \put(226.5,85){\scriptsize$ \gamma $}
  \put(47,55.5) {\scriptsize$ \beta $}
  \put(26,85)   {\scriptsize$ \alpha $} 
  \put(28,71)   {\scriptsize$ p $}
  \put(23,110)  {\scriptsize$ l$}
  \put(230,71)  {\scriptsize$ q $}
  \put(223,110) {\scriptsize$ l$}
  \epicture22 \labl{eq:Bmat-def} \xlabp 05 
If instead the inverse braiding is used one obtains the inverse
braid matrices $\BBmat^{-(ijk)l}_{}$.

The braid matrices can be expressed in terms of the fusing
matrices $\FF$ from (I:2.36) and the braid matrices $\RR$ from (I:2.41) as
  \be
  \Bmat{i\,j}klpq\alpha\beta\gamma\delta = \sum_{\mu,\nu}
  \Rm kip\beta\mu\, \F{j\,k}ilpq\alpha\mu\delta\nu\, \R iql\nu\gamma\, .
  \labl{eq:B=RFR}
This can be verified by first applying an inverse $\RR$-move to the morphism 
$\beta\iN\Hom(U_i\oti U_k,U_p)$ on the left hand side of \erf{eq:Bmat-def},
then performing an $\FF$-move, and finally an $\RR$-move applied to the 
resulting morphism in $\Hom(U_q\oti U_i,U_l)$.

Setting $k\eq0$ in \erf{eq:Bmat-def} and comparing to (I:2.41),
and recalling also convention (I:2.33), one arrives at the relation
  \be
  \R ijk\alpha\beta = \Bmat{i\,j}0kij\alpha{}\beta{} .
  \labl{eq:R=B}
Note that by convention the couplings in the one-dimensional morphism
spaces involving the tensor unit $U_0\eq\one$ are not spelt out explicitly.

Furthermore, using (I:2.42) the relation \erf{eq:B=RFR} is easily inverted:
  \be
  \F{i\,j}klpq\alpha\beta\gamma\delta = \sum_{\mu,\nu}
  \R kjp\beta\mu\, \Bmat{k\,i}jlpq\alpha\mu\nu\gamma\, \Rm qkl\nu\delta .
  \labl{eq:F=RBR}
Thus one can obtain both $\RR$ and $\FF$ from the braid matrices $\BBmat$. One 
may wonder whether it is possible to recover the twists $\theta_k$ and the 
quantum dimensions $\dim(U_k)$ as well. For this to be the case, we need to 
assume that the quantum dimensions are all positive. Then we can take 
the absolute value of (I:2.45) so as to find
  \be
  \dim(U_k) = \big| \Fs k{\bar k}kk00\, \Rs -{\bar k}k0 \big|^{-1} .
  \labl{eq:dimk=FR}
The twist is then directly given by (I:2.45),
  \be
  \theta_k = \dim(U_k)\, \Fs k{\bar k}kk00\, \Rs -{\bar k}k0 .
  \labl{eq:thetak=dimFR}

Up to equivalence, a modular tensor category is determined by the
list of isomorphism classes of simple objects $\{ U_k \,|\,k{\in}\II\}$,
the dimensions $\N ijk$ of the morphism spaces
$\Hom(U_i{\otimes}U_j,U_k)$, and the numbers
  \be
  \dim(U_k) \,,\quad
  \theta_k \,,\quad \R ijk\alpha\beta \,,\quad
  \F{i\,j}klpq\alpha\beta\gamma\delta \,.
  \labl{eq:modcat-data}
The considerations above show that, upon the assumption that the
quantum dimensions are positive, instead of \erf{eq:modcat-data} it is 
sufficient to give the numbers
  \be
  \Bmat{i\,j}klpq\alpha\beta\gamma\delta \,.
  \ee

\subsubsection*{Braid matrices from four-point conformal blocks}

The braid matrices also appear upon analytic continuation of the standard 
four-point conformal blocks. In short, one takes the standard four-point 
block and exchanges the field insertions at the points 2 and 3 via analytic 
continuation. The result can be expanded in terms of standard blocks, and 
the coefficients occurring in that expansion are the braid matrices. 
A more precise description is as follows.

Define a family $E^c(t)$ of extended Riemann surfaces over $\complex 
\,{\cup}\, \{ \infty \}$ by taking $E^c(t)$ to have field insertions
  \be
  (1,[\varphi_1],\SCA_i) \,,\qquad
  (w_t,[\varphi_2^t],\SCA_j) \,,\qquad
  (z_t,[\varphi_3^t],\SCA_k) \,,\qquad
  (4,[\varphi_4],\SCA_{\bar l}) ) \,,  \ee
where $\varphi_1(\zeta)\eq \zeta{+}1$, $\varphi_4(\zeta)\eq \zeta{+}4$,
  \be
  \varphi_2^t(\zeta) = \zeta - \Tfrac12\, \eE^{-\pi\ii t} + \Tfrac52 \,, \qquad
  \varphi_3^t(\zeta) = \zeta + \Tfrac12\, \eE^{-\pi\ii t} + \Tfrac52 \,,
  \ee
and $w_t \eq \varphi_2^t(0)$, $z_t \eq \varphi_3^t(0)$. It follows in particular 
that $E^c(0) \eq E^c_{0,4}[i,j,k,\bar l]$. For $E^c(1)$ we must also
take into account the ordering of the marked points, resulting in
$P_{(23)}(E^c(1)) \eq E^c_{0,4}[i,k,j,\bar l]$, where $(23)$ denotes
the transposition that exchanges $2$ and $3$. Transporting the conformal block
$\beta^{\mu\nu}[i,j,k,\bar l;p;\mu,\nu] \iN \calh^c(E^c_{0,4})$ along
$\gamma_{E^c}^{}$ from $0$ to $t$ results in the block
$\beta^{\mu\nu}_t \iN \calh^c(E^c(t))$ given by
  \be
  \beta^{\mu\nu}_t(v_1,v_2,v_3,v_4)
  = \langle 0 |\, \VO 0{\bar l}l{v_4}4 \,
  \VO {l,\nu}kp{v_3}{z_t} \, \VO {p,\mu}ji{v_2}{w_t} \,
  \VO {i}i0{v_1}{1} \, |0\rangle  \ee
with $z_t \eq \Tfrac12\, \eE^{-\pi\ii t} + \Tfrac52$ and $w_t \eq{-}\Tfrac12\, 
\eE^{-\pi\ii t} z + \Tfrac52$. Since $\beta^{\mu\nu}_1$ is in the space 
$\calh^c(E^c(1))$ we can apply to it the map $\calp_{(23)}{:}~ 
\calh^c(E^c(1)) \To \calh^c(E^c_{0,4}[i,k,j,\bar l])$ and expand the result
in the standard basis for the latter space,
  \beaa
  (\calp_{(23)} \beta^{\mu\nu}_1)(u_1,u_2,u_3,u_4) \!\!
  &= \beta^{\mu\nu}_1(u_1,u_3,u_2,u_4)
  \\{}\\[-.8em] &\displaystyle
  = \sum_q \sum_{\rho,\sigma} \Bmat{jk}ilpq\nu\mu\rho\sigma\,
  \beta_{0,4}[i,k,j,\bar l;q;\sigma\rho](u_1,u_2,u_3,u_4)\,.
  \eear \labl{eq:Bmat-conf}
By definition of the braiding in the representation category
$\calc \eq {\rm Rep}(\CA)$ of the chiral algebra, the matrices 
$\Bsmat ijkl{}{}$ which appear in \erf{eq:Bmat-conf}
are the same as those defined in \erf{eq:Bmat-def} through a
relation between morphisms in $\calc$, see e.g.\ \cite{huan24} for details.

\subsubsection*{Compatibility of $\Alpha$ and braiding}

Again one obtains a family $E(t)$ of extended surfaces by applying $\ff$ to the 
family $E^c(t)$. For the associated cobordism $\M_E$, composed with a cobordism 
$B_{0,4}$ describing the standard basis of $\calh(E_{0,4})$, one finds
  \begin{eqnarray} \begin{picture}(420,175)(10,0)
  \put(140,0)   {\Includeourbeautifulpicture 118 }
  \put(140,0)     {\setlength{\unitlength}{.38pt}
                  \put(243,262){\scriptsize$ i $} 
                  \put(381,262){\scriptsize$ j $} 
                  \put(527,262){\scriptsize$ k $} 
                  \put(671,262){\scriptsize$ \overline l $} 
                  \put(360,406){\scriptsize$ p $} 
                  \put(438,459){\scriptsize$ l $} 
                  \put(317,373){\scriptsize$ \mu $} 
                  \put(402,437){\scriptsize$ \nu $} 
                  \put(254,39){\scriptsize$ 1 $} 
                  \put(396,39){\scriptsize$ 2 $} 
                  \put(539,39){\scriptsize$ 3 $} 
                  \put(680,39){\scriptsize$ 4 $} 
                  \setlength{\unitlength}{1pt}}
  \put(0,125)   {$ Z(\M_E) \cir Z(B_{0,4}[i,j,k,{\bar l};p;\mu,\nu]) ~= $}
  \end{picture} \nonumber
  \\[1.2em] \dsty
  =\; \sum_q \sum_{\rho,\sigma} \Bmat{jk}ilpq\nu\mu\rho\sigma \,
  Z(B_{0,4}[i,k,j,\bar l;q;\sigma,\rho]) \;.~~~
  \label{eq:Al-braid-cobord}  
  \end{eqnarray} \Xlabp 118
We can now verify the relation \erf{eq:Alpha-inter} for the braiding on a basis 
$\beta_{0,4}[i,j,k,\bar l;p;\mu,\nu]$ of the space $\calh^c(E^c_{0,4}[i,j,k,\bar l])$ 
as follows:
  \beaa
  Z(\M_E) \cir \Alpha( \beta_{0,4}[i,j,k,\bar l;p;\mu,\nu] ) \!\!
  &= Z(\M_E) \cir Z(B_{0,4}[i,j,k,{\bar l};p;\mu,\nu]) 1
  \\{}\\[-.6em] &\displaystyle
  = \sum_q \sum_{\rho,\sigma} \Bmat{jk}ilpq\nu\mu\rho\sigma \,
      Z(B_{0,4}[i,k,j,\bar l;q;\sigma,\rho ] ) 1
  \\{}\\[-.8em] &\displaystyle
  = \sum_q \sum_{\rho,\sigma}
    \Bmat{jk}ilpq\nu\mu\rho\sigma \,
      \Alpha( \beta_{0,4}[i,k,j,\bar l;q;\sigma,\rho] )
  \\{}\\[-.8em] 
  &= \Alpha \cir \calp_{(23)} ( \beta^{\mu\nu}_1 )
  = \Alpha \cir U_{\gamma_{E^c}^{}} ( \beta_{0,4}[i,j,k,\bar l;p;\mu,\nu] )\,.
  \eear\ee
Here the first and third step just use the definition of
$\Alpha$ in \erf{eq:Alpha-on-seRS}, the second and fourth step 
are the equalities in \erf{eq:Al-braid-cobord} and \erf{eq:Bmat-conf},
respectively, and the final step amounts to the definition 
$\beta^{\mu\nu}_1 \eq U_{\gamma_{E^c}^{}} \beta_{0,4}[i,j,k,\bar l;p;\mu,\nu]$
together with the first of the properties \erf{eq:Alpha-prop}.

\subsubsection*{Consistency of $\Alpha$} 

Let us now sketch how compatibility with the braiding implies
\erf{eq:Alpha-inter}. Let $\gamma$ be a closed path in $\hat\calm_{0,n}$ 
starting and ending at $[E^c_{0,n}]$. $\gamma^c$ can always be written as the 
composition of two paths $\gamma_1$ and $\gamma_2$, of which $\gamma_1$ only 
changes the local coordinates but leaves the insertion points fixed, while 
$\gamma_2$ only moves the insertion points, with all local coordinates 
remaining of the form $\varphi_k(z) \eq z{+}p_k$.

The paths $\gamma_1$ and $\gamma_2$ are closed already by themselves, and 
hence can be treated separately. For paths of type $\gamma_1$ relation 
\erf{eq:Alpha-inter} holds by an argument similar to the one used in the 
first example above. A path of type $\gamma_2$ furnishes an element of the 
braid group with $n$ strands, which has the special property that the strand 
starting at position $k$ also ends at position $k$. The braid group is generated
by moves $g_{k,k+1}$ that exchange two neighbouring strands. It is thus 
sufficient to verify the relation $\Alpha \cir U_{\gamma_{E^c}^{}} \eq Z(\M_E) 
\cir \Alpha$ for such elements. This in turn follows by a similar argument as 
used for the four-point conformal block above, taking also into account that 
the braid matrices $\Bsmat ijkl{}{}$ control the behaviour of the intertwiners 
themselves under analytic continuation, not only that of the four-point block,
  \be
  \VO {l,\nu}kp{v_3}{z} \, \VO {p,\mu}ji{v_2}{w} 
  = \sum_q \sum_{\rho,\sigma} \Bmat{j\,k}ilpq\nu\mu\rho\sigma\,
  \VO {l,\rho}jq{v_2}{w} \, \VO {q,\sigma}ki{v_3}{z} \,,
  \ee
where the \lhs\ is understood as analytic continuation from $|w|\,{<}\,|z|$ 
to $|w|\,{>}\,|z|$, taking $w$ clockwise around $z$.


\sect{The fundamental correlation functions}\label{sec:cf}

Ultimately, we would like to obtain the correlators of a CFT as functions
of the world sheet moduli and of the field insertions. To this end we must
combine the results from the previous sections with additional information
about the world sheet.
In particular, for non-zero central charge the correlators $C$ do depend
explicitly on the world sheet metric $g$ and not only on the conformal
equivalence class of metrics, see e.g.\ \cite{gawe13}. Thus, to find the
value of a correlator it is not enough to know the conformal structure of
the world sheet, but rather the metric itself is relevant.
In sections \ref{sec:cf-bnd3pt}\,--\,\ref{sec:cf-1defect-xcap} we will treat 
certain ratios of correlators in which the Weyl anomaly cancels.\,%
  \footnote{~These ratios, in turn, are obtained from {\em unnormalised
  correlators\/} $C(\,\cdots)$, as opposed to the (normalised) expectation
  values $\langle\, \cdots \rangle$ which obey $\langle\, \one \,\rangle \eq
  1$. It is the correlators $C$ which appear in factorisation constraints,
  and which give, e.g., the torus and annulus partition functions.}

\subsection{Riemannian world sheets}\label{sec:ws-db-cmf}

In sections \ref{sec:fi} and \ref{sec:rc} we were concerned
with the 3-d TFT, and accordingly only needed the world sheet as
a topological manifold. To obtain the correlator as a function,
rather than as a vector in the TFT state space, we need the
world sheet metric. To distinguish the two settings, we will use the terms 
{\em topological world sheet\/} and {\em Riemannian world sheet\/}.

\subsubsection*{Definition}

By a Riemannian word sheet $\X^g$ we mean a compact,
not necessarily orientable, two-manifold with
metric $g$ which may have non-empty boundary, defects and field insertions.

In section \ref{sec:fi} we found that on the topological world sheet the fields,
boundaries and defects are labelled by equivalence classes of certain tuples.
The same holds true for insertions on the Riemannian world
sheet, but the quantities entering the tuples are slightly
different. We will only describe the tuples and refrain from
giving the equivalence relations explicitly. They take a similar form as in 
the topological case, and will not be needed for the applications below.

Define a {\em local bulk coordinate germ\/} $[f]$ to be a germ of injective 
functions $f{:}\; D_\delta \To \X^g$ of an arbitrarily small disk $D_\delta\eq
\{ z\iN\complex \,\big|\,|z|\,{<}\,\delta \}$ to $\X^g$ such that $f(0)$ lies 
in the interior of $\X^g$. On the disk $D_\delta$ we take the standard metric 
induced by $\complex \cong \reals^2$. The maps $f$ are required to be conformal, 
i.e.\ satisfy $f^* g \eq \Omega(x,y) ({\rm d}x^2 \,{+}\, {\rm d}y^2)$.
Define further a {\em local boundary coordinate germ} $[d]$ 
to be a germ of injective functions $d{:}\; H_\delta \To \X^g$ 
of an arbitrarily small semidisk $H_\delta \eq 
\{ z\iN\complex \,\big|\,|z|\,{<}\,\delta\;{\rm and}\;{\rm Im}(z)\,{\ge}\,0 \}$ 
to $\X^g$ such that $d(x)\iN\partial \X^g$ iff $x\iN\reals$. Again on $H_\delta$
we take the standard metric and require the map $d$ to be conformal.

Since the definition of local coordinate germs depends only on the
conformal structure on $\X^g$, we denote fields on $\X^g$ with
a superscript $c$ to distinguish them from their topological counterparts in 
section \ref{sec:fi}. The possible field insertions on $\X^g$ are as follows.
\ITemize
\item[\Nxt]
A {\em bulk field\/} $\vPhi^c$ is a tuple
  \be
  \vPhi^c = (i,j,\phi,[f])\,,
  \ee
where $i,j\iN\II$ label simple objects, $\phi$ is a morphism
in $\HomAA(U_i\,{\otimes^+}A\,{\otimes^-}\,U_j,A)$,
and $[f]$ is a local bulk coordinate germ.
\item[\Nxt]
A {\em boundary field\/} $\hatpsi^c$ is a tuple
  \be
  \hatpsi^c = (M,N,\RCA,\psi,[d])\,,
  \ee
where $M$ and $N$ are left $A$-modules, $\RCA$ is an object of $\calc$, 
$\psi\iN\HomA(M\Oti \RCA,N)$, and $[d]$ is a local boundary coordinate germ.
\item[\Nxt]
A {\em defect field\/} $\Thetha^c$ is a tuple
  \be
  \Thetha^c = (\X,\orr_2(X),Y,\orr_2(Y),i,j,\vartheta,[f])\,,
  \ee
where $X$, $Y$ are $A$-bimodules, $\orr_2(X)$ is
a local orientation of $\X^g$ around the defect segment
labelled by $X$, and similarly for $\orr_2(Y)$. Further, $i,j\iN\II$ and
$\vartheta \iN \HomAA(U_i\,{\otimes^+}\,Z\,{\otimes^-}\,U_j,Z')$,
where $Z, Z'$ are defined as in \erf{eq:defect-ZZ-XY}. $[f]$ is a local bulk 
coordinate germ which must fulfill the condition that $f(x)$ lies on the 
defect circle if $x$ lies on the real axis.
\end{itemize}
As an additional datum, the boundary and bulk insertions are ordered, i.e.\ 
if there are $m$ boundary fields and altogether $n$ bulk and defect fields,
we have to pick maps from $\{1,2,\dots,m\}$ to the set of boundary
fields and from $\{1,2,\dots,n\}$ to the set of bulk/defect fields.
Thinking of bulk fields as a special type of defect fields, we write
  \be
  \hatpsi^c_k = (M_k,N_k,\RCA_k,\psi_k,[d_k]) \qquad {\rm and} \qquad
  \Thetha^c_k = (\X_k,\orr_2(X)_k,Y_k,\orr_2(Y)_k,i_k,j_k,\vartheta_k,[f_k])
  \ee
for the $k$th boundary and defect field, respectively, and their defining data.
The ordering is needed because below we construct an extended Riemann surface
from $\X^g$, and in that situation the marked points from an ordered set.

\subsubsection*{The complex double of $\X^g$}

From a Riemannian world sheet $\X^g$ we can obtain an
extended Riemann surface $\hat\X^g$, the {\em complex double of} $\X^g$.
It is constructed as follows.

If there are no field insertions on $\X^g$, then $\hat \X^g$ is
simply the orientation bundle over $\X^g$, divided by the usual
equivalence relation, i.e.
  \be
  \hat \X^g = {\rm Or}(\X^g) /{\sim}
  \qquad {\rm with}~~ (x,\orr_2) \sim (x,-\orr_2) 
  \quad{\rm for}~~ x\iN\partial \X^g \,.
  \ee
Since a metric (or just a conformal structure) together with an orientation 
defines a complex structure, we obtain a complex structure
on ${\rm Or}(\X^g)$. The equivalence relation leading from ${\rm Or}(\X^g)$ to 
$\hat \X^g$ respects this complex structure. In terms of
charts it amounts to identifying a subset of the upper 
half plane with the conjugate subset of the lower half plane along 
the real axis. Thus $\hat \X^g$ is a compact Riemann surface without boundary.

If in addition there are field insertions on $\X^g$, one obtains an
ordered set of marked points on $\hat \X^g$ as follows.
\ITemize
\item[\Nxt] 
Each boundary field $\hatpsi^c \eq (M,N,\RCA,\psi,[d])$ gives rise to a single
marked point $(p,[\varphi],\RCA)$ on $\hat \X^g$, where
  \be
  \varphi(z) = \begin{cases}
  [d(z),\orr_2(d)] & {\rm for}~ {\rm Im}(z) \,{\ge}\, 0 \,, \\[.3em]
  [d(z^*),-\orr_2(d)] & {\rm for}~ {\rm Im}(z) \,{\le}\, 0 \,. \end{cases}
  \labl{eq:com-double-bnd}
Here $\orr_2(d)$ denotes the local orientation of $\X^g$ obtained
form the push-forward of the standard orientation on $\complex$ via $d$. 
One can verify that $\varphi$ is an injective analytic function from 
a small disk in $\complex$ to $\hat \X^g$.
The insertion point $p$ is equal to $\varphi(0)$. Finally, if $\hatpsi^c$
has label $k$ from the ordering of fields, we also assign the number $k$ 
to the marked point \erf{eq:com-double-bnd}.
\item[\Nxt] 
Each bulk field $\vPhi^c \eq (i,j,\phi,[f])$ or defect field
$\Thetha^c \eq (X,\orr_2(X),Y,\orr_2(Y),i,j,\vartheta,[f])$ 
gives rise to two marked points $(p_i,[\varphi_i],\SCA_i)$ and 
$(p_j,[\varphi_j],\SCA_j)$ on $\hat \X^g$. Here 
  \be
  \varphi_i(z) = [f(z),\orr_2(f)]  \qquad{\rm and}\qquad
  \varphi_j(z) = [f(z^*),-\orr_2(f)]
  \labl{eq:com-double-bulk}
are analytic functions, and $p_{i,j} \eq \varphi_{i,j}(0)$. If the 
bulk\,/\,defect field has label $k$  from the ordering on $\X^g$, the two marked
points \erf{eq:com-double-bulk} are assigned the numbers $m\,{+}\,2k\,{-}\,1$ 
and $m\,{+}\,2k$, respectively, with $m$ the number of boundary fields.
\end{itemize}
We also need to fix a Lagrangian submodule $\lambda^c\,{\subset}\,H_1(\hat\X^g,
\zet)$. This is again done via the inclusion $\iota{:}~X^g \To \MX$ (here $\MX$ 
is the connecting manifold for $\X^g$, considered as a topological manifold), 
which gives rise to a map $\iota_*{:}~ H_1(\hat\X^g,\zet) \To H_1(\MX,\zet)$.
We take $\lambda^c$ to be the unique Lagrangian submodule that contains 
${\rm ker}(\iota_*)$.

\subsubsection*{Holomorphic factorisation}

Given a Riemannian world sheet $\X^g$, the correlation function
$C(\X^g)$ of the CFT for that world sheet is a linear function
  \be
  C(\X^g) : \quad
  \bigotimes_{k=1}^m \RCA_k \otimes
  \bigotimes_{l=1}^n\, (\SCA_{i_l} \oti \SCA_{j_l})
  \,\longrightarrow\, \complex \,.
  \labl{eq:C-function}
Any correlation function must solve the chiral Ward identities coming from the 
holomorphic (and anti-holomorphic) fields in the theory, in particular those 
coming from fields in the chiral algebra. This leads to the principle of 
holomorphic factorisation \cite{witt39}, which can be simply formulated 
as the statement that
  \be
  C(\X^g) \in \calh^c( \hat \X^g ) \,,
  \ee
i.e.\ that the correlator for $\X^g$ is an element of the space of
conformal blocks on the complex double of $\X^g$.

\subsubsection*{Metric dependence}

Consider a Riemannian world sheet $\X^g$. 
Under a Weyl transformation $g \,{\mapsto}\, g' \eq \eE^{\gamma} g$
for some $\gamma{:}~ \X^g \To \reals$, the correlator changes by a factor:
  \be
  C(\X^{g'}) = \eE^{c S[\gamma]}\, C(\X^g) \,,
  \labl{eq:Weyl-anom}
where $c$ is the central charge and $S[\gamma]$ is the 
(suitably normalised) Liouville action, see e.g.\ \cite{gawe13} for details. 

If one works with ratios of correlators, the prefactor can be cancelled 
in the following manner. Let $\dot\X^g$ be the Riemannian world
sheet obtained from $\X^g$ by removing all field insertions and
labelling all boundaries and defects by the algebra $A$. Then
$C(\dot\X^g)$ is just a complex number, rather than a function
as in \erf{eq:C-function}, but it shows the same behaviour 
\erf{eq:Weyl-anom} under
Weyl transformations of the metric as $C(\X^g)$. It follows that the ratio
  \be
  C(\X^g) / C(\dot\X^g) 
  \labl{eq:corr-ratio}
only depends on the conformal equivalence class of the metric $g$.
It is these ratios, rather than the correlators themselves, that we will 
present below for the fundamental correlators.

\subsubsection*{The topological world sheet and its double}

Let us define an operation $\tildeF$ which assigns to a Riemannian world
sheet $\X^g$ a topological world sheet $\X \eq \tildeF(\X^g)$ as follows.

If there are no field insertions on $\X^g$, then $\X$ is obtained from
$\X^g$ by just forgetting the metric. If in addition there
are field insertions on $\X^g$, then the corresponding insertions on the
topological world sheet $\X$ are obtained as follows.
  \ITemize
  \item[\Nxt] 
A boundary field  $\hatpsi^c \eq (M,N,V,\psi,[d])$ on $\X^g$ gives an insertion
$\hatpsi \eq (M,N,V,\psi,p,[\gamma])$ on $\X$, where
$p \eq d(0)$ and $\gamma(t) \eq d(t)$.
  \item[\Nxt] 
A bulk field $\vPhi^c \eq (i,j,\phi,[f])$ on $\X^g$ gives an insertion
$\vPhi \eq (i,j,\phi,p,[\gamma],\orr_2(p))$ on $\X$, where $p \eq f(0)$ and 
$\gamma(t) \eq f(t)$.  The local orientation $\orr_2(p)$ of $\X^g$ is given 
by $\orr_2(f)$, i.e.\ the push-forward of the standard orientation
of $\complex$ via $f$.
\\[3pt]
Similarly, for a defect field $\Thetha^c \eq (X,\orr_2(X),Y,\orr_2(Y),i,j,
\vartheta,[f])$  on $\X^g$ one gets an insertion $\Thetha^c \eq 
(X,\orr_2(X),Y,\orr_2(Y),i,j,\vartheta,p,[\gamma],\orr_2(p))$ on $\X$.
  \end{itemize}
With these definitions one can verify that the diagram
  \be
  \begin{array}{rcccl}
  & \X^g & \stackrel\tildeF\longmapsto & \X
  \\[1pt]
  &\top&&\top \\[-8pt] &\downarrow&&\downarrow
  \\[2pt]
  & \hat\X^g & \stackrel\ff\longmapsto & \hat\X
  \end{array}
  \ee
with $\ff$ the map from extended Riemann surfaces to extended surfaces 
defined in section \ref{sec:cb-link}, commutes.

The TFT construction describes the correlator as a state of the TFT,
$C(\X) \iN \calh(\hat \X)$. In order to obtain the correlator 
$C(\X^g) \iN \calh^c( \hat \X^g )$ we thus need an isomorphism
  \be \Alpha :\quad \calh^c( \hat \X^g ) \stackrel \cong \to
  \calh(\hat \X) \,.
  \ee
Since the metric $g$ does not appear in the topological construction, 
$\Alpha$ should have some dependence on $g$. 
Here we are interested in correlators (or rather, the ratios
\erf{eq:corr-ratio}) on the upper half plane and on the complex plane
(together with the point at infinity). In both cases we consider
the standard metric\,\footnote{%
  ~This makes both world sheets non-compact, but they are still conformally
  equivalent to compact world sheets. We could have chosen
  a compact metric instead, but since the ratios \erf{eq:corr-ratio}
  are invariant under Weyl-transformations this makes no difference.}
${\rm d}x^2+{\rm d}y^2$. For this metric we take the isomorphism $\Alpha$ to be
the one defined in section \ref{sec:cb-link}.

\subsection{Boundary three-point function on the upper half plane}
\label{sec:cf-bnd3pt}

Let $A$ be a symmetric special Frobenius algebra in $\calc$. Take the 
Riemannian world sheet $\X^g$ to be the upper half plane together with three
field insertions
  \be
  \hatpsi_1 = (N,M,\SCA_i,\psi_1,[d_1])\,,\quad
  \hatpsi_2 = (K,N,\SCA_j,\psi_2,[d_2])\,,\quad
  \hatpsi_3 = (M,K,\SCA_k,\psi_3,[d_3])
  \ee
on its boundary. Here $M, N, K$ are left $A$-modules, the three fields
are inserted at positions $x_i$ satisfying $0\,{<}\,x_1\,{<}\,x_2\,{<}\,x_3$, 
and the local boundary coordinate germs $[d_n]$ around these points are given 
by arcs $d_n(z) \eq x_n{+}z$.  
Note that $\tildeF(\X^g)$ gives precisely the 
topological world sheet displayed in figure \erp 24.

The complex double of $\X^g$ is the extended Riemann surface consisting of
the Riemann sphere $\hat \X^g\eq\mathbb{P}^1$, parametrised
as $\complex\,{\cup}\,\{\infty\}$, and the three marked points
$(x_1, [\varphi_1] , \SCA_i)$, $(x_2, [\varphi_2] , \SCA_j)$ and 
$(x_3, [\varphi_3] , \SCA_k)$, with the local coordinates given by
$\varphi_n(z) \eq x_n{+}z$, $n\eq1,2,3$. The correlator $C(\X^g)$ is an element 
of the space of conformal blocks $\calh^c(\hat \X^g)$. A basis of this space 
is provided by the three-point blocks \erf{eq:3pt-block}. One can thus 
write the correlator as a linear combination
  \be
  C(\X^g)(u,v,w) = \sum_{\delta=1}^{\N ij{\bar k}} 
  c(M \hatpsi_1 N \hatpsi_2 K \hatpsi_3 M)_\delta \,
     \langle 0|\, \VO 0k{\bar k}w{x_3}\,
             \VO {\bar k,\delta}jiv{x_2}\, \VO ii0u{x_1} \,|0\rangle \,,
  \labl{eq:uhp-3pt-block}
where $u\iN\SCA_i$, $v\iN\SCA_j$ and $w\iN\SCA_k$.
The constants $c(M \hatpsi_1 N \hatpsi_2 K \hatpsi_3 M)_\delta$
are precisely those determined by the TFT-analysis in \erf{eq:bnd-3pt-rib}.
This follows from applying the map $\Alpha$ to both sides of 
\erf{eq:uhp-3pt-block}, which results in \erf{eq:uhp-3pt-cobord}.
Altogether, for the ratio \erf{eq:corr-ratio} we find, 
using the explicit form \erf{eq:3pt-block} of the three-point block, 
  \be  \!\begin{array}{lr}\dsty
  \frac{C(\X^g)(u,v,w)}{C(\dot\X^g)} \!\! &\dsty
  = \Big( \sum_{\delta=1}^{\N ij{\bar k}} 
  \frac{c(M \hatpsi_1 N \hatpsi_2 K \hatpsi_3 M)_\delta}{
  \dim(A)}\, B^\delta_{kji}(w,v,u) \Big) \,
  (x_3{-}x_2)^{\Delta_i(u)-\Delta_j(v)-\Delta_k(w)} ~~
  \\{}\\[-.8em] &\dsty
  (x_3{-}x_1)^{\Delta_j(v)-\Delta_i(u)-\Delta_k(w)}\,
  (x_2{-}x_1)^{\Delta_k(w)-\Delta_i(u)-\Delta_j(v)} .
  \eear\labl{eq:3bnd-corr}
The vectors $u\iN\SCA_i$, $v\iN\SCA_j$ and $w\iN\SCA_k$ tell 
us which fields of the infinite-dimensional representation spaces of the
chiral algebra are inserted at the points $x_3\,{>}\,x_2\,{>}\,x_1$, and
$\Delta_i(u)$, $\Delta_j(v)$, $\Delta_k(w)$ are their conformal weights.
Note also that $C(\dot\X^g) \eq \dim(A)$. The constants $c_\delta$ have
been expressed in a basis in \erf{eq:3bnd-inv-basis};
for the Cardy case they are given by \erf{eq:3bnd-basis-cardy}.

\dt{Remark}
Even though the correlator of three boundary
fields is built directly from a single conformal block, rather than from
a bilinear combination of blocks like the correlator of three bulk
fields, it is {\em not\/}, in general, an analytic function of the
insertion points. As an example, take the correlator of two
equal boundary fields\,\footnote{%
  ~This is zero in general, in particular if $i\,{\neq}\,\bar\imath$.
  But there are many examples, like the Virasoro minimal models, where it
  can be non-zero.}
$\hatpsi^c$, which for $x_2\,{>}\,x_1$ is given by
  \be
  C = c(M\hatpsi M\hatpsi M)\, B_{ii}(u,u)\, (x_2\,{-}\,x_1)^{-2\Delta_i(u)}
  .  \ee
For $x_1\,{>}\,x_2$ one finds the same answer up to an exchange 
$x_1\,{\leftrightarrow}\, x_2$, so that for all values of $x_1, x_2$ we can
write
  \be
  C = c(M\hatpsi M\hatpsi M)\, B_{ii}(u,u)\,
  \big|x_2\,{-}\,x_1\big|^{-2\Delta_i(u)} , \ee
which is not analytic in $x_1$ and $x_2$ (unless $\Delta_i\iN\zet$, which is 
a very special case). It is thus not quite appropriate to say that
the theory on the boundary is chiral.

\dtl{Remark}{MNKA}
Consider the case $M\eq N\eq K\eq A$, i.e.\ on each boundary segment the 
boundary condition is given by $A$, regarded as a left module over itself. 
Using the Frobenius reciprocity relation $\HomA(A\oti U, A) \,{\cong}\,
\Hom(U,A)$ (see e.g.\ proposition I:4.12), the elements of
$\HomA(A\oti \SCA_k, A)$ can be expressed as
  \be
  \psi = m \circ (\id_A \oti b^A_{k,\alpha})\,,
  \ee
where $m$ is the multiplication morphism of $A$ and
$b^A_{k,\alpha}$ denotes the basis \erf{pic77} of $\Hom(\SCA_k,A)$. 
The ribbon graph \erf{eq:bnd-3pt-rib} then takes the form
  \bea \begin{picture}(380,133)(10,46)
  \put(130,0)   {\Includeourbeautifulpicture 28a }
  \put(310,0)   {\Includeourbeautifulpicture 28b }
  \put(163,144)    {\scriptsize $A$}
  \put(180,104)    {\scriptsize $A$}
  \put(195,123)    {\scriptsize $A$}
  \put(249,60)     {\scriptsize $i$}
  \put(206,59)     {\scriptsize $j$}
  \put(217,17.3)   {\scriptsize $\bar k$}
  \put(190,27)     {\scriptsize $k$}
  \put(230.8,41)   {\scriptsize $\delta$}
  \put(253,77)     {\scriptsize $\alpha$}
  \put(213,77)     {\scriptsize $\beta$}
  \put(189,65)     {\scriptsize $\gamma$}
  \put(349,143)    {\scriptsize $A$}
  \put(330,100)    {\scriptsize $A$}
  \put(380,100)    {\scriptsize $A$}
  \put(307,116)    {\scriptsize $A$}
  \put(382.4,84)   {\scriptsize $\alpha$}
  \put(343,84)     {\scriptsize $\beta$}
  \put(318.2,84)   {\scriptsize $\gamma$}
  \put(373.5,67)   {\scriptsize $i$}
  \put(341.5,67)   {\scriptsize $j$}
  \put(346,17)     {\scriptsize $\bar k$}
  \put(319,27)     {\scriptsize $k$}
  \put(359.5,41)   {\scriptsize $\delta$}
  \put(0,86)       {$ c(A \hatpsi_1 A \hatpsi_2 A \hatpsi_3 A)_\delta \;= $} 
  \put(275,86)     {$ = $} 
  \epicture29 \labl{eq:bnd-3pt-rib-A} \xlabp 28
In the second equality one uses associativity of the multiplication, the
transformations (I:3.49), as well as the specialness relation 
$m\cir\Delta\eq\id_A$. One can now substitute the expression of the
multiplication in a basis as in (I:3.7) to evaluate the ribbon invariant
\erf{eq:bnd-3pt-rib-A}. This yields
  \be
  c(A \hatpsi_1 A \hatpsi_2 A \hatpsi_3 A)_\delta
  = \sum_{\mu,\nu} m_{j\beta,i\alpha}^{\bar k \mu; \delta}\,
  m_{k\gamma,\bar k\mu}^{0\nu}\, \big(\eps \cir b^A_{0,\nu}\big) \,.
  \labl{eq:C-m-OPE}
Using the OPE of boundary fields as defined in 
(I:3.11) to evaluate the correlator
$\langle 0|\hatpsi_3 \hatpsi_2 \hatpsi_1 |0\rangle$
and comparing the result to \erf{eq:C-m-OPE} gives
  \be
  m_{j\beta,i\alpha}^{\bar k \mu; \delta} =
  C_{j\beta,i\alpha}^{\bar k \mu; \delta} \,.  \ee
This is nothing but the relation (I:3.14) that was already established
in section I:3.2. In words it says that the boundary OPE on the boundary
condition labelled by $A$ is equal to the multiplication on the algebra $A$.

\subsection{One bulk and one boundary field on the upper half plane}
\label{sec:cf-1bulk1bnd}

The next correlator we consider is that of one bulk field and
one boundary field on the upper half plane. 
Let $\X^g$ be the upper half plane with field insertions
  \be
  \vPhi^c = (i,j,\phi,[f]) \qquad {\rm and} \qquad
  \hatpsi^c = (M,M,\SCA_k,\psi,[d]) \,,
  \ee
where $i,j,k\iN\II$ label simple objects, $\phi\iN\HomAA(\SCA_i\otiP A\otim 
\SCA_j,A)$ and $\psi\iN\HomA(M\oti \SCA_k,M)$. Further, $[f]$ is a local 
bulk coordinate germ and $[d]$ a local boundary coordinate germ, with
$f(\zeta)\eq z\,{+}\,\zeta$ and $d(\zeta) \eq s\,{+}\,\zeta$ for
$z$ in the upper half plane and $s$ on the real axis.
In fact we take $z\eq x\,{+}\,\ii y$ and $s\,{>}\,x\,{>}\,0$.

The complex double of $\X^g$ is the extended Riemann surface $\hat \X^g$ 
consisting of the Riemann sphere $\complex\,{\cup}\,\{\infty\}$ and the three 
marked points $(s, [\varphi_1] , \SCA_k)$, $(x{+}\ii y, [\varphi_2] , \SCA_i)$ 
and $(x{-}\ii y, [\varphi_3], \SCA_j)$, with the local coordinates given by 
$\varphi_1(\zeta) \eq s\,{+}\,\zeta$, $\varphi_2(\zeta) \eq x{+}\ii y\,{+}\,
\zeta$ and $\varphi_3(\zeta) \eq x{-}\ii y\,{+}\,\zeta$. We select a basis 
$\{ \beta_\delta \}$ in the space of conformal blocks $\calh^c(\hat \X^g)$ as
  \beaa
  \beta_\delta \!\!& := \langle 0|\, \VO 0k{\bar k}w{z_3}\,
             \VO {\bar k,\delta}iju{z_2}\, \VO jj0v{z_1} \,|0\rangle 
  \,\Big|_{z_1=x{-}\ii y , z_2=x{+}\ii y , z_3=s}~~
  \\[15pt]\displaystyle
  & \;= B^\delta_{kij}(w,u,v)\,
  \eE^{\frac{\pi \ii}{2}(\Delta_k(w)-\Delta_i(u)-\Delta_j(v))}\,
  \big(s{-}x{+}\ii y\big)^{2(\Delta_i(u)-\Delta_j(v))}
  \\[7pt]\displaystyle&\hspace*{10em}\times
  \big( (s{-}x)^2+y^2 \big)^{-\Delta_i(u)+\Delta_j(v)-\Delta_k(u)}
  \big(2y\big)^{\Delta_k(w)-\Delta_i(u)-\Delta_j(v)} ,
  \eear\ee
where in the first line it is understood that the points $z_{1,2,3}$ are taken 
to their present position from the standard block \erf{eq:standard-block} with 
$z_n\eq n$ via continuation along the contours indicated in figure 
\erf{eq:Bxys}.  In particular, the connection to $B_\delta$ in \erf{eq:Bxys}
is $\Alpha(\beta_\delta) \eq Z(B(x,y,s)_\delta, \emptyset,\hat\X)$.
Applying $\Alpha^{-1}$ to both sides of \erf{eq:1B1b-TFT} thus yields
  \be
  C(\X^g) = \sum_{\delta} c(\vPhi ; M \hatpsi)_\delta \,\beta_\delta \,,
  \labl{eq:blk-bnd-corr-const}
where the constants $c(\vPhi; M \hatpsi)_\delta$ 
are given by the ribbon invariant \erf{eq:1bulk-1bnd-sc}.

We conclude that the correlator ratio for one bulk field 
and one boundary field on the upper half plane is of the form
  \bea  
  \!\!\!\! \dsty \frac{C(\X^g)(w,u,v)}{C(\dot\X^g)} = \dsty 
  \Big( \sum_{\delta=1}^{\N ij{\bar k}}
  \frac{c(\phi; M \hatpsi)_\delta}{\dim(A)} \, 
  B^\delta_{kij}(w,u,v)\,
  \eE^{\frac{\pi \ii}{2}(\Delta_k(w)-\Delta_i(u)-\Delta_j(v))} \Big)
  \\{}\\[-.7em]  \hspace*{3.9em} \dsty
  \big(s{-}x{+}\ii y\big)^{2(\Delta_i(u)-\Delta_j(v))} 
  \big( (s{-}x)^2\,{+}\,y^2 \big)^{-\Delta_i(u)+\Delta_j(v)-\Delta_k(u)}
  \big(2y\big)^{\Delta_k(w)-\Delta_i(u)-\Delta_j(v)} ,
  \eear\labl{eq:1bulk-1bnd}
where $u \oti v \iN \SCA_i \oti \SCA_j$ gives the bulk field and
$w \iN \SCA_k$ the boundary field.

\dtl{Remark}{rem:1bulk-uhp'}
The special case of one bulk field on the upper half plane without any boundary 
field insertion (compare remark \ref{rem:1bulk-uhp}) is obtained by setting 
$\hatpsi \eq (M,M,\one,\id_M,[g])$ as well as $w\eq |0\rangle \iN \SCA_0$. 
Then \erf{eq:1bulk-1bnd} becomes
  \be
  \frac{C(\X^g)(u,v)}{C(\dot\X^g)} = \frac{c(\phi; M )}{\dim(A)} \, 
  B_{ij}(u,v)\, \eE^{- \pi \ii \Delta_i(u)} \big(2y\big)^{-2\Delta_i(u)} .
  \labl{eq:1bulk}
In \erf{eq:1bulk} the two-point block \erf{eq:2pt-block} appears,
which vanishes unless $j\eq\bar\imath$ and $\Delta_i(u)\eq\Delta_j(v)$. Note 
that in the Cardy case, the phase in \erf{eq:1bulk} combines with the factor 
$t_i$ in \erf{eq:Cardy-1disc} to a sign $(-1)^{\Delta_i(u)-\Delta_i}$.

\subsection{Three bulk fields on the complex plane} \label{sec:cf-3bulk}

We now turn to the correlator of three bulk fields 
on the complex plane, i.e.\ $\X^g$ is $\complex \,{\cup}\, \{\infty\}$ together
with three field insertions
  \be
  \vPhi_1^c = (i,j,\phi_1,[f_1]) \,,\qquad
  \vPhi_2^c = (k,l,\phi_2,[f_2]) \,,\qquad
  \vPhi_3^c = (m,n,\phi_3,[f_3]) \,.
  \ee
Here $i,j,k,l,m,n \iN \II$ label simple objects; $\phi_1$ is an element of
$\HomAA(\SCA_i\otiP A\otim \SCA_j,A)$,
and similarly one has $\phi_2\iN\HomAA(\SCA_k\otiP A\otim \SCA_l,A)$
and $\phi_3\iN\HomAA(\SCA_m\otiP A\otim \SCA_n,A)$.
The three fields are inserted at the positions $z_1$, $z_2$, $z_3$
with local bulk coordinate germs $[f_{1,2,3}]$ 
given by $f_{1,2,3}(\zeta) \eq z_{1,2,3} \,{+}\, \zeta$. 

The complex double of $\X^g$ is the extended Riemann surface $\hat \X^g$ 
given by two copies of the Riemann sphere $\complex\,{\cup}\,\{\infty\}$
together with six marked points, three on each connected component.
Recall that we have chosen an orientation of $\X^g$ from the outset.
Let $\hat \X^g_+$ be the connected component of $\hat \X^g$ that has the
same orientation as $\X^g$, and $\hat \X^g_-$ be the component with
opposite orientation. We have $\hat \X^g_\pm \,{\cong}\, \mathbb{P}^1$.
The three marked points on the component $\hat \X^g_+$ 
are $(z_1, [\varphi_1] , \RCA_i)$, $(z_2, [\varphi_2] , \RCA_k)$ and
$(z_3, [\varphi_3] , \RCA_m)$, while on $\hat \X^g_-$ the marked points are
$(z^*_1, [\tilde\varphi_1] , \RCA_j)$, $(z^*_2, [\tilde\varphi_2] , \RCA_l)$,
$(z^*_3, [\tilde\varphi_3] , \RCA_n)$. Here $z^*\eq x\,{-}\,\ii y$ 
is the point complex conjugate to $z\eq x\,{+}\,\ii y$.
The local coordinates are given by $\varphi_i(\zeta) \eq z_i\,{+}\,\zeta$
and $\tilde\varphi_i(\zeta) \eq z_i^*\,{+}\,\zeta$, for $i\eq1,2,3$.
We select a basis $\{ \beta_{\mu\nu} \}$ 
in $\calh^c(\hat \X^g)$ via the cobordism \erf{eq:3bulk-bas},
  \be
  \Alpha(\beta_{\mu\nu}) = Z(B(z_1,z_2,z_3)_{\mu\nu},\emptyset,\hat \X)\,, 
  \labl{eq:3bulk-block}
i.e.\ $\beta_{\mu\nu}$ is a product of two three-point blocks 
\erf{eq:3pt-block}. Applying $\Alpha^{-1}$ to both sides
of \erf{Z3bul} results in
  \be
  C(\X^g) = \sum_{\mu,\nu} c(\vPhi_1 \vPhi_2 \vPhi_3)_{\mu\nu}\,
  \beta_{\mu\nu} \,,
  \labl{eq:3bulk-corr-const}
with $c(\vPhi_1 \vPhi_2 \vPhi_3)_{\mu\nu}$ given by the ribbon invariant
\erf{eq:3bulk-sc-aux}.

To find the explicit functional dependence of the correlator on the insertion 
points, we use that the conformal block \erf{eq:3bulk-block} is a product of 
two three-point blocks. Rearranging the individual factors
in a suitable way we arrive at the expression
  \beaa \displaystyle
  \frac{C(\X^g)(v_i,...\,,v_n)}{C(\dot\X^g)} \!\!& \dsty
  = \sum_{\mu=1}^{\N ik{\bar m}} \sum_{\nu=1}^{\N jl{\bar n}} 
  \frac{S_{0,0} \, c(\vPhi_1 \vPhi_2 \vPhi_3)_{\mu\nu}}{  \dim(A)}\,
  B^\mu_{mki}(v_m,v_k,v_i)~B^\nu_{nlj}(v_n,v_l,v_j) 
  \\{}\\[-.7em] & \displaystyle~~ \times\,
  |z_3-z_2|^{w_1-w_2-w_3}\, |z_3-z_1|^{w_2-w_1-w_3}\, |z_2-z_1|^{w_3-w_1-w_2}
  \\[7pt]  & \displaystyle~~ \times\,
  \exp\!\big(\, \ii \varphi_{32} (s_1{-}s_2{-}s_3) +
    \ii \varphi_{31} (s_2{-}s_1{-}s_3) +
    \ii \varphi_{21} (s_3{-}s_1{-}s_2) \,\big)
  \eear\labl{eq:3bulk}
for the correlator ratio for three bulk fields.
Here $\varphi_{ij} \eq {\rm arg}(z_i{-}z_j)$, and we abbreviated
  \be\begin{array}{llll}
  w_1 \!\!\!&= \Delta_i(v_i)+\Delta_j(v_j) \,,\qquad
  w_2 \!\!\!&= \Delta_k(v_k)+\Delta_l(v_l) \,,\qquad
  w_3 \!\!\!&= \Delta_m(v_m)+\Delta_n(v_n) \,,~~ \\[7pt]
  s_1 \!\!\!&= \Delta_i(v_i)-\Delta_j(v_j) \,,\qquad
  s_2 \!\!\!&= \Delta_k(v_k)-\Delta_l(v_l) \,,\qquad
  s_3 \!\!\!&= \Delta_m(v_m)-\Delta_n(v_n) \,.
  \eear\ee
The function \erf{eq:3bulk} is a continuous (but not analytic)
function of the insertion points $z_1$, $z_2$, $z_3$. It does
not have branch cuts, because $s_{1,2,3} \iN \zet$ which, in turn,
follows from the fact that a bulk field transforming in the
chiral/antichiral representation $i,j$ exists only if
$Z(A)_{ij} \,{\neq}\, 0$. The matrix $Z(A)_{ij}$ commutes with
the matrix $\hat T_{kl} \eq \delta_{k,l} \theta_k^{-1}$ (this is part
of modular invariance, see theorem I:5.1), implying that
$Z(A)_{ij}\eq (\theta_i / \theta_j)\, Z(A)_{ij}$, i.e.\ 
$\theta_i\eq\theta_j$ for $Z(A)_{ij} \,{\neq}\, 0$.

\subsection{Three defect fields on the complex plane}
\label{sec:cf-3defect}

As already seen in section \ref{sec:rc-3defect}, the
calculation for three defect fields differs only slightly from the one
for three bulk fields. $\X^g$ is now the complex plane with field insertions
  \bea
  \Thetha^c_1 = (X,\orr_2,Y,\orr_2,i,j,\vartheta_1,[f_1])\,,
  \qquad
  \Thetha^c_2 = (Y,\orr_2,Z,\orr_2,k,l,\vartheta_2,[f_2])\,,
  \\{}\\[-.7em]
  \Thetha^c_3 = (Z,\orr_2,X,\orr_2,m,n,\vartheta_3,[f_3])\,.
  \eear\ee
Here $\orr_2$ is the standard orientation of the complex plane, $i,j,k,l,m,n 
\iN \II$ label simple objects, and the morphisms $\vartheta_{1,2,3}$ are elements 
of the relevant spaces of bimodule morphisms:
  \bea
  \vartheta_1 \in \HomAA(\SCA_i\otiP X\otim \SCA_j,Y) \,,
  \qquad
  \vartheta_2 \in \HomAA(\SCA_k\otip Y\otim \SCA_l,Z) \,,~~
  \\{}\\[-.7em]
  \vartheta_3 \in \HomAA(\SCA_m\otiP Z\otim \SCA_n,X) \,.
  \eear\ee
The three fields are inserted at $z_1$, $z_2$ and $z_3$, with local bulk 
coordinate germs $[f_{1,2,3}]$ given by $f_{1,2,3}(\zeta) \eq 
z_{1,2,3}\,{+}\,\zeta$. 

Here it is important to note that in \erf{eq:3defect-geom}
we have chosen the defect circle to run parallel to the
real axis in a neighbourhood of the defect field insertions
$z_i$. Otherwise the local bulk coordinate germs
$[f_i]$ must be modified so as to assure the condition that
$f_i(x)$ lies on the defect if $x$ lies on the real axis.
We further take $|z_3|\,{>}\,|z_2|\,{>}\,|z_1|$ as indicated in figure
\erf{eq:3defect-geom}.

The complex double of $\X^g$ is the same as for the case of three
bulk fields. Correspondingly the conformal blocks needed to express the 
correlator \erf{eq:3defect-geom} are the same, too, and we can expand
  \be
  C(\X^g) = \sum_{\mu,\nu} c(X,\Thetha_1,Y,\Thetha_2,Z,\Thetha_3,X)_{\mu\nu}\,
  \beta_{\mu\nu} \,,
  \labl{eq:3defect-corr-const}
with blocks $\beta_{\mu\nu}$ as defined in \erf{eq:3bulk-block}.
The ribbon invariant for the coefficient is given by \erf{eq:3defect-rib},
as calculated in section \ref{sec:rc-3defect}.
In the same manner as we arrived at the explicit form of the correlator 
\erf{eq:3bulk} of three bulk fields, for three defect fields we then find
  \beaa\dsty
  \frac{C(\X^g)(v_i,{...},v_n)}{C(\dot\X^g)} \!\!\!&\dsty
  = \sum_{\mu=1}^{\N ik{\bar m}} \sum_{\nu=1}^{\N jl{\bar n}}
  \frac{S_{0,0} \, c(X,\Thetha_1,Y,\Thetha_2,Z,\Thetha_3, X)_{\!\mu\nu\,}^{}}
     { \dim(A) } 
  B^\mu_{mki}(v_m,v_k,v_i)\, B^\nu_{nlj}(v_n,v_l,v_j)  
  \\{}\\[-.6em] &\displaystyle~~\times\,
  |z_3{-}z_2|^{w_1-w_2-w_3}\, |z_3{-}z_1|^{w_2-w_1-w_3}\,
  |z_2{-}z_1|^{w_3-w_1-w_2}
  \\[7pt]&  \displaystyle~~\times
  \exp\!\big(\, \ii \varphi_{32} (s_1{-}s_2{-}s_3) +
    \ii \varphi_{31} (s_2{-}s_1{-}s_3) +
    \ii \varphi_{21} (s_3{-}s_1{-}s_2) \,\big) \,,
  \\[-1.7em]{}
  \eear\labl{eq:3defect}
where $\varphi_{ij}\eq {\rm arg}(z_i{-}z_j)$ and
  \be\begin{array}{llll}
  w_1 \!\!&= \Delta_i(v_i)+\Delta_j(v_j) \,,\quad
  w_2 \!\!&= \Delta_k(v_k)+\Delta_l(v_l) \,,\quad
  w_3 \!\!&= \Delta_m(v_m)+\Delta_n(v_n) \,,\\[7pt]
  s_1 \!\!&= \Delta_i(v_i)-\Delta_j(v_j) \,,\quad
  s_2 \!\!&= \Delta_k(v_k)-\Delta_l(v_l) \,,\quad
  s_3 \!\!&= \Delta_m(v_m)-\Delta_n(v_n) \,.
  \eear\ee
In contrast to the correlator of three bulk fields,
the expression \erf{eq:3defect} is {\em not\/} necessarily
single valued in the coordinates $z_1 , z_2 , z_3$, since the numbers
$s_k$ do not have to be integers. This is to be expected. Indeed,
if, for example, in the setup \erf{eq:3defect-geom} we take
the field at $z_2$ around the field at $z_1$ we must deform
the defect circle in order to avoid taking the field $\Thetha^c_2$ across
the defect. In this way we end up with an arrangement of
defect lines different from the one we started with.

\subsection{One bulk field on the cross cap}
\label{sec:cf-1bulk-xcap}

Next we consider the correlator of one bulk field on the cross cap. This 
surface is non-orientable; accordingly we take $A$ to be a Jandl algebra. 
Analogously as in section \ref{sec:rc-1bulk-xcap} the world sheet
$\X^g$ is given by $\complex / \sigrp$ with $\sigrp$ the anti-holomorphic
involution $\sigrp(\zeta) \eq {-}1/{\zeta^*}$. The bulk insertion is
$\vPhi^c \eq (i,j,\phi,[f])$ with $f(\zeta) \eq z \,{+}\, \zeta$,
$i,j\iN\II$, and $\phi\iN\HomAA(\SCA_i\otiP A\otim \SCA_j,A)$.
The complex double of $\X^g$ is the Riemann sphere
$\complex \cup \{\infty\}$ together with the two marked points
$(z,[\varphi],\RCA_i)$ and $(\sigrp(z),[\tilde\varphi],\RCA_j)$.
The holomorphic coordinate germs are given by
$\varphi(\zeta) \eq z\,{+}\,\zeta$ and 
$\tilde\varphi(\zeta) \eq \sigrp(\varphi(\zeta^*)) \eq {-}(z^*{+}\zeta)^{-1}$.

The space $\calh^c(\hat \X^g)$ is the space of two-point blocks
on the Riemann sphere and hence one-dimensional, provided
$j\eq\bar\imath$. It is spanned by the block
  \be
  \beta_2 = \langle 0 |  \,
  \VOs^0_{i,\bar\imath}
    \!\big(R\big(\kappa_0(\varphi)\big)^{-1} v_i \,;\, z\big)\,
  \VOs^{\bar\imath}_{\bar\imath,0}
    \!\big(R\big(\kappa_0(\tilde\varphi)\big)^{-1} 
    v_{\bar\imath} \,;\,-\Frac1{z^*}\big)\,
  |0\rangle\,, \labl{eq:xcap-block}
where we also made explicit the dependence on the local
coordinates as in \erf{eq:H0n-conf-block}. In the correlators treated in the 
previous sections, all operators $R(\,\cdots)^{-1}$ were just the identity,
owing to the local coordinates taking the simple form 
$\zeta\,{ \mapsto}\, p{+}\zeta$. In the present case this is only true
for $R\big(\kappa_0(\varphi)\big)^{-1}$. The correlator is thus given by
  \be
  C(\X^g)(v_i,v_{\bar\imath}) = c(\vPhi^c) \, \langle 0 |\,
  \VOs^0_{i,\bar\imath} \!\big( v_i \,;\, z\big)\,
  \VOs^{\bar\imath}_{\bar\imath,0}
    \!\big(R\big(\kappa_0(\tilde\varphi)\big)^{-1} 
    v_{\bar\imath} \,;\,-\Frac1{z^*}\big) 
  |0\rangle \,. \labl{eq:xcap-corr-pre}
The constant $c(\vPhi^c)$ is given by the ribbon invariant \erf{eq:xcap-cobord},
as follows from applying $\Alpha$ to both sides of \erf{eq:xcap-corr-pre}
and comparing with \erf{eq:xcap-1bulk-expand}. This result also uses the
equality $\Alpha(\beta_2) \eq Z(B_2,\emptyset,\hat \X)1$, with $B_2$ given by
\erf{eq:xcap-aux-1}.

We conclude that the correlator ratio for one bulk field on the cross cap is
  \be
  \frac{C(\X^g)(v_i,v_{\bar\imath})}{C(\dot\X^g)} 
  = \frac{S_{0,0}\,c(\vPhi^c)}{
  {\Gamma^{\phantom I\!\!\!}}^\sigma(\one,\eps)}\,
  B_{i\bar\imath}(v_i, R\big(\kappa_0(\tilde\varphi)\big)^{-1}
    v_{\bar\imath})\, \big(z\,{+}\,\Frac1{z^*}\big)^{-2\Delta_i(v_i)} .
  \ee
Here $v_i \oti v_{\bar\imath} \iN \SCA_i \oti \SCA_{\bar\imath}$ specifies the 
state representing the bulk field. 
Note that for the identity bulk insertion, in \erf{chatphic} we have
$\phi'\eq\eps$, the counit of $A$, so that $C(\dot\X^g)\eq
{\Gamma^{\phantom I\!\!\!}}^\sigma(\one,\eps) / S_{0,0}$. Using
\erf{eq:P-def} and (II:3.90) leads to the expression
  \be
  \Gamma^\sigma(\one,\eps) = \sum_{a \In A} \frac{t_a P_{0,a}}{\sigma(a)} \,.
  \ee
Suppose now further that $v_{\bar\imath}$ is a Virasoro-primary state. Then
$R\big(\kappa_0(\tilde\varphi)\big)^{-1} v_{\bar\imath} \eq (z^*)^{-2\Delta_
{\bar\imath}(v_{\bar\imath})}v_{\bar\imath}$, so that in this case
  \be
  \frac{C(\X^g)(v_i,v_{\bar\imath})}{C(\dot\X^g)} 
  = \frac{S_{0,0}\,c(\vPhi^c)}{{\Gamma^{\phantom I\!\!\!}}^\sigma(\one,\eps)}\,
  B_{i\bar\imath}(v_i,v_{\bar\imath})\,\big(1\,{+}\,|z|^2\big)^{-2\Delta_i(v_i)}
  .  \labl{eq:1B-xcap}
The constant $c(\vPhi^c)$ is evaluated in \erf{chatphic} and (II:3.110);
in the Cardy case it takes the from \erf{eq:cardy-1bulk-xcap}.

\subsection{One defect field on the cross cap}
\label{sec:cf-1defect-xcap}

To obtain the correlator of a defect field on the cross cap one 
essentially repeats the calculation in the previous section. Again,
$\X^g \eq \complex / \sigrp$, this time with the insertion of a defect
field, given by $\Thetha^c \eq (X,-\orr_2,X,\orr_2,i,j,\vartheta,[f])$ with 
$f(\zeta) \eq z \,{+}\, \zeta$ and $\vartheta \iN \HomAA(\SCA_i\otiP X^s 
\otim \SCA_j,X)$. The choice of $f$ implies that the defect runs parallel 
to the real axis at the insertion point of the defect field.
The orientations $\pm\orr_2$ of the neighbourhood of $X$
are obtained as follows. Let $\orr_2$ be the local orientation around
$f(0)$ induced by $[f]$. The orientation of $X$ is obtained by transporting
$\orr_2$ along the defect to the right. When passing through the identification
circle, the orientation gets reversed, so that one arrives at the
insertion point $f(0)$ with $-\orr_2$ from the left, see figure \Erp 112.

The space $\calh^c(\hat \X^g)$ is again spanned by the single
block \erf{eq:xcap-block}; the correlator is 
  \be
  C(\X^g) = c(X,\Thetha^c)\, 
  \langle 0 |  \VOs^0_{i,\bar\imath\,} \!\big( v_i \,; z\big)
  \VOs^{\bar\imath}_{\bar\imath,0\,}
    \!\big(R\big(\kappa_0(\tilde\varphi)\big)^{-1} 
    v_{\bar\imath} \,;\,-\Frac1{z^*}\big) |0\rangle \,,
  \labl{eq:xcap-defect-corr-pre}
where $c(X,\Thetha^c)$ is given by the ribbon invariant \erf{eq:xcap-def-inv}, 
as calculated in section \ref{sec:rc-1defect-xcap}.
Taking the states $v_i\iN\SCA_i$ and $v_{\bar\imath}\iN\SCA_{\bar\imath}$
describing the defect field to be Virasoro-primaries, altogether the ratio 
of correlators for one defect field on the cross cap becomes
  \be
  \frac{C(\X^g)(v_i,v_{\bar\imath})}{C(\dot\X^g)} 
  = \frac{S_{0,0} \, c(X,\Thetha^c)}{\Gamma^\sigma_{}(\one,\eps)}\,
  B_{i\bar\imath}(v_i,v_{\bar\imath}) \,
  \big(1\,{+}\,|z|^2\big)^{-2\Delta_i(v_i)} .
  \labl{eq:1D-xcap}
In the Cardy case, the constant
$c(X,\Thetha^c)$ is given by \erf{eq:1def-Xcap-bas-Cardy}.

 \newpage

\newcommand\wb{\,\linebreak[0]} \def\wB {$\,$\wb}
 \newcommand\Bi       {\bibitem}
 \renewcommand\J[5]     {{\sl #5\/}, {#1} {#2} ({#3}) {#4} }
 \renewcommand\K[6]     {{\sl #6\/}, {#1} {#2} ({#3}) {#4}~\,[#5]}
 \newcommand\PhD[2]   {{\sl #2\/}, Ph.D.\ thesis (#1)}
 \newcommand\Mast[2]  {{\sl #2\/}, Master's thesis (#1)}
 \newcommand\Prep[2]  {{\sl #2\/}, pre\-print {#1}}
 \newcommand\BOOK[4]  {{\sl #1\/} ({#2}, {#3} {#4})}
 \newcommand\inBO[7]  {{\sl #7\/}, in:\ {\sl #1}, {#2}\ ({#3}, {#4} {#5}), p.\ {#6}}
 \newcommand\iNBO[7]  {{\sl #7\/}, in:\ {\sl #1} ({#3}, {#4} {#5}) }
 \newcommand\Erra[3]  {\,[{\em ibid.}\ {#1} ({#2}) {#3}, {\em Erratum}]}
 \def\jf    {J.\ Fuchs}
 \def\dim   {dimension}  
 \def\adma  {Adv.\wb Math.}
 \def\anop  {Ann.\wb Phys.}
 \def\aspm  {Adv.\wb Stu\-dies\wB in\wB Pure\wB Math.}
 \def\coia  {Com\-mun.\wB in\wB Algebra}
 \def\comp  {Com\-mun.\wb Math.\wb Phys.}
 \def\cpma  {Com\-pos.\wb Math.}
 \def\fiic  {Fields\wB Institute\wB Commun.}
 \def\foph  {Fortschritte\wB d.\wb Phys.}
 \def\ijmp  {Int.\wb J.\wb Mod.\wb Phys.\ A}
 \def\jams  {J.\wb Amer.\wb Math.\wb Soc.}
 \def\jgap  {J.\wb Geom.\wB and\wB Phys.}
 \def\jhep  {J.\wb High\wB Energy\wB Phys.}
 \def\jomp  {J.\wb Math.\wb Phys.}
 \def\jopa  {J.\wb Phys.\ A}
 \def\josp  {J.\wb Stat.\wb Phys.}
 \def\lemp  {Lett.\wb Math.\wb Phys.} 
 \def\maan  {Math.\wb Annal.}
 \def\mpla  {Mod.\wb Phys.\wb Lett.\ A}
 \newcommand\nqma[2] {\inBO{Non-perturbative QFT Methods and Their Applications}
            {Z.\ Horv\'ath and L.\ Palla, eds.} \WS\Si{2001} {{#1}}{{#2}} }
 \newcommand\nqft[2] {\inBO{Nonperturbative Quantum Field Theory}
            {G.\ 't Hooft et al., eds.} \PL\NY{1988} {{#1}}{{#2}} }
 \def\nuci  {Nuovo\wB Cim.}
 \def\nupb  {Nucl.\wb Phys.\ B}
 \newcommand\phgt[2] {\inBO{Physics, Geometry, and Topology}
            {H.C.\ Lee, ed.} \PL\NY{1990} {{#1}}{{#2}} }
 \def\phla  {Phys.\wb Lett.\ A}
 \def\phlb  {Phys.\wb Lett.\ B}
 \def\phrl  {Phys.\wb Rev.\wb Lett.}
 \def\phrp  {Phys.\wb Rep.}
 \def\phrd  {Phys.\wb Rev.\ D}
 \def\phre  {Phys.\wb Rev.\ E}
 \def\pnas  {Proc.\wb Natl.\wb Acad.\wb Sci.\wb USA}
 \def\prtp  {Progr.\wb Theor.\wb Phys.}
 \newcommand\qfsm[2] {\inBO{\Q Fields and Strings: A Course for Mathematicians}
            {P.\ Deligne et al., eds.} \AMS\PR{1999} {{#1}}{{#2}} }
 \def\rvmp  {Rev.\wb Math.\wb Phys.}
 \def\sebo  {S\'emi\-naire\wB Bour\-baki}
 \def\sjnp  {Sov.\wb J.\wb Nucl.\wb Phys.}
 \def\trgr  {Trans\-form.\wB Groups}
 \def\zfpc  {Z.\wb Phy\-sik C}
 \def\AMS    {{American Mathematical Society}}
 \def\BIR    {{Birk\-h\"au\-ser}}
 \def\PL     {{Plenum Press}}
 \def\SV     {{Sprin\-ger Ver\-lag}}
 \def\WS     {{World Scientific}}
 \def\Bo     {{Boston}} 
 \def\PR     {{Providence}}
 \def\Si     {{Singapore}}
 \def\NY     {{New York}}

\end{document}